# THÈSE DE DOCTORAT

Soutenue à Aix-Marseille Université
le 7 octobre 2020 par

## Minh Duy TRUONG

Théorie et modélisation numérique des résonances photoniques :

Expansion modale quasinormale
Applications en théorie électromagnétique

**Discipline**
Physique et Sciences de la Matière

**Spécialité**
Optique, Photonique et Traitement d'Image

**École doctorale**
ED 352 PHYSIQUE ET SCIENCES DE LA MATIERE

**Laboratoire/Partenaires de recherche**
Aix Marseille Université
CNRS
Institut Fresnel

**Composition du jury**

| | |
|---|---|
| André NICOLET<br>PR. Aix Marseille Université | Directeur de thèse |
| Philippe LALANNE<br>DR. Labo LP2N | Co-directeur de thèse |
| Christophe GEUZAINE<br>PR. University of Liège | Rapporteur |
| Stéphane LANTERI<br>DR. INRIA | Rapporteur |
| Guillaume DEMÉSY<br>MC. Aix Marseille Université | Examinateur |
| Christophe SAUVAN<br>CR. Laboratoire Charles Fabry | Examinateur |
| Sven BURGER<br>DR. Zuse Institute Berlin | Examinateur |
| Maryna KACHANOVSKA<br>CR. INRIA | Examinatrice |

Je soussigné, Minh Duy TRUONG, déclare par la présente que le travail présenté dans ce manuscrit est mon propre travail, réalisé sous la direction scientifique de André NICOLET et Philippe LALANNE, dans le respect des principes d'honnêteté, d'intégrité et de responsabilité inhérents à la mission de recherche. Les travaux de recherche et la rédaction de ce manuscrit ont été réalisés dans le respect à la fois de la charte nationale de déontologie des métiers de la recherche et de la charte d'Aix-Marseille Université relative à la lutte contre le plagiat.

Ce travail n'a pas été précédemment soumis en France ou à l'étranger dans une version identique ou similaire à un organisme examinateur.

Fait à Marseille le 01/07/2020
Minh Duy TRUONG

# Résumé

L'idée de l'expansion modale en électromagnétisme découle de la recherche sur les résonateurs électromagnétiques, qui jouent un rôle essentiel dans les développements en nanophotonique, et qui incluent les microcavités optiques, les lasers à semi-conducteurs, les nanorésonateurs plasmoniques, les métamatériaux optiques, les nanotechnologies à ADN et de traitement de l'information quantique. Tous les résonateurs électromagnétiques partagent une propriété commune : ils possèdent un ensemble discret de fréquences spéciales qui apparaissent comme des pics dans les spectres de diffusion, appelés modes résonants.

Ces modes de résonance régissent les interactions entre les résonateurs électromagnétiques et la lumière. Cela conduit à l'hypothèse que la réponse optique des résonateurs est le résultat de l'excitation de chaque état de résonance physique dans le système : sous l'excitation d'impulsions externes, ces modes de résonance sont initialement chargés, puis libèrent leur énergie qui contribue aux réponses optiques totales des résonateurs. Ces modes de résonance avec des fréquences complexes sont connus dans la littérature sous le nom de modes quasinormaux (QNM). Mathématiquement, ces QNM correspondent à des solutions du problème aux valeurs propres des équations de Maxwell sans source. Dans le cas où la structure optique des résonateurs est dans un domaine spatial non borné et que les milieux sont dispersifs (éventuellement anisotropes et non réciproques), cela nécessite de résoudre des problèmes aux valeurs propres non linéaires (en fréquence) et non hermitiens.

Dans les résonateurs physiques réalistes, il y a un certain degré de dissipation de l'énergie vers l'environnement. Cet effet entraîne un élargissement des pics dans les spectres et se caractérise généralement par la quantité appelée facteur de qualité Q. Ce facteur est proportionnel au temps de confinement dans une unité de la période optique : Plus le facteur Q est élevé, plus il faut de temps pour qu'une excitation initiale du résonateur se dissipe, soit par rayonnement vers l'environnement, soit par absorption dans le matériau. Par conséquent, les modes de résonance physique ont des fréquences complexes, où la dissipation est indiquée par la partie imaginaire des fréquences. Dans la littérature, ce mode est connu sous le nom de mode quasinormal (QNM), où le préfixe "quasi-" a été utilisé pour distinguer les QNM à fréquence complexe dans un système non conservatif (avec un facteur Q élevé) des modes normaux d'un système conservatif. D'un point de vue mathématique, ces modes résonants (ou QNMs) correspondent aux solutions du problème de la valeur propre des équations de Maxwell sans source. Dans la pratique, lorsque les milieux sont très dispersifs et que la structure optique des résonateurs n'est pas limitée, il faut résoudre des problèmes de valeur propre non linéaires et non Hermitiens.

En conséquence, tout le problème se résume à l'étude de la théorie spectrale pour les opérateurs électromagnétiques de Maxwell dans des structures bornées ou illimitées où la perméabilité et la permittivité peuvent être dispersives, anisotropes et même éventuellement non réciproques : L'interaction entre la lumière et les résonateurs peut se traduire par la question de la représentation spectrale des ondes électromagnétiques. Puisque les opérateurs de Maxwell peuvent être non linéaires et non hermitiens, leurs solutions propres, c'est-à-dire les QNM, n'héritent pas immédiatement de nombreuses propriétés intéressantes de représentation spectrale dans le cadre mathématique bien connu développé pour les problèmes de valeurs propres hermitiennes. Par exemple, ils ne peuvent pas être normalisés de manière standard par la norme qui est souvent utilisée pour les problèmes hermitiens. C'est la raison pour laquelle pendant longtemps, la littérature a contourné cette difficulté en considérant la dissipation d'énergie comme une perturbation d'un système conservateur.

Par la suite, il y a eu de nombreuses percées dans la recherche des propriétés spectrales des QNM. L'objectif ultime est le formalisme de l'expansion modale quasinormale (QNM) qui peut être utilisé pour modéliser les propriétés optiques des résonateurs sur la base de l'état de résonance naturel. Pour ceux qui s'intéressent à l'histoire de cette recherche, voici un bref aperçu de la littérature existante : La théorie de la perturbation des QNM a été présentée par Lai, Leung, et al dans [1], et plus tard ils ont également traité de la complétude dans [2-4]. Lee et al. ont également traité de la complétude du QNM et des expansions du tenseur Green électromagnétique [5] ainsi que de la théorie des perturbations [6]. Ils ont également mentionné le cas de la théorie des perturbations dégénérées dans [7]. Dans [8-10], les problèmes à une et trois dimensions ont été traités par Muljarov et al. en utilisant des expansions directes dans un sous-espace de QNM qu'ils ont appelé "expansion d'état résonant". Doost et al. ont appliqué l'expansion QNM du tenseur de Green à des dalles, des cylindres et des sphères dans [11-13] respectivement. Mansuripur et al. ont examiné l'intégralité des expansions QNM dans des sphères fabriquées à partir de matériaux dispersifs dans [14]. [15] a utilisé une expansion QNM du tenseur de Green pour modéliser des guides d'ondes planaires avec une incidence oblique de la lumière. Dans [16], les auteurs ont suggéré l'utilisation d'une expansion de l'état de résonance basée sur les QNM d'une sphère en combinaison avec l'équation de Dyson pour calculer les propriétés de diffusion des résonateurs généraux. En calcul numérique, diverses méthodes ont été employées pour traiter le caractère non-borné des structures. Celles-ci comprennent une analyse rigoureuse des ondes couplées pour les structures périodiques [17] et les techniques appelées Couches Parfaitement Adaptées (PML) [18,19], où les QNM sont normalisées par une intégration de volume qui s'étend à travers la PML avant d'être utilisée dans l'expansion modale. Le formalisme de l'expansion a ensuite été développé et systématisé par Yan et al. dans [20] en trois dimensions. Une approche similaire utilisant le même ensemble de modes contenant les modes QNM dominants et les modes PML a également été mentionnée dans [21]. De nombreux auteurs différents ont également abordé indépendamment ces méthodes de normalisation dans la littérature récente [9,22-25]. Enfin, il convient également de mentionner la dérivation de l'expansion QNM basée sur le théorème de Keldysh par Zolla et al. [26].

Toutes les études ci-dessus et l'état actuel de la recherche sur l'expansion modale des résonateurs électromagnétiques sont résumés dans [27]. Cependant, certains concepts mathématiques subtils des propriétés spectrales des opérateurs de Maxwell sont obscurcis par les caractéristiques physiques et doivent être revus. L'objectif de cette thèse est d'aborder de manière systématique et compréhensible, certains des problèmes les plus importants du sujet. Pour ce faire, nous allons supprimer la couche de camouflage liée aux propriétés optiques des résonateurs électromagnétiques et simplifier le problème de l'excitation résonante pour la réduire à l'étude spectrale des opérateurs de Maxwell non linéaires. Dans un souci de clarté et de simplicité, la thèse n'approfondira pas trop les preuves mathématiques, mais ne traitera que des propriétés spectrales les plus fondamentales des opérateurs et les utilisera pour tirer des conclusions physiques importantes. Ce faisant, nous révélerons un nouvel aperçu de la théorie spectrale appliquée aux équations de Maxwell : Il n'existe pas seulement une formule, mais une famille continue de formules d'expansion QNM (voir chapitre 4 et chapitre 5 pour plus de détails), qui peuvent être appliquées à des matériaux dispersifs, anisotropes, et même non réciproques.

Le manuscrit a été rédigé non seulement comme un document officiel pour l'obtention d'un doctorat, mais il se veut également un petit guide pour ceux qui s'intéressent aux QNM, en particulier, et à la théorie spectrale non linéaire, en général. En particulier, les propriétés théoriques de l'opérateur non-linéaire obtenues dans la thèse peuvent être appliquées en dehors du cadre de la photonique, par exemple, dans l'étude des QNMs en relativité générale contraignant les perturbations autour d'une solution de trou noir [28]. Dans le même temps, les modélisations numériques de la thèse peuvent être considérées comme des exemples d'application de l'expansion modale en calcul numérique. Elle souligne également plusieurs difficultés dans la construction des réponses optiques des systèmes électromagnétiques basées sur les modes résonnants.

Le cadre général de cette thèse comprend les parties suivantes :

**Partie 1 :** est consacrée à l'introduction du concept de mode quasi-normal (QNM) et rappelle quelques éléments de théorie électromagnétique de base. Il est également conçu pour aider le lecteur à mieux comprendre la motivation qui sous-tend l'étude de l'expansion modale dans les résonateurs électromagnétiques. En particulier, la recherche de fréquences de résonance est associée aux problèmes de valeurs propres de l'opérateur des équations de Maxwell. Ce faisant, nous introduisons le concept d'opérateurs non-linéaires et non-hermitiens comme sujet principal de l'étude. La fin de ce chapitre soulève la question du rôle de la normalisation appliquée aux QNM et aborde les problèmes critiques concernant la non-unicité du formalisme d'expansion.

**Partie 2 :** Cette partie est le cœur de la thèse et contient 4 chapitres consacrés à la théorie spectrale en électromagnétisme. L'objectif principal de cette partie est de formaliser l'équation de l'expansion QNM des opérateurs de Maxwell. Pour ce faire, nous développerons des théorèmes spectraux allant de simples opérateurs linéaires auto-adjoints à des opérateurs non linéaires compliqués. Les formules finales de expansion des opérateurs non linéaires sont conservées sous la forme la plus générale possible. Cela élargit la possibilité d'appliquer les formules d'expansion modale à d'autres opérateurs non linéaires, en dehors du cadre des opérateurs de Maxwell et des problèmes électromagnétiques.

**Partie 3 :** Cette partie comporte 3 chapitres. Nous y mettons en pratique la théorie de l'expansion en QNM dérivée dans les parties précédentes. Cela comprend l'utilisation de l'analyse par éléments finis pour étudier la modélisation mathématique des structures optiques. La modélisation est implémentée via le package open source Onelab / Gmsh / GetDP. Le concept de couches parfaitement adaptées (PML) est utilisé pour gérer les structures non bornées spatialement. Nous y présentons l'utilisation des fonctions rationnelles pour exprimer la dépendance de la permittivité diélectrique vis-à-vis de la fréquence (ainsi que son prolongement analytique pour les valeurs complexes de la fréquence) et représenter la dispersion du matériau dans les opérateurs de Maxwell. Nous étudions deux des principaux problèmes numériques des simulations fondées sur l'expansion modale : les formules d'expansion divergent autour des pôles de la fonction de permittivité et des résonances plasmoniques apparaissent pour certaines valeurs critiques.

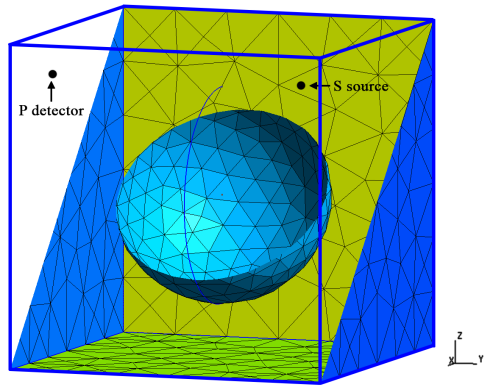

La géométrie 3D et le maillage d'une boîte carrée fermée contenant une sphère.

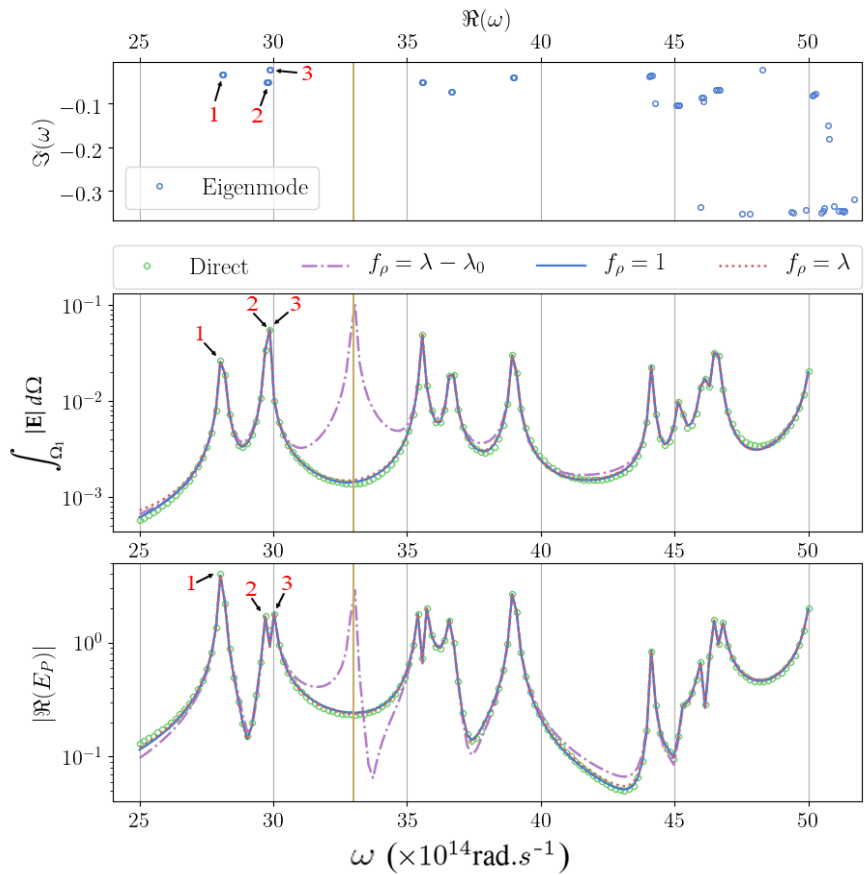

L'application de l'extension QNM dans la boîte fermée :
(TOP) Spectre des fréquences propres complexes.
(MIDDLE) La valeur moyenne du module du champ électrique à l'intérieur de l'objet
$\Omega_1 : \int_{\Omega_1} |\mathbf{E}| \, \Omega$.
(BOTTOM) La magnitude de la partie réelle du champ $\mathbf{E}$ au point de détection
$(x_P, y_P, z_P) : |\text{Re}\{\mathbf{E}(x_P, y_P, z_P)\}|$.

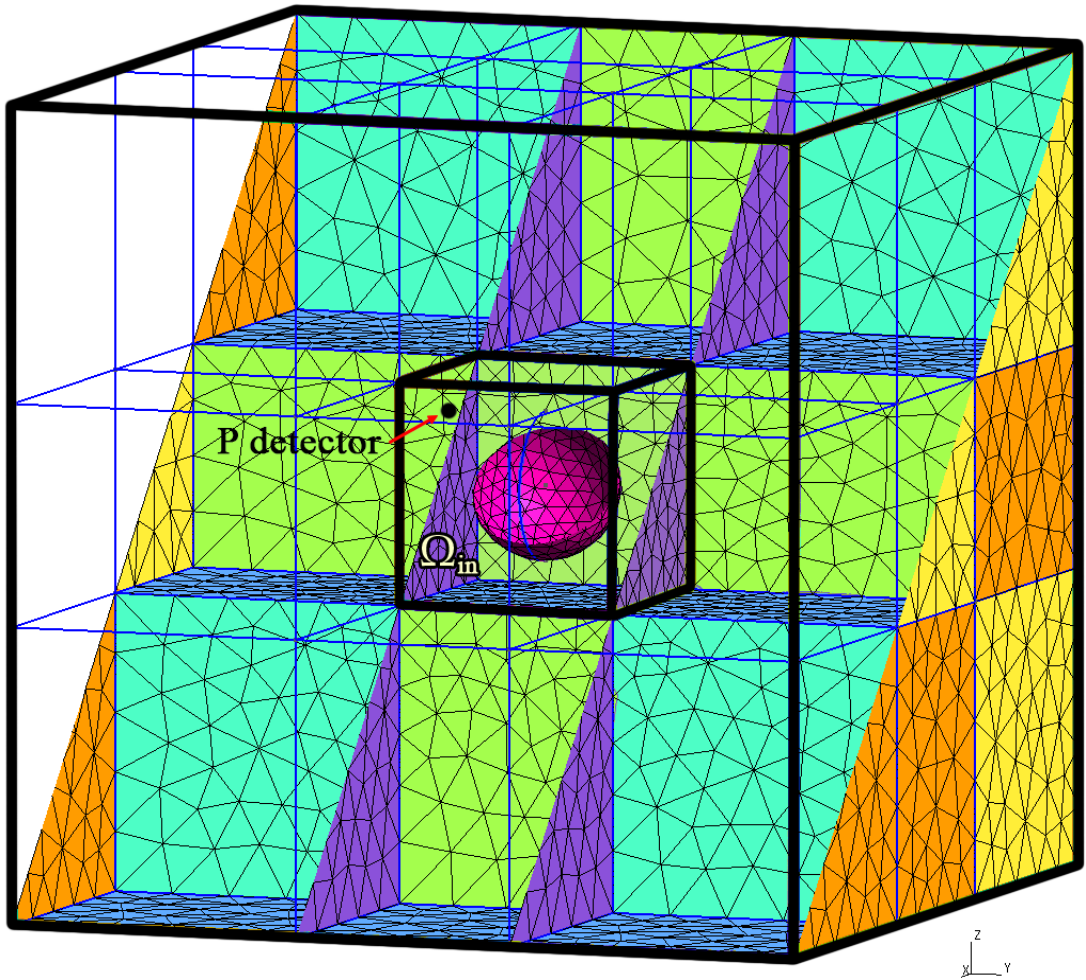

La géométrie 3D et le maillage de la sphère dans un domaine ouvert.

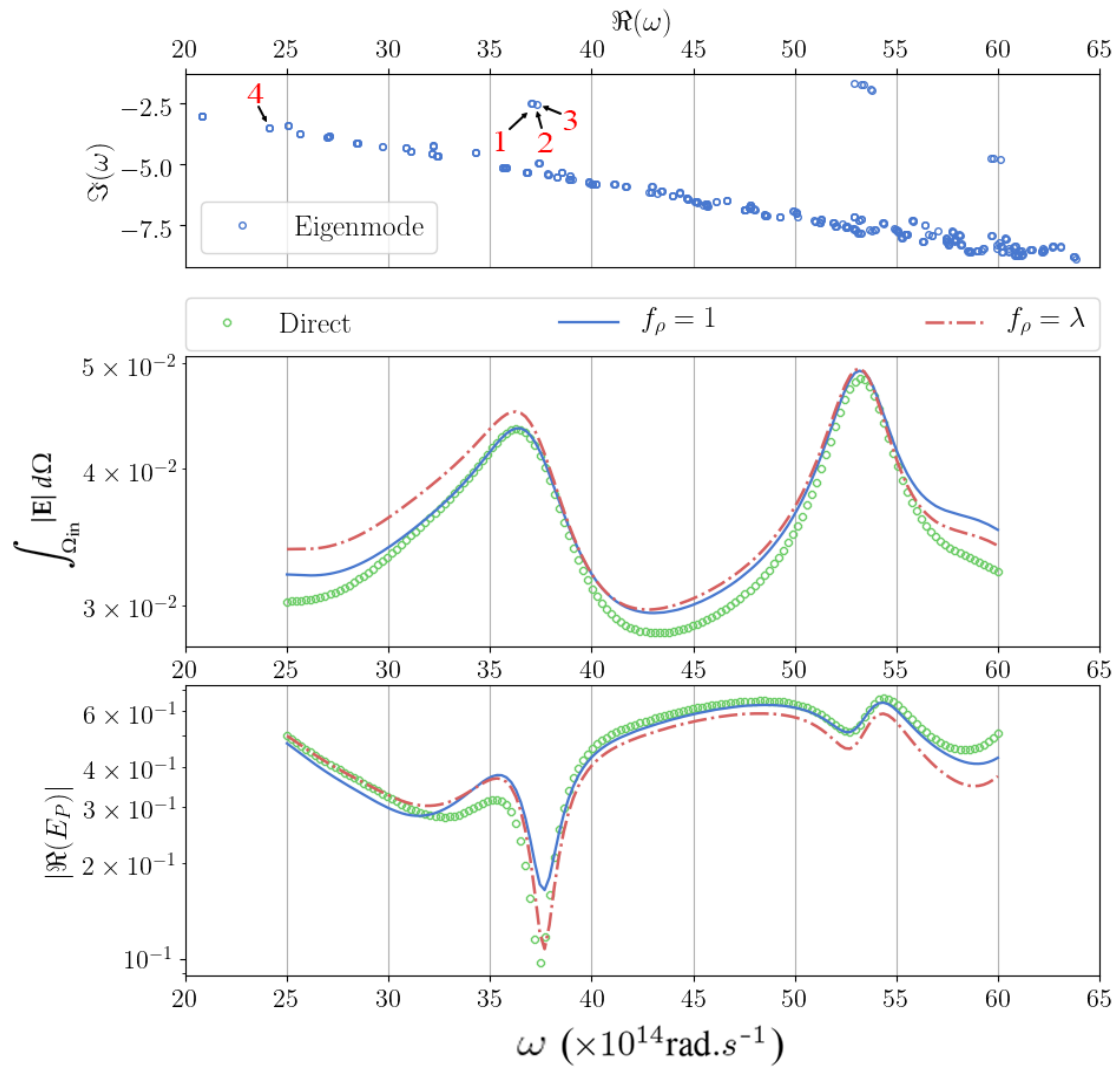

L'application de l'expansion QNM pour une sphère dans un espace ouvert.
(TOP) Spectre de fréquences propres complexes.
(MIDDLE) Valeur moyenne du champ électrique en dehors des PMLs : $\int_{\Omega_{\text{in}}} |\mathbf{E}| \, d\Omega$.
(BOTTOM) La magnitude de la partie réelle du champ $\mathbf{E}$ au point de détection $(x_P, y_P, z_P)$ : $|\text{Re}\{\mathbf{E}(x_P, y_P, z_P)\}|$.

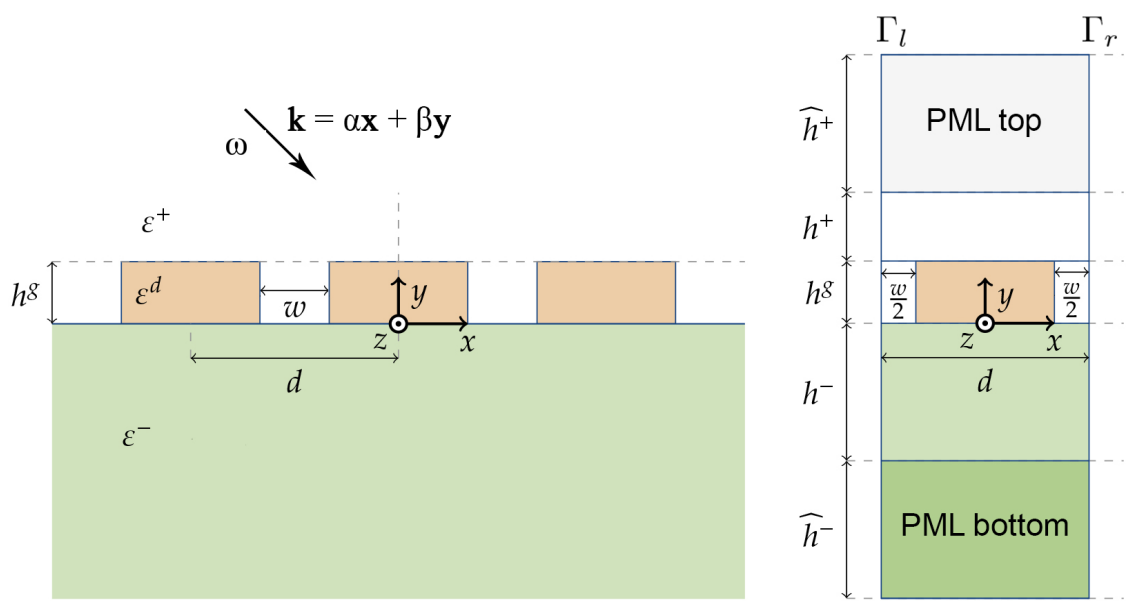

La structure géométrique 2-D pour le problème des réseaux de diffraction.

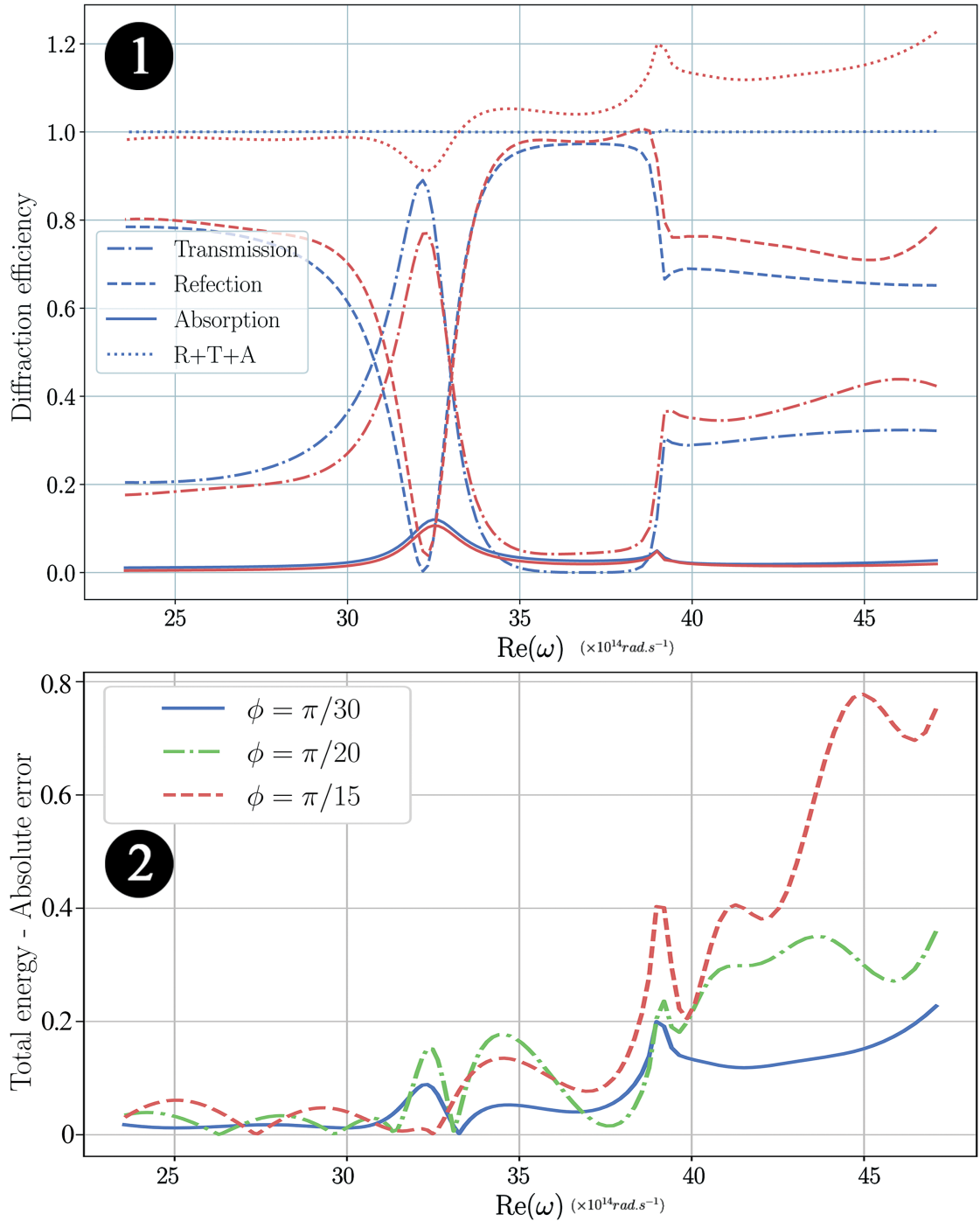

L'application de l'expansion QNM pour la structure du réseau de diffraction :

1/ Efficacité de la diffraction calculée à $\alpha = 0$ : Les lignes bleues se réfèrent au calcul direct, les lignes rouges sont reconstruites par l'expansion DQNM à partir de 700 modes avec $\phi = \pi/30$.

2/ Erreurs absolues sur l'énergie totale ($R + T + A$ dans la sous-figure supérieure) reconstruites à partir de 700 modes pour différentes valeurs $\phi$.

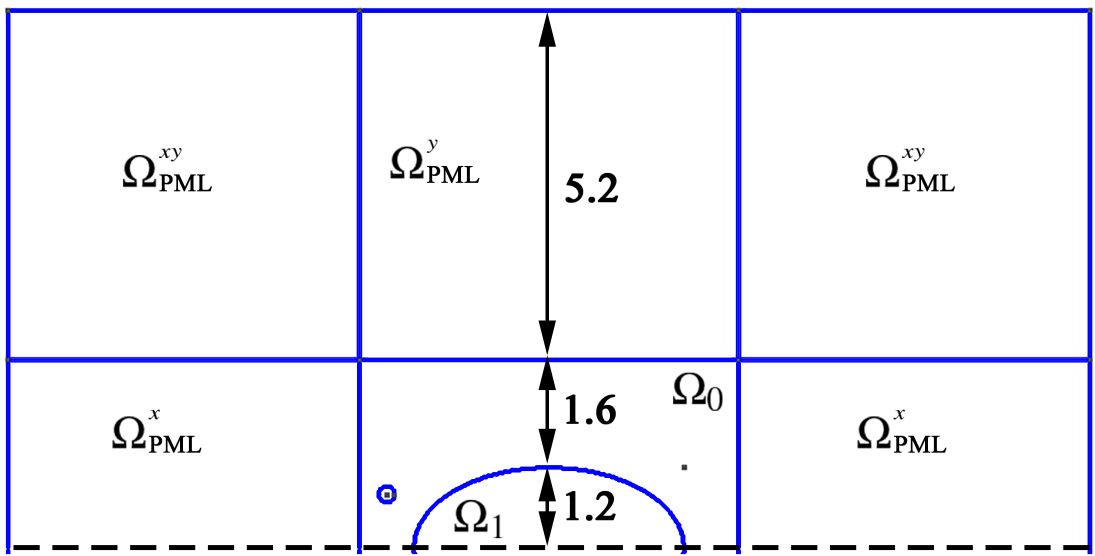

La moitié supérieure de la géométrie de la structure elliptique (en silicium) dans un espace ouvert.

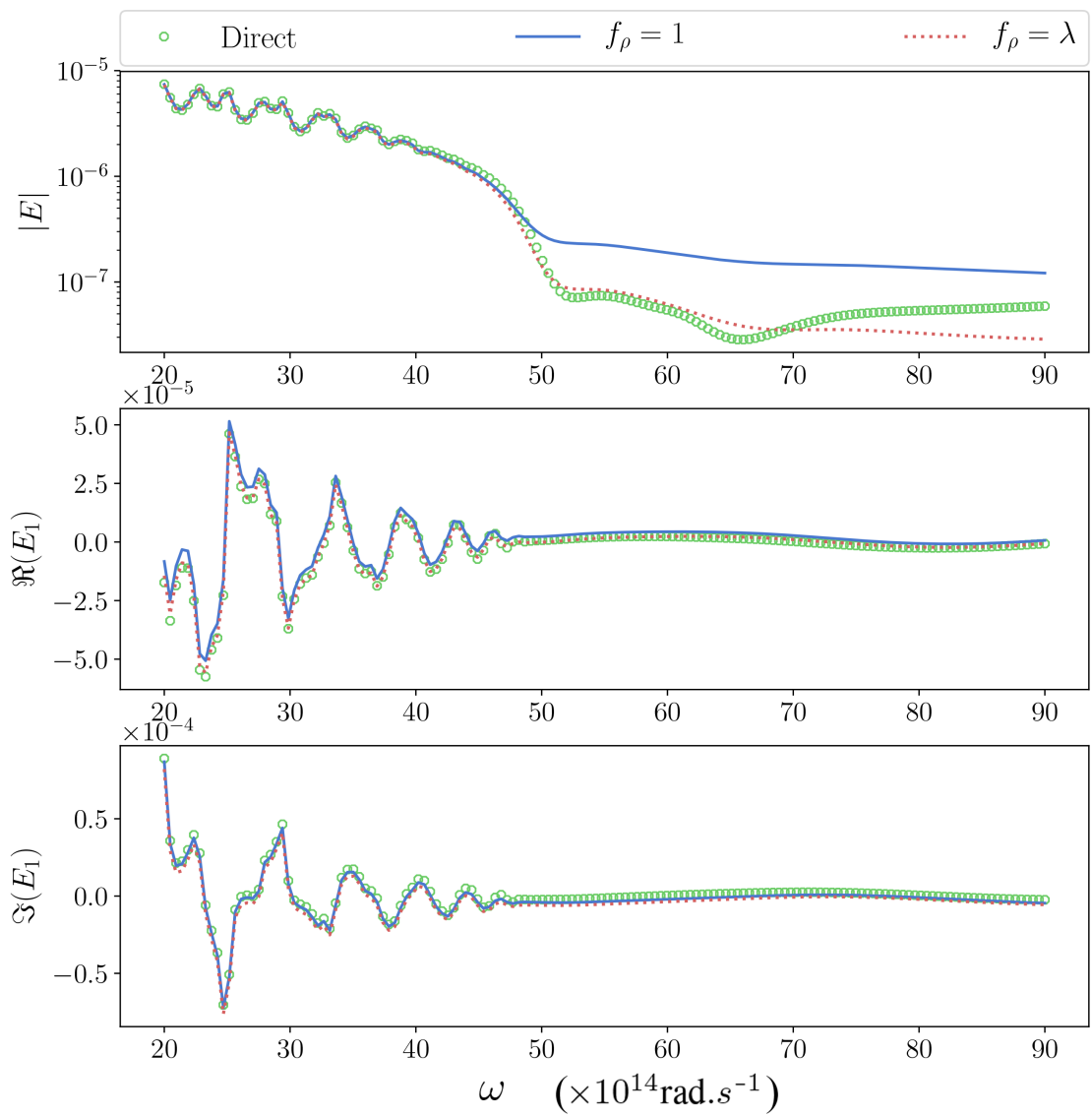

L'application de l'expansion QNM pour la structure elliptique (en silicium avec une permittivité à 4 pôles) dans un espace ouvert :
(TOP) Intégrale sur $\Omega_1$ de la norme du champ magnétique $\int_{\Omega_1} |\mathbf{E}| d\Omega$.
(MOYEN et INFÉRIEUR) Partie réelle et imaginaire du champ électrique calculée au point de détection.

**Theory and numerical modeling of photonic resonances:**

# Quasinormal Modal Expansion Applications in Electromagnetics

Minh Duy TRUONG

October 26, 2020

Fresnel-AMU

# Acknowledgements

Firstly, I would like to thank my supervisor Prof. André Nicolet for guiding me through the entire process of being a PhD student and writing this thesis: All the way from choosing me for this PhD to figuring out the last minute details, his input has been invaluable.

To my co-supervisor Philippe Lalanne, for the lesson to become a good researcher.

I would also like to thank the whole ATHENA team: Prof. Frédéric Zolla, Guillaume Demésy, for helping me with all the theoretical and numerical problems I encountered. Without their input, both comprehending the simple beauty of math and debugging the numerical implementation would have taken much much longer.

To my friends and family for helping me keep the motivation up throughout these years.

Endless gratitude to my best friend Plamen Dimitrov, for being my trustworthy mental support and wisdom, ironically even without acknowledgment of his own dedication.

Finally, to my parents and my wife Do Ngoc Tram, FOR EVERYTHING, nothing less.

# Preface

The idea of the modal expansion in electromagnetics is derived from the research of electromagnetic resonators, which plays an essential role in developments in nanophotonics, ranging from microwave resonators, via optical microcavities and semiconductor lasers, to plasmonic nanoresonators and optical metamaterials, from DNA nanotechnologies to quantum information processing. All of the electromagnetic resonators share a common property: they possess the discrete set of special frequencies that show up as peaks in scattering spectra, which are called resonant modes.

It is not so long until these resonant modes are recognized to dictate the interaction between electromagnetic resonators and light. This leads to an intuitive belief that the optical response of a given structure of resonators is the synthesis of the excitation of each physical-resonance-state in the system: Under the excitation of external pulses, these resonant modes are initially loaded, then release their energy which contributes to the total optical responses of the resonators.

In realistic physical resonators, there is a certain degree of dissipation of energy to the environment. This effect results in a broadening of the peaks in the spectra and is usually characterized by the quantity so-called quality factor Q. This factor is proportional to the confinement time in a unite of the optical period: The higher the Q factor, the longer it takes for an initial excitation of the resonator to dissipate away, either through radiation to the environment or absorption in the material. As a result, the physical resonant modes have complex frequencies, where the dissipation is indicated by the imaginary part of the frequencies. This is known in the literature as the Quasinormal Mode (QNM), where the prefix 'quasi' has been used to distinguish the QNMs with complex-valued frequency in a non-conservative system (with a high Q-factor) from the normal modes of a conservative system.

From a mathematical point of view, these resonant modes (or QNMs) correspond to solutions of the eigenvalue problem of source-free Maxwell's equations. In the practice where the media are highly dispersive and the optical structure of resonators is unbounded, this requires solving non-linear, non-Hermitian eigenvalue problems.

As a result, the whole problem boils down to the study of the spectral theory for electromagnetic Maxwell operators in bounded or unbounded structures where the permeability and permittivity can be dispersive, anisotropic, and even possibly non-reciprocal: The interaction between light and resonators can be translated to the question of the spectral representation of the electromagnetic waves. Since the Maxwell operators can be non-linear and non-Hermitian, their eigensolutions, i.e. QNMs, do not immediately inherit many nice properties of spectral representation in the well-known mathematical framework developed for Hermitian eigenvalue problems. For example, they cannot be normalized in a standard fashion by the norm that is often used for Hermitian problems. This is the reason why for a long time, the literature has bypassed this difficulty by considering energy dissipation as a perturbation of a conservative system.

After that, there are many breakthroughs in the research of spectral properties of QNMs. The ultimate goal is the Quasinormal Modal expansion (QNM expansion) formalisms which can be used to model the optical properties of resonators on a natural resonance-state basis. For those interested in the history of research, here is a short overview of the existing literature: The perturbation theory for QNMs was presented by Lai, Leung, et al in [1], and later they also have treated the completeness in [2–4]. Lee et al. also discussed the QNM completeness and electromagnetic Green tensor expansions [5] as well as perturbation theory [6]. They also mentioned the case of degenerate perturbation theory in [7]. In [8–10] both one and three-dimensional problems were treated by Muljarov et al. using direct expansions in a subspace of QNMs which they called the 'resonant state expansion'. Doost et al. have applied the QNM expansion of the Green tensor into slabs, cylinders, and spheres in [11–13] respectively. Mansuripur et al. discussed the completeness of QNM expansions in spheres made from dispersive materials in [14]. [15] used a QNM expansion of the Green tensor to model planar waveguides with an oblique incidence of light. In [16], the authors suggested the use of a resonant state expansion based on the QNMs of a sphere in combination with the Dyson equation to calculate the scattering properties of general resonators. In numerical computation, a variety of methods have been employed to handle the unboundedness of the structures. These include rigorous coupled-wave analysis for periodic structures [17] and the techniques so-called Perfectly Matched Layers (PMLs) [18, 19], where the QNMs are normalized by a volume integration which extended through the PML before being used in the modal expansion. The expansion formalism was then further developed and systematized by Yan et al. in [20] into three dimensions. A similar approach using the same set of modes containing the dominant QNMs and the PML modes was also mentioned in [21]. Many different authors also independently addressed these normalization methods in the recent literature [9, 22–25]. Finally, it should also mention the attempt to derive the QNM expansion based on the Keldysh theorem by Zolla et al. [26].

All of the above studies and the current state of research of modal expansion in electromagnetic resonators is summarized in [27]. However, some subtle mathematical concepts of spectral properties of Maxwell operators are clouded by physical characteristics and need to be revisited. The aim of this thesis is to tackle in a systematic and understandable way, some of the most important issue of the topic. In order to do that, we will remove the camouflage layer related to the optical properties of electromagnetic resonators and simplify the problem of resonant excitation to become the spectral study of the non-linear Maxwell operators. For the sake of clarity and simplicity, the thesis will not go too deep into mathematical proofs, but will instead discuss only the most basic spectral properties of operators and use them to draw important physical conclusions. By doing that, we will reveal a new insight on spectral theory when applied to Maxwell's equations: There exists not just one, but a continuous family of QNM expansion formulas (see Chapter 4 and Chapter 5 for more details), which can be applied to dispersive, anisotropic, and even non-reciprocal materials.

The manuscript not only has been written as an official document to obtain a doctorate degree but also aims to be a short guidebook for those who are interested in QNMs, in particular, and non-linear spectral theory, in general. In particular, the theoretical properties of the non-linear operator obtained in the thesis can be applied outside the framework of photonics, for example, in the study of QNM in general relativity constraining perturbations around a black hole solution [28]. At the same time, the numerical modelings in the thesis can be seen as examples in the application of modal expansion in numerical computation (The source code of these examples is freely available at `https://github.com/truongminhduy`). It also points out several difficulties in constructing the optical responses of the electromagnetic systems based on resonant modes.

The general frame of this thesis encompasses the following parts:

Part 1 We introduce the definition of Quasinormal Mode and recall some basic electromagnetic theories. It is also designed to help readers better understand the motivation behind the study of modal expansion in electromagnetic resonators. The most important role of this part is to introduce the concept of non-linear and non-Hermitian operators as the main subject of the study.

Part 2 This part is the heart of the thesis and contains 4 chapters. The main goal of this part is to formalize the equation of QNM expansion of Maxwell operators. In order to do that, we will develop spectral theorems from simple linear self-adjoint operators to complicated non-linear operators. The final expansion formulas of non-linear operators are kept in the most general form possible. This expands the ability to apply the modal expansion formulas on other non-linear operators, outside the framework of Maxwell operator and electromagnetic problems. In particular, the QNM expansion formalism is deduced for general infinite-dimensional operators, which refer to continuous Maxwell's equations. In numerical computations, these infinite-dimensional operators will be replaced with 'discretized' operators, i.e. finite-dimensional matrices, by using numerical discretization methods defined on a finite mapped space (see for instance the Finite Element analysis in Chapter 6).

Part 3 There are 3 chapters in this part. We will put into practice the theory of the QNM expansion derived in the previous part. This includes using the Finite Element Analysis to study the mathematical modeling of optical structures. The modeling is implemented through the open-source package Onelab/Gmsh/GetDP (`http://onelab.info`) [29]. The concept of Perfectly Matched Layers (PMLs) will be used to handle the unbounded structures. Finally, we mention the multi-pole rational functions of the permittivity as a way to impose the dispersion of the material into the Maxwell operators. This leads to two main difficulties of simulation of the modal expansion: The expansion formula diverges around the poles and the plasmonic resonances (the corner problem).

*Duy*

# Contents

# List of Figures

# List of Expansion Formalisms

# INTRODUCTION

# Quasinormal Mode and modal expansion

# 1

This chapter is intended as an attempt to provide a brief exposition of the Quasinormal Mode (QNM) as well as to motivate the reader to the concept of modal expansion.

## 1.1 Quasinormal Mode

In order to understand the Quasinormal Mode, it is better to start with a more intuitive and understandable concept: Resonance. In any vibration-like phenomenon, in general, and electromagnetic waves, in particular, resonance is of uttermost importance. In optics, it refers to the effect when small changes in the incident field result in significant changes in the diffracted field. And frequencies at which the response amplitude is a relative maximum are known as resonant frequencies. The applications of resonant frequency can vary from string vibration in musical instruments and acoustics, from which the term 'resonance' indeed is originated, to electrical resonance, i.e the exchange electric and magnetic energy between the capacitors and inductors in a circuit, to the optical resonator in lasers, where the electromagnetic waves are trapped in forms of a standing wave by mirrors.

In the context of optical resonators, i.e. optical cavities, it is important to distinguish 2 cases:

▶ Closed cavities, where the electromagnetic fields are fully confined in a finite region of space.
▶ Open cavities, when the field are not strictly confined and can leak to the whole universe [27] .

In the case of closed cavities with specific boundary conditions, for example with perfectly conducting walls, the energy can not escape outside and the system is conservative. Thus, the resonant frequencies are real and called 'normal'. On the other hand, in open and non-conservative systems, i.e open cavities, it is clear that the resonant oscillations can not be reserved and will decay with respect to time. In particular, the system would still oscillate with the resonant frequencies but with an exponentially decaying envelope which characterizes the damping rate. We can witness the same phenomenon in an acoustic spherical Helmholtz resonator in Figure 1.1.

The loss of energy in open cavities can come from the absorption of the materials or the energy radiation into the environment. This energy leakage can be characterized by a figure of merit of resonators: the quality factor. The quality factor, or Q-factor, is a dimensionless parameter describing how damped a resonance is, which is equivalent to the resonant bandwidth relative to its center frequency. To put it simply, the higher the Q-factor, the lower the energy leaks compared to the energy stored in the resonator.

[27]: Lalanne et al. (2018), 'Light Interaction with Photonic and Plasmonic Resonances'

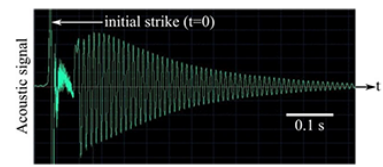

**Figure 1.1:** Illustration of the concept of QNM with an acoustic spherical Helmholtz resonator with a volume of $3000\,cm^3$, a cylindrical neck of length 8 cm, and a cross-sectional area of $8.3\,cm^2$. Oscillogram recorded by a microphone placed inside the resonator, initially knocked by the palm of one hand. For t > 0.1 s, the acoustic signal oscillates at the resonance frequency and exponentially decays in time, i.e. gradually releases its energy through dissipation [27].

When the energy leakage is relatively small, it is expected that certain concepts from closed systems can be extended to open cases with a satisfactory degree of accuracy and physical intuition. From there, the resonant modes adopt a new name 'Quasinormal Mode', which is originated from the research of blackhole for historical reasons.

One of the natural characteristics of 'Quasinormal Modes' (QNMs) is their complex frequencies. In particular, the QNMs are often shown to be proportional to $\exp(-i\omega't)\exp(-\omega''t)$ where $\omega'$ gives the resonant frequency and $\omega''$ refers to the linewidth of the resonance. In other words, the modes of open resonators are usually time-harmonic fields with a complex frequency:

$$\omega = \omega' - i\omega''.$$

The complex frequency also implies that the QNMs are no longer of finite energy and possess a finite lifetime while growing exponentially in space at infinity.

In order to better understand the properties of QNMs, it is instructive to study one of the simplest resonator: a 1-D model of dielectric barrier (the Fabry-Perot resonator). But before that, it is worth recalling some prerequisite knowledge in electromagnetics.

## Wave equations

In the framework of classical electrodynamics, the electromagnetic fields in a source-free region $\Omega$ are adequately described by 4 Maxwell's equations:

$$\nabla \times \mathbf{E}(\mathbf{r}, t) = -\frac{\partial \mathbf{B}(\mathbf{r}, t)}{\partial t} \qquad \nabla \cdot \mathbf{D}(\mathbf{r}, t) = 0$$

$$\nabla \times \mathbf{H}(\mathbf{r}, t) = \frac{\partial \mathbf{D}(\mathbf{r}, t)}{\partial t} \qquad \nabla \cdot \mathbf{B}(\mathbf{r}, t) = 0.$$

with appropriate boundary conditions.[1] The constitutive relations for materials are simply given by:

$$\mathbf{H} = \mu_0^{-1}\mathbf{B} - \mathbf{M} \quad \text{and} \quad \mathbf{D} = \varepsilon_0\mathbf{E} + \mathbf{P}_e. \tag{1.1}$$

where $\mathbf{M}$ and $\mathbf{P}_e$ are the magnetization and polarization vectors respectively.

For all the numerical examples in this thesis, the materials present a negligible magnetization. On the other hand, the electrical displacement $\mathbf{D}$ is influenced not only by the electric field $\mathbf{E}$ but also by the electric polarization of the materials, characterized by vector $\mathbf{P}$.[2]

$$\mathbf{P}_e(\mathbf{r}, t) = \frac{\varepsilon_0}{2\pi}\int_{-\infty}^{+\infty} \chi_e(t - \tau)\mathbf{E}(\mathbf{r}, \tau)\, d\tau.$$

With the assumption that the electromagnetic fields are time-harmonic, it is possible to express Maxwell's equations in the frequency domain via the following Fourier transform convention:[3]

1: The boundary conditions of Maxwell's equations will be discussed in details in Chapter 2 and Chapter 3.

2: see Chapter 8 for more details of the causality properties of this polarization.

3: It is worth pointing out that although the following conventions are expressed in the classical sense, we have to understand all the Fourier transform in a more general form as the operator such that (see [30] for more details):

$$\left\langle \mathcal{F}_{t\to\omega}\{f(t)\}(\omega), \phi(\omega) \right\rangle := \left\langle f(t), \phi(t) \right\rangle$$

$$\left\langle \mathcal{F}_{\omega\to t}^{-1}\{\hat{f}(\omega)\}(t), \phi(t) \right\rangle := \left\langle \hat{f}(\omega), \phi(\omega) \right\rangle$$

(1.2)

where the brackets stand for integration in the whole real line and $\phi(\omega)$ is the Gaussian test function.

$$\hat{f}(\omega) := \frac{1}{2\pi} \int_{t \in \mathbb{R}} f(t) \exp(i\omega t)\, dt$$

$$f(t) := \int_{t \in \mathbb{R}} \hat{f}(\omega) \exp(-i\omega t)\, dt. \tag{1.3}$$

Then, we obtain:

$$\nabla \times \hat{\mathbf{E}}(\mathbf{r}, \omega) = i\omega \hat{\mathbf{B}}(\mathbf{r}, \omega) \qquad \nabla \cdot \hat{\mathbf{D}}(\mathbf{r}, \omega) = 0$$

$$\nabla \times \hat{\mathbf{H}}(\mathbf{r}, \omega) = -i\omega \hat{\mathbf{D}}(\mathbf{r}, \omega) \qquad \nabla \cdot \hat{\mathbf{B}}(\mathbf{r}, \omega) = 0. \tag{1.4}$$

with frequency-dependent constitutive relations given by:

$$\hat{\mathbf{D}}(\mathbf{r}, \omega) = \boldsymbol{\varepsilon}(\mathbf{r}, \omega)\hat{\mathbf{E}}(\mathbf{r}, \omega) \quad \text{and} \quad \hat{\mathbf{B}}(\mathbf{r}, \omega) = \boldsymbol{\mu}(\mathbf{r}, \omega)\hat{\mathbf{H}}(\mathbf{r}, \omega) \tag{1.5}$$

where $\boldsymbol{\varepsilon}(\mathbf{r}, \omega)$ and $\boldsymbol{\mu}(\mathbf{r}, \omega)$ are second-order tensors depending on the space and frequency.

Since the focus of our research is the frequency domain, it is of convenience drop the 'hat' notation on the top of the letters representing the electromagnetic field. Similarly, the notation of the space-dependence will also be omitted. For example, $\hat{\mathbf{E}}(\mathbf{r}, \omega) \to \mathbf{E}$.

Then, after some manipulations, we can obtain the following wave equations for electric and magnetic fields:

$$-\nabla \times \left(\boldsymbol{\mu}^{-1}(\omega)\nabla \times \mathbf{E}\right) + \omega^2 \boldsymbol{\varepsilon}(\omega)\mathbf{E} = 0$$

$$-\nabla \times \left(\boldsymbol{\varepsilon}^{-1}(\omega)\nabla \times \mathbf{H}\right) + \omega^2 \boldsymbol{\mu}(\omega)\mathbf{H} = 0.$$

Since $\varepsilon$ (resp. $\mu$) is a matrix, it is more precise to write $\boldsymbol{\varepsilon}^{-1}$ (resp. $\boldsymbol{\mu}^{-1}$) instead of $\frac{1}{\varepsilon}$ (resp. $\frac{1}{\mu}$).

It is important to keep in mind that the electromagnetic fields in the previous equations also need to satisfy the divergence conditions, i.e. $\nabla \cdot \boldsymbol{\varepsilon} \mathbf{E} = 0$ and $\nabla \cdot \boldsymbol{\mu} \mathbf{H} = 0$. Indeed, these wave equations of electromagnetic will be the main subject of our research throughout this thesis.

## 1-D dielectric barrier

Now, let's take a look on a classical example of optical resonators: 1-D model of a dielectric barrier with refractive index $n_p(\omega)$ embedded in a homogeneous background with refractive index $n_0$. The barrier has a width of L and is centered on the origin $O$. In other words, the geometry of the system is divided into 3 spatial domains (see Figure 1.2):

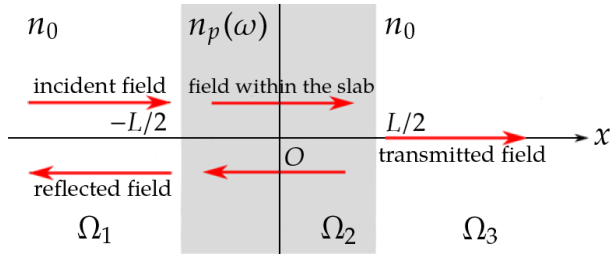

**Figure 1.2:** The geometrical setup of a slab of frequency dispersive material embedded in the medium $n_0$. The red arrows represent the plane waves that form the QNMs.

▶ The superstratum $\Omega_1$ with the permittivity $\varepsilon_1$ for $x < -L/2$,
▶ the slab $\Omega_2$ the permittivity $\varepsilon_2$ for $-L/2 < x < L/2$,

▶ the substratum $\Omega_3$ the permittivity $\varepsilon_3$ for $L/2 < x$.

The slab is illuminated by an electric field which is linearly polarized along the $y$-axis and propagates along the $x$-direction. This implies normal incidence and the electric field can be written such that:

$$\mathbf{E} = E(x, \omega)\mathbf{e}_y. \tag{1.6}$$

By substituting (1.6) into the wave equation for electric fields $-\nabla \times \left(\boldsymbol{\mu}^{-1}(\omega)\nabla \times \mathbf{E}\right) + \omega^2 \boldsymbol{\varepsilon}(\omega)\mathbf{E} = 0$, we obtain a simple version of second-order ordinary differential equation:

$$\frac{d^2 E}{dx^2} + \beta^2(\omega)E = 0, \tag{1.7}$$

with $\beta^2(\omega) = \omega^2 \mu(\omega)\varepsilon(\omega)$. The wave number $\beta(\omega)$ can be computed by taking the square root of $\omega^2 \mu(\omega)\varepsilon(\omega)$. But since the frequency is not necessarily real, $\beta^2(\omega)$ must be viewed as a complex-valued function. And the computation of the square root of a complex-valued function remains to be elucidated with caution (see [31] for more details). Thus, for the sake of simplicity, we will neglect the dispersion property of the slab $n_p(\omega) \to n_p$. Then, the wave number can be simply described as:

$$\beta(\omega) = \omega\sqrt{\mu\varepsilon} = \frac{\omega n}{c}$$

where $c = \frac{1}{\sqrt{\varepsilon_0 \mu_0}}$ stands for the velocity of light in vacuum and $n$ is the refractive index depending on the space:

$$n = \left\{ \begin{array}{ll} n_p & \text{for} \quad x < -L/2 \text{ or } L/2 < x \\ n_0 & \text{for} \quad -L/2 < x < L/2 \end{array} \right. .$$

For simplicity, we will adopt the following notation:

$$\beta(\omega) = \left\{ \begin{array}{ll} \beta_1 = \omega n_0 & \text{in } \Omega_1 \\ \beta_2 = \omega n_p & \text{in } \Omega_2 \\ \beta_3 = \omega n_0 & \text{in } \Omega_3 \end{array} \right. .$$

From here, it is straightforward to solve the differential equation (1.7) to obtain the following solution:

$$E = \left\{ \begin{array}{ll} \underbrace{A_1^+ \exp(i\beta_1 x)}_{\text{reflected field}} + \underbrace{A_1^- \exp(-i\beta_1 x)}_{\text{incident field}} & \text{in } \Omega_1 \\[2ex] \underbrace{A_2^+ \exp(i\beta_2 x) + A_2^- \exp(-i\beta_2 x)}_{\text{field within the slab}} & \text{in } \Omega_2 \\[2ex] \underbrace{A_3^+ \exp(i\beta_3 x) + A_3^- \exp(-i\beta_3 x)}_{\text{transmitted field}} & \text{in } \Omega_3 \end{array} \right. , \tag{1.8}$$

where $A_i^\phi$ with $\phi = \{+, -\}$ and $i = \{1, 2, 3\}$ refer to the magnitude of the propagating (or counter-propagating) plane wave in the domain $\Omega_i$. In particular, according to our Fourier transform convention, the signal '$-$' indicates the wave propagates along the $x$-direction and vice versa. For example, the term $A_1^- \exp(-i\beta_1 x)$ can identified as the incident field. All the fields are depicted by the red arrows in Figure 1.2.

The next job is to clarify the value of $A_i^\varphi$. From the outgoing wave condition, i.e. Sommerfeld radiation condition [32], it is easily seen that $A_3^+ = 0$. Then by considering the interface conditions [33] , according to [31], the Fresnel coefficients can be expressed as follows:

$$A_1^- = \frac{1}{4} \left[ (1-p)(1-q)\exp(i\beta_2 L) + (1+p)(1+q)\exp(-i\beta_2 L) \right] A_3^-$$

$$A_1^+ = \frac{1}{4} \left[ (1+p)(1-q)\exp(i\beta_2 L) + (1-p)(1+q)\exp(-i\beta_2 L) \right] A_3^-$$

$$A_2^- = \frac{1}{2}(1+q)A_3^-$$

$$A_2^+ = \frac{1}{2}(1-q)A_3^-.$$

where $p = \beta_2/\beta_1$ and $q = \beta_3/\beta_2$.

As we know, the QNMs are the resonant modes of the system where there is no source. Thus, finding QNMs implies setting the incident field to zero, i.e. $A_1^- = 0$, which is equivalent to the following equations:

$$\exp(2i\beta_2 L) = G,$$

where

$$G = -\frac{(1+p)(1+q)}{(1-p)(1-q)} = \frac{(\beta_2+\beta_1)(\beta_2+\beta_3)}{(\beta_2-\beta_1)(\beta_2+\beta_3)} = \frac{(n_p+n_0)^2}{(n_p-n_0)^2}.$$

It follows that the QNM frequencies, i.e. resonant frequencies, are given by:

$$\frac{\omega_m n_p}{c} = \beta_2 = \frac{1}{L} \left[ m\pi - \frac{i}{2}\ln(|G|) \right].$$

It is clear that $|G| > 1$, which implies that $\Im(\omega_m) < 0$. For this particular choice of refractive index, the real part of the QNM frequencies are evidently spaced equally by $L$, and all QNM frequencies have the same negative imaginary part. The complex frequency spectrum is displayed in the bottom panel of Figure 1.3 where the QNM frequencies are shown by dark spots. It is also worth pointing out that the decrease of $|G|$ yields a decrease of the imaginary part of the complex frequencies $\Im(\omega_m)$. It is consistent with the intuitive prediction that the light remains trapped in the cavity for longer times when the ratio of diffraction indices $n_p/n_0$ increases.

Figure 1.4 shows an example of a QNM field in the vicinity of the resonator for the case $m = 4$. In particular, the QNMs are represented by standing waves inside the resonator. This is partly similar to the case of the closed cavity made from perfect reflectors at each side. The difference is that, for open cavities, the fields indeed propagate away from the resonator at both sides. At the same time, their magnitude clearly increases in the direction away from the resonator, which is a general feature of QNMs.

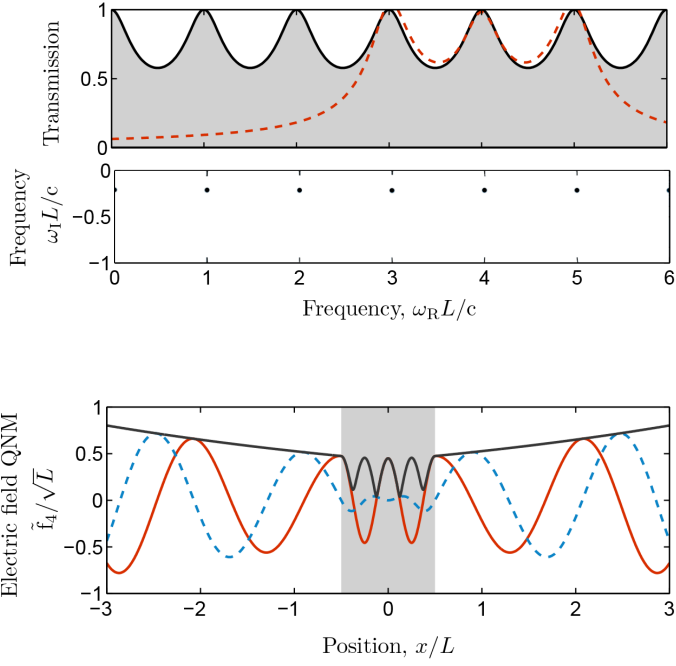

**Figure 1.3:** (TOP) Transmission through the dielectric barrier as a function of frequency for the case of $n_p = \pi$ and $n_0 = 1$. Dashed red curve shows the approximate transmission calculated using three QNMs. (BOTTOM) The complex spectrum: The discrete QNM frequencies are indicated by dark spots in the lower half of the complex plane [34].

**Figure 1.4:** The electric field QNM with $m = 4$. Red solid and blue dashed curves show the real and imaginary parts of the field, respectively, and the black curve stands for the absolute value, which increases (exponentially) as a function of distance from the resonator. Gray shading indicates the extent of the dielectric barrier [34].

## The eigenvalue problem

In the previous subsection, we illustrate how to find the QNMs in a simple 1-D model of the dielectric barrier. The idea behind the process is quite simple:

> Given an optical system, find the resonant frequencies such that the electromagnetic oscillations can sustain for a certain period of time without additional source.

From the mathematical point of view, the previous statement is equivalent to solving the eigenvalue problems of the given electromagnetic system. In fact, if source-free Maxwell's equations can be represented by an electromagnetic operator (for instance $\mathscr{L}(\omega)$), these resonant modes, i.e. QNMs, (and their corresponding resonant frequencies) are nothing more than the eigenvectors (and their corresponding eigenvalues) of such operator.

Before going to the next topic, it is worth spending some time reflecting on the issues that arise when solving the eigenvalue problems of open systems:

1. *Which mathematical operators should be chosen to represent the given electromagnetic problems?*
   Indeed, there is no unique way to construct the eigenvalue problem, and its corresponding mathematical operator, from Maxwell's equations (1.4). For example, (1.4) can be rewritten by the following eigenvalue problem:

   $$\mathscr{L}_1(\omega_n)\mathbf{E}_n = 0$$

   where the operator $\mathscr{L}_1(\omega)$ is given by:[4]

4: We leave the complete definition of electromagnetic operators for the next part of the thesis.

$$\mathscr{L}_1(\omega) = -\nabla \times \left( \boldsymbol{\mu}^{-1}(\omega) \nabla \times \cdot \right) + \omega^2 \boldsymbol{\varepsilon}(\omega)$$

Another expression of (1.4) yields:

$$\mathscr{L}_2(\omega_n) \begin{pmatrix} \mathbf{E}_n \\ \mathbf{H}_n \end{pmatrix} = 0$$

where

$$\mathscr{L}_2(\omega) = \begin{pmatrix} 0 & i\boldsymbol{\varepsilon}^{-1}(\omega)\nabla\times \\ -i\boldsymbol{\mu}^{-1}(\omega)\nabla\times & 0 \end{pmatrix} - \omega \begin{pmatrix} 1 & 0 \\ 0 & 1 \end{pmatrix} \qquad (1.9)$$

It is clear that although the natures of two operators $\mathscr{L}_1(\omega)$ and $\mathscr{L}_2(\omega)$ are different,[5] they both share the same set of eigenvalues, i.e resonant frequencies, as well as QNMs.

Since the eigensolutions (QNMs) remain intact despite the choice of mathematical operators, there must not be any further restriction on the construction of these operators. In fact, the electromagnetic operator should be built in accordance with the user's own purposes as long as the physical properties of QNMs are transformed into the spectral characteristics of their mathematical operators. This gives us an idea that, instead of studying the physical quantities of the given open system, it is better to focus on the intrinsic nature of its mathematical operator $\mathscr{L}(\omega)$. And one can not talk about the nature of the operator without mentioning two main characteristics: the non-linearity and non-self-adjointness.

5: For example, if the materials are reciprocal, the matrix $\mathscr{L}_1(\omega)$ is symmetric while $\mathscr{L}_2(\omega)$ are not.

▶ **Non-self-adjointness**: In the case of closed cavities, the system is conservative and its associated operators are shown to be self-adjoint (or Hermitian). On the other hand, for open cavities, the system would be open and non-conservative, and the associated mathematical operators would not be Hermitian in the usual sense. And as we know, the eigenfunctions of non-self-adjoint operators, i.e. QNMs, possess complex eigenvalues, i.e. eigenfrequencies $\omega_n$.

▶ **non-linearity**: It is easily seen that both operators $\mathscr{L}_1(\omega)$ and $\mathscr{L}_2(\omega)$ in the previous example are non-linear with respect to the variable $\omega$. It results from the fact that the permittivity $\boldsymbol{\varepsilon}$ and permeability $\boldsymbol{\mu}$ depend on the frequency (see Chapter 8 for more details). As a consequence, find QNMs requires to solve non-linear eigenvalue problems.

2. *How to normalize the eigensolutions (QNMs)?*
   Since the magnitude of eigen-fields is not fixed and can be determined up to non-zero complex factors, it is of convenience to normalize the QNMs. Unfortunately, the non-self-adjointness of our electromagnetic operators raises a delicate question of the QNM normalization. In physics, it is a habit to process the eigenvector normalization according to some important physical quantities. For example, in classical textbooks [35], the electric fields $\mathbf{E}_n$ in a closed domain $\Omega_0$ are often normalized based on the electric energy in the volume $\Omega_0$ such as:

$$\int_{\Omega_0} \varepsilon |\mathbf{E}_n|^2 d\Omega.$$

Unfortunately, according to the out-going wave boundary condi-tions, QNMs exponentially diverge in space. Moreover, it is not an easy task to take integration over an infinite region of space. In other words, QNMs are no longer finite energy and the standard normalization based on energy consideration can no longer be applied. The key issue is the necessity to cope with open systems and diverging fields [27, 35]. Different authors have tried to address this problem of QNM normalization but a general approach of normalization methods is still a debated topic in the recent literature [9, 18, 24, 25] .

It is easy to notice that a part of the problem stems from the fact that people tend to come up with a universal formula for the normalization; while the form of the electromagnetic operator $\mathscr{L}(\omega)$ could vary in accordance with the purpose of use. Perhaps the deep root of the problem is that the normalization is often associated with fixed physical quantities, such as mode volume, instead of adapting to change of the constructed operators, based on rigorous mathematical analysis. In order to solve the problem at its root, it is important to re-question the nature of normalization: What is the mathematical motivation behind the standardization of eigensolutions? Or is it genuinely necessary to normalize QNMs?

In short, the two previous issues imply that the best thorough way to understand the nature of QNMs in open resonators is to leave aside the physical intuition and to look for the core platform of QNM by systematically studying the eigenvalue problems for non-linear, non-Hermitian operators.

## 1.2 Quasinormal Modal expansion (QNM expansion)

In order to understand the motivation behind the QNM expansion, let's revisit the example of the 1-D dielectric barrier in the previous section. The top panel of Figure 1.3 depicts the transmission through the barrier as a function of frequency. It is easy to recognize a number of distinct peaks with unity transmission, which coincides with the position of the real part of the frequency of QNMs in the complex plane. These are the well-known Fabry-Perot resonances. Figure 1.3 proves that the transmission property of the slab is clearly influenced by the location of QNMs. In order to clarify this prediction, an approximate transmission, which is calculated using three QNMs such that $3 \leq \Re(\omega_m)L \leq 5$, is represented by dashed red curve in the top panel of Figure 1.3. Unsurprisingly, we witness a local agreement between the exact transmission and the approximation around the considered QNMs.

From here, it is tempting to draw a conclusion that the optical response of the structure (which, in this case, the transmission property) is determined by the synthesis of all the QNMs of the system. The explanation is as follows: Under the excitation of external pulses, the resonant modes (QNMs) are initially loaded, then release their energy which, in turn, contributes to the macroscopic optical property of the resonators. This physical intuition results in a naive but elegant idea that, similar to the geometrical projection, the resonator responses (for example the scattered electromagnetic fields) can be expanded (or decomposed) into a vector space formed by the QNMs.

In fact, the concept of the spectral decomposition of electromagnetic waves is not a new idea. It originated from quantum mechanics [36, 37] and has been theoretically studied in electromagnetics since the seventies [38]. Further developments of QNM expansion have been well documented in the review [27]. In that paper, the authors classified two different approaches to the formalism of QNM expansion:

▶ **Orthogonality-decomposition approach:** The main feature of this approach is the orthogonality relation between the QNMs. Moreover, the technique requires to derive the final expansion formula from the Lorentz reciprocity theorem [18, 20, 39] .

▶ **Residue-decomposition approach:** It relies on the theory of functions of complex variable. The idea of this technique is built around the concept of the Green tensor $G(\mathbf{r}, \mathbf{r}', \omega)$, the solution of the following equation:

$$-\nabla \times \left(\mu^{-1}\nabla \times G(\mathbf{r}, \mathbf{r}', \omega)\right) + \omega^2 \varepsilon G(\mathbf{r}, \mathbf{r}', \omega) = \mathbf{I}\delta(\mathbf{r} - \mathbf{r}').$$

By Mittag-Leffler's theorem, the Green tensor $G(\mathbf{r}, \mathbf{r}', \omega)$ can be expanded as a series of poles (see [8, 9, 13, 40] for more details):

$$G(\mathbf{r}, \mathbf{r}', \omega) = \sum_n \frac{\text{Res}[G(\mathbf{r}, \mathbf{r}', \omega_n)]}{\omega - \omega_n},$$

where $\omega_n$ stand for the resonant frequencies while $\text{Res}[G(\mathbf{r}, \mathbf{r}', \omega_n)]$ refers to the residue of $G(\mathbf{r}, \mathbf{r}', \omega_n)$. Once the residues of the Green tensor are known, it is straightforward to deduce the expansion formula.

Each of the above QNM expansion approach has its own pros and cons. And it is inevitable to have certain concerns regarding the nature of the QNM expansion formalism:

1. *The 'left' eigenvalue problem*

   It is important to note that most of the attempts to derive the QNM expansion formulas are done under the assumption that the media are reciprocal (often isotropic and homogeneous). For example, the Lorentz reciprocity theorem (from the Orthogonality-decomposition approach) only holds for the condition that media are reciprocal. Unfortunately, in a wide range of optical applications, both permittivity and permeability are not only dispersive but also anisotropic, and possibly non-reciprocal. This urges more systematic approaches on the QNM expansion formalization of the general form of the electromagnetic operator $\mathscr{L}(\omega)$: Perhaps it is necessary to revisit some mathematical concepts in spectral theory to find out what is missing.

   In fact, we will point out, in Chapter 3, that the lack of research on the QNM expansion for non-reciprocal media comes from the neglect of the concept of 'left' eigenvector. In particular, given the operator $\mathscr{L}(\omega)$, the 'left' and 'right' eigenvectors ($\langle \mathbf{w}_n |$ and $| \mathbf{v}_n \rangle$ respectively) and their corresponding eigenvalues $\omega_n$ are defined as follows:[6]

6: see Chapter 3 for further explanation of the following notations.

$$\langle \mathbf{w}_n | \mathscr{L}(\omega_n) = 0 \quad \text{and} \quad \mathscr{L}(\omega_n) | \mathbf{v}_n \rangle = 0.$$

The 'right' eigenvectors $\mathbf{v}_n$, which are simply the solutions of the 'common' eigenvalue problem, can be directly associated with the 'physical' resonant modes. On the other hand, it is hard to intuitively explain the meaning of the 'left' eigenvectors $\mathbf{w}_n$ or associate them with available physical quantities. It raises the question about the importance of the contribution of these 'left' eigenvectors on the modal expansion. Or to put it simply, why the 'left' eigenvectors are vital in some cases (with non-reciprocal media) but negligible in other casese? What is the relation between the 'left' and 'right' eigenvectors?

2. *Normalization*

   Similar to the previous section, the precise way to normalize the QNMs is still an open question considering that the structures are unbounded and the materials are dispersive and non-reciprocal. This also evokes a central question if the QNM normalization is genuinely necessary? If yes what is the role of the normalization in the general picture of modal expansion?

3. *Overcompleteness*

   It is easily seen that there are many different 'final' formulas for QNM expansion in the literature. Part of the reason is that there are mainly 2 different approaches to formalize the QNM expansion. In [27], the authors have attempted to bridge the apparent gap between these two different frameworks. They showed that it is possible to derive the residue of the Green tensor $\text{Res}[G(\mathbf{r}, \mathbf{r}', \omega_n)]$ used in the residue-decomposition approach from Lorentz-reciprocity arguments. But it doesn't completely solve the mystery of the non-uniqueness of the QNM expansions. Another explanation is the freedom of choice in the construction of electromagnetic operators (as we showed in the previous section) and linearization [41] , which leads to different expansions which are specific to each choice of operator. It raises the question of what is the relation between these formulas? It is also reasonable to wonder which expansion (or decomposition) provide the most accurate prediction of the optical responses of resonators? In fact, these questions are still debated topic recently.

   Or perhaps we have been asking the wrong questions. Perhaps all the formulas are correct and provide the same exact representation of the spectral properties of open resonators. In fact, since the constructed electromagnetic operator $\mathscr{L}(\omega)$ is non-linear, its eigensolutions, i.e. QNMs, can be proved to be not linear independent. Thus, the eigenbasis spaned by these QNMs (if it exists) is overcomplete [2] . This could be the premise for a new insight on spectral theory when applied to Maxwell's equations: There exists not just one, but a continuous family of QNM expansions.

In the rest of this thesis, we will try to answer all of the questions posed above through research and analysis of the spectral theory of non-linear and non-self-adjoint operators. And by that way, we hope to reveal subtle insights and the beauty of the spectral representation of the optical responses on the resonant-state-basis.

# SPECTRAL THEOREM FOR ELECTROMAGNETICS

# Spectral theorem for self-adjoint operators

# 2

In photonics, the interaction of electromagnetic field (light) and matter heavily relies on the concept of resonant mode of the structure. Under external excitation, these modes are initially loaded, and then release their energy which, in turn, reflects the optical properties of the system. Thus, it is intuitively believed that we can take advantages of these resonances with the help of modal expansion technique in order to discover the physical properties of our considered system. In fact, the study of modal expansion of natural resonance, i.e. Quasinormal Mode, is recently of fundamental interest in photonics. (It is highly recommended to see [27] and the references given there before reading this thesis.)

Mathematically, if the given system can be represented by an electromagnetic operator, these resonant modes are nothing more than the eigensolutions of such operator. Therefore, to study the modal expansion requires to deeply understand the spectral theorem of electromagnetic operators.

The aim of this chapter in particular, and the spectral theorem part in general, is to study in a very systematic way the spectral properties in electromagnetics and to mathematically verify modal expansion formulas obtained recently. This requires to revisit the most basic concepts of the Fourier expansion of linear operators. For deeper discussions associated with the eigenvalue problem of linear operators, we refer the reader to appendix A.

Before going further, it is worth pointing out that all the electromagnetic operators mentioned in this thesis are 'linear' in terms of their action on the elements of the vector space (according to Definition A.2.1). At the same time, these operators are also functions that can be linear or non-linear with respect to the variable $\lambda$. For the sake of clarity, from now on, the linearity of an operator will simply be associated with its dependence on the variable $\lambda$. For example, the operator $L(\lambda)$ is called linear if it can be expressed as a linear function with respect to $\lambda$ namely $L(\lambda) = A + \lambda B$. Similarly, the definition of the polynomial and rational operator are stated in Definition 4.2.1 and Definition 5.0.1 respectively.

## 2.1 Spectral properties of linear operators

Given the linear operator $L(\lambda) = A + \lambda B$ in the domain $D_L$, such that $A, B : H \rightarrow H$ be self-adjoint (or symmetric) linear operators, $\langle \mathbf{x}, A\mathbf{x} \rangle \neq 0$ or $\langle \mathbf{x}, B\mathbf{x} \rangle \neq 0, \forall \mathbf{x} \neq 0$ in $D_L$, the eigenfunctions $\mathbf{v}_n$ such that $L(\lambda_n)\mathbf{v}_n = 0$ on a Hilbert space $H$ form an orthonormal basis.[1] Equipped with such eigenbasis, our purpose is to deduce the eigenfunction expansion for the solution of $L(\lambda)\mathbf{u} = \mathbf{S}$ with corresponding boundary conditions.

1: All the notations and definitions used in this chapter can be found in appendix A.

**Theorem 2.1.1** (Fourier expansion) *Given the operator $L(\lambda) = A + \lambda B$ where $A, B : H \longrightarrow H$ self-adjoint such that $\langle x, Ax \rangle \neq 0$ or $\langle x, Bx \rangle \neq 0, \forall x \neq 0$ in $D_L$, the solution $\boldsymbol{u}$ of the boundary value problem $L(\lambda)\boldsymbol{u} = \boldsymbol{S}$ can be rewritten as generalized Fourier expansion on the orthonormal basis $\{\boldsymbol{v}_n\}$, solutions of the generalized eigenvalue problem $L(\lambda)\boldsymbol{v}_n = (A + \lambda_n B)\boldsymbol{v}_n = 0$:*

$$u = \sum_{n=1}^{\infty} \frac{1}{\lambda - \lambda_n} \frac{\langle v_n, S \rangle}{\langle v_n, B v_n \rangle} v_n. \tag{2.1}$$

It is obvious that the expansion (2.1) does not exist if $\lambda = \lambda_n$.

*Proof:* By Definition A.1.1 of the expansion of $\mathbf{u}$ onto $\{\mathbf{v}_n\}$, we have

$$L(\lambda)\mathbf{u} = L(\lambda)\sum_{n=1}^{\infty} \langle \mathbf{v}_n, \mathbf{u} \rangle \mathbf{v}_n = \sum_{n=1}^{\infty} \langle \mathbf{v}_n, \mathbf{u} \rangle (A + \lambda B)\mathbf{v}_n = \mathbf{S}.$$

Since $\mathbf{v}_n$ are eigenfunctions of the operator $L(\lambda_n)\mathbf{v}_n = (A + \lambda_n B)\mathbf{v}_n = 0$, we can replace the term $A\mathbf{v}_n$ in the previous equation by $(\lambda_n B)\mathbf{v}_n$, which yields:

$$\sum_{n=1}^{\infty} \langle \mathbf{v}_n, \mathbf{u} \rangle (-\lambda_n + \lambda) B\mathbf{v}_n = \mathbf{S}.$$

Taking the inner product of both sides with $\mathbf{v}_m$ and exploiting orthonormality from Theorem A.3.2 $(\lambda_m - \lambda_n)\langle \mathbf{v}_n, B\mathbf{v}_m \rangle = 0$ (keep in mind that B is a self-adjoint operator) lead to:

$$(\lambda - \lambda_n)\langle \mathbf{v}_n, \mathbf{u} \rangle \langle \mathbf{v}_n, B\mathbf{v}_n \rangle = \langle \mathbf{v}_n, \mathbf{S} \rangle.$$

This results in

$$\langle \mathbf{v}_n, \mathbf{u} \rangle = \frac{1}{\lambda - \lambda_n} \frac{\langle \mathbf{v}_n, \mathbf{S} \rangle}{\langle \mathbf{v}_n, B\mathbf{v}_n \rangle},$$

or

$$u = \sum_{n=1}^{\infty} \frac{1}{\lambda - \lambda_n} \frac{\langle \mathbf{v}_n, \mathbf{S} \rangle}{\langle \mathbf{v}_n, B\mathbf{v}_n \rangle} \mathbf{v}_n. \tag{2.2}$$

The sum $\sum_{n=1}^{\infty}$ runs from one to infinity as the space has infinite dimensions.

## 2.2 Fourier expansion for electromagnetics

In this section, we will try to apply Theorem 2.1.1 in a bounded, non-dispersive, non-dissipative electromagnetic problem. Consider Maxwell's equations applied to a finite domain $\Omega$ bounded by a perfectly conducting, closed surface $\Gamma$:

$$\nabla \times \mathbf{H}(\mathbf{r}) + i\omega\varepsilon(\mathbf{r})\mathbf{E}(\mathbf{r}) = \mathbf{J}$$
$$\nabla \times \mathbf{E}(\mathbf{r}) - i\omega\mu(\mathbf{r})\mathbf{H}(\mathbf{r}) = 0 \tag{2.3}$$
$$n \times \mathbf{E}|_{\Gamma} = 0$$

Our goal is to rewrite the previous problem into a system containing only self-adjoint operators.

One possible option is to set:

$$A = i \begin{pmatrix} 0 & -\nabla \times \\ \nabla \times & 0 \end{pmatrix} \quad \text{and} \quad B = \begin{pmatrix} \varepsilon & 0 \\ 0 & \mu \end{pmatrix} \tag{2.4}$$

with the domains:

$$D_L \equiv \{\mathbf{u} : \mathbf{u}, \nabla \times \mathbf{u} \in L^2(\Omega)^6, n \times \mathbf{u} = 0\}$$

If we choose and $\lambda = \omega$, $\mathbf{u} = \begin{pmatrix} \mathbf{E} \\ \mathbf{H} \end{pmatrix}$ and $\mathbf{S} = -i \begin{pmatrix} \mathbf{J} \\ 0 \end{pmatrix}$, (2.4) is equivalent to the following equation:

$$L(\lambda)\mathbf{u} = (A + \lambda B)\mathbf{u} = \mathbf{S}. \tag{2.5}$$

The resonant states of the system are the eigensolutions of the generalized linear eigenvalue problem $(A + \lambda_n B)\mathbf{v}_n = 0$ or

$$\left[ i \begin{pmatrix} 0 & -\nabla \times \\ \nabla \times & 0 \end{pmatrix} + \omega_n \begin{pmatrix} \varepsilon & 0 \\ 0 & \mu \end{pmatrix} \right] \begin{pmatrix} \mathbf{E}_n \\ \mathbf{H}_n \end{pmatrix} = \begin{pmatrix} 0 \\ 0 \end{pmatrix} \tag{2.6}$$

The permittivity $\varepsilon$ and permeability $\mu$ in the very simple case of linear, homogeneous, isotropic materials are scalar $\varepsilon$ and $\mu$. [2] In the case that the media are anisotropic, the value of $\varepsilon$ and $\mu$ are represented as matrices:

$$\chi = \begin{pmatrix} \chi_{xx} & \chi_{xy} & \chi_{xz} \\ \chi_{yx} & \chi_{yy} & \chi_{yz} \\ \chi_{zx} & \chi_{zy} & \chi_{zz} \end{pmatrix} \quad \chi = \{\varepsilon, \mu\}.$$

And in order for the operators $B$ to be self-adjoint, the matrices $\varepsilon$ and $\mu$ must also be self-adjoint. [3] Moreover, the operator is positive definite.

As the same time, the operator $A$ is self-adjoint since the matrix $A = i \begin{pmatrix} 0 & -\nabla \times \\ \nabla \times & 0 \end{pmatrix}$ is also self-adjoint.

From the self-adjointness of operators $A$ and $B$, we can extract 2 properties from the eigenvalue problem (2.6):

▶ By Theorem A.3.1 , we can show that the resonance frequencies $\lambda_n = \omega_n$ are real $\omega_n \in \mathbf{R}$.
▶ By Theorem A.3.2, eigenfunctions $\mathbf{v}_n$ of $L(\lambda)$ corresponding to different eigenvalues are orthogonal: $\langle \mathbf{v}_n, B\mathbf{v}_m \rangle = 0$ for $m \neq n$.

Then, by applying Proposition A.3.5, we know that $\{\mathbf{v}_n\}$ form an orthogonal basis on $H = D_L$ such that $\mathbf{u} = \sum_n \langle \mathbf{v}_n, \mathbf{u} \rangle \mathbf{v}_n$ for $\forall \mathbf{u} \in D_L$.

Theorem 2.1.1 shows us exactly how the term $\langle \mathbf{v}_n, \mathbf{u} \rangle$ look like:

$$u = \sum_{n=1}^{\infty} \frac{1}{\lambda - \lambda_n} \frac{\langle \mathbf{v}_n, \mathbf{S} \rangle}{\langle \mathbf{v}_n, B\mathbf{v}_n \rangle} \mathbf{v}_n.$$

We remind that $\lambda_n = \omega_n$ are eigen-frequencies, i.e. the resonant frequencies of operator $L(\omega)$ with the corresponding eigenfunctions $\mathbf{v}_n = \begin{pmatrix} \mathbf{E}_n \\ \mathbf{H}_n \end{pmatrix}$

The inner product $\langle \mathbf{v}_n, B\mathbf{v}_n \rangle$ can be explicated as follows:

$$\langle \mathbf{v}_n, B\mathbf{v}_n \rangle = \left\langle \begin{pmatrix} \mathbf{E}_n \\ \mathbf{H}_n \end{pmatrix}, \begin{pmatrix} \boldsymbol{\varepsilon} & 0 \\ 0 & \boldsymbol{\mu} \end{pmatrix} \begin{pmatrix} \mathbf{E}_n \\ \mathbf{H}_n \end{pmatrix} \right\rangle \tag{2.7}$$

$$= \int_\Omega \left[ \overline{\mathbf{E}_n} \cdot (\boldsymbol{\varepsilon} \mathbf{E}_n) + \overline{\mathbf{H}_n} \cdot (\boldsymbol{\mu} \mathbf{H}_n) \right] d\Omega. \tag{2.8}$$

Finally, with the expression of the numerator product:

$$\langle \mathbf{v}_n, \mathbf{S} \rangle = -i \int_\Omega \overline{\mathbf{E}_n} \cdot \mathbf{J} \, d\Omega,$$

we achieve the Fourier expansion for electric fields:

$$\mathbf{E} = \sum_{n=1}^{\infty} \frac{1}{\omega - \omega_n} \frac{-i \int_\Omega \overline{\mathbf{E}_n} \cdot \mathbf{J} d\Omega}{\int_\Omega \left[ \overline{\mathbf{E}_n} \cdot (\boldsymbol{\varepsilon} \mathbf{E}_n) + \overline{\mathbf{H}_n} \cdot (\boldsymbol{\mu} \mathbf{H}_n) \right] d\Omega} \mathbf{E}_n. \tag{2.9}$$

*Expansion of electric fields for self-adjoint operators 1*

We also notice some interesting properties for the 'electric energy' and 'magnetic energy' in the denominator of (2.9).

**Remark 2.2.1** If the matrix $\boldsymbol{\mu}$ is diagonal, it is possible to prove that $\int_\Omega \overline{\mathbf{E}_n} \cdot (\boldsymbol{\varepsilon} \mathbf{E}_n) d\Omega = \int_\Omega \overline{\mathbf{H}_n} \cdot (\boldsymbol{\mu} \mathbf{H}_n) d\Omega$.

*Proof:* First, we have to remind that From Maxwell's equations it is easy to see that:

$$\boldsymbol{\varepsilon} \mathbf{E} = -\frac{\nabla \times \mathbf{H}}{i\omega}$$

$$\nabla \times \overline{\mathbf{E}} = -i\omega \boldsymbol{\mu} \overline{\mathbf{H}}.$$

The second equality is achieved by taking the complex conjugate of Maxwell's equation and the fact that $\boldsymbol{\mu}$ contains only real components. [4] The 'electric energy' term can be rewritten as:

4: Since our system is non dissipative.

$$\int_\Omega \overline{\mathbf{E}_n} \cdot (\boldsymbol{\varepsilon} \mathbf{E}_n) d\Omega = \int_\Omega \overline{\mathbf{E}_n} \cdot \left( -\frac{\nabla \times \mathbf{H}}{i\omega} \right) d\Omega$$

$$= \int_\Omega \left( -\frac{\nabla \times \overline{\mathbf{E}_n}}{i\omega} \right) \cdot \mathbf{H} \, d\Omega + \int_\Gamma (n \times \overline{\mathbf{E}}) \cdot \mathbf{H} \, dS$$

$$= \int_\Omega \boldsymbol{\mu} \overline{\mathbf{H}} \cdot \mathbf{H} \, d\Omega = \int_\Omega \overline{\mathbf{H}} \cdot (\boldsymbol{\mu} \mathbf{H}) \, d\Omega,$$

where the boundary term $\int_\Gamma (n \times \overline{\mathbf{E}}) \cdot \mathbf{H} \, dS$ vanishes due to appropriate boundary conditions. [5]

5: The boundary conditions will be discussed in details in section Section 3.3

If Remark 2.2.1 holds, the inner product $\langle \mathbf{v}_n, B\mathbf{v}_n \rangle$ can be expressed solely in terms of electric fields:

$$\langle \mathbf{v}_n, B\mathbf{v}_n \rangle = 2 \int_\Omega \overline{\mathbf{E}_n} \cdot (\boldsymbol{\varepsilon} \mathbf{E}_n) \, d\Omega.$$

### "Normalization"

Given an eigenvalue $\tilde{\lambda}$, the set $\tilde{V} = \{\mathbf{v} : L(\tilde{\lambda})\mathbf{v} = 0\}$, which is the union of the zero vector with the set of all eigenvectors associated with $\tilde{\lambda}$, is called the eigenspace or characteristic space of $L(\lambda)$ associated with $\tilde{\lambda}$. In particular, if $\mathbf{v}_1$ and $\mathbf{v}_2$ are eigenvectors of $L(\lambda)$ associated with eigenvalue $\tilde{\lambda}$: $\mathbf{v}_1, \mathbf{v}_2 \in \tilde{V}$ then $\mathbf{v}_1 + \mathbf{v}_2$ and $a\mathbf{v}_1$ are either zero or eigenvectors of $L(\lambda)$ associated with $\tilde{\lambda}$ for $\forall a \in \mathbb{C}$.

Henceforward, we assume that the geometric multiplicity of all the eigenvalues $\lambda_n$ equals 1, i.e. each eigenvalue possesses only one eigenvector. [6] Even with the previous assumption, the eigenfunctions (i.e. eigenvectors $\mathbf{v}_n$) are still not strictly determined:

> **Remark 2.2.2** If $\mathbf{v}_n$ is an eigenfunction of $L(\lambda_n)\mathbf{v}_n = (A + \lambda_n B)\mathbf{v}_n = 0$, then so is $a\mathbf{v}_n$ for $\forall a \in \mathbb{C}$.

If we replace $\mathbf{v}_n$ by $a\mathbf{v}_n$ the Fourier expansion of $\mathbf{u}$ (2.1) still remain unchanged. Thus, it is possible to express (2.1) as:

$$u = \sum_{n=1}^{\infty} \frac{1}{\lambda - \lambda_n} \langle \mathbf{v}_n, \mathbf{S} \rangle \mathbf{v}_n, \tag{2.10}$$

if we set $\langle \mathbf{v}_n, B\mathbf{v}_n \rangle = 1$.

This kind of maneuver is called 'normalization'.[27]

In (2.7) the term $\langle \mathbf{v}_n, B\mathbf{v}_n \rangle$ can be seen as the sum of electric and magnetic energy. Thus it is easy to misunderstand that the normalization is done through energy standardization of the whole field. It is true that the energy of a closed, non-dissipative, non-dispersive electromagnetic systems is conserved but that is not the mathematical reason why we should normalize the eigenvectors in terms of energy. The reality is that we only try to build a general eigenvalue problem with $L(\lambda)\mathbf{v}_n = (A + \lambda B)\mathbf{v}_n$ such that both $A$ and $B$ are self-adjoint operators and $B$ is positive definite in order to apply Theorem 2.1.1. [7]

In fact, there are a lot of way to derive a matrix system of equations from Maxwell's equations. Unfortunately, none of them give us two self-adjoint operators $A$ and $B$, where $B$ is positive definite in order to have Theorem 2.1.1.

For example (see for instance [20] ), we can choose $\lambda = \omega$ and

$$A = \begin{pmatrix} 0 & -i\boldsymbol{\varepsilon}^{-1}\nabla\times \\ i\boldsymbol{\mu}^{-1}\nabla\times & 0 \end{pmatrix} \quad \text{and} \quad B = \begin{pmatrix} 1 & 0 \\ 0 & 1 \end{pmatrix}$$

with the domains:

$$D_L \equiv \{\mathbf{u} : \mathbf{u}, \nabla \times \mathbf{u} \in L^2(\Omega)^6, n \times \mathbf{u}|_\Gamma = 0\}.$$

It is clear that the operator $B$ is self-adjoint and positive definite. Unfortunately, the operator A is not self-adjoint. The inner product give us different equation $\langle \mathbf{v}_n, B\mathbf{v}_n \rangle = \int_\Omega \left[ \overline{\mathbf{E}_n} \cdot (\mathbf{E}_n) + \overline{\mathbf{H}_n} \cdot (\mathbf{H}_n) \right] d\Omega$. However, we can't use this inner product to normalize the eigenfunctions because Theorem 2.1.1 no longer holds here.

6: In practice, it is rare for the eigenvalue to degenerate. Even if the geometric multiplicity of the eigenvalue is analytically greater than 1, the numerical computation still results in several eigenvalues which are approximately equal.

7: In other words, the 'normalization', if exists, is only a mathematical consequence of the inner product created by self-adjoint operators, rather than based on a specific physical quantity such as energy.

It is interesting to see that the restriction of Theorem 2.1.1 leads us to the normalization over the electromagnetic energy. From physical point of view, we know the electromagnetic energy of a self-adjoint system is conserved. Therefore, by constructing mathematically self-adjoint operators, we unintentionally guarantee the conservation of energy. The fact that the inner product $\langle \mathbf{v}_n, B\mathbf{v}_n \rangle$ of our normalization has a form of energy should be seen as mathematical consequences of our process, instead of causes.

## 2.3 The non-uniqueness of Fourier expansion

We will close this chapter by constituting another operator equation for (2.3). In particular, it is possible to rephrase (2.3) into an electric wave equation:

$$-\nabla \times \left( \boldsymbol{\mu}^{-1} \nabla \times \mathbf{E} \right) + \omega^2 \boldsymbol{\varepsilon} \mathbf{E} = -i\omega \mathbf{J} \qquad (2.11)$$
$$n \times \mathbf{E}|_\Gamma = 0.$$

Then, we can set two operators $A$ and $B$ as follows:

$$A = -\nabla \times \left( \boldsymbol{\mu}^{-1} \nabla \times \cdot \right) \quad \text{and} \quad B = \boldsymbol{\varepsilon},$$

with the domain $D_L \equiv \{ \mathbf{u} : \mathbf{u}, -\nabla \times (\mu^{-1} \nabla \times \mathbf{u}) \in L^2(\Omega)^3, n \times \mathbf{u}|_\Gamma = 0 \}$.

Seting $\lambda = \omega^2$ as the eigenvalue and $\mathbf{S} = -i\omega \mathbf{J}$, we obtain again $L(\lambda)\mathbf{u} = (A + \lambda B)\mathbf{u} = \mathbf{S}$.

The eigensolutions of operator $L(\lambda)$ are given by:

$$-\nabla \times \left( \boldsymbol{\mu}^{-1} \nabla \times \mathbf{E}_n \right) + \omega_n^2 \boldsymbol{\varepsilon} \mathbf{v}_n = 0. \qquad (2.12)$$

It is clear that $A$ and $B$ are self-adjoint while $B$ is positive definite. Thus, similar to Section 2.2, the following statements hold:

▶ By Theorem A.3.1, the resonance frequencies $\lambda_n = \omega_n^2$ are real.
▶ By Theorem A.3.2, eigenvectors $\mathbf{v}_n = \mathbf{E}_n$ form an orthogonal basis such that: $\langle \mathbf{v}_n, B\mathbf{v}_m \rangle = 0$ for $m \neq n$.

Now, we can use Theorem 2.1.1 to expand the solution $\mathbf{u}$ as follows:

$$\mathbf{u} = \sum_{n=1}^{\infty} \frac{1}{\lambda - \lambda_n} \frac{\langle \mathbf{v}_n, \mathbf{S} \rangle}{\langle \mathbf{v}_n, B\mathbf{v}_n \rangle} \mathbf{v}_n,$$

or

$$\mathbf{E} = \sum_{n=1}^{\infty} \frac{1}{\omega^2 - \omega_n^2} \frac{-i\omega \int_\Omega \overline{\mathbf{E}_n} \cdot \mathbf{J} \, d\Omega}{\int_\Omega \overline{\mathbf{E}_n} \cdot (\boldsymbol{\varepsilon} \mathbf{E}_n) \, d\Omega} \mathbf{E}_n. \qquad (2.13)$$

*Expansion of electric fields for self-adjoint operators 2*

As discussed in Section 2.2, we can also rewrite the previous Fourier expansion as:

$$\mathbf{E} = \sum_{n=1}^{\infty} \left[ \frac{-i\omega}{\omega^2 - \omega_n^2} \left( \int_{\Omega} \overline{\mathbf{E}_n} \cdot \mathbf{J} \, d\Omega \right) \right] \mathbf{E}_n$$

after "normalize" the eigenfunctions by setting $\int_{\Omega} \overline{\mathbf{E}_n} \cdot (\varepsilon \mathbf{E}_n) \, d\Omega = 1$

From a mathematical point of view, it is tempting to show that Fourier expansions (2.9) and (2.13) are similar. Although, on the first glance, the factor between 2 expansions are quite different, we can discover that

- ▶ If $\omega_n$ is the eigenvalue corresponding to eigen functions $\mathbf{v}_n$ for the eigenproblem (2.12), then so is $-\omega_n$. For the sake of convenience, let's denote $\omega_{-n} := -\omega_n$.
- ▶ The sum $\sum$ in (2.13) is different from the one in (2.9): The former consider only the eigenvalues $\omega_n^2 > 0$; while the latter indeed counts double: all the positive and negative-real-part eigenvalues $\omega_n$.

In order to compare between two equations, we have to make some modification to standardize the sum. [8]

Let's start from (2.13):

$$\mathbf{E} = \sum_{n=1}^{\infty} \frac{-i\omega}{\omega^2 - \omega_n^2} \frac{\int_{\Omega} \overline{\mathbf{E}_n} \cdot \mathbf{J} \, d\Omega}{\int_{\Omega} \overline{\mathbf{E}_n} \cdot (\varepsilon \mathbf{E}_n) \, d\Omega} \mathbf{E}_n$$

$$= \sum_{n=1}^{\infty} \left( \frac{-i\omega}{(\omega - \omega_n)(\omega + \omega_n)} \right) \frac{\int_{\Omega} \overline{\mathbf{E}_n} \cdot \mathbf{J} \, d\Omega}{\int_{\Omega} \overline{\mathbf{E}_n} \cdot (\varepsilon \mathbf{E}_n) \, d\Omega} \mathbf{E}_n$$

$$= \begin{cases} \sum_{n=1}^{\infty} \dfrac{-i}{2} \left( \dfrac{1}{\omega - \omega_n} + \dfrac{1}{\omega + \omega_n} \right) \dfrac{\int_{\Omega} \overline{\mathbf{E}_n} \cdot \mathbf{J} \, d\Omega}{\int_{\Omega} \overline{\mathbf{E}_n} \cdot (\varepsilon \mathbf{E}_n) \, d\Omega} \mathbf{E}_n \\[4mm] \sum_{n=1}^{\infty} \dfrac{-i\omega}{2} \left( \dfrac{1}{\omega_n(\omega - \omega_n)} + \dfrac{1}{-\omega_n(\omega + \omega_n)} \right) \dfrac{\int_{\Omega} \overline{\mathbf{E}_n} \cdot \mathbf{J} \, d\Omega}{\int_{\Omega} \overline{\mathbf{E}_n} \cdot (\varepsilon \mathbf{E}_n) \, d\Omega} \mathbf{E}_n \end{cases}$$

We remind the fact that in (2.12), one eigenfunction $\mathbf{E}_n$ has two eigenvalues $\omega_n$ and $\omega_{-n}$. [9] This helps us to rewrite (2.13) with a new sum:

$$\begin{cases} \displaystyle\sum_{n=-\infty}^{\infty} \frac{-i}{2(\omega - \omega_n)} \frac{\int_{\Omega} \overline{\mathbf{E}_n} \cdot \mathbf{J} \, d\Omega}{\int_{\Omega} \overline{\mathbf{E}_n} \cdot (\varepsilon \mathbf{E}_n) \, d\Omega} \mathbf{E}_n & \text{(2.14a)} \\[5mm] \displaystyle\sum_{n=-\infty}^{\infty} \frac{-i\omega}{2\omega_n(\omega - \omega_n)} \frac{\int_{\Omega} \overline{\mathbf{E}_n} \cdot \mathbf{J} \, d\Omega}{\int_{\Omega} \overline{\mathbf{E}_n} \cdot (\varepsilon \mathbf{E}_n) \, d\Omega} \mathbf{E}_n & \text{(2.14b)} \end{cases}$$

where (2.14a) differs from (2.14b) by a factor of $\dfrac{\omega_n}{\omega}$.

If $\mu$ is a diagonal matrix, using Remark 2.2.1, we can see immediately that expression (2.14a) gives the same result as (2.9),[10] which confirms our prediction.

The question appear is what could explain the second expression (2.14b). Deriving from (2.13), we know that (2.14b) must be a legit Fourier expansion. This can only be understood by the fact that the eigenvalue problem (2.12) is not linear, but indeed quadratic with respect to $\omega$. In particular, the system of linear eigenvalue problems containing electric and magnetic fields can be considered as a single higher-order eigenvalue problem. And, we have to keep in mind the fact that the modal expansion for non-linear eigenvalue problem is not unique.[11]

11: Pay attention that this does not contrast to the theory of the existence and uniqueness of the Fourier expansion onto an eigenbasis.

# Spectral theorem for non-self-adjoint operators

# 3

In the previous chapter, the Fourier expansion is constructed based on the self-adjointness of operators. Up until now, we still haven't come out of the framework of the classical modal expansion. Unfortunately, as we can see from Section 2.2, it is not a simple task to build an eigenvalue system solely from self-adjoint operators. In addition, in physics, some systems are simply not self-adjoint because their total energy is not conserved. In particular, for electromagnetics, it is the case where the dissipation of materials is taken into account. This urges the need for new eigen-decomposition technique for eigenvalue problem containing non-self-adjoint operators. Such technique is known in literature [27] as Quasinormal-mode expansion. The name 'Quasinormal mode' is derived from the fact that the eigen-functions are no longer orthogonal to each other because of the lack of self-adjoint-ness.

The work in this chapter is motivated by [21].

## 3.1 The left eigenvectors - Eigentriplet

In this chapter, we introduce a concept which is rarely mentioned in detail in physics and often assimilated with its popular counterpart 'right' eigenvectors: The 'left' eigenvectors.

### Left eigenvectors

Firstly, for simplicity of notation, let's introduce the 'Bra-ket' vector: The 'ket' vector $|\mathbf{v}\rangle$ denotes a vector in an vector space $H$. In particular, we can simple right $\mathbf{v} = |\mathbf{v}\rangle$. On the other hand, the 'bra' $\langle\mathbf{w}|$ represents the vector in dual space $H' = H$. The conjugate transpose (also called Hermitian conjugate) of a bra is the corresponding ket and vice versa:

$$\langle\mathbf{v}|^\dagger = |\mathbf{v}\rangle.$$

We set that $\mathbf{v}^\dagger = (\overline{\mathbf{v}})^\mathsf{T}$ where $\mathbf{v}^\mathsf{T}$ denotes the transpose and $\overline{\mathbf{v}}$ stands for the complex conjugated of $\mathbf{v}$.

*Example:* In finite dimensional spaces, considering the vector space $\mathbb{C}^k$, a ket can be identified with a column vector, and a bra as a row vector:

$$\langle\mathbf{w}| = (\overline{w_1}\ \overline{w_2}\ \overline{w_3}\ \dots\ \overline{w_k}) \qquad |\mathbf{v}\rangle = \begin{pmatrix} v_1 \\ v_2 \\ v_3 \\ \dots \\ v_k \end{pmatrix}.$$

Thus, the inner product on Hilbert space $H$ is now equivalent to an identification between the space of 'kets' and that of 'bras' in the 'bra-ket' notation:

$$\langle \mathbf{w}, \mathbf{v} \rangle = \langle \mathbf{w} | \mathbf{v} \rangle.$$

Now, we can define the left eigenvectors as follows

**Definition 3.1.1** *Given two operators $A, B : H \longrightarrow H$ with the domain $D_L$, the vector $w_n \in D_L$ such that $w_n \neq 0$ is called the left eigenvector of the operator $L(\lambda) = A + \lambda B$ with a corresponding eigenvalue $\lambda_n \in \mathbb{C}$ if*

$$\langle w_n | L(\lambda_n) = \langle w_n | (A + \lambda_n B) = 0. \tag{3.1}$$

### Spectral properties of left eigen-solutions

To solve the left eigenvalue problem is simply to find the eigen-solutions of the adjoint operator $L^\dagger(\lambda)$:

$$L^\dagger(\lambda_n) | \mathbf{w}_n \rangle = (A^\dagger + \overline{\lambda_n} B^\dagger) | \mathbf{w}_n \rangle = 0. \tag{3.2}$$

We are interested in finding the relations between the 'left' (3.2) and 'right' eigen-solutions (3.3).

$$L(\lambda_n) | \mathbf{v}_n \rangle = (A + \lambda_n B) | \mathbf{v}_n \rangle = 0. \tag{3.3}$$

Pay attention that $A^\dagger$ denotes the Hermitian conjugate (or adjoint) of the operator $A$. Only in the case where $A$ can be expressed as a square matrix, $A^\dagger$ becomes the notation of conjugate transpose matrix.

**Lemma 3.1.1** *Given $A, B$ self-adjoint (or symmetric) operators, the left and right eigensolutions of the operator $L(\lambda) = A + \lambda B$ such that:*

$$\langle w_m | L(\lambda_m) = 0 \quad L(\lambda_n) | v_n \rangle = 0$$

*are equivalent: $(\lambda_m, w_m) = (\lambda_n, v_n)$.*

**Proof:** It is trivial given $A^\dagger = A$ and $B^\dagger = B$.[1]

It it easily seen that if $A, B$ are not self-adjoint operators, there is no guarantee that the 'left' eigenvectors equal to the right ones. The question arises as to whether the connection between left and right eigenvalues is still preserved. The answer is yes but only in finite-dimensional spaces. In particular, the relation between the left and right eigenvalues is determined by the diagonalizability of the operators $L(\lambda)$.

Consider a finite-dimensional subspace $V$ of the Hilbert space $H$, (3.2) and (3.3) can be formulated as eigenvalue problems of matrices.[2] In this case, the relation between the 'left' and 'right' eigenvalues is established through the following theorem.

1: We remind that writing $A^\dagger = A$ implies $\langle \mathbf{w}, A\mathbf{v} \rangle = \langle A\mathbf{w}, \mathbf{v} \rangle$ and $D_A = D_{A^\dagger}$

2: One typical example of such technique is the Finite Element Method which allows us to replace the infinite-dimensional linear problem into finite-dimensional, see Chapter 6 for more details.

**Lemma 3.1.2** *Let $V \subset H$ be a finite-dimensional nonzero inner product spaces. Then $\tilde{\lambda}$ is an eigenvalue of $L(\lambda)$ if and only if $\overline{\tilde{\lambda}}$ is an eigenvalue of $L^\dagger(\lambda)$.*

***Proof:*** The theorem is proved by contradiction. First, suppose that $\tilde{\lambda}$ is not an eigenvalue of $L(\lambda)$. This implies that $(A + \lambda B)$ is invertible, and so there exists a linear map $S$ such that

$$S(A + \tilde{\lambda}B) = (A + \tilde{\lambda}B)S = I,$$

where $I$ is the identity matrix. By the adjoint of both sides of the equation above, we get:

$$(A^\dagger + \overline{\tilde{\lambda}}B^\dagger)S^\dagger = S^\dagger(A^\dagger + \overline{\tilde{\lambda}}B^\dagger) = I^\dagger = I$$

The previous equation implies that $L^\dagger(\tilde{\lambda}) = A^\dagger + \overline{\tilde{\lambda}}B^\dagger$ is invertible, which concludes that $\overline{\tilde{\lambda}}$ is not an eigenvalue of $L^\dagger(\lambda)$. Similarly, we can show that if $\overline{\tilde{\lambda}}$ is not an eigenvalue of $L^\dagger(\lambda)$, then $\tilde{\lambda}$ is not an eigenvalue of $L(\lambda)$; which completes the proof.

By Lemma 3.1.2, we see that in finite-dimensional spaces, the left and right eigenvalue problems share the same set of eigenvalues, even though their eigenvectors are different. In infinite dimensional spaces, theorem no longer holds. Indeed, the left and right eigenvalues are only equal to each other under the condition that the operator $L(\lambda)$ is diagonalizable. For the sake of simplicity, without going into details, we assume that all the operators are diagonalizable in the remainder of this thesis.

In order to comprehend the diagonalizability of operators in infinite-dimensional space H, let's consider a finite-dimensional subspace $V$ of $H$, which is much more practical for numerical applications

> **Lemma 3.1.3** *A square $n \times n$ matrix L over a finite-dimensional subspace $V$ is called diagonalizable if there exists an invertible matrix P such that:*
>
> $$L \in V^{n \times n} \text{ diagonalizable } \iff \exists P, P^{-1} \in V^{n \times n} : D = P^{-1}LP \text{ diagonal}$$

The column vectors of $P$ are 'right' eigenvectors $\mathbf{v}_n$ of $L$, corresponding to eigenvalues given by diagonal entries of $D$. As the same time, the row vectors of $P^{-1}$ are the 'left' eigenvectors $\mathbf{w}_n$ of $L$. If the matrix $L$ is Hermitian, $P^{-1} = P^*$, we have a 1-1 correspondence between 'right' eigenvectors, i.e. the basis, and 'left' eigenvectors, i.e. the dual basis.

Theoretically, it is possible to obtain the set of 'left' eigenvectors from the 'right' counterpart by reversing the matrix $P$. Unfortunately, when the size of the matrix $L$ becomes larger, finding the inverse matrix turns out to be a more difficult task than solving the 'left' eigenvalue problem $\langle \mathbf{w}_n | L(\lambda_n) = 0$.

## Bi-orthogonality

Since $A$, $B$ are no longer self-adjoint, we can't use Theorem A.3.2 to prove the orthogonality of the eigenvectors $\mathbf{v}_n$, solutions of the 'right' eigenvalue problem $(A + \lambda_n B)\mathbf{v}_n = 0$. Therefore, the orthogonality property in the case of non-self-adjoint operators is replaced by the bi-orthogonality as follows:

**Lemma 3.1.4** (Bi-orthogonality) *Give two operators* $A, B : H \to H$*, the left and right eigenvectors* $\langle w_m |$ *and* $|v_n\rangle$ *(respectively) of operator* $L(\lambda)$ *corresponding to*

$$\langle w_m | L(\lambda_m) = 0 \quad L(\lambda_n) | v_n\rangle = 0, \tag{3.4}$$

*satisfy the bi-orthogonality relationships:*

$$(\lambda_m - \lambda_n)\langle w_m, B v_n\rangle = 0. \tag{3.5}$$

**Proof:** From $\langle \mathbf{w}_m | L(\lambda_m) = 0$ and $L(\lambda_n) | \mathbf{v}_n\rangle = 0$, it is clear that $\langle \mathbf{w}_m, L(\lambda_m) \mathbf{v}_n\rangle = \langle \mathbf{w}_m, L(\lambda_n) \mathbf{v}_n\rangle = 0$. This implies that $\lambda_m \langle \mathbf{w}_m, B \mathbf{v}_n\rangle = \lambda_n \langle \mathbf{w}_m, B \mathbf{v}_n\rangle$ and completes the proof.

Although the eigenvectors $\mathbf{v}_n$ are not orthogonal, by Lemma 3.1.4, they can be proved nevertheless linearly independent.

**Proposition 3.1.5** *If the 'right' eigenvectors* $v_n$ *of* $L(\lambda)$ *are nondegenerate, they are linearly independent, and thus span the Hilbert space* $H$ *such that:*

$$u = \sum_{n=1}^{\infty} \langle w_n, u\rangle v_n. \tag{3.6}$$

*for* $\forall u \in H$*.*

**Proof by contradiction:** Let's assume that the set $\{\mathbf{v}_n\}$ are linearly dependent. Then there exists a set of numbers $c_n$ such that $\sum_n |c_n| \neq 0$ and that

$$\sum_n c_n \mathbf{v}_n = 0.$$

Then, we see that, for each $m$, $\sum_n c_n \langle \mathbf{w}_m, B \mathbf{v}_n\rangle = 0$. By Lemma 3.1.4, we have $\langle \mathbf{w}_m, B \mathbf{v}_n\rangle \neq 0$ only if $m \neq n$, which implies $c_n \langle \mathbf{w}_m, B \mathbf{v}_n\rangle = 0$. It follows that $c_m = 0$ for any $m$ we choose, contradicting the hypothesis. This concludes that the nondegenerate eigenvectors $\mathbf{v}_n$ span the Hilbert space $H$.

## 3.2 Spectral properties of non-self-adjoint linear operator

Now, we have enough tools to derive the quasinormal modal expansion for non-self-adjoint linear operator.

**Lemma 3.2.1** *Given the operator* $L(\lambda) = A + \lambda B$*, the solution* $u$ *of the boundary value problem* $L(\lambda)u = S$ *can be expressed as quasinormal modal expansion based on the set of right eigenvectors* $\{v_n\}$*, solutions of the 'right' eigenvalue problem* $L(\lambda_n)|v_n\rangle = (A + \lambda_n B)|v_n\rangle = 0$*:*

$$u = \sum_{n=1}^{\infty} \frac{1}{\lambda - \lambda_n} \frac{\langle w_n, S\rangle}{\langle w_n, B v_n\rangle} v_n, \tag{3.7}$$

Similar to Theorem 2.1.1 the expression (3.7) fails if $\lambda = \lambda_n$. Fortunately, the situation where $\lambda = \lambda_n$ rarely happens in practice. In the case of electromagnetics, the resonant frequencies, i.e. eigenvalues $\lambda_n$, are complex; while the value of $\lambda$, as a physical quantity, must be real. Therefore, we don't have to worry about the singularity where $\lambda = \lambda_n$.

$$\langle w_m | L(\lambda_m) = \langle w_m | (A + \lambda_m B) = 0. \tag{3.8}$$

**Proof:** By Proposition 3.1.5 and the assumption that $\{\mathbf{v}_n\}$ are non-degenerate, we can expand the solution $\mathbf{u}$ onto the set of 'right' eigenvector $\{\mathbf{v}_n\}$ as follows:

$$L(\lambda)\mathbf{u} = L(\lambda) \sum_{n=1}^{\infty} \langle \mathbf{w}_n, \mathbf{u} \rangle \mathbf{v}_n = \sum_{n=1}^{\infty} \langle \mathbf{w}_n, \mathbf{u} \rangle (A + \lambda B)\mathbf{v}_n = \mathbf{S}.$$

Since $\mathbf{v}_n$ are eigenfunctions of the operator $L(\lambda_n)\mathbf{v}_n = (A + \lambda_n B)\mathbf{v}_n = 0$, we can replace the term $A\mathbf{v}_n$ in the previous equation by $(\lambda_n B)\mathbf{v}_n$, which yields:

$$\sum_{n=1}^{\infty} \langle \mathbf{w}_n, \mathbf{u} \rangle (-\lambda_n + \lambda)B\mathbf{v}_n = \mathbf{S}$$

By taking the inner product of both sides with $\mathbf{w}_m$ and exploiting the bi-orthogonality from Lemma 3.1.4 $(\lambda_m - \lambda_n)\langle \mathbf{w}_m, B\mathbf{v}_n \rangle = 0$ lead to:

$$(\lambda - \lambda_n)\langle \mathbf{w}_n, \mathbf{u} \rangle \langle \mathbf{w}_n, B\mathbf{v}_n \rangle = \langle \mathbf{w}_n, \mathbf{S} \rangle$$

This results in

$$\langle \mathbf{w}_n, \mathbf{u} \rangle = \frac{1}{\lambda - \lambda_n} \frac{\langle \mathbf{w}_n, \mathbf{S} \rangle}{\langle \mathbf{w}_n, B\mathbf{v}_n \rangle}$$

or

$$\mathbf{u} = \sum_{n=1}^{\infty} \frac{1}{\lambda - \lambda_n} \frac{\langle \mathbf{w}_n, \mathbf{S} \rangle}{\langle \mathbf{w}_n, B\mathbf{v}_n \rangle} \mathbf{v}_n$$

## 3.3 Breaking the self-adjointness

Before proceeding further with Lemma 3.2.1, it is important to figure out how exactly the self-adjointness no longer hold for certain cases in electromagnetics. In order to see the issue, let's go back to the second order electric wave equation (see Section 2.3) with the source $\mathbf{S} = -i\omega \mathbf{J}$:

$$L(\lambda)\mathbf{u} = (A + \lambda B)\mathbf{u} = \mathbf{S}, \tag{3.9}$$

where we set

$$A = -\nabla \times \left( \mu^{-1}\nabla \times \cdot \right) \quad \text{and} \quad B = \varepsilon$$

and the domain

▶ with homogeneous Dirichlet boundary condition:

$$D_L^D \equiv \left\{ \mathbf{u} : \mathbf{u}, -\nabla \times (\mu^{-1}\nabla \times \mathbf{u}) \in L^2(\Omega)^3, \mathbf{n} \times \mathbf{u}|_\Gamma = 0 \right\},$$

▶ with homogeneous Neumannn boundary condition:

$$D_L^N \equiv \left\{ \mathbf{u} : \mathbf{u}, -\nabla \times (\mu^{-1}\nabla \times \mathbf{u}) \in L^2(\Omega)^3, \mathbf{n} \times (\mu^{-1}\nabla \times \mathbf{u})|_\Gamma = 0 \right\}.$$

We remind that $L(\lambda)$ is a linear operator with respect to $\lambda$ where $\lambda = \omega^2$.

The next step is to compute explicitly the adjoint operator $L^\dagger(\lambda)$. As in the previous chapter, we try to use integration by parts. Given two vectors

$\mathbf{x} \in D_L^D$ or $D_L^N$ and $\mathbf{y} \in H$ we have:

$$\langle \mathbf{y}, L(\lambda)\mathbf{x} \rangle = \int_\Omega \overline{\mathbf{y}} \cdot \left[ \nabla \times \left( \boldsymbol{\mu}^{-1} \nabla \times \mathbf{x} \right) + \omega^2 \boldsymbol{\varepsilon} \mathbf{x} \right] \, d\Omega$$

$$= \int_\Omega \left[ \nabla \times \left( \boldsymbol{\mu}^{-\mathsf{T}} \nabla \times \overline{\mathbf{y}} \right) + \omega^2 \boldsymbol{\varepsilon}^{\mathsf{T}} \overline{\mathbf{y}} \right] \cdot \mathbf{x} \, d\Omega + \text{b.t.}$$

$$= \underbrace{\int_\Omega \overline{\left[ \nabla \times \left( \boldsymbol{\mu}^{-*} \nabla \times \mathbf{y} \right) + \omega^2 \boldsymbol{\varepsilon}^* \mathbf{y} \right]} \cdot \mathbf{x} \, d\Omega}_{\langle L^\dagger(\lambda)\mathbf{y}, \mathbf{x} \rangle} + \text{b.t.}$$

where the boundary term

$$\text{b.t.} = \int_\Gamma \left[ (\boldsymbol{\mu}^{-\mathsf{T}} \nabla \times \overline{\mathbf{y}}) \cdot (\mathbf{n} \times \mathbf{x}) - (\mathbf{n} \times \overline{\mathbf{y}}) \cdot (\boldsymbol{\mu}^{-1} \nabla \times \mathbf{x}) \right] \, dS$$

is obtained using $\int_\Omega \overline{\mathbf{y}} \cdot (\nabla \times \mathbf{x}) \, d\Omega = \int_\Omega (\nabla \times \overline{\mathbf{y}}) \cdot \mathbf{x} \, d\Omega - \int_{\partial \Omega} (\overline{\mathbf{y}} \times \mathbf{x}) \cdot \mathbf{n} \, dS$
and $\overline{\mathbf{y}} \cdot (\boldsymbol{\varepsilon} \mathbf{x}) = (\boldsymbol{\varepsilon}^{\mathsf{T}} \overline{\mathbf{y}}) \cdot \mathbf{x}$.

The domain $D_L$ is chosen such that the boundary term b.t., i.e. the conjunct,[3] vanishes:

▶ The homogeneous Dirichlet boundary condition

$$D_{L^\dagger}^D \equiv \{ \mathbf{u} : \mathbf{u}, -\nabla \times (\boldsymbol{\mu}^{-1} \nabla \times \mathbf{u}) \in L^2(\Omega)^6, n \times \mathbf{u}|_\Gamma = 0 \}$$

▶ The homogeneous Neumann boundary condition

$$D_{L^\dagger}^N \equiv \{ \mathbf{u} : \mathbf{u}, -\nabla \times (\boldsymbol{\mu}^{-1} \nabla \times \mathbf{u}) \in L^2(\Omega)^6, n \times (\boldsymbol{\mu}^{-1} \nabla \times \mathbf{u})|_\Gamma = 0 \}$$

There are 2 cases where the operators $A$, $B$ does not satisfy the condition of self-adjointness:

▶ The matrices $\boldsymbol{\varepsilon}$ and $\boldsymbol{\mu}$ are not Hermitian.
▶ The domain of $D_L$ is modified such that $D_L \neq D_{L^\dagger}$, which results from the change of boundary conditions.

Firstly, let's take a look at the matrices $\boldsymbol{\varepsilon}$ and $\boldsymbol{\mu}$. In this chapter, we exclude that case where $\boldsymbol{\varepsilon}$ and $\boldsymbol{\mu}$ depend on frequency $\omega$, which is associated to dispersive materials. If the media have dissipative characteristic,[4] the energy of the system is absorbed and not conserved, which break the self-adjointness characteristic. In addition, the imaginary part of value of $\boldsymbol{\varepsilon}$ and $\boldsymbol{\mu}$ could also modify boundary conditions in the domain of operators.

Secondly, we will point out in next subsection, that there exists some specific boundary conditions cause $D_L \neq D_{L^\dagger}$, and, in turn, break the self-adjoint-ness.

For the sake of clarity, the superscript $^{-\mathsf{T}}$ stands for the transposition of the inverse matrix. The superscript $*$ represents the Hermitian conjugate of matrix $\boldsymbol{\varepsilon}^* = \overline{\boldsymbol{\varepsilon}^{\mathsf{T}}}$, while $*$ denotes the Hermitian conjugate of the inverse matrix. We have to be careful distinguishing the concept of adjoint tensor (conjugate transpose of matrix) and adjoint operator.

3: For more details about the conjunct and adjoint operator, we refer the reader to appendix A and [42].

4: In reality, due to the property of causality, the tensor of permittivity and permeability must be Hermitian symmetry $\overline{\boldsymbol{\varepsilon}}(i\omega) = \boldsymbol{\varepsilon}(\overline{i\omega})$ and $\overline{\boldsymbol{\mu}}(i\omega) = \boldsymbol{\mu}(\overline{i\omega})$. Therefore, the imaginary part of $\boldsymbol{\varepsilon}$ and $\boldsymbol{\mu}$ can't exist without the presence of frequency $\omega$. Roughly speaking, the media can't be dissipative without being time-dispersive. The situation of time-dispersion (frequency-dependent characteristic of materials) will be explored in Chapter 4, Chapter 5. Meanwhile, in this chapter, we only tackle the dissipation (non-self-adjointness) of the operator without the presence of time-dispersion, by assuming the tensors $\boldsymbol{\varepsilon}$ and $\boldsymbol{\mu}$ are complex and independent of frequency.

## Open (unbounded) space

The problem arises when we try to extend our domain of interest $\Omega$ into open space, and impose the outgoing wave condition on the boundary:

$$D_L^O \equiv \left\{ \mathbf{u} : \mathbf{u}, -\nabla \times (\boldsymbol{\mu}^{-1} \nabla \times \mathbf{u}) \in L^2(\Omega)^6, \right.$$
$$\left. \lim_{r \to \infty} r \left( \boldsymbol{\mu}^{-1/2} \nabla \times \mathbf{u} - i\omega \hat{\mathbf{r}} \boldsymbol{\varepsilon}^{1/2} \mathbf{u} \right) = 0 \right\}. \quad (3.10)$$

It is worth noting that when $r \to \infty$, the media are usually isotropic and homogeneous (for example air). Thus, the permittivity and permeability at infinity in 3.10 can be expressed as scalar quantities: $\varepsilon$ and $\mu$ respectively.

We may notice that in this situation, the integral $\int_\Omega$ in the definition of inner product $\langle \cdot, \cdot \rangle$ becomes improper and must be taken over the whole open space. (see Figure 3.1)

In this case, even if the media are lossless, the operator $A$ is not self-adjoint due to the imposition of the outgoing radiation condition. It is easily seen that $D_L^O \neq D_{L^\dagger}^O$. Indeed, the domain $D_{L^\dagger}^O$ of the adjoint operator imposes the incoming wave condition of the function:

$$D_{L^\dagger}^O = \left\{ \mathbf{u} : \mathbf{u}, -\nabla \times (\boldsymbol{\mu}^{-1} \nabla \times \mathbf{u}) \in L^2(\Omega)^6, \right.$$
$$\left. \lim_{r \to \infty} r \left( \boldsymbol{\mu}^{-1/2} \nabla \times \mathbf{u} + i\omega \boldsymbol{\varepsilon}^{1/2} \mathbf{u} \right) = 0 \right\} \equiv D_L^I, \quad (3.11)$$

## Fourier expansion for continuous spectrum

In fact, the open space not only breaks the self-adjoint-ness but also creates the issue of continuous spectrum. In particular, by extending the boundary of our domain of interest $\Omega$ to infinite $r \to \infty$, improper eigen-solutions appear.

***Example:*** Consider the operator $A(\lambda) = \dfrac{d^2}{dx^2} + \lambda$ with the domain $D_A = \{ u : u \in L^2(0, \infty), u(0) = 0 \}$. It is possible to point out a solution $u$ such that:

$$\frac{d^2}{dx^2} \sin(\omega x) + \omega^2 \sin(\omega x) = 0. \quad (3.12)$$

Unfortunately, $u = \sin(\omega x)$ is not a proper eigenvector of the operator $A(\lambda)$ because $\sin(\omega x) \notin L^2(0, \infty)$:

$$\int_0^\infty \sin^2(\omega x)\, dx = \int_0^{2\pi/\omega} \sin^2(\omega x)\, dx + \sum_{n=1}^\infty \int_{2\pi(n-1)/\omega}^{2\pi(n+1)/\omega} \sin^2(\omega x)\, dx \to \infty$$

Hence, the eigenvector $u = \sin(\omega x)$ and eigenvalue $\lambda = \omega$ are improper. To be more precise, there is no proper eigenvector for the operator $A(\lambda) = \dfrac{d^2}{dx^2} + \lambda$ with the given condition $D_A$.

As showed in the previous example and discussed in appendix A, proper eigenfunctions (for point spectrum) may fail to exist. Therefore, functions of interest has to be expanded in the continuous spectrum of improper eigen-solutions. In the case of electromagnetics, these improper eigen-solutions belong to the resolvent set of the operator, which from a physical perspective, are not directly related to physical resonances of our optical structure. However, mathematically, they still play an important role in spectral expansions and must be involved into our Fourier expansion.

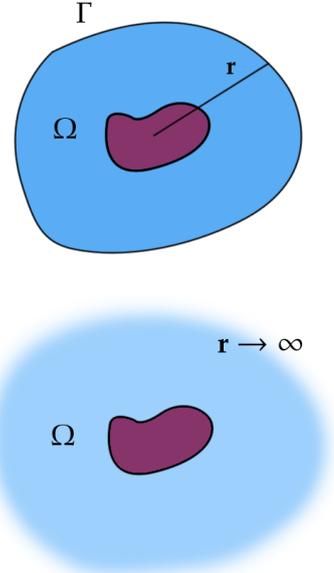

**Figure 3.1:** Open space (down) versus closed space (up).

In general, the 'left' and 'right' improper eigenvalue problem for operator $L(\lambda) = A + \lambda B$ can be expressed as follows:

$$\langle \mathbf{w}_\nu | L(\lambda_\nu) = 0 \qquad L(\lambda_\nu) | \mathbf{v}_\nu \rangle = 0 \tag{3.13}$$

The bi-orthogonality will also be extended to the continuous spectrum:

$$(\lambda_\nu - \lambda_\nu)\langle \mathbf{w}_\nu, B\mathbf{v}_\nu \rangle = 0$$
$$(\lambda_\nu - \lambda_n)\langle \mathbf{w}_\nu, B\mathbf{v}_n \rangle = 0$$
$$(\lambda_n - \lambda_\nu)\langle \mathbf{w}_n, B\mathbf{v}_\nu \rangle = 0,$$

where the discrete proper eigen-triplet $(\lambda_n, \langle \mathbf{w}_n |, |\mathbf{v}_n \rangle)$ denoted with Latin index stands for proper eigen-solutions. This distinguishes from the continuous improper spectrum using Greek index.

We follow [2, 5] in assuming that the spectrum of Maxwell operators in open uniform backgrounds is complete. Then, the expansion for **u** in an open space $\Omega$ can be formalized as follows:

$$\mathbf{u} = \sum_n \langle \mathbf{w}_n, \mathbf{u} \rangle \mathbf{v}_n + \int_\nu \langle \mathbf{w}_\nu, \mathbf{u} \rangle \mathbf{v}_\nu$$

Finally, following the construction in Section 3.2 and Lemma 3.2.1, we can obtain the spectral expansion for non-self-adjoint linear operator with continuous spectrum (see [42] for more details).

**Lemma 3.3.1** *Given the non-self-adjoint operator* $L(\lambda) = A + \lambda B$, *the solution* $\boldsymbol{u}$ *of the boundary value problem* $L(\lambda)\boldsymbol{u} = \boldsymbol{S}$ *can be expressed as quasinormal modal expansion based on the set of 'right' proper* $\{\boldsymbol{v}_n\}$ *and improper eigenvectors* $\{\boldsymbol{v}_\nu\}$, *solutions of the 'right' eigenvalue problems* $L(\lambda_n)|\boldsymbol{v}_n\rangle = 0$ *and* $L(\lambda_\nu)|\boldsymbol{v}_\nu\rangle = 0$ :

$$\boldsymbol{u} = \sum_{n=1}^\infty \frac{1}{\lambda - \lambda_n} \frac{\langle \boldsymbol{w}_n, \boldsymbol{S} \rangle}{\langle \boldsymbol{w}_n, B\boldsymbol{v}_n \rangle} \boldsymbol{v}_n + \int_\nu \frac{1}{\lambda - \lambda_n} \frac{\langle \boldsymbol{w}_\nu, \boldsymbol{S} \rangle}{\langle \boldsymbol{w}_\nu, B\boldsymbol{v}_n \rangle} \boldsymbol{v}_\nu, \tag{3.14}$$

*where* $w_n$ *and* $w_\nu$ *are eigensolutions of the 'left' proper and improper eigenvalue problems respectively:*

$$\langle \boldsymbol{w}_n | L(\lambda_n) = 0 \qquad \langle \boldsymbol{w}_\nu | L(\lambda_\nu) = 0. \tag{3.15}$$

Despite the existence of continuous spectrum, in practice, Lemma 3.3.1 is rarely used. The reason is that, for the sake of computation, it is more pragmatic to truncate infinite Hilbert space into finite dimension, and limit the computational domain.[5] This transforms the integral $\int_\nu$ in (3.14) into a discrete finite sum. Since, the topic of analytically continuous spectrum exceeds the scope of this paper,[6] henceforth we restrict our attention on the discrete spectrum.

5: The procedure of truncation the domain of interest $\Omega$ by using Perfectly Matched Layer (PML) in Finite Element will be discussed in Chapter 7.

6: We are interested in studying the 'proper' resonant state of optical systems and numerical applications of spectral theory.

## 3.4 Reciprocal media

In this section, we will study the application of Lemma 3.2.1 in electromagnetics. For non-self-adjoint linear operators, the quasinormal modal expansion requires to compute not only the 'right' eigen-solutions but also the 'left' eigenvectors (solving the adjoint eigenvalue problem). Fortunately, for most cases in electromagnetics, there is a connection between the 'left' and 'right' eigen-fields, which implies we only have to solve the spectral problem once to obtain the entire set of eigen-solutions. In particular, we will focus our study on reciprocal materials. From a mathematical point of view, this is equivalent to say that the permittivity and permeability matrices are symmetric:

$$\boldsymbol{\varepsilon}^{\mathsf{T}} = \boldsymbol{\varepsilon} \qquad \boldsymbol{\mu}^{\mathsf{T}} = \boldsymbol{\mu}$$

> **Lemma 3.4.1** *Given the operator $L(\omega^2) = -\nabla \times \left(\boldsymbol{\mu}^{-1}\nabla \times \cdot\right) + \omega^2 \boldsymbol{\varepsilon}$ with Dirichlet, Neumann or outgoing wave boundary condition, the 'left' and 'right' eigenvectors*
>
> $$\langle w_n | L(\omega_n^2) = 0 \qquad L(\omega_n^2) | v_n \rangle = 0 \qquad (3.16)$$
>
> *are the complex conjugate of each other if the materials are reciprocal.*

We come back to the electric wave equation with the boundary conditions (3.9) in Section 3.3:

$$L(\lambda)\mathbf{u} = (A + \lambda B)\mathbf{u} = \mathbf{S}$$

with $A \equiv -\nabla \times \left(\boldsymbol{\mu}^{-1}\nabla \times \cdot\right)$, $B \equiv \boldsymbol{\varepsilon}$, $\mathbf{S} = -i\omega \mathbf{J}$ and $\lambda = \omega^2$.

The 'left' eigen-vectors are achieved by solving the adjoint eigenvalue problem:

$$L^{\dagger}(\lambda_n)\mathbf{w}_n = -\nabla \times \left(\boldsymbol{\mu}^{-*}\nabla \times \mathbf{w}_n\right) + \overline{\lambda_n}\boldsymbol{\varepsilon}^* \mathbf{w}_n = 0. \qquad (3.17)$$

If we assume that the permittivity and permeability are represented by symmetric tensors $\boldsymbol{\varepsilon}^{\mathsf{T}} = \boldsymbol{\varepsilon}$ and $\boldsymbol{\mu}^{\mathsf{T}} = \boldsymbol{\mu}$, we have:

$$\overline{L^{\dagger}(\lambda_n)\mathbf{w}_n} = -\nabla \times \left(\boldsymbol{\mu}^{-1}\nabla \times \overline{\mathbf{w}_n}\right) + \lambda_n \boldsymbol{\varepsilon}\overline{\mathbf{w}_n} = 0. \qquad (3.18)$$

which implies $\mathbf{v}_n = \overline{\mathbf{w}_n}$.

We notice again that the concept of symmetric tensor is quite different from symmetric operator.

It is tempted to claim that the 'left' eigenvectors $\mathbf{w}_n$ are simply the complex conjugate of their 'right' counterparts $\mathbf{v}_n$, but we have to be careful. In order to for the previous property to be true, we must verify if $\overline{\mathbf{w}_n} \in D_L$. This process is done by taking the complex conjugate of $D_{L^{\dagger}}$ and then comparing with $D_L$.

Pay attention that $\overline{\mathbf{w}_n} \in D_L$ is not relevant to the self-adjointness charateristic of operator.

For the cases of homogeneous Dirichlet boundary and homogeneous Neumann boundary conditions, the process is straightforward:

▶ Dirichlet

$$\overline{\mathbf{w}_n} \in \left\{\overline{\mathbf{u}} : \overline{\mathbf{u}}, -\nabla \times \left(\boldsymbol{\mu}^{-1}\nabla \times \overline{\mathbf{u}}\right) \in L^2(\Omega)^6, n \times \overline{\mathbf{u}}|_{\Gamma} = 0\right\} = D_L^D,$$

▶ Neumann

$$\overline{\mathbf{w}_n} \in \left\{ \overline{\mathbf{u}} : \overline{\mathbf{u}}, -\nabla \times (\boldsymbol{\mu}^{-1}\nabla \times \overline{\mathbf{u}}) \in L^2(\Omega)^6, \, n \times (\boldsymbol{\mu}^{-1}\nabla \times \overline{\mathbf{u}}) \,|_\Gamma = 0 \right\} = D_L^N.$$

Then, by Lemma 3.2.1, we can derive the quasinormal modal expansion for electric fields from the operator $L(\omega^2) = -\nabla \times (\boldsymbol{\mu}^{-1}\nabla \times \cdot) + \omega^2 \boldsymbol{\varepsilon}$ as follows:

$$\mathbf{E} = \sum_{n=1}^{\infty} \frac{-i\omega}{\omega^2 - \omega_n^2} \frac{\int_\Omega \overline{\mathbf{E}_{ln}} \cdot \mathbf{J} \, d\Omega}{\int_\Omega \overline{\mathbf{E}_{ln}} \cdot (\boldsymbol{\varepsilon}\mathbf{E}_{rn}) \, d\Omega} \mathbf{E}_{rn}. \tag{3.19}$$

The final step is to use Lemma 3.4.1 to simplify (3.19) to obtain:

$$\mathbf{E} = \sum_{n=1}^{\infty} \frac{-i\omega}{\omega^2 - \omega_n^2} \frac{\int_\Omega \mathbf{E}_{rn} \cdot \mathbf{J} \, d\Omega}{\int_\Omega \mathbf{E}_{rn} \cdot (\boldsymbol{\varepsilon}\mathbf{E}_{rn}) \, d\Omega} \mathbf{E}_{rn}. \tag{3.20}$$

*Expansion of electric fields for symmetric electromagnetic matrices 1*

## Bloch-Floquet quasi-periodicity conditions

In this subsection, we will show a small example proving that it is not an easy task to enforce the complex conjugate relation between the 'left' and 'right' eigenvectors, even if the matrix operator is symmetric: Bloch-Floquet quasiperiodicity conditions.

The situation occurs when we have a periodic structure whose boundaries $\Gamma$ contains with two parallel lines denoted by $\Gamma_l$, $\Gamma_r$ separated by a width $d$ (see Figure 3.2), and factorized by the quasi-periodicity real coefficient $\mathbf{k} \in \mathbb{R}^3$. By definition, both the 'left' and 'right' eigenvectors must be imposed by Bloch wave:

$$\mathbf{w}_n^{\mathbf{k}}(\mathbf{r}) = \mathbf{w}_{\#n}^{\mathbf{k}}(\mathbf{r}) \exp(i\mathbf{k} \cdot \mathbf{r}) \qquad \mathbf{v}_n^{\mathbf{k}}(\mathbf{r}) = \mathbf{v}_{\#n}^{\mathbf{k}}(\mathbf{r}) \exp(i\mathbf{k} \cdot \mathbf{r}) \tag{3.21}$$

where $\mathbf{w}_{\#n}^{\mathbf{k}}$ and $\mathbf{v}_{\#n}^{\mathbf{k}}$ are $d$-periodic functions along $\mathbf{k}$.

Consider the operator $L(\omega^2) = -\nabla \times (\boldsymbol{\mu}^{-1}\nabla \times \cdot) + \omega^2 \boldsymbol{\varepsilon}$ with the domain:

$$\mathbf{u} \in D_L^P \equiv \left\{ \mathbf{u} : \mathbf{u}, -\nabla \times (\boldsymbol{\mu}^{-1}\nabla \times \overline{\mathbf{u}}) \in L^2(\Omega)^6, \mathbf{u}^{\mathbf{k}}(\mathbf{r}) = \mathbf{u}_\#^{\mathbf{k}}(\mathbf{r}) \exp(i\mathbf{k} \cdot \mathbf{r}) \right\}.$$

where $\mathbf{u}_\#$ is a d-periodic function.

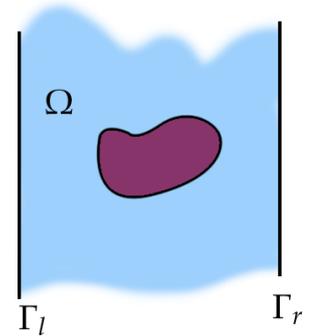

**Figure 3.2:** A periodic structure

Then, it is clear that $\overline{\mathbf{w}_n^{\mathbf{k}}(\mathbf{r})} = \overline{\mathbf{w}_{\#n}^{\mathbf{k}}(\mathbf{r})} \exp(-i\mathbf{k} \cdot \mathbf{r}) \notin D_L^P$, which implies that Lemma 3.4.1 fails to hold in this case. In particular, the equality $\mathbf{w}_n^{\mathbf{k}} = \overline{\mathbf{v}_n^{\mathbf{k}}}$ only holds when the dephasing term $\exp(i\mathbf{k} \cdot \mathbf{r})$ vanishes, i.e. $\mathbf{k} = 0$.

When the quasi-periodicity coefficient $\mathbf{k} \neq 0$, we have $\mathbf{w}_n^{\mathbf{k}} \neq \overline{\mathbf{v}_n^{\mathbf{k}}}$. But this doesn't mean that there is no connection between the 'left' and 'right' quasi-periodic eigenvectors. On the other hand, it is possible to show that:

**Remark 3.4.1** Under the Bloch condition, the 'left' eigenvectors are the complex conjugate of their 'right' counterpart with negative quasi-periodicity coefficient $\mathbf{k}$.

$$\mathbf{w}_n^{\mathbf{k}} = \overline{\mathbf{v}_n^{-\mathbf{k}}}$$

***Proof:*** [7] If $\mathbf{v}^{\mathbf{k}} = \mathbf{v}_{\#}^{\mathbf{k}} \exp(i\mathbf{k}\cdot\mathbf{r})$ is the 'right' eigenvector of $L(\lambda)$ as follows:

$$L(\lambda)\mathbf{v}^{\mathbf{k}} = -\nabla\times\left(\boldsymbol{\mu}^{-1}\nabla\times\mathbf{u}^{\mathbf{k}}\right) + \lambda\boldsymbol{\varepsilon}\mathbf{v}^{\mathbf{k}} = 0,$$

then, $\mathbf{v}_{\#}^{\mathbf{k}}$ will satisfy the following equation:

$$L_{\#}^{\mathbf{k}}(\lambda)\mathbf{v}_{\#}^{\mathbf{k}} \equiv -(i\mathbf{k}+\nabla)\times\left(\boldsymbol{\mu}^{-1}(i\mathbf{k}+\nabla)\times\mathbf{v}_{\#}^{\mathbf{k}}\right) + \lambda\boldsymbol{\varepsilon}\mathbf{v}_{\#}^{\mathbf{k}} = 0.$$

The previous equation can be achieved by simply replacing the operator $\nabla\times$ by $(i\mathbf{k}+\nabla)\times$ when going from $\mathbf{v}^{\mathbf{k}}$ to $\mathbf{v}_{\#}^{\mathbf{k}}$. It is true using the fact that:

$$\begin{aligned}
\nabla\times\mathbf{v}^{\mathbf{k}} &= \nabla\times\left(\mathbf{v}_{\#}^{\mathbf{k}}\exp(i\mathbf{k}\cdot\mathbf{r})\right) \\
&= (i\mathbf{k}\times\mathbf{v}_{\#}^{\mathbf{k}})\exp(i\mathbf{k}\cdot\mathbf{r}) + (\nabla\times\mathbf{v}_{\#}^{\mathbf{k}})\exp(i\mathbf{k}\cdot\mathbf{r}) \\
&= \left((i\mathbf{k}+\nabla)\times\mathbf{v}_{\#}^{\mathbf{k}}\right)\exp(i\mathbf{k}\cdot\mathbf{r})
\end{aligned}$$

Next, let us consider the similar problem with Floquet-Bloch boundary conditions but this time with opposite quasi-periodicity coefficient $-\mathbf{k}$. In this case, the 'right' eigenvector is expressed as: $\mathbf{v}^{-\mathbf{k}} = \mathbf{v}_{\#}^{-\mathbf{k}}\exp(-i\mathbf{k}\cdot\mathbf{r})$. Therefore, under the same argument in the previous case and replacing $\nabla\times$ by $(-i\mathbf{k}+\nabla)\times$, $\mathbf{v}_{\#}^{-\mathbf{k}}$ must satisfy the following equation:

$$L_{\#}^{-\mathbf{k}}(\lambda)\mathbf{v}_{\#}^{-\mathbf{k}} \equiv -(-i\mathbf{k}+\nabla)\times\left(\boldsymbol{\mu}^{-1}(-i\mathbf{k}+\nabla)\times\mathbf{v}_{\#}^{-\mathbf{k}}\right) + \lambda\boldsymbol{\varepsilon}\mathbf{v}_{\#}^{-\mathbf{k}} = 0.$$

The next step is to show that, under appropriate boundary conditions, the complex conjugate of the operator $L_{\#}^{-\mathbf{k}}(\lambda)$ is in fact the adjoint operator of $L_{\#}^{\mathbf{k}}(\lambda)$. Indeed, given two vectors $\mathbf{x}, \mathbf{y} \in H$, it is possible to prove:[8]

$$\langle\mathbf{y}, L_{\#}^{\mathbf{k}}(\lambda)\mathbf{x}\rangle = \langle\overline{L_{\#}^{-\mathbf{k}}(\lambda)}\mathbf{y}, \mathbf{x}\rangle + \text{b.t.}$$

The boundary term b.t. in the previous equation is given as follows:

$$\text{b.t.} = \int_{\Gamma}(\mathbf{n}\times\overline{\mathbf{y}})\cdot\left(\boldsymbol{\mu}^{-1}(i\mathbf{k}+\nabla)\times\mathbf{x}\right) - \left(\boldsymbol{\mu}^{-\mathsf{T}}(i\mathbf{k}\times\overline{\mathbf{y}})\right)\cdot(\mathbf{n}\times\mathbf{x})dS$$

In particular, the boundary term $\Gamma$ except $\Gamma_l$ and $\Gamma_r$ is considered to vanish due to appropriate boundary conditions (For example, Dirichlet/Neumann or even outgoing wave conditions). The problem remains on the contribution of the surface integrals on $\Gamma_l$ and $\Gamma_r$. We notice that the normal vector $\mathbf{n}$ indeed has opposite directions along $\Gamma_l$ and $\Gamma_r$. As a consequence, the integrals of $\Gamma_l$ and $\Gamma_r$ have the same absolute values but are opposite in signs. This makes the whole boundary term vanish.

As a result, we can conclude that the operator $L_{\#}^{\mathbf{k}}(\lambda)$ and $L_{\#}^{-\mathbf{k}}(\lambda)$ share the same set of eigenvalues $\lambda_n$; and the 'left' eigenvector $\mathbf{w}_{\#}^{\mathbf{k}}$ of $L_{\#}^{\mathbf{k}}(\lambda)$ must satisfy the following equation:

$$\mathbf{w}_{\#}^{\mathbf{k}} = \overline{\mathbf{v}_{\#}^{-\mathbf{k}}}, \tag{3.22}$$

which implies that $\mathbf{w}^{\mathbf{k}} = \overline{\mathbf{v}^{-\mathbf{k}}}$ and completes our proof.

7: For the sake of simplicity, in this proof, we will drop the index $n$ and the space variable $\mathbf{r}$: $\mathbf{v}_n^{\mathbf{k}}(\mathbf{r}) = \mathbf{v}^{\mathbf{k}}$.

8: using $\mathbf{y}\cdot(i\mathbf{k}\times\mathbf{x}) = (-i\mathbf{k}\times\mathbf{y})\cdot\mathbf{x}$ and $\mathbf{y}\cdot(\nabla\times\mathbf{x}) = (\nabla\times\mathbf{y})\cdot\mathbf{x} + \nabla\cdot(\mathbf{y}\times\mathbf{x})$.

## Lagrangian 'normalization'

Previously, we tried to apply quasinormal modal expansion on the operator $L(\omega^2)\mathbf{u} = -\nabla \times (\boldsymbol{\mu}^{-1}\nabla \times \mathbf{u}) + \omega^2 \boldsymbol{\varepsilon}\mathbf{u}$, which is a quadratic operator with respect to $\omega$. It is of interest to verify what would happen if we build a system of first order operator with respect to $\omega$.

In particular, we reconsider Maxwell's equations applied to a finite domain $\Omega$ bounded by a perfectly conducting, closed surface $\Gamma$:

$$\nabla \times \mathbf{H}(\mathbf{r}) + i\omega\boldsymbol{\varepsilon}\mathbf{E}(\mathbf{r}) = \mathbf{J}$$
$$\nabla \times \mathbf{E}(\mathbf{r}) - i\omega\boldsymbol{\mu}\mathbf{H}(\mathbf{r}) = 0 \qquad (3.23)$$
$$n \times \mathbf{E}|_\Gamma = 0.$$

Our task now is to construct the operator matrices[9] $A, B$ such that $L(\lambda)\mathbf{u} = (A + \lambda B)\mathbf{u} = \mathbf{S}$. Unlike Section 2.2, there is no requirement that the matrices $A, B$ need to be Hermitian [10] in order for the operators to be self-adjoint. Thus, there are unlimited choices to combine electric and magnetic fields $E$ and $H$ respectively. For example, we can set:

$$A = i \begin{pmatrix} 0 & -\nabla\times \\ \nabla\times & 0 \end{pmatrix} \quad \text{and} \quad B = \begin{pmatrix} \boldsymbol{\varepsilon} & 0 \\ 0 & \boldsymbol{\mu} \end{pmatrix}, \qquad (3.24)$$

or

$$A = \begin{pmatrix} 0 & -i\boldsymbol{\varepsilon}^{-1}\nabla\times \\ i\boldsymbol{\mu}^{-1}\nabla\times & 0 \end{pmatrix} \quad \text{and} \quad B = \begin{pmatrix} 1 & 0 \\ 0 & 1 \end{pmatrix},$$

as what we did in Section 2.2.

The problem is that solving the adjoint eigenvalue problem

$$L^\dagger(\lambda_n)\mathbf{w}_n = (A^\dagger + \overline{\lambda_n}B^\dagger)\mathbf{w}_n = 0$$

will be a complicated task. Therefore, it would be better to take advantage of the reciprocity of materials, and construct symmetric matrices $A, B$. Indeed, it is reasonable to choose the operator $L(\lambda) = A + \lambda B$ as follows:

$$A = -i \begin{pmatrix} 0 & \nabla\times \\ \nabla\times & 0 \end{pmatrix} \quad \text{and} \quad B = \begin{pmatrix} \boldsymbol{\varepsilon} & 0 \\ 0 & -\boldsymbol{\mu} \end{pmatrix}, \qquad (3.25)$$

with the domain:

$$D_L^D \equiv \left\{ \mathbf{u} : \mathbf{u}, \nabla \times \mathbf{u} \in L^2(\Omega)^6, n \times \mathbf{u}|_\Gamma = 0 \right\},$$

or

$$D_L^N \equiv \left\{ \mathbf{u} : \mathbf{u}, \nabla \times \mathbf{u} \in L^2(\Omega)^6, n \times \left( \boldsymbol{\mu}^{-1}\nabla \times \mathbf{u} \right)|_\Gamma = 0 \right\}.$$

Thus, (3.23) can be rewritten as follows:

$$\left[ -i \begin{pmatrix} 0 & \nabla\times \\ \nabla\times & 0 \end{pmatrix} + \omega \begin{pmatrix} \varepsilon & 0 \\ 0 & -\mu \end{pmatrix} \right] \begin{pmatrix} \mathbf{E} \\ \mathbf{H} \end{pmatrix} = \begin{pmatrix} -i\mathbf{J} \\ 0 \end{pmatrix}. \qquad (3.26)$$

Then, the 'right' spectral problem will become: Finding eigenvectors

$\mathbf{v}_n = \begin{pmatrix} \mathbf{E}_{rn} \\ \mathbf{H}_{rn} \end{pmatrix}$ and their corresponding eigenvalues $\lambda_n = \omega_n$ such that

$$(A + \lambda_n B)\mathbf{v}_n := \left[ -i \begin{pmatrix} 0 & \nabla \times \\ \nabla \times & 0 \end{pmatrix} + \omega_n \begin{pmatrix} \varepsilon & 0 \\ 0 & -\mu \end{pmatrix} \right] \begin{pmatrix} \mathbf{E}_{rn} \\ \mathbf{H}_{rn} \end{pmatrix} = \begin{pmatrix} 0 \\ 0 \end{pmatrix}. \quad (3.27)$$

At a first glance, (3.27) is just a small changed version of (3.24). But we should recognize that in (3.27) if $\varepsilon^{\mathsf{T}} = \varepsilon$ and $\mu^{\mathsf{T}} = \mu$, then $A^{\mathsf{T}} = A$ and $B^{\mathsf{T}} = B$, which implies that $\mathbf{w}_n = \overline{\mathbf{v}_n}$ as discussed in the previous subsection.

By building a symmetric matrix of $L(\lambda) = A + \lambda B$, the reciprocity of the material results in the wonderful fact that the 'left' eigenvectors $\mathbf{w}_n = \begin{pmatrix} \mathbf{E}_{ln} \\ \mathbf{H}_{ln} \end{pmatrix}$ are the complex conjugate of their 'right' counterparts:

$$\begin{pmatrix} \mathbf{E}_{ln} \\ \mathbf{H}_{ln} \end{pmatrix} = \begin{pmatrix} \overline{\mathbf{E}_{rn}} \\ \overline{\mathbf{H}_{rn}} \end{pmatrix}. \quad (3.28)$$

As a final step, by Lemma 3.2.1, we can extend the solution of (3.26) on the set of quasinormal modes:

$$u = \sum_{n=1}^{\infty} \frac{1}{\lambda - \lambda_n} \frac{\langle \mathbf{w}_n, \mathbf{S} \rangle}{\langle \mathbf{w}_n, B\mathbf{v}_n \rangle} \mathbf{v}_n,$$

where the inner product in the denominator needs to be derived as follows:

$$\langle \mathbf{w}_n, B\mathbf{v}_n \rangle = \left\langle \begin{pmatrix} \mathbf{E}_{ln} \\ \mathbf{H}_{ln} \end{pmatrix}, \begin{pmatrix} \varepsilon & 0 \\ 0 & -\mu \end{pmatrix} \begin{pmatrix} \mathbf{E}_{rn} \\ \mathbf{H}_{rn} \end{pmatrix} \right\rangle$$
$$= \int_{\Omega} \left[ \overline{\mathbf{E}_{ln}} \cdot (\varepsilon \mathbf{E}_{rn}) - \overline{\mathbf{H}_{ln}} \cdot (\mu \mathbf{H}_{rn}) \right] d\Omega.$$

Thenceforth, the quasinormal modal expansion for electric fields is deduced as follows:

$$\begin{pmatrix} \mathbf{E} \\ \mathbf{H} \end{pmatrix} = \sum_{n=1}^{\infty} \frac{1}{\omega - \omega_n} \frac{-i \int_{\Omega} \overline{\mathbf{E}_{ln}} \cdot \mathbf{J} \, d\Omega}{\int_{\Omega} \left[ \overline{\mathbf{E}_{ln}} \cdot (\varepsilon \mathbf{E}_{rn}) - \overline{\mathbf{H}_{ln}} \cdot (\mu \mathbf{H}_{rn}) \right] d\Omega} \begin{pmatrix} \mathbf{E}_{rn} \\ \mathbf{H}_{rn} \end{pmatrix}.$$

Substituting (3.28) into the previous equation yields:

$$\boxed{\mathbf{E} = \sum_{n=1}^{\infty} \frac{1}{\omega - \omega_n} \frac{-i \int_{\Omega} \mathbf{E}_{rn} \cdot \mathbf{J} \, d\Omega}{\int_{\Omega} \left[ \mathbf{E}_{rn} \cdot (\varepsilon \mathbf{E}_{rn}) - \mathbf{H}_{rn} \cdot (\mu \mathbf{H}_{rn}) \right] d\Omega} \mathbf{E}_{rn}.} \quad (3.29)$$

*Expansion of electric fields for symmetric electromagnetic matrices 2*

$$\boxed{\mathbf{H} = \sum_{n=1}^{\infty} \frac{1}{\omega - \omega_n} \frac{-i \int_{\Omega} \mathbf{E}_{rn} \cdot \mathbf{J} \, d\Omega}{\int_{\Omega} \left[ \mathbf{E}_{rn} \cdot (\varepsilon \mathbf{E}_{rn}) - \mathbf{H}_{rn} \cdot (\mu \mathbf{H}_{rn}) \right] d\Omega} \mathbf{H}_{rn}.} \quad (3.30)$$

*Expansion of magnetic fields for symmetric electromagnetic matrices 1*

Even though, (3.29) looks very similar to (2.9) without the complex conjugate, we have to keep in mind that:

▶ (2.9) only applies for the self-adjoint operators, while in this section the operator matrix $L(\lambda)$ is symmetric.
▶ The eigenvalues $\omega_n$ in (2.9) only accept real value, meanwhile the eigenvalues in (3.29) can be complex.
▶ The quasinormal modal expansion (3.29) can be applied for unbounded space, which is impossible for the self-adjoint case (2.9).

It is worth reminding that for an eigenvalue $\tilde{\lambda}$, there exists an eigenspace or characteristic space of $L(\lambda)$ associated with $\tilde{\lambda}$: $\tilde{V} = \{\mathbf{v} : L(\tilde{\lambda})\mathbf{v} = 0\}$. In particular, if $\mathbf{v}_1$ and $\mathbf{v}_2$ are eigenvectors of $L(\lambda)$ associated with eigenvalue $\tilde{\lambda}$: $\mathbf{v}_1, \mathbf{v}_2 \in \tilde{V}$ then $\mathbf{v}_1 + \mathbf{v}_2$ and $a\mathbf{v}_1$ are either zero or eigenvectors of $L(\lambda)$ associated with $\tilde{\lambda}$ for $\forall a \in \mathbb{C}$. Hence, the 'right' eigenvectors $\begin{pmatrix} \mathbf{E}_{rn} \\ \mathbf{H}_{rn} \end{pmatrix}$ can be determined up to non-zero complex factors $\alpha_r \begin{pmatrix} \mathbf{E}_{rn} \\ \mathbf{H}_{rn} \end{pmatrix}$. Since (3.29) remains unchanged with respect to $\alpha_r$, it is possible to choose $\begin{pmatrix} \mathbf{E}_{rn} \\ \mathbf{H}_{rn} \end{pmatrix}$ in such a way that the denominator of (3.29) equals 1:

$$\langle \mathbf{w}_n, B\mathbf{v}_n \rangle = \int_\Omega \left[ \mathbf{E}_{rn} \cdot (\boldsymbol{\varepsilon}\mathbf{E}_{rn}) - \mathbf{H}_{rn} \cdot (\boldsymbol{\mu}\mathbf{H}_{rn}) \right] d\Omega = 1. \tag{3.31}$$

The previous expression is known as 'normalization' with respect to Lagrangian 'energy' (see [18, 27] ).

In particular, (3.31) is similar to the formula proposed in [18] when the media are reciprocal and non-dispersive. In [18], the normalization is derived from the Lorentz reciprocity, which implies the utilization of reciprocal media, i.e. symmetric matrices. Similarly, in this section, by linearizing Maxwell's equations (3.23) in a way such that preserves the symmetry of the matrices $A$, $B$, we unintentionally enforce the Lorentz reciprocity and end up with a resembling 'Lagrangian' normalization.

From this subsection, it is vital to see that the 'normalization' (3.31) is solely a consequence of our choice of linearization of Maxwell's equations (3.23). In other words, the formula of normalization is not unique and must depend on the way which the operators $A$, $B$ are constructed.

## Non-uniqueness

As the previous chaper, it is worth remarking that both (3.29) and (3.20) provide the same formula for quasinormal modal expansion:

> **Remark 3.4.2** Given symmetric matrices $\boldsymbol{\varepsilon}$ and $\boldsymbol{\mu}$ , it is possible to prove that $\int_\Omega \mathbf{E}_{rn} \cdot (\boldsymbol{\varepsilon}\mathbf{E}_{rn}) d\Omega = - \int_\Omega \mathbf{H}_{rn} \cdot (\boldsymbol{\mu}\mathbf{H}_{rn}) d\Omega$

*Proof:* From Maxwell's equations, we have :

$$\boldsymbol{\varepsilon}\mathbf{E}_{rn} = -\frac{\nabla \times \mathbf{H}_{rn}}{i\omega}$$

$$\nabla \times \mathbf{E}_{rn} = i\omega\boldsymbol{\mu}\mathbf{H}_{rn}$$

Then the following deduction is true:

$$
\int_{\Omega} \mathbf{E}_{rn} \cdot (\varepsilon \mathbf{E}_n) \, d\Omega = \int_{\Omega} \mathbf{E}_{rn} \cdot \left( -\frac{\nabla \times \mathbf{H}_{rn}}{i\omega} \right) d\Omega
$$

$$
= -\int_{\Omega} \left( \frac{\nabla \times \mathbf{E}_{rn}}{i\omega} \right) \cdot \mathbf{H}_{rn} \, d\Omega + \text{b.t.}
$$

$$
= -\int_{\Omega} \mu \mathbf{H}_{rn} \cdot \mathbf{H}_{rn} \, d\Omega = -\int_{\Omega} \mathbf{H}_{rn} \cdot (\mu \mathbf{H}_{rn}) \, d\Omega,
$$

Where the boundary term b.t. $= \int_{\Gamma} (n \times \mathbf{E}_{rn}) \cdot \mathbf{H} \, dS$ can be eliminated thanks to appropriate boundary conditions, i.e. Dirichlet/Neumann or outgoing wave conditions. [11]

By Remark 3.4.2, we can see that the quasinormal modal expansion (3.20) indeed implies (3.29). Similarly to Chapter 2, from (3.20) we have:

$$
\mathbf{E} = \sum_{n=1}^{\infty} \frac{-i\omega}{\omega^2 - \omega_n^2} \frac{\int_{\Omega} \mathbf{E}_{rn} \cdot \mathbf{J} \, d\Omega}{\int_{\Omega} \mathbf{E}_{rn} \cdot (\varepsilon \mathbf{E}_{rn}) \, d\Omega} \mathbf{E}_{rn}
$$

$$
= \sum_{n=1}^{\infty} \left( \frac{-i\omega}{(\omega - \omega_n)(\omega + \omega_n)} \right) \frac{\int_{\Omega} \mathbf{E}_{rn} \cdot \mathbf{J} \, d\Omega}{\int_{\Omega} \mathbf{E}_{rn} \cdot (\varepsilon \mathbf{E}_{rn}) \, d\Omega} \mathbf{E}_{rn}
$$

$$
= \begin{cases} \sum_{n=1}^{\infty} \dfrac{-i}{2} \left( \dfrac{1}{\omega - \omega_n} + \dfrac{1}{\omega + \omega_n} \right) \dfrac{\int_{\Omega} \mathbf{E}_{rn} \cdot \mathbf{J} \, d\Omega}{\int_{\Omega} \mathbf{E}_{rn} \cdot (\varepsilon \mathbf{E}_{rn}) \, d\Omega} \mathbf{E}_{rn} \\[2ex] \sum_{n=1}^{\infty} \dfrac{-i\omega}{2} \left( \dfrac{1}{\omega_n(\omega - \omega_n)} + \dfrac{1}{-\omega_n(\omega + \omega_n)} \right) \dfrac{\int_{\Omega} \mathbf{E}_{rn} \cdot \mathbf{J} \, d\Omega}{\int_{\Omega} \mathbf{E}_{rn} \cdot (\varepsilon \mathbf{E}_{rn}) \, d\Omega} \mathbf{E}_{rn} \end{cases}
$$

The following step is implemented by rewriting the sum over all the positive and negative-real-part eigenvalues $\omega_n$ and $\omega_{-n}$.

$$
\begin{cases} \displaystyle\sum_{n=-\infty}^{\infty} \frac{-i}{2(\omega - \omega_n)} \frac{\int_{\Omega} \mathbf{E}_{rn} \cdot \mathbf{J} \, d\Omega}{\int_{\Omega} \mathbf{E}_{rn} \cdot (\varepsilon \mathbf{E}_{rn}) \, d\Omega} \mathbf{E}_{rn} & \text{(3.32a)} \\[2ex] \displaystyle\sum_{n=-\infty}^{\infty} \frac{-i\omega}{2\omega_n(\omega - \omega_n)} \frac{\int_{\Omega} \mathbf{E}_{rn} \cdot \mathbf{J} \, d\Omega}{\int_{\Omega} \mathbf{E}_{rn} \cdot (\varepsilon \mathbf{E}_{rn}) \, d\Omega} \mathbf{E}_{rn} & \text{(3.32b)} \end{cases}
$$

The first result (3.32a) gives us the modal expansion (3.29) by using Remark 3.4.2. In addition, the second formula (2.14b) implies the non-uniqueness of the modal expansion, which happens when the operator $L(\omega^2)$ is a quadratic function with respect to eigenvalues $\omega_n$.

## Magnetic fields expansion: Mutatis mutandis

It is worth pointing out that in (3.29) and (3.30), the expansions of the electric and magnetic fields share the same expansion coefficients, i.e. excitation coefficients. The question is raised: Whether this is purely the consequence of our choice of operator following (3.25) or the expansion formula based on the magnetic fields is indeed the same as the electric counterpart. In order to answer the question, let's consider the magnetic operator $L(\lambda)$ such as:

$$L(\lambda)\mathbf{u} = (A + \lambda B)\mathbf{u} = \mathbf{S},$$

where $A \equiv -\nabla \times (\varepsilon^{-1}\nabla \times \cdot)$, $B \equiv \boldsymbol{\mu}$, $\mathbf{S} \equiv -\nabla \times (\varepsilon^{-1}\mathbf{J})$, $\mathbf{u} \equiv \mathbf{H}$ and $\lambda = \omega^2$, satisfies the Maxwell's wave equation for magnetic fields.

Then, by Lemma 3.2.1, we can derive the quasinormal modal expansion for magnetic fields from the operator $L(\omega^2) = -\nabla \times (\varepsilon^{-1}\nabla \times \cdot) + \omega^2\boldsymbol{\mu}$ as follows:

$$\mathbf{H} = \sum_{n=1}^{\infty} \frac{1}{\omega^2 - \omega_n^2} \frac{\int_{\Omega} \overline{\mathbf{H}_{ln}} \cdot \mathbf{S}\, d\Omega}{\int_{\Omega} \overline{\mathbf{H}_{ln}} \cdot (\boldsymbol{\mu}\mathbf{H}_{rn})\, d\Omega} \mathbf{H}_{rn}. \qquad (3.33)$$

By Lemma 3.4.1, we can draw the following expansion:

$$\mathbf{H} = \sum_{n=1}^{\infty} \frac{1}{\omega^2 - \omega_n^2} \frac{\int_{\Omega} \mathbf{H}_{rn} \cdot \mathbf{S}\, d\Omega}{\int_{\Omega} \mathbf{H}_{rn} \cdot (\boldsymbol{\mu}\mathbf{H}_{rn})\, d\Omega} \mathbf{H}_{rn}. \qquad (3.34)$$

*Expansion of magnetic fields for symmetric electromagnetic matrices 2*

with $\mathbf{S} = -\nabla \times (\varepsilon^{-1}\mathbf{J})$.

Our goal is to rewrite the expansion coefficient part

$$\frac{1}{\omega^2 - \omega_n^2} \frac{\int_{\Omega} \mathbf{H}_{rn} \cdot \mathbf{S}\, d\Omega}{\int_{\Omega} \mathbf{H}_{rn} \cdot (\boldsymbol{\mu}\mathbf{H}_{rn})\, d\Omega}$$

in the previous equation in terms of electric fields. In particular, the following mathematical transformation is true:

$$\begin{aligned}
\int_{\Omega} \mathbf{H}_{rn} \cdot \mathbf{S}\, d\Omega &= -\int_{\Omega} \mathbf{H}_{rn} \cdot \nabla \times (\varepsilon^{-1}\mathbf{J})\, d\Omega \\
&= -\int_{\Omega} (\nabla \times \mathbf{H}_{rn}) \cdot (\varepsilon^{-1}\mathbf{J})\, d\Omega + \text{b.t.} \\
&= -\int_{\Omega} (\varepsilon^{-1}\nabla \times \mathbf{H}_{rn}) \cdot \mathbf{J}\, d\Omega = i\omega \int_{\Omega} \mathbf{E}_{rn} \cdot \mathbf{J}\, d\Omega, \quad (3.35)
\end{aligned}$$

where the boundary term b.t. $= \int_{\Gamma} (\mathbf{n} \times \mathbf{H}) \cdot (\varepsilon^{-1}\mathbf{J})\, dS$ can vanish thanks to appropriate boundary conditions. The third line of the previous equation is obtained using $\varepsilon = \varepsilon^{\mathsf{T}}$ and $-i\omega\varepsilon\mathbf{E}_{rn} = \nabla \times \mathbf{H}_{rn}$.

From (3.35) and Remark 3.4.2, (3.34) becomes:

$$\mathbf{H} = \sum_{n=1}^{\infty} \frac{-i\omega}{\omega^2 - \omega_n^2} \frac{\int_{\Omega} \mathbf{E}_{rn} \cdot \mathbf{J}\, d\Omega}{\int_{\Omega} \mathbf{E}_{rn} \cdot (\varepsilon\mathbf{E}_{rn})\, d\Omega} \mathbf{H}_{rn}. \qquad (3.36)$$

From (3.36), it is easily seen that the expansion coefficients of (3.34) is similar to (3.20). As a consequence, following the same argument with the previous subsection, (3.34) is indeed the general version of expansion formula of (3.30).

In conclusion, the subsection demonstrated the fact that no matter if the Quasinormal-mode expansions for the electromagnetic operators are developed in terms of electric/magnetic fields or both of them, all expansions can be converted to the same formula. Therefore, from now on, we will focus only on the expansion of electric fields: The magnetic formulation can be derived according to exactly the same guidelines.

# Spectral theorem for polynomial operators

# 4

## 4.1 Motivation

In the last two chapters, we discover the non-uniqueness of modal expansion for electromagnetics, even for non-dispersive materials . Intuitively, we know that this property results from the fact the electromagnetics operator is either a quadratic function with respect to eigenvalues (eigenfrequencies $\omega_n$), or a system of two linear operators. Even so, we still don't know rigorously the number of possible formulas for modal expansion in electromagnetics. This requires a systematical study in this chapter.

The second issue appears when we impose the dissipation of the materials through the permittivity and permeability. According to the Kramers-Kronig relations, it is only possible with the presence of time-dispersion. The reason of this is partly explained in [43]. To put the problem simple, let's consider the polarization vector $\mathbf{P}$. The causality principle is imposed via the constitutive relation between $\mathbf{P}$ and electric field $\mathbf{E}$, represented by the following equation:

$$\sum_j a_j \frac{\partial^j \mathbf{P}}{\partial t^j} = \sum_k b_k \frac{\partial^k \mathbf{E}}{\partial t^k} \tag{4.1}$$

Since the previous equation refers to real physical problems, all the coefficients $a_j$, $b_k$ must be real. The frequency-dependent electric susceptibility $\chi(\omega)$ is given by carrying out the following Fourier transform:

$$f(\omega) = \int_{\mathbb{R}} f(t) \exp(-i\omega t)\, dt,$$

into (4.1).[1]

$$\left(\sum_j a_j(-i\omega)^j\right)\mathbf{P}(\omega) = \left(\sum_k b_k(-i\omega)^k\right)\mathbf{E}(\omega)$$

The previous equation implies that the electric susceptibility $\chi(i\omega)$, such that $\mathbf{P}(\omega) = \chi(i\omega)\mathbf{E}(\omega)$, must be a function with respect to $i\omega$ where all the coefficients are real; and so the permittivity. From a mathematical point of view, the previous statement is equivalent to claim that the permittivity matrix $\boldsymbol{\varepsilon}$ is Hermitian symmetric $\overline{\boldsymbol{\varepsilon}}(i\omega) = \boldsymbol{\varepsilon}(\overline{i\omega})$. Similarly, we apply the same argument for permeability to obtain $\overline{\boldsymbol{\mu}}(i\omega) = \boldsymbol{\mu}(\overline{i\omega})$.

The two previous issues urge us to extend our study to highly dispersive media by setting permittivity and permeability as frequency-dependent functions $\boldsymbol{\varepsilon}(i\omega)$, $\boldsymbol{\mu}(i\omega)$. From the mathematical perspective, it implies our linear operators $A$, $B$ now depending on $\lambda$: $L(\lambda) = A(\lambda) + \lambda B(\lambda)$ where $\lambda = i\omega$. This forces us to extend our spectral study to non-linear problem with $L(\lambda) = f(\lambda)$ where $f(\lambda)$ is the function of linear operators.

1: Keep in mind that the Fourier transform still maintains the causality property.

The first problem of non-linear operators is that the eigenvalue problem is no longer linear, which is not easy to solve both analytically and numerically. Fortunately, a powerful software library specialized for non-linear eigen-solvers is currently under development: SLEPc [44] , which we will study in Chapter 6. Since both Theorem 2.1.1 and Lemma 3.2.1 are no longer hold in this situation, our next step is try to derive a new modal expansion formula for non-linear operators. We start from the most simplest case of function operator: the polynomial operator.

The ideas in this chapter are inspired by the Athena team's internal reports from Prof. F. Zolla, G. Demésy, and A. Nicolet.

## 4.2 Polynomial operators

**Definition 4.2.1** *A polynomial operator* $\mathcal{P}_L^N(\lambda) : H \to H$ *with the domain* $D_P$ *is defined as:*

$$\mathcal{P}_L^N(\lambda) = \sum_{i=0}^{N} \lambda^i L_i$$

*with* $N$ *standing for the highest order of polynomial. The operator* $\mathcal{P}_L^N(\lambda)$ *should be understood as a function of complex variable* $\lambda$*, where its coefficients are given by linear operators* $L_i$*.*

As discussed in the previous section, it is more reasonable to choose $\lambda = i\omega$ to demonstrate the causality of our physical problem.[2] As a consequence, it is obvious that all the operators $L_i$ should be real.

2: Instead of $\lambda = \omega$ like in Chapter 2 and Chapter 3

The spectral problem requests to search for eigen-triplets $(\lambda_n, \langle \mathbf{w}_n |, | \mathbf{v}_n \rangle)$ of operator $\mathcal{P}_L^N(\lambda)$ as the solutions of the 'left' and 'right' polynomial eigen-problems as follows (respectively):

$$\langle \mathbf{w}_n | \mathcal{P}_L^N(\lambda_n) = 0 \qquad \mathcal{P}_L^N(\lambda_n) | \mathbf{v}_n \rangle = 0, \qquad (4.2)$$

where $\lambda_n$ are the eigenvalues and $\langle \mathbf{w}_n |, | \mathbf{v}_n \rangle$ represent the corresponding left and right eigenvectors.

In particular, the left eigen-solutions $\mathbf{w}_n$ are given by solving the adjoint eigenvalue problem:[3]

$$\left( \sum_{i=0}^{N} \left( \overline{\lambda_n} \right)^i L_i^\dagger \right) \mathbf{w}_n = 0, \qquad (4.3)$$

where $L_i^\dagger$ is the adjoint operator of $L_i$: $\langle L_i^\dagger \mathbf{x}, \mathbf{y} \rangle = \langle \mathbf{x}, L_i \mathbf{y} \rangle$ for $\forall \mathbf{x}, \mathbf{y} \in D_P$.

3: We remind that without any further assumption of symmetry/self-adjointness, the 'left' and 'right' polynomial eigenvalue problems have to be solved separately.

For the sake of simplicity, we assume the operator $\mathcal{P}_L^N(\lambda)$ is diagonalizable. In finite dimensional space, the assumption means for each of the eigenvalues, the algebraic multiplicity equals the geometric multiplicity. Roughly speaking, we will consider that all the eigenvalues are semi-simple.

We can also make a stronger assumption that all the eigenvalues are simple.[4] This requires that $\left. \dfrac{\partial \mathcal{P}_L^N(\lambda)}{\partial \lambda} \right|_{\lambda_n} \mathbf{v}_n \neq 0$.

4: A simple eigenvalue has the multiplicity equal to 1.

**Modal expansion**

Similar to the two previous chapters, we expect that the 'right' polynomial eigen-solutions can span a Hilbert space $H$. This means that any solution of the following non-homogeneous problem such that:

$$\mathscr{P}_L^N(\lambda)\mathbf{u} = \mathbf{S}.$$

with the source $\mathbf{S}$, can be linearly decomposed onto the set of 'right' eigenvectors $\{\mathbf{v}_n\}$:

$$\mathbf{u} = \sum_{n=1}^{\infty} \langle \mathbf{w}_n, \mathbf{u} \rangle \mathbf{v}_n.$$

## 4.3 System of linear operators

**Motivation**

Since we lack necessary mathematical tools to formulate the modal expansion equations for non-linear operators, we must find a way to make use of what we already have in Lemma 3.2.1. Fortunately, from Section 2.3 and Section 3.4, we know that a second-order eigenvalue problem is similar to the system of $2 \times 2$ matrix whose elements are linear operators. Intuitively, this implies that we can rewrite a N-order polynomial eigenvalue problems (4.2) into a $N \times N$ matrices whose indices are given by only linear operators as follows:

$$\langle w_n | (\mathscr{A} + \lambda_n \mathscr{B}) = 0 \qquad (\mathscr{A} + \lambda_n \mathscr{B}) | v_n \rangle = 0, \tag{4.4}$$

where $\mathscr{A}$ and $\mathscr{B}$ are two $N \times N$ 'big' operator matrices, i.e. matrices whose elements are linear operators. The mathematical maneuvers in this section indeed follow the technique of reduction of a high-order differential problem to a set of lower-order differential equations.

The main idea is to deduce the quasinormal modal expansion for polynomial operator, by using Lemma 3.2.1 for two operator $\mathscr{A}$ and $\mathscr{B}$. And, our first job is try to write explicitly how two matrices $\mathscr{A}$ and $\mathscr{B}$ are.

**The right eigenvalue problem**

In order to obtain two 'big' operator matrices $\mathscr{A}$ and $\mathscr{B}$, we have to firstly clarify the 'long' right eigenvectors $v \equiv v_n$ as follows:[5]

$$v := \begin{pmatrix} \mathbf{v}^0 \\ \mathbf{v}^1 \\ \vdots \\ \vdots \\ \mathbf{v}^{N-1} \end{pmatrix} = \begin{pmatrix} \mathbf{v} \\ \lambda^1 \mathbf{v} \\ \vdots \\ \vdots \\ \lambda^{N-1} \mathbf{v} \end{pmatrix}$$

*Example:* If $\mathbf{v} \in L^2(\Omega)^3$ then $v \in L^2(\Omega)^{3N}$

The right eigen-problem $\mathcal{P}_L^N(\lambda)|\mathbf{v}\rangle = 0$ can be rewritten as follows:

$$\begin{cases} L_0\mathbf{v}^0 + L_1\mathbf{v}^1 + \ldots + L_{N-1}\mathbf{v}^{N-1} & + & \lambda L_N\mathbf{v}^{N-1} & = & 0 \\ \mathbf{v}^1 & - & \lambda\mathbf{v}^0 & = & 0 \\ \mathbf{v}^2 & - & \lambda\mathbf{v}^1 & = & 0 \\ & & & \vdots & \\ \mathbf{v}^{N-1} & - & \lambda\mathbf{v}^{N-2} & = & 0 \end{cases}$$

Then, the previous system can be abbreviated as the following matrix equation with the new vector $v$:

$$(\mathcal{A} + \lambda\mathcal{B})\, v = 0 \tag{4.5}$$

with

$$\mathcal{A} = \begin{pmatrix} L_0 & L_1 & \ldots & \ldots & L_{N-1} \\ 0 & I & \ldots & & 0 \\ \vdots & \ddots & \ddots & & \vdots \\ \vdots & & & & \vdots \\ 0 & \ldots & \ldots & 0 & I \end{pmatrix}, \quad \mathcal{B} = \begin{pmatrix} 0 & \ldots & \ldots & \ldots & L_N \\ -I & 0 & \ldots & & 0 \\ \vdots & \ddots & \ddots & & \vdots \\ \vdots & & & & \vdots \\ 0 & \ldots & \ldots & -I & 0 \end{pmatrix}$$

The final result gives us a system of simple linear equation (expressed via matrices $\mathcal{A}$ and $\mathcal{B}$) which is converted from the right polynomial eigen-problem (4.2): Find eigen-solution $(\lambda, v)$ satisfying eq.(4.5).

## The 'left' eigenvalue problem

The next step is to rewrite the 'left' eigenvalue problem $\langle \mathbf{w}_n|\mathcal{P}_L^N(\lambda_n) = 0$ as a new system of linear operators as what've done for the 'right' counterpart. However, instead of deducing the matrices directly from $\langle \mathbf{w}_n|\mathcal{P}_L^N(\lambda_n) = 0$, it is much easier to start from opposite direction with the 'left' eigenvalue problem of operator $(\mathcal{A} + \lambda\mathcal{B})$:

$$\langle w|(\mathcal{A} + \lambda\mathcal{B}) = 0 \quad \text{or} \quad (\mathcal{A}^\dagger + \overline{\lambda}\mathcal{B}^\dagger)|w\rangle = 0. \tag{4.6}$$

By default, the 'left' eigenvector $w$ is represented by:

$$w := \begin{pmatrix} \mathbf{w}^0 \\ \mathbf{w}^1 \\ \vdots \\ \vdots \\ \mathbf{w}^{N-1} \end{pmatrix}, \tag{4.7}$$

where the forms of $\mathbf{w}^0, \mathbf{w}^1, \ldots, \mathbf{w}^{N-1}$ are still unknown. [6] Our main goal in the section is to derive explicitly the equation of each vector $\mathbf{w}^p$, with an interger $p \in \{0, 1, 2, \ldots, N\}$.

We emphasize that in (4.6), the adjoint operator matrices $\mathcal{A}^\dagger$ and $\mathcal{B}^\dagger$ must be defined as adjoint operators:[7]

$$\langle \mathcal{A}^\dagger x, y \rangle = \langle x, \mathcal{A}y \rangle \qquad \langle \mathcal{B}^\dagger x, y \rangle = \langle x, \mathcal{B}y \rangle \tag{4.8}$$

[6]: Unlike the right eigenvector $v$, hitherto, there is no direct connection between $w$ and the left eigensolution of (4.2) $\mathbf{w}_n$.

[7]: This can't be done by simply take the conjugate transpose of $\mathcal{A}$ and $\mathcal{B}$.

with $x := \begin{pmatrix} \mathbf{x}^0 \\ \vdots \\ \mathbf{x}^{N-1} \end{pmatrix}$ and $y := \begin{pmatrix} \mathbf{y}^0 \\ \vdots \\ \mathbf{y}^{N-1} \end{pmatrix}$.

Now, the most pressing issue is to find out the form of adjoint operator of $\mathscr{A}$ and $\mathscr{B}$. Given a matrix $\mathscr{M}$ whose components are operator $M_{ij}$, the following property holds:

$$\langle x, \mathscr{M}y \rangle = \sum_{i,j} \langle \mathbf{x}_j, M_{ij}\mathbf{y}_j \rangle = \sum_{i,j} \langle M_{ij}^\dagger \mathbf{x}_j, \mathbf{y}_j \rangle$$

The second equality comes from the property of adjoint operator $\langle M^\dagger \mathbf{x}, \mathbf{y} \rangle = \langle \mathbf{x}, M\mathbf{y} \rangle$ for $\forall \mathbf{x}, \mathbf{y} \in D_P$.

At the same time, we know that the adjoint of $\mathscr{M}$ is defined such that $\langle x, \mathscr{M}y \rangle = \langle \mathscr{M}^\dagger x, y \rangle$. This implies that the adjoint operator $\mathscr{M}^\dagger$ is the transpose of the matrix whose indices are given by adjoint operators $M_{ij}^\dagger$.

From the last property, we now know the explicit forms of $\mathscr{A}^\dagger$ and $\mathscr{B}^\dagger$:

$$\mathscr{A}^\dagger = \begin{pmatrix} L_0^\dagger & 0 & \dots & \dots & 0 \\ L_1^\dagger & I & \dots & & 0 \\ \vdots & \ddots & \ddots & & \vdots \\ \vdots & & & & \vdots \\ L_{N-1}^\dagger & \dots & \dots & 0 & I \end{pmatrix}, \quad \mathscr{B}^\dagger = \begin{pmatrix} 0 & -I & \dots & \dots & 0 \\ 0 & 0 & -I & \dots & 0 \\ \vdots & \ddots & \ddots & & \vdots \\ \vdots & & & & -I \\ L_N^\dagger & \dots & \dots & 0 & 0 \end{pmatrix}$$

By substituting the previous equations into (4.6), we can obtain a system of linear eigenvalue problems:

$$\begin{cases} L_0^\dagger \mathbf{w}^0 & - & \overline{\lambda}\mathbf{w}^1 & = & 0 \\ L_1^\dagger \mathbf{w}^0 + \mathbf{w}^1 & - & \overline{\lambda}\mathbf{w}^2 & = & 0 \\ & & \vdots & & \\ L_{N-1}^\dagger \mathbf{w}^0 + \mathbf{w}^{N-1} & + & \overline{\lambda}(L_N^\dagger)\mathbf{w}^0 & = & 0 \end{cases}$$

To solve the problem, let us try to multiply the first line by $\left(\overline{\lambda}\right)^0$, the second by $\left(\overline{\lambda}\right)^1$, so forth up to the last line by $\left(\overline{\lambda}\right)^{N-1}$. Then by adding all the equations in the system, we get:

$$\left(\overline{\lambda}\right)^{N-1} L_{N-1}^\dagger \mathbf{w}^0 + \sum_{i=0}^{N-2} \left(\overline{\lambda}\right)^i L_i^\dagger \mathbf{w}^0 + \left(\overline{\lambda}\right)^N L_N^\dagger \mathbf{w}^0 = 0.$$

The previous equation is equivalent to the following formula:

$$\sum_{i=0}^{N} \left(\overline{\lambda}\right)^i L_i^\dagger \mathbf{w}^0 = \mathscr{P}_L^{N,\dagger}(\overline{\lambda})\mathbf{w}^0 = 0,$$

which is exactly the left eigenvalue problem (4.2):

As a result, it is reasonable to set $\mathbf{w}_0 = \mathbf{w}$. Then, it naturally follows that $\mathbf{w}^1 = \frac{1}{\overline{\lambda}} L_0^\dagger \mathbf{w}^0 = \frac{1}{\overline{\lambda}} L_0^\dagger \mathbf{w}$. By iterating, we find out the general formula for each component $\mathbf{w}^k$

$$\mathbf{w}^k = \sum_{i=0}^{k-1} \frac{1}{\overline{\lambda}^{k-1}} L_i^\dagger \mathbf{w}. \tag{4.9}$$

## 4.4 Spectral properties of operator matrices

### Bi-orthogonality

Similar to Section 3.1 of non-self-adjoint linear operator, we try to derive the bi-orthogonality relation between 'left' and 'right' eigenvectors $v_n$ and $w_m$.[8]

> **Lemma 4.4.1** (Bi-orthogonality for operator matrices) *Give two operator matrices $\mathscr{A}, \mathscr{B}$, the left and right eigenvectors $\langle w_m|$ and $|v_n\rangle$ (respectively) corresponding to*
>
> $$\langle w_m|(\mathscr{A} + \lambda_m \mathscr{B}) = 0 \quad (A + \lambda_n \mathscr{B})|v_n\rangle = 0, \qquad (4.10)$$
>
> *satisfy the bi-orthogonality relationships:*
>
> $$(\lambda_m - \lambda_n)\langle w_m, \mathscr{B}v_n\rangle = 0. \qquad (4.11)$$

The proof is similar to the one in Lemma 3.1.4. By combining (4.5) and (4.6), we obtain the following result:

$$\langle w_m, (\mathscr{A} + \lambda_m \mathscr{B})v_n\rangle = \langle w_m, (A + \lambda_n \mathscr{B})v_n\rangle = 0.$$

Since the matrices $\mathscr{A}, \mathscr{B}$ do not depend on eigen-values $\lambda_n$, it follows that:

$$\lambda_m \langle w_m, \mathscr{B}v_n\rangle = \lambda_n \langle w_m, \mathscr{B}v_n\rangle.$$

which completes the proof.

### Modal expansion

In order to apply Lemma 3.2.1 for the operator $(\mathscr{A} + \lambda \mathscr{B})$, the final intermediate step is to express the non-homogeneous equation (4.4) in terms of matrices:

$$(\mathscr{A} + \lambda \mathscr{B})\, u = \mathscr{S}. \qquad (4.12)$$

where 'vectors' $u$ and $\mathscr{S}$ are defined as

$$u := \begin{pmatrix} \mathbf{u} \\ \lambda \mathbf{u} \\ \vdots \\ \vdots \\ \lambda^{N-1}\mathbf{u} \end{pmatrix}, \quad \mathscr{S} := \begin{pmatrix} \mathbf{S} \\ 0 \\ \vdots \\ \vdots \\ 0 \end{pmatrix}, \qquad (4.13)$$

where the vector $\mathbf{u}$ is the solution of $\mathscr{P}_L^N(\lambda)\mathbf{u} = \mathbf{S}$.

Finally, by theorem Lemma 3.2.1 and Lemma 4.4.1, we obtain:

> **Lemma 4.4.2** (Quasinormal modal expansion for operator matrices) *Given the operator $\mathscr{A} + \lambda \mathscr{B}$, the solution $u$ of the boundary value problem*

> *$(\mathcal{A} + \lambda\mathcal{B})u = \mathcal{S}$ can rewrite as quasinormal modal expansion based on the set of right eigenvectors $\{v_n\}$, solutions of the 'right' eigenvalue problem $(A + \lambda_n B)|v_n\rangle = 0$*
>
> $$u = \sum_{n=1}^{\infty} \frac{1}{\lambda - \lambda_n} \frac{\langle w_n, \mathcal{S} \rangle}{\langle w_n, \mathcal{B}v_n \rangle} v_n, \qquad (4.14)$$
>
> *where $w_n$ are eigensolutions of the left eigenvalue problem*
>
> $$\langle w_n | (\mathcal{A} + \lambda_n \mathcal{B}) = 0.$$

## 4.5 Spectral properties of polynomial operator

### Sesquilinear products

In the previous section, we just derive the equation of the polynomial quasinormal modal expansion (4.14). Unfortunately, it is written in terms of 'sesquilinear products' with 'long' vectors $v_n$ and $w_n$ and 'big' operator matrices $\mathcal{A}$ and $\mathcal{B}$. In order to complete the formalism, it is sufficient to express these 'sesquilinear products' in respect of our 'normal' linear operator $L_i$ for $i \in \{0, 1, \dots, N\}$.

The process for 'sesquilinear product' in the numerator of (4.14) is straightforward. Since all the components of $\mathcal{S}$ are zero except the first one, it is easy to check that:

$$\langle w_n, \mathcal{S} \rangle = \langle \mathbf{w}_n, \mathbf{S} \rangle. \qquad (4.15)$$

What remains is to compute explicitly the 'sesquilinear product' $\langle w_n, \mathcal{B}v_n \rangle$. In order to do that, we need the explicit forms of two 'vectors' $w_n$ and $\mathcal{B}v_n$:

$$w_n = \begin{pmatrix} \mathbf{w}_n \\ \frac{1}{\lambda_n} L_0^{\dagger} \mathbf{w}_n \\ \frac{1}{(\lambda_n)^2} L_0^{\dagger} \mathbf{w}_n + \frac{1}{\lambda_n} L_1^{\dagger} \mathbf{w}_n \\ \vdots \\ \sum_{i=0}^{N-2} \frac{1}{(\lambda_n)^{(N-1-i)}} L_i^{\dagger} \mathbf{w}_n \end{pmatrix} \qquad \mathcal{B}v_n = \begin{pmatrix} L_N \lambda_n^{N-1} \mathbf{v}_n \\ -\mathbf{v}_n \\ -\lambda_n \mathbf{v}_n \\ \vdots \\ -\lambda_n^{N-2} \mathbf{v}_n \end{pmatrix}.$$

The term $\langle w_n \mathcal{B}v_n \rangle$ then becomes:

$$\langle w_n \mathcal{B}v_n \rangle = \lambda_n^{N-1} \langle \mathbf{w}_n, L_N \mathbf{v}_n \rangle - \frac{1}{\lambda_n} \langle \mathbf{w}_n, L_0 \mathbf{v}_n \rangle$$

$$- \frac{1}{\lambda_n} \langle \mathbf{w}_n, L_0 \mathbf{v}_n \rangle - \langle \mathbf{w}_n, L_1 \mathbf{v}_n \rangle - \dots - \sum_{i=0}^{N-2} \lambda_n^{i-1} \langle \mathbf{w}_n L_i \mathbf{v}_n \rangle$$

$$= \langle \mathbf{w}_n, \mathcal{DP}_L(\lambda_n) \mathbf{v}_n \rangle. \qquad (4.16)$$

After the second equal sign, we use the notation $\mathcal{DP}_L(\lambda_n)$ to shorten the equation.

$$\mathscr{D}\mathscr{P}_L(\lambda_n) := \lambda_n^{N-1}L_N - \lambda_n^{N-3}L_{N-2} + \ldots$$
$$+ (3-N)\lambda_n L_2 + (2-N)L_1 + (1-N)\lambda_n^{-1}L_0$$

Adding $\frac{1}{\lambda}(N-1)\mathscr{P}_L^N(\lambda_n)\mathbf{v}_n$ to $\mathscr{D}\mathscr{P}_L(\lambda_n)\mathbf{v}_n$ yields :

<div style="text-align: right">We remind that $\mathscr{P}_L^N(\lambda_n)\mathbf{v}_n = 0$</div>

$$\mathscr{D}\mathscr{P}_L(\lambda_n) = L_1 + 2\lambda_n^1 L_2 + 3\lambda^2 L_3 + \ldots$$
$$+ (N-1)\lambda_n^{(N-2)}L_{(N-1)} + N\lambda_n^{(N-1)}L_N$$
$$= \left(\mathring{\mathscr{P}}_L^N(\lambda_n)\right)\mathbf{v}_n \tag{4.17}$$

where $\mathring{\mathscr{P}}_L^N(\lambda_n) := \dfrac{d}{d\lambda}\mathscr{P}_L^N(\lambda)\Big|_{\lambda=\lambda_n}$ is the complex derivative.

Finally, by combining (4.16) and (4.17), we achieve the following equality:

$$\boxed{\langle w_n, \mathscr{B}v_n\rangle = \langle \mathbf{w}_n, \mathring{\mathscr{P}}_L^N(\lambda_n)\mathbf{v}_n\rangle.} \tag{4.18}$$

## The quasinormal modal expansion for polynomial operators

Now we have enough tool to derive the modal expansion for polynomial operator. By substituting (4.15) and (4.18) into (4.14), we get:

$$u = \sum_n \frac{1}{\lambda - \lambda_n} \frac{\langle \mathbf{w}_n, \mathbf{S}\rangle}{\langle \mathbf{w}_n, \mathring{\mathscr{P}}_L^N(\lambda_n)\mathbf{v}_n\rangle} v_n. \tag{4.19}$$

It is interesting to find out that the previous equation is not a unique expansion formula, but indeed implies N sub-formulas. In order to see that, firstly we recall that the vectors $u$ and $v_n$ possess $N$ different components:

$$v_n = \begin{pmatrix} \mathbf{v}_n \\ \lambda_n^1 \mathbf{v}_n \\ \vdots \\ \vdots \\ \lambda_n^{N-1}\mathbf{v}_n \end{pmatrix} \qquad u = \begin{pmatrix} \mathbf{u} \\ \lambda^1 \mathbf{u} \\ \vdots \\ \vdots \\ \lambda^{N-1}\mathbf{u} \end{pmatrix}.$$

For the $(\rho + 1)$ component of $u$ and $v_n$, (4.19) implies that:

$$\mathbf{u} = \sum_n \left(\frac{\lambda_n}{\lambda}\right)^\rho \frac{1}{\lambda - \lambda_n} \frac{\langle \mathbf{w}_n, \mathbf{S}\rangle}{\langle \mathbf{w}_n, \mathring{\mathscr{P}}_L^N(\lambda_n)\mathbf{v}_n\rangle} \mathbf{v}_n,$$

with $\rho \in \{0, 1, 2, \ldots, N-1\}$.

**Lemma 4.5.1** *Given a polynomial operator $\mathscr{P}_L^N(\lambda)$ with eigentriplets $(\lambda_n, \langle w_n|, |v_n\rangle)$ such that $\langle w_n|\mathscr{P}_L^N(\lambda_n) = 0$ and $\mathscr{P}_L^N(\lambda_n)|v_n\rangle = 0$, the solution $u$ of the non-homogeneous problem with the source $S$: $\mathscr{P}_L^N(\lambda)u_s = S$*

*can be expanded in the form of the following quasinormal modal expansion formula:*

$$u = \sum_n \left(\frac{\lambda_n}{\lambda}\right)^\rho \frac{1}{\lambda - \lambda_n} \frac{\langle w_n, S \rangle}{\langle w_n, \dot{\mathscr{P}}_L^N(\lambda_n) v_n \rangle} v_n, \qquad (4.20)$$

*where $\rho$ is an integer such that $\rho \in \{0, 1, \ldots, N-1\}$, and $\dot{\mathscr{P}}_L^N(\lambda_n)$ stands for the derivative operator: $\dot{\mathscr{P}}_L^N(\lambda_n) := \left. \dfrac{d}{d\lambda} \mathscr{P}_L^N(\lambda) \right|_{\lambda = \lambda_n}$.*

## Non-uniqueness

At first glance, from Lemma 4.5.1, we may believe that there are $N$ different modal expansion formulas for operator $\mathscr{P}_L^N(\lambda_n)$. But the equation (4.20) indeed implies that the number of expansion formulas for $\forall \mathbf{u} \in H$ is infinite.

In fact, by Lemma 4.5.1, all the following formulas are equal:

$$\mathbf{u} = \sum_n \left(\frac{\lambda_n}{\lambda}\right)^0 \frac{1}{\lambda - \lambda_n} \frac{\langle \mathbf{w}_n, \mathbf{S} \rangle}{\langle \mathbf{w}_n, \dot{\mathscr{P}}_L^N(\lambda_n) \mathbf{v}_n \rangle} \mathbf{v}_n$$

$$= \sum_n \left(\frac{\lambda_n}{\lambda}\right)^1 \frac{1}{\lambda - \lambda_n} \frac{\langle \mathbf{w}_n, \mathbf{S} \rangle}{\langle \mathbf{w}_n, \dot{\mathscr{P}}_L^N(\lambda_n) \mathbf{v}_n \rangle} \mathbf{v}_n$$

$$= \ldots$$

$$= \sum_n \left(\frac{\lambda_n}{\lambda}\right)^{N-1} \frac{1}{\lambda - \lambda_n} \frac{\langle \mathbf{w}_n, \mathbf{S} \rangle}{\langle \mathbf{w}_n, \dot{\mathscr{P}}_L^N(\lambda_n) \mathbf{v}_n \rangle} \mathbf{v}_n.$$

The previous set of equations leads to:[9]

$$\mathbf{u} = \sum_n \frac{a_0 + a_1\lambda_n + a_2\lambda_n^2 + \ldots + a_{N-1}\lambda_n^{N-1}}{a_0 + a_1\lambda + a_2\lambda^2 + \ldots + a_{N-1}\lambda^{N-1}} \frac{1}{\lambda - \lambda_n} \frac{\langle \mathbf{w}_n, \mathbf{S} \rangle}{\langle \mathbf{w}_n, \dot{\mathscr{P}}_L^N(\lambda_n) \mathbf{v}_n \rangle}.$$

where $a_0, a_1, a_2, \ldots, a_{N-1}$ are arbitrary complex coefficients.

For the sake of simplicity, we rewrite the previous formula as follows:

$$\mathbf{u} = \sum_n \frac{f_\rho(\lambda_n)}{f_\rho(\lambda)} \frac{1}{\lambda - \lambda_n} \frac{\langle \mathbf{w}_n, \mathbf{S} \rangle}{\langle \mathbf{w}_n, \dot{\mathscr{P}}_L^N(\lambda_n) \mathbf{v}_n \rangle} \qquad (4.21)$$

where $f_\rho(\lambda)$ is an arbitrary polynomial degree $\rho$: $\rho \in \{0, 1, \ldots, N-1\}$.

**Lemma 4.5.2** (The quasinormal modal expansion for polynomial operators) *Given a polynomial operator $\mathscr{P}_L^N(\lambda)$ with eigentriplets $(\lambda_n, \langle w_n|, |v_n\rangle)$ such that:*

$$\langle w_n | \mathscr{P}_L^N(\lambda_n) = 0 \qquad \mathscr{P}_L^N(\lambda_n) | v_n \rangle = 0, \qquad (4.22)$$

*the solution $u$ of the non-homogeneous problem with the source $S$:*

$$\mathscr{P}_L^N(\lambda) u_s = S,$$

*can be expanded in form of the following quasinormal modal expansion*

9: It is straightforward considering $\dfrac{a}{b} = \dfrac{c}{d} = \dfrac{a+c}{b+d}$.

*formula:*

$$\boldsymbol{u} = \sum_n \frac{f_\rho(\lambda_n)}{f_\rho(\lambda)} \frac{1}{\lambda - \lambda_n} \frac{\langle \boldsymbol{w}_n, \boldsymbol{S} \rangle}{\langle \boldsymbol{w}_n, \dot{\mathcal{P}}_L^N(\lambda_n) \boldsymbol{v}_n \rangle} \boldsymbol{v}_n, \qquad (4.23)$$

*where $f_\rho(\lambda)$ is an arbitrary polynomial function degree $\rho$ such that $\rho \in \{0, 1, \dots, N-1\}$,*

*$\dot{\mathcal{P}}_L^N(\lambda_n)$ stands for the complex derivative operator: $\dot{\mathcal{P}}_L^N(\lambda_n) := \frac{d}{d\lambda} \mathcal{P}_L^N(\lambda) \Big|_{\lambda = \lambda_n}$*

In order to fully comprehend the non-uniqueness of Lemma 4.5.2, let's reduce our infinite-dimensional spaces to finite-dimensional one which is more intuitive. In particular, we will study the spectral property of operator in $V$ a finite-dimensional subspace of the infinite-dimensional functional space $H$. [10] First, in the finite-dimensional space V, let's consider the case of linear operator $L(\lambda) = A + \lambda B$.

10: In the finite-dimensional space $V$, all the spectral results we obtain for infinite-dimensional operator still hold. The only difference is that all the operators now can be expressed by their matrix representation.

**Example 4.5.1** Given two $M \times M$ matrices $A, B : V \to V$, by Lemma 3.2.1, the solution $\boldsymbol{u} \in V$ of the boundary value problem $(A + \lambda B)\boldsymbol{u} = \boldsymbol{S}$ can rewrite as quasinormal modal expansion based on the set of 'left' and 'right' eigenvectors $\langle \boldsymbol{w}_n|(A + \lambda_n B) = 0$ and $(A + \lambda_n B)|\boldsymbol{v}_n \rangle = 0$ respectively:

$$\boldsymbol{u} = \sum_{n=1}^{M} \frac{1}{\lambda - \lambda_n} \frac{\langle \boldsymbol{w}_n, \boldsymbol{S} \rangle}{\langle \boldsymbol{w}_n, B \boldsymbol{v}_n \rangle} \boldsymbol{v}_n. \qquad (4.24)$$

In the previous example, assuming all the eivenvalues are simple, we will have $M$ number of independent (but not orthogonal) 'right' eigenvectors $\boldsymbol{v}_n$ solution of $(A + \lambda B)\boldsymbol{v}_n = 0$. Thus, the sum in (4.24), naturally, takes into account all $M$ eigenvectors $\boldsymbol{v}_n$ with $n \in \{1, 2, 3, \dots, M\}$. Since the set of 'right' eigenvectors is linear independent and span the whole complete $M$-dimensional space, the modal expansion is unique.

In fact, even if the matrices $A, B$ are non self-adjoint, deriving the uniqueness of the modal expansion for the linear operator $L(\lambda) = A + \lambda B$ is still a straightforward process from Lemma 3.2.1. However, when moving from linear operator to non-linear operator, the quasinormal modal expansion must be treated with care.

**Example 4.5.2** Consider the polynomial operator $\mathcal{P}_L^N(\lambda) = \sum_{i=0}^{N} \lambda^i L_i$ where $L_i$ are $M \times M$ matrices $L_i : V \to V$ with eigentriplets $(\lambda_n, \langle \boldsymbol{w}_n|, |\boldsymbol{v}_n \rangle)$. The quasinormal modal expansion for $\boldsymbol{u} \in V$ the solution of the boundary value problem $\mathcal{P}_L^N(\lambda)\boldsymbol{u} = \boldsymbol{S}$ can be written as follows:

$$\boldsymbol{u} = \sum_{n=1}^{N \times M} \frac{f_\rho(\lambda_n)}{f_\rho(\lambda)} \frac{1}{\lambda - \lambda_n} \frac{\langle \boldsymbol{w}_n, \boldsymbol{S} \rangle}{\langle \boldsymbol{w}_n, \dot{\mathcal{P}}_L^N(\lambda_n) \boldsymbol{v}_n \rangle} \boldsymbol{v}_n. \qquad (4.25)$$

To solve the spectral problem for operator $\mathcal{P}_L^N$ implies to find all the eigenvalues $\lambda_n$ and corresponding eigenvectors $\boldsymbol{v}_n \in V$ such that $\mathcal{P}_L^N(\lambda_n)\boldsymbol{v}_n = 0$.[11] It is interesting to find out that since the operator is a polynomial function with respect to $\lambda$, we have not only $M$ number of eigen-solutions like in Example 4.5.1 but indeed $N \times M$ eigenvalues $\lambda_n$ with $N \times M$ corresponding eigenvectors $\boldsymbol{v}_n$. This implies that the set of $N \times M$ components of eigenvectors $\boldsymbol{v}_n$ is 'over-complete' to span the $M$-dimensional space V.[12] Hence, the matrices $\mathscr{A}$ and $\mathscr{B}$ are

11: We assume that all the eigenvalues are simple.

12: In other words, the set of eigenvectors $\boldsymbol{v}_n$ is not linear independent.

indeed $(N \times M) \times (N \times M)$. Thus the sum in (4.25) run through $N \times M$ eigenvectors.

> **Lemma 4.5.3** *Since the set of eigen-space is over-complete and eigenvectors are not linear independent, the uniqueness of quasinormal modal expansion in Lemma 4.5.2 fails to hold.*

# Spectral theorem for rational operators

<div style="text-align: right">**5**</div>

In Chapter 4, we've already successfully established the quasinormal modal expansion for polynomial operators. Unfortunately as mentioned in chapter Chapter 3 and Chapter 4, the frequency-dependence of the permittivity is generally expressed in terms of rational functions (such as the Drude model) rather than simple polynomials. As a result, instead of polynomial operators, the electromagnetics operators usually contain rational functions of the complex variable $\lambda$, which will be called rational operators in this chapter:

> **Definition 5.0.1** *A rational operator $\mathcal{R}_L^N(\lambda) : H \to H$ is defined as follows:*
>
> $$\mathcal{R}_L^N(\lambda) := \sum_{i=0}^N \mathcal{R}_i(\lambda) L_i, \tag{5.1}$$
>
> *where $L_i$ are linear operators and $\mathcal{R}_i(\lambda)$ stand for the rational functions with respect to $\lambda$:*
>
> $$\mathcal{R}_i = \frac{n_i(\lambda)}{d_i(\lambda)}. \tag{5.2}$$
>
> *The numerator and denominator of rational function $\mathcal{R}_i(\lambda)$ are polynomials of degree $deg(n_i)$ and $deg(d_i)$ respectively.*

The idea of this chapter is to rewrite the rational operator $\mathcal{R}_L^N(\lambda)$ as a polynomial operator $\mathcal{P}_L^N(\lambda)$ in order to utilise Lemma 4.5.2. Then we can formalize the quasinormal modal expansion for rational operators and complete the study of spectral properties for non-linear and non-self-adjoint operators.

## 5.1 Spectral properties of rational operators

### Rational operator $\mathcal{R}_L^N(\lambda)$

For the sake of simplicity, the rational operator $\mathcal{R}_L^N(\lambda)$ is re-written as follows:

$$\mathcal{R}_L^N(\lambda) = \frac{\mathcal{N}_L^{N_N}(\lambda)}{\mathcal{D}^{N_D}(\lambda)}. \tag{5.3}$$

where the values of $N_N$ and $N_D$ are given by (5.4).

In what follows, $\mathcal{N}_L^{N_N}(\lambda)$ is set to be a polynomial operator degree $N_N$:

$$\mathcal{N}_L^{N_N}(\lambda) = \mathcal{P}_L^{N_N}(\lambda) = \sum_{j=0}^{N_N} \lambda^j L_j,$$

where $L_j$ refer to linear operators.

On the other hand, the denominator $D^{N_D}(\lambda)$ is described as a polynomial degree $N_D$ of the variable $\lambda$:

$$\mathscr{D}^{N_D}(\lambda) = \sum_{k=0}^{N_D} a_k \lambda^k,$$

with $a_k$ are complex numbers.

It is easy to check that if $\text{root}(d_i) \neq \text{root}(d_j)$ [1] for $\forall i, j \in \{0, 1, \dots, N\}$, we have:

$$N_N = \sup_i \left( \deg(n_i) + \sum_{h \neq i} \deg(d_h) \right) \tag{5.4}$$
$$N_D = \sum_i \deg(d_i).$$

## Spectral properties

Using (5.3), it is possible to re-write the rational operator $\mathscr{R}_L^N(\lambda)$ in terms of polynomial operators. In particular, by multiplying $\mathscr{R}_L^N(\lambda)$ by $\mathscr{D}^{N_D}(\lambda)$, we have:

$$\begin{cases} \mathscr{R}_L^N(\lambda)|\mathbf{u}\rangle = |\mathbf{S}\rangle \\ \mathscr{R}_L^N(\lambda_n)|\mathbf{v}_n\rangle = 0 \\ \langle\mathbf{w}_n|\mathscr{R}_L^N(\lambda_n) = 0 \end{cases} \implies \begin{cases} \mathscr{N}_L^{N_N}(\lambda)|\mathbf{u}\rangle = \mathscr{D}^{N_D}(\lambda)|\mathbf{S}\rangle \\ \mathscr{N}_L^{N_N}(\lambda_n)|\mathbf{v}_n\rangle = 0 \\ \langle\mathbf{w}_n|\mathscr{N}_L^{N_N}(\lambda_n) = 0 \end{cases}. \tag{5.5}$$

The relation in (5.5) is a right implication '$\Rightarrow$' instead of a equivalence '$\Leftrightarrow$' because we have to take into account the value of $\lambda$ where $\mathscr{D}^{N_D}(\lambda) = 0$

Then, by applying Lemma 4.5.2 for polynomial operators to the operator $\mathscr{N}^{N_N}(\lambda)$, we obtain:

$$\mathbf{u} = \sum_n \frac{f_{\rho_1}(\lambda_n)}{f_{\rho_1}(\lambda)} \frac{1}{\lambda - \lambda_n} \frac{\langle\mathbf{w}_n, \mathscr{D}^{N_D}(\lambda)\mathbf{S}\rangle}{\left\langle\mathbf{w}_n, \dot{\mathscr{N}}_L^{N_N}(\lambda_n)\mathbf{v}_n\right\rangle} \mathbf{v}_n, \tag{5.6}$$

where $f_{\rho_1}(\lambda)$ is an arbitrary polynomial degree $\rho_1$ such as $\rho_1 \in \{0, 1, \dots, N_N - 1\}$; and $N_N$ is the degree of polynomial operator $\mathscr{N}_L^{N_N}(\lambda)$.

Although, the expansion (5.6) has already provided the spectral formula to our problem, it is worth investigating the equation further in order to retrieve the operator $\mathscr{R}_L^N(\lambda)$.

Let's remind that $\mathscr{N}_L^{N_N}(\lambda) = \mathscr{R}_L^N(\lambda)\mathscr{D}^{N_D}(\lambda)$. This leads to:

$$\dot{\mathscr{N}}_L^{N_N}(\lambda_n)\mathbf{v}_n = \left(\dot{\mathscr{D}}^{N_D}(\lambda_n)\mathscr{R}_L^N(\lambda_n)\right)\mathbf{v}_n + \left(\mathscr{D}^{N_D}(\lambda_n)\dot{\mathscr{R}}_L^N(\lambda_n)\right)\mathbf{v}_n$$
$$= \left(\mathscr{D}(\lambda_n)\dot{\mathscr{R}}_L^N(\lambda_n)\right)\mathbf{v}_n.$$

In the previous equation, the term $\left(\dot{\mathscr{D}}^{N_D}(\lambda_n)\mathscr{R}_L^N(\lambda_n)\right)\mathbf{v}_n$ vanishes since the eigenvalue problem $\mathscr{R}_L^N(\lambda_n)\mathbf{v}_n = 0$.

Then (5.6) can be expressed as follows:

$$\mathbf{u} = \sum_n \frac{f_{\rho_1}(\lambda_n)}{f_{\rho_1}(\lambda)} \frac{1}{\lambda - \lambda_n} \frac{\langle \mathbf{w}_n, \mathscr{D}^{N_D}(\lambda)\mathbf{S} \rangle}{\langle \mathbf{w}_n, \mathscr{D}^{N_D}(\lambda_n)\dot{\mathscr{R}}_L^N(\lambda_n)\mathbf{v}_n \rangle} \mathbf{v}_n.$$

It is worth noting that $\mathscr{D}^{N_D}(\lambda)$ is only a simple polynomial of variable $\lambda$, and can be moved out of bra-ket notation to obtain:

$$\mathbf{u} = \sum_n \frac{f_{\rho_1}(\lambda_n)\mathscr{D}^{N_D}(\lambda)}{f_{\rho_1}(\lambda)\mathscr{D}^{N_D}(\lambda_n)} \frac{1}{\lambda - \lambda_n} \frac{\langle \mathbf{w}_n, \mathbf{S} \rangle}{\langle \mathbf{w}_n, \dot{\mathscr{R}}_L^N(\lambda_n)\mathbf{v}_n \rangle} \mathbf{v}_n. \tag{5.7}$$

where $f_{\rho_1}(\lambda)$ is an arbitrary polynomial degree up to $\rho_1 = N_N - 1$.

On account of the arbitrariness of function $f_{\rho_1}(\lambda)$, we choose a particular $f_{\rho_1}(\lambda)$ as follows:

$$f_{\rho_1}(\lambda) = g_{\rho_2}(\lambda)\mathscr{D}^{N_D}(\lambda). \tag{5.8}$$

We emphasize that (5.8) holds under the hypothesis that the degree of polynomial $f_{\rho_1}(\lambda)$ is greater than the one of $\mathscr{D}^{N_D}(\lambda)$: $\rho_1 \geq N_D$. [2]

It follows that $g_{\rho_2}(\lambda)$ must be an arbitrary polynomial degree $\rho_2$ such as $\rho_2 = \rho_1 - N_D$. This implies that $\rho_2 \in \{0, 1, \ldots, N_N - N_D - 1\}$.

Finally, substituting (5.8) into (5.7) yields the quasinormal modal expansion for rational operators:

**Lemma 5.1.1** *Consider a rational operator* $\mathscr{R}_L(\lambda) : H \longrightarrow H$ *such that:*

$$\mathscr{R}_L(\lambda) = \frac{\mathscr{N}_L^{N_N}(\lambda)}{\mathscr{D}^{N_D}(\lambda)} \tag{5.9}$$

*with eigentriplets* $(\lambda_n, \langle \mathbf{w}_n|, |\mathbf{v}_n\rangle)$ *satisfying* $\langle \mathbf{w}_n|\mathscr{R}_L(\lambda_n) = 0$ *and* $\mathscr{R}_L(\lambda_n)|\mathbf{v}_n\rangle = 0$. *The solution* $\mathbf{u}$ *of the non-homogeneous problem with the source* $\mathbf{S}$:

$$\mathscr{R}_L(\lambda)\mathbf{u}_s = \mathbf{S}$$

*can be expanded in form of the quasinormal modal expansion formula followed:*

$$\mathbf{u} = \sum_n \frac{f_\rho(\lambda_n)}{f_\rho(\lambda)} \frac{1}{\lambda - \lambda_n} \frac{\langle \mathbf{w}_n, \mathbf{S} \rangle}{\langle \mathbf{w}_n, \dot{\mathscr{R}}_L(\lambda_n)\mathbf{v}_n \rangle} \mathbf{v}_n \tag{5.10}$$

*where* $f_\rho(\lambda)$ *is an arbitrary polynomial degree* $\rho$ *with* $\rho \in \{0, 1, \ldots, N_N - N_D - 1\}$, $\dot{\mathscr{R}}_L(\lambda_n)$ *stands for the derivative operator:* $\dot{\mathscr{R}}_L(\lambda_n) := \frac{\partial \mathscr{R}_L}{\partial \lambda}(\lambda)\Big|_{\lambda = \lambda_n}$

## 5.2 Dispersive quasinormal expansion for electromagnetics

The aim of this section is to deduce the quasinormal expansion for electromagnetic problem with dispersive media, by utilising Lemma 5.1.1.

Given a space $\Omega$ with ideal boundary, let's consider the following Maxwell equations:

$$\nabla \times \mathbf{H}(\mathbf{r}) + i\omega \boldsymbol{\varepsilon}(\omega, \mathbf{r})\mathbf{E}(\mathbf{r}) = \mathbf{J} \qquad (5.11)$$

$$\nabla \times \mathbf{E}(\mathbf{r}) - i\omega \boldsymbol{\mu}(\omega, \mathbf{r})\mathbf{H}(\mathbf{r}) = 0$$

Unlike Chapter 2 and Chapter 3, the matrices $\boldsymbol{\varepsilon}(\omega, \mathbf{r})$ and $\boldsymbol{\mu}(\omega, \mathbf{r})$ now depend on both space $\mathbf{r}$ and frequency $\omega$. As explained in Chapter 8, the frequency dependence of the permittivity and permeability must be expressed as a rational function with respect to $i\omega$. Thus, it is better henceforth to denote $\boldsymbol{\varepsilon}(\lambda)$ and $\boldsymbol{\mu}(\lambda)$ where $\lambda = i\omega$ to reflect the causality property. [3]

If we denote:

$$L_1 = \begin{pmatrix} 0 & \nabla\times \\ \nabla\times & 0 \end{pmatrix} \quad \text{and} \quad L_2 = \begin{pmatrix} \boldsymbol{\varepsilon}(\lambda) & 0 \\ 0 & -\boldsymbol{\mu}(\lambda) \end{pmatrix}, \qquad (5.12)$$

Maxwell's equation (5.11) will become:

$$(L_1 + \lambda L_2)\mathbf{u} = \mathbf{S}, \qquad (5.13)$$

with $\mathbf{u} = \begin{pmatrix} \mathbf{E} \\ \mathbf{H} \end{pmatrix}$ and $\mathbf{S} = \begin{pmatrix} \mathbf{J} \\ 0 \end{pmatrix}$.

The next goal is to expand the solution $\mathbf{u}$ of non-homogeneous problem (5.13) as a linear combination of the 'left' and 'right' eigen-fields $\mathbf{w}_n = \begin{pmatrix} \mathbf{E}_{ln} \\ \mathbf{H}_{ln} \end{pmatrix}$ and $\mathbf{v}_n = \begin{pmatrix} \mathbf{E}_{rn} \\ \mathbf{H}_{rn} \end{pmatrix}$ respectively:

$$\langle \mathbf{w}_n | (L_1 + \lambda_n L_2) = 0 \qquad (L_1 + \lambda_n L_2)|\mathbf{v}_n\rangle = 0,$$

Unfortunately, it is important to see that at the moment, the operator $\mathcal{NL}(\lambda) = L_1 + \lambda L_2$ is not a rational operator. The reason is that $\boldsymbol{\varepsilon}(\lambda)$ and $\boldsymbol{\mu}(\lambda)$ are not simple scalars function, but indeed matrices. Our task now is to modify and put the operator $\mathcal{NL}(\lambda)$ in the rational form.

The idea comes from the fact that as matrices, $\boldsymbol{\varepsilon}(\lambda)$ and $\boldsymbol{\mu}(\lambda)$ can be rewritten as the following sum:

$$\boldsymbol{\chi}(\lambda) = \begin{pmatrix} \chi_{xx}(\lambda) & \chi_{xy}(\lambda) & \chi_{xz}(\lambda) \\ \chi_{yx}(\lambda) & \chi_{yy}(\lambda) & \chi_{yz}(\lambda) \\ \chi_{zx}(\lambda) & \chi_{zy}(\lambda) & \chi_{zz}(\lambda) \end{pmatrix} = \sum_{i,j \in \{x,y,z\}} \chi_{ij}(\lambda) \left( \mathcal{F}_{ij} \right), \qquad (5.14)$$

with $\chi = \{\varepsilon, \mu\}$. $\left( \mathcal{F}_{ij} \right) : L^2(\Omega)^3 \to L^2(\Omega)^3$ represents the matrix whose all the elements are zero, except the $i, j$-element given by the identity operator $\mathcal{F} : L^2(\Omega) \to L^2(\Omega)$.

For example, if we assume that the materials are isotropic, the operator $L_2$ in (5.12) can be expressed as follows:

$$L_2 = \begin{pmatrix} \boldsymbol{\varepsilon}(\lambda) & 0 \\ 0 & -\boldsymbol{\mu}(\lambda) \end{pmatrix} = \varepsilon(\lambda) \begin{pmatrix} \mathcal{F} & 0 \\ 0 & 0 \end{pmatrix} - \mu(\lambda) \begin{pmatrix} 0 & 0 \\ 0 & \mathcal{F} \end{pmatrix},$$

where $\varepsilon(\lambda)$ and $\mu(\lambda)$ are scalar rational function.

Then it is easy to see that the operator:

$$\mathcal{NL}(\lambda) = \begin{pmatrix} 0 & \nabla \times \\ \nabla \times & 0 \end{pmatrix} + \lambda \varepsilon(\lambda) \begin{pmatrix} \mathcal{I} & 0 \\ 0 & 0 \end{pmatrix} - \lambda \mu(\lambda) \begin{pmatrix} 0 & 0 \\ 0 & \mathcal{I} \end{pmatrix}, \qquad (5.15)$$

is the rational operator.

We emphasize that although the media are isotropic in the previous example, the operator $\mathcal{NL}(\lambda)$ can still be written as the rational form even for an-isotropic, non-reciprocal and dispersive media.

Then, by Lemma 5.1.1 for rational operator, the quasinormal modal expansion for operator $\mathcal{NL}(\lambda)$ is deduced:

$$\mathbf{u} = \sum_n \frac{f_\rho(\lambda_n)}{f_\rho(\lambda)} \frac{1}{\lambda - \lambda_n} \frac{\langle \mathbf{w}_n, \mathbf{S} \rangle}{\langle \mathbf{w}_n, \mathcal{NL}_L(\lambda_n)\mathbf{v}_n \rangle} \mathbf{v}_n, \qquad (5.16)$$

where the sesquilinear product $\langle \mathbf{w}_n, \mathcal{NL}_L(\lambda_n)\mathbf{v}_n \rangle$ is computed as follows:

$$\langle \mathbf{w}_n, \dot{\mathcal{NL}}_L(\lambda_n)\mathbf{v}_n \rangle =$$
$$\int_\Omega \left[ \overline{\mathbf{E}_{ln}} \cdot \left( (\lambda_n \varepsilon(\lambda_n))' \mathbf{E}_{rn} \right) - \overline{\mathbf{H}_{ln}} \cdot \left( (\lambda_n \boldsymbol{\mu}(\lambda_n))' \mathbf{H}_{rn} \right) \right] d\Omega. \qquad (5.17)$$

where $f'(\lambda_n) \coloneqq \left. \dfrac{\partial f(\lambda)}{\partial \lambda} \right|_{\lambda = \lambda_n}$.

The final step is to identify the value of $\rho$ in (5.16). This requires to find out the value of $N_D$ and $N_N$ from the operator $\mathcal{NL}(\lambda)$:

$$\mathcal{NL}(\lambda) = \frac{\mathcal{N}_L^{N_N}(\lambda)}{\mathcal{D}_D^N(\lambda)}.$$

**Proposition 5.2.1** *If the frequency-dependence of permittivity and permeability is expressed as a rational function, the order of polynomial of the numerator and denominator of such rational function are equal.*

The reason behind Proposition 5.2.1 will be mentioned in Chapter 8.

Using Proposition 5.2.1 and (5.4), it is straightforward to verify that

$$N_N = N_D + 1, \qquad (5.18)$$

which concludes that the highest order of the polynomial $f_\rho(\lambda)$ for the operator $(L_1 + \lambda L_2)$ given by (5.12) is zero $\rho = 0$. This implies that with our choose of operator (5.12), the expansion (5.16) is unique.

In conclusion, the dispersive quasinormal modal expansion for electric field $\mathbf{E}$ solution of the Maxwell's equation (5.11) is formalized as follows:

$$\mathbf{E} = \sum_n \frac{1}{\lambda - \lambda_n} \frac{\int_\Omega \overline{\mathbf{E}_{ln}} \cdot \mathbf{J} \, d\Omega}{\int_\Omega \left[ \overline{\mathbf{E}_{ln}} \cdot \left( (\lambda_n \varepsilon(\lambda_n))' \mathbf{E}_{rn} \right) - \overline{\mathbf{H}_{ln}} \cdot \left( (\lambda_n \boldsymbol{\mu}(\lambda_n))' \mathbf{H}_{rn} \right) \right] d\Omega} \mathbf{E}_{rn}, \qquad (5.19)$$

with $\lambda = i\omega$.

Following the same argument in Section 3.4, (5.19) can be simplified if the media are reciprocal. In particular, if the matrices of $\boldsymbol{\varepsilon}$ and $\boldsymbol{\mu}$ are symmetric, the 'left' eigenvectors will be equal to the 'right' counterpart:

$$\begin{pmatrix} \mathbf{E}_{ln} \\ \mathbf{H}_{ln} \end{pmatrix} = \begin{pmatrix} \overline{\mathbf{E}_{rn}} \\ \overline{\mathbf{H}_{rn}} \end{pmatrix}. \tag{5.20}$$

Then our quasinormal modal expansion will become:

$$\mathbf{E} = \sum_n \frac{1}{\lambda - \lambda_n} \frac{\int_\Omega \mathbf{E}_{rn} \cdot \mathbf{J} \, d\Omega}{\int_\Omega \left[ \mathbf{E}_{rn} \cdot ((\lambda_n \boldsymbol{\varepsilon}(\lambda_n))' \mathbf{E}_{rn}) - \mathbf{H}_{rn} \cdot ((\lambda_n \boldsymbol{\mu}(\lambda_n))' \mathbf{H}_{rn}) \right] \, d\Omega} \mathbf{E}_{rn}, \tag{5.21}$$

*Expansion of electric fields for non-linear operators 1: Lagrangian form*

which is similar to the 'Lagrangian' formulas obtained in [10, 18, 20] .

It is worth remembering that the procedures of scaling eigenvectors, and formulating the spectral expansion are not fixed and depend on our choice of linearization of Maxwell's equations. These procedures must be seen as a consequence obtained through mathematical rigor, rather than the standardization based on any conventional physical concept. For example, we can set:

$$L_1 = \begin{pmatrix} 0 & \nabla \times \\ -\nabla \times & 0 \end{pmatrix} \quad \text{and} \quad L_2 = \begin{pmatrix} \boldsymbol{\varepsilon}(\lambda) & 0 \\ 0 & \boldsymbol{\mu}(\lambda) \end{pmatrix},$$

in order to obtain the 'energy form' normalization and expansion, similarly to [45] . In this case, even if the media are non-reciprocal, Lemma 5.1.1 still holds.

## 5.3 Non uniqueness

After the demonstration of dispersive quasinormal modal expansion in the last chapter, we may wonder what happens if another different choice of operators is considered. Therefore, let's try to apply Lemma 5.1.1 once again, but this time with the electric wave operator:

$$\mathcal{N}\mathcal{L}(\lambda) = \nabla \times \left( \boldsymbol{\mu}^{-1}(\lambda) \nabla \times \cdot \right) + \lambda^2 \boldsymbol{\varepsilon}(\lambda), \tag{5.22}$$

with $\lambda = i\omega$.

Hence, the boundary Maxwell's equations (5.11) will be equivalent to:

$$\mathcal{N}\mathcal{L}(\lambda)\mathbf{u} = \mathbf{S}, \tag{5.23}$$

where $\mathbf{u}$ stands for the electric field $\mathbf{E}$ and the source is given by $\mathbf{S} = \lambda \mathbf{J}$.

The 'right' eigen-solutions $\mathbf{v}_n = \mathbf{E}_{rn}$ of the operator is given by:

$$\mathcal{N}\mathcal{L}(\lambda_n)\mathbf{v}_n = \nabla \times \left( \boldsymbol{\mu}^{-1}(\lambda_n) \nabla \times \mathbf{v}_n \right) + \lambda_n^2 \boldsymbol{\varepsilon}(\lambda_n)\mathbf{v}_n = 0,$$

while the 'left' eigenvectors satisfy the following equation:[4]

4: As we've already derived in Section 3.4.

$$\mathscr{NL}(\lambda_n)\mathbf{w}_n = \nabla \times \left(\boldsymbol{\mu}^{-*}(\lambda_n)\nabla \times \mathbf{v}_n\right) + \lambda_n^2 \boldsymbol{\varepsilon}^*(\lambda_n)\mathbf{w}_n = 0. \qquad (5.24)$$

As discussed in the previous section, through some appropriate transformation, the operator $\mathscr{NL}(\lambda)$ can be written as a rational operator. Then, by Lemma 5.1.1, the quasinormal modal expansion for electric fields appears to be:

$$\mathbf{E} = \sum_n \frac{f_\rho(\lambda_n)}{f_\rho(\lambda)} \frac{1}{\lambda - \lambda_n} \frac{\langle \mathbf{E}_{ln}, \lambda \mathbf{J}\rangle}{\langle \mathbf{E}_{ln}, \mathscr{NL}_L(\lambda_n)\mathbf{E}_{rn}\rangle} \mathbf{E}_{rn}$$

where the inner product in the denominator can be computed explicitly as follows:

$$\langle \mathbf{E}_{ln}, \mathscr{NL}_L(\lambda_n)\mathbf{E}_{rn}\rangle =$$

$$\int_\Omega \left[\overline{\mathbf{E}_{ln}} \cdot \left(\nabla \times ((\boldsymbol{\mu}^{-1}(\lambda_n))'\nabla \times \mathbf{E}_{rn})\right) + \overline{\mathbf{E}_{ln}} \cdot \left((\lambda_n^2 \boldsymbol{\varepsilon}(\lambda))'\mathbf{E}_{rn}\right)\right] d\Omega.$$

What remains is to figure out the value of $N_D$ and $N_N$ of the operator $\mathscr{NL}(\lambda)$ given by (5.22). In particular, by Proposition 5.2.1, we can show that:

$$N_N = N_D + 2,$$

which implies $\rho \in \{0, 1\}$.

Finally, we achieve the final expansion for electric field on the eigen-triplets $(\lambda_n, \langle\mathbf{E}_{ln}|, |\mathbf{E}_{rn}\rangle)$:

$$\mathbf{E} = \sum_n \frac{f_\rho(\lambda_n)}{f_\rho(\lambda)} \frac{\lambda}{\lambda - \lambda_n} \frac{\int_\Omega \overline{\mathbf{E}_{ln}} \cdot \mathbf{J}}{\int_\Omega \left[\overline{\mathbf{E}_{ln}} \cdot \left(\nabla \times ((\boldsymbol{\mu}^{-1}(\lambda_n))'\nabla \times \mathbf{E}_{rn})\right) + \overline{\mathbf{E}_{ln}} \cdot \left((\lambda_n^2 \boldsymbol{\varepsilon}(\lambda))'\mathbf{E}_{rn}\right)\right] d\Omega} \mathbf{E}_{rn}. \qquad (5.25)$$

*Expansion of electric fields for non-linear operators 2*

where $f_\rho(\lambda)$ is any arbitrary polynomial such that $\rho \in \{0, 1\}$. This means $f_\rho(\lambda)$ would take the form $f_\rho(\lambda) = \alpha + \lambda\beta$ with $\forall \alpha, \beta \in \mathbf{C}$. As a consequence, there are indeed infinite amount of expansion formulas for the electric field $\mathbf{E}$ of the operator $\mathscr{NL}(\lambda) = \nabla \times \left(\boldsymbol{\mu}^{-1}(\lambda)\nabla \times \cdot\right) + \lambda^2 \boldsymbol{\varepsilon}(\lambda)$

Before ending this section, it is worth verifying the relation between (5.21) and (5.25).

**Proposition 5.3.1** *By choosing* $f_\rho(\lambda) = \lambda$, *(5.25) implies (5.21).*

Set $f_\rho(\lambda) = \lambda$, it directly follows that:

$$\mathbf{E} = \sum_n \frac{1}{\lambda - \lambda_n} \underbrace{\frac{\int_\Omega \overline{\mathbf{E}_{ln}} \cdot \mathbf{J}\, d\Omega}{\lambda_n^{-1}\int_\Omega \left[\overline{\mathbf{E}_{ln}} \cdot \left(\nabla \times (\boldsymbol{\mu}^{-1}(\lambda_n))'\nabla \times \mathbf{E}_{rn}\right) + \overline{\mathbf{E}_{ln}} \cdot \left((\lambda_n^2 \boldsymbol{\varepsilon}(\lambda_n))'\mathbf{E}_{rn}\right)\right] d\Omega}}_{=D} \mathbf{E}_{rn}.$$

Let us denote by $D$ the denominator part of the previous equation as the above notation:

$$D := \int_\Omega \left[ \overline{\mathbf{E}_{ln}} \cdot \left( \lambda_n^{-1} \nabla \times (\mu^{-1}(\lambda_n))' \nabla \times \mathbf{E}_{rn} \right) \right.$$
$$\left. + \overline{\mathbf{E}_{ln}} \cdot \left( \lambda_n^{-1} (\lambda_n^2 \varepsilon(\lambda_n))' \mathbf{E}_{rn} \right) \right] d\Omega.$$

Doing integration by parts yields:

$$D = \int_\Omega \left( \nabla \times \overline{\mathbf{E}_{ln}} \right) \cdot \left[ \lambda_n \left( \mu^{-1}(\lambda_n) \right)' \nabla \times \mathbf{E}_{rn} \right]$$
$$+ \overline{\mathbf{E}_{ln}} \cdot \left[ \lambda_n^{-1} \left( \lambda_n^2 \varepsilon(\lambda_n) \right)' \mathbf{E}_{rn} \right] d\Omega + \text{b.t.},$$

where the boundary term b.t. vanishes thanks to the boundary condition.

The following step is to apply the matrix derivative formula $(\mu(\lambda_n)^{-1})' = -\mu(\lambda_n)^{-1}\mu'(\lambda_n)\mu(\lambda_n)^{-1}$ to the first term of D to get:

$$D = \int_\Omega - \left( \mu^{-\mathsf{T}}(\lambda_n) \nabla \times \overline{\mathbf{E}_{ln}} \right) \cdot \left[ \lambda_n^{-1} \mu'(\lambda_n) \mu(\lambda_n) \nabla \times \mathbf{E}_{rn} \right]$$
$$+ \overline{\mathbf{E}_{ln}} \cdot \left[ \lambda_n^{-1} \left( \lambda_n^2 \varepsilon(\lambda_n) \right)' \mathbf{E}_{rn} \right] d\Omega.$$

From Maxwell equations for both $\mathbf{E}_{ln}$ and $\mathbf{E}_{rn}$:

$$\begin{cases} \nabla \times \overline{\mathbf{E}_{ln}} & = \lambda_n \mu^{\mathsf{T}}(\lambda_n) \overline{\mathbf{H}_{ln}} \\ \nabla \times \mathbf{E}_{rn} & = \lambda_n \mu(\lambda_n) \mathbf{H}_{rn} \end{cases}, \qquad (5.26)$$

The first equation for eletromagnetic 'left' eigenfield in (5.26) is deduced by taking the complex conjugate of (5.24).

we have:

$$D = \int_\Omega -\overline{\mathbf{H}_{ln}} \cdot \left( \lambda_n \mu'(\lambda_n) \mathbf{H}_{rn} \right) + \overline{\mathbf{E}_{ln}} \cdot \left[ \lambda_n^{-1} \left( \lambda_n^2 \varepsilon(\lambda_n) \right)' \mathbf{E}_{rn} \right] d\Omega$$
$$= \int_\Omega -\overline{\mathbf{H}_{ln}} \cdot \left( \lambda_n \mu'(\lambda_n) \mathbf{H}_{rn} \right)$$
$$+ \underbrace{\overline{\mathbf{E}_{ln}} \cdot (\varepsilon(\lambda_n) \mathbf{E}_{rn})}_{D_2} + \overline{\mathbf{E}_{ln}} \cdot \left[ (\lambda_n \varepsilon(\lambda_n))' \mathbf{E}_{rn} \right] d\Omega. \qquad (5.27)$$

The second line is computed using the formula $(fg)' = f'g + fg'$

Let $D_2$ denote the second term of (5.27). From (5.26), the following deduction is true:

$$D_2 := \int_\Omega \overline{\mathbf{E}_{ln}} \cdot (\varepsilon(\lambda_n) \mathbf{E}_n) \, d\Omega = -\int_\Omega \overline{\mathbf{E}_{ln}} \cdot \left( \lambda_n^{-1} \nabla \times \mathbf{H}_{rn} \right) d\Omega$$
$$= -\int_\Omega \left( \lambda_n^{-1} \nabla \times \overline{\mathbf{E}_{ln}} \right) \cdot \mathbf{H}_{rn} \, d\Omega + \text{b.t.}$$
$$= -\int_\Omega \left( \mu^{\mathsf{T}}(\lambda_n) \mathbf{H}_{ln} \right) \cdot \mathbf{H}_{rn} \, d\Omega = \int_\Omega \mathbf{H}_{ln} \cdot \left( \mu(\lambda_n) \mathbf{H}_{rn} \right) d\Omega.$$

where the boundary term b.t. $= \int_\Gamma (n \times \mathbf{E}_{ln}) \cdot \mathbf{H}_{rn} \, dS$ can be omitted thanks to appropriate boundary conditions

The final step is to substitute $D_2$ into (5.27) in order to obtain:

$$D = \int_\Omega -\overline{\mathbf{H}_{ln}} \cdot (\lambda_n \boldsymbol{\mu}'(\lambda_n) \mathbf{H}_{rn}) - \overline{\mathbf{H}_{ln}} \cdot (\mu(\lambda_n) \mathbf{H}_{rn}) + \overline{\mathbf{E}_{ln}} \cdot \left[ (\lambda_n \boldsymbol{\varepsilon}(\lambda_n))' \mathbf{E}_{rn} \right] d\Omega$$

$$= \int_\Omega -\overline{\mathbf{H}_{ln}} \cdot \left[ (\lambda_n \boldsymbol{\mu}(\lambda_n))' \mathbf{H}_{rn} \right] + \overline{\mathbf{E}_{ln}} \cdot \left[ (\lambda_n \boldsymbol{\varepsilon}(\lambda_n))' \mathbf{E}_{rn} \right] d\Omega$$

The second equality is established by combining the magnetic terms together, which also proves that (5.25):

$$\mathbf{E} = \sum_n \frac{f_P(\lambda_n)}{f_P(\lambda)} \frac{\lambda}{\lambda - \lambda_n} \frac{\int_\Omega \overline{\mathbf{E}_{ln}} \cdot \mathbf{J}}{\int_\Omega \left[ \overline{\mathbf{E}_{ln}} \cdot \left( \nabla \times ((\boldsymbol{\mu}^{-1}(\lambda_n))' \nabla \times \mathbf{E}_{rn}) \right) + \overline{\mathbf{E}_{ln}} \cdot \left( (\lambda_n^2 \boldsymbol{\varepsilon}(\lambda))' \mathbf{E}_{rn} \right) \right] d\Omega} \mathbf{E}_{rn}$$

implies (5.21):

$$\mathbf{E} = \sum_n \frac{1}{\lambda - \lambda_n} \frac{\int_\Omega \mathbf{E}_{rn} \cdot \mathbf{J} \, d\Omega}{\int_\Omega \left[ \mathbf{E}_{rn} \cdot ((\lambda_n \boldsymbol{\varepsilon}(\lambda_n))' \mathbf{E}_{rn}) - \mathbf{H}_{rn} \cdot \left( (\lambda_n \boldsymbol{\mu}(\lambda_n))' \mathbf{H}_{rn} \right) \right] \, d\Omega} \mathbf{E}_{rn}.$$

## 5.4 The Keldysh theorem

In this chapter, with the assumption that all the eigenvalues are simple, we have succeeded formalize the spectral expansion for 'diagonalizable' rational operators Lemma 5.1.1. In practice, it is much more useful to reduce our electromagnetic problem from a infinite-dimensional function space to a finite matrix level. In particular, most of all the numerical computations in photonics are executed in finite-dimensional spaces. [5]

In finite-dimensional spaces, on the path of seeking the formulation of spectral expansion, it is worth noticing the existence of Keldysh's theorem. (The construction in this subsection is due to [26, 46] ).

For the sake of simplicity, we will state below a simple version of the Keldysh theorem for the operators $\mathcal{T}(\lambda)$:

> **Theorem 5.4.1** (Keldysh theorem) *Given a domain $D \subset \mathbb{C}$ and an interger $m > 1$, let $C \subset D$ be a compact subset, let $\mathcal{T}$ be a matrix-valued function $\mathcal{T} : D \to \mathbb{C}^{m \times m}$ analytic in $D$ and let $n(C)$ denote the number of eigenvalues of $\mathcal{T}$ in $C$. Let $\lambda_k : k = 1, \ldots, n(C)$ denote these eigenvalues and suppose that all of them are simple. Let $\langle \boldsymbol{w}_k |$ and $| \boldsymbol{v}_k \rangle$ denote their 'left' and 'right' eigenvectors:*
>
> $$\langle \boldsymbol{w}_k | \mathcal{T}(\lambda_k) = 0 \quad \mathcal{T}(\lambda_k) | \boldsymbol{v}_k \rangle = 0$$
>
> *Then there is a neighborhood $U$ of $C$ and a matrix-valued analytic function $R : U \to \mathbb{C}^{m \times m}$ such that the resolvent $\mathcal{T}(z)^{-1}$ can be written as*
>
> $$\mathcal{T}(z)^{-1} = \sum_{k=1}^{n(C)} \frac{1}{(z - \lambda_k)} \frac{| \boldsymbol{w}_k \rangle \langle \boldsymbol{v}_k |}{\langle \boldsymbol{w}_k, \dot{\mathcal{T}}(\lambda_k) \boldsymbol{v}_k \rangle} + R(z)$$
>
> *for all $z \in U \backslash \{ \lambda_1, \ldots, \lambda_{n(C)} \}$*

Pay attention that if $\mathcal{T}(\lambda)$ is a matrix-valued strictly proper rational function, $R(z) = 0$.

Our purpose is to deduce the spectral expansion for the operator $\mathcal{T}(\lambda)$ using Theorem 5.4.1. By default, let's denote $\mathbf{u}$ is the solution of the following non-homogeneous problem with the source $\mathbf{S}$:

$$\mathcal{T}(\lambda)\mathbf{u} = \mathbf{S}. \tag{5.28}$$

By applying Theorem 5.4.1 to the operator $\mathcal{T}(\lambda)$, we obtain:

$$\mathcal{T}^{-1}(\lambda) = \sum_n \frac{1}{(\lambda - \lambda_n)} \frac{|\mathbf{v}_n\rangle \langle \mathbf{w}_n|}{\langle \mathbf{w}_n, \dot{\mathcal{T}}(\lambda_n)\mathbf{v}_n \rangle}.$$

Then, by having the operator $\mathcal{T}^{-1}(\lambda)$ act upon $|\mathbf{u}\rangle$ and due to the fact that $\mathbf{u} = \mathcal{T}^{-1}(\lambda)\mathbf{S}$, we achieve the following mathematical formula of spectral expansion:

$$|\mathbf{u}\rangle = \sum_n \frac{1}{(\lambda - \lambda_n)} \frac{\langle \mathbf{w}_n, \mathbf{S} \rangle}{\langle \mathbf{w}_n, \dot{\mathcal{T}}(\lambda_n)\mathbf{v}_n \rangle} |\mathbf{v}_n\rangle, \tag{5.29}$$

which is indeed the expansion for non-linear operator in Lemma 5.1.1.

It is worth noting that the expansion based on Keldysh's theorem, i.e. Theorem 5.4.1, misses the factor $\dfrac{f_\rho(\lambda_n)}{f_\rho(\lambda)}$ in (5.10). Indeed, (5.29) is a simplified version of (5.10) when $f_\rho = 1$. Nevertheless, without such factor, the Keldysh theorem doesn't reflect the non-uniqueness of expansion of non-linear operators.

## 5.5  Tuning out a resonant mode

Thus far, we have already mentioned many different examples of the over-completeness of the set of eigen-solutions of non-linear operators. Still, it is easy to overlook or misunderstand the counter-intuitive-ness of the non-uniqueness of quasinormal modal expansion. Therefore, the last section of this chapter will be devoted to discuss one very interesting property of our non-linear spectral expansion:

> **Remark 5.5.1**  It is possible to tune out the contribution of any resonant mode, i.e eigen-solution from the quasinormal modal expansion formulation of non-linear operators.

Although being counter-intuitive, Remark 5.5.1 is mathematically correct. In order to fully understand it, let's reconsider the quasinormal modal expansion for electric fields in (5.25):

$$\mathbf{E} = \sum_n \frac{f_\rho(\lambda_n)}{f_\rho(\lambda)} \frac{\lambda}{\lambda - \lambda_n} \frac{\int_\Omega \overline{\mathbf{E}_{ln}} \cdot \mathbf{J}}{\int_\Omega \left[ \overline{\mathbf{E}_{ln}} \cdot \left(\nabla \times ((\mu^{-1}(\lambda_n))'\nabla \times \mathbf{E}_{rn})\right) + \overline{\mathbf{E}_{ln}} \cdot \left((\lambda_n^2 \varepsilon(\lambda))'\mathbf{E}_{rn}\right) \right] d\Omega} \mathbf{E}_{rn},$$

where $f_\rho(\lambda)$ can be any arbitrary function such that $f_\rho(\lambda) = \alpha + \lambda\beta$, with $\forall \alpha, \beta \in \mathbf{C}$ and $|\alpha|+|\beta| \neq 0$. Then, among $n$ eigenvalues $\lambda_n$, let's choose specifically an eigenvalue $\tilde{\lambda}$ with a corresponding eivenvector $\tilde{\mathbf{E}}$. The issue arise if we choose the function $f_\rho(\lambda)$ as follows:

$$\alpha = -\tilde{\lambda}\beta. \tag{5.30}$$

Then the factor $f_p(\lambda_n)$ of (5.25) equal 0 when $\lambda_n = \tilde{\lambda}$. This implies that at the eigenvalue $\tilde{\lambda}$, the resonant mode, i.e. the eigenvector $\tilde{\mathbf{E}}$ has no contribution to the expansion of the electric field $\mathbf{E}$.

Moreover, the non-uniqueness of Dispersive Quasinormal Modal (DQNM) expansion also raises a question of which formula of $f_p(\lambda)$ we should choose to perform the expansion. We notice that the expansion for electric field (5.25) will explode at the rate $\frac{1}{\lambda - \lambda_n}$ when $\lambda$ moves close to $\lambda_n$. In practice, this singularity never occurs because $\lambda_n$ are complex resonant frequencies while $\lambda$ only accepts imaginary values. Unfortunately, the problem will raise if we choose, for example, $f_p(\lambda) = \lambda - \tilde{\lambda}$ where $\tilde{\lambda}$ is in the domain of interest. Then the expansion (5.25) will explode at the rate $\frac{1}{(\lambda - \tilde{\lambda})}$ when $\lambda$ moves close to $\tilde{\lambda}$. In fact, we will numerically demonstrate in the next section that the expansion (5.25) is no longer accurate in the vicinity of $\tilde{\lambda}$.

In the beginning of our study of spectral theorem of electromagnetic operators, it is expected to decompose the electromagnetic field of an optical system onto a set of resonant quasinormal modes, i.e. eigenvectors. It is based on an intuitive belief that the optical response of the given structure is the synthesis of the excitation of each physical-resonance state in the system. Unfortunately, that belief has to be reconsidered and developed to the new level of non-linearity. In particular, Remark 5.5.1 allows us to turn off the contribution of any resonant mode no matter how big its magnitude is. This results from the fact that all the quasinormal modes of the system are no longer linear-independent. As a result, modifying any eigen-modes can deeply affect the rest of the spectrum, which in turn, reflects on the spectral expansion of non-linear operators.

The dependence of eigen-solutions of non-linear operators also raises the difficulty in terms of numerical calculations. Technically, in practice, it is compulsory to reduce our infinite-dimensional spaces to finite-dimensional one. Moreover, because of the limitation of computational volume, the sum in the spectral expansion must be also reduced as much as possible. In specific, the most vital question is how to select eigen-modes considering all of them are dependent. As we soon discover in the third part of this thesis, the issue of convergence and well-posedness of spectral expansion is still a big question with no proper answer.

# DETAILS IN NUMERICAL TECHNIQUES

# Finite Element Analysis | 6

In the previous part of this thesis, we have already gone through a profound study of spectral properties for electromagnetic operators. In particular, we reveal the possibility to expand the electromagnetic fields into an infinite sum of eigenstates, solutions of the eigenvalue problem for electromagnetic operators. Based on the spectral characteristics of so-called rational operators, we have theoretically developed the Dispersive Quasinormal Modal (DQNM) Expansion for electromagnetic fields in an anisotropic and dissipative and even dispersive media.

The role of the third (final) part of this thesis is none other than to demonstrate the correctness of conclusions drawn from the previous chapter in practice. This can be done directly by explicitly solving the eigenvalue problem of the chosen continuous infinite-dimensional operator, then analytically taking the infinite summation. Unfortunately, there seem to be two major obstacles: Firstly, justifying the value of the infinite series is not an easy task. Most importantly, the mathematical models of most electromagnetic problems are so complicated that an analytical (closed-form) solution is practically not available. A common solution is to convert the continuous problem into a discrete problem which can be solved explicitly using only arithmetic operations which can be executed by computers.

Among numerical methods, Finite Element Analysis (FEA) has been already well developed and played an important role in studying electromagnetics. The most direct advantages of FEA are the flexibility in applying on a complex mathematical model as well as its comprehensibility: It provides intuitive and physical results, including some which might have been hidden in an analytical approach. In addition, FEA also provides us a useful tool to handle our obstacle of infinity: Our infinite-dimensional operators could be transformed into a finite-dimensional subspace where the expansion is finite.

The main role of this chapter is to make a general introduction of FEA for the electromagnetic wave. This will pave the way for the numerical application of Dispersive Quasinormal Modal (DQNM) Expansion in electromagnetics and play a role as a platform on which later chapters will be constructed. All issues will be discussed around the well-posedness of our finite model and its effects on the implementation of the modal expansion. The second function of this chapter is to introduce the reader to a free finite element solver GetDP [47, 48] , which can be equipped with the eigensolver package SLEPc (Scalable Library for Eigenvalue Problem Computations)[44]. The tools from this software will be widely used throughout the thesis.

## 6.1 Finite Element Analysis for electromagnetics

In this section, we will explore the application of FEA in solving electromagnetic problems. For deeper discussions associated with the general concepts of FEA, we refer the reader to appendix B.

### Electric Vector problem

In order to demonstrate the application and characteristics of Finite Element Analysis, let's consider the following vector problem for electric fields in a closed domain $\Omega$:

$$\nabla \times \left(\boldsymbol{\mu}^{-1}(\lambda)\nabla \times \mathbf{E}\right) + \lambda^2 \boldsymbol{\varepsilon}(\lambda)\mathbf{E} = \mathbf{S} := \lambda \mathbf{J}, \qquad (6.1)$$

with the homogeneous Dirichlet boundary condition $\mathbf{n} \times \mathbf{E}|_{\partial\Omega} = 0$.

By adapting the notation from the previous chapter, we write the wave operator of this equation as follows:

$$\mathscr{L}_E(\lambda) = \nabla \times \left(\boldsymbol{\mu}^{-1}(\lambda)\nabla \times \cdot\right) + \lambda^2 \boldsymbol{\varepsilon}(\lambda). \qquad (6.2)$$

It is worth reminding that if the media are not reciprocal, i.e. the permittivity and permeability matrices are symmetric, the operator $\mathscr{L}_E(\lambda)$ is not self-adjoint. As the same time, the solution $\mathbf{E}$ must belong to the domain of the operator:

$$D^D_{\mathscr{L}_E} \equiv \left\{\mathbf{E} : \mathbf{E}, \nabla \times (\boldsymbol{\mu}^{-1}\nabla \times \mathbf{E}) \in L^2(\Omega)^3, \mathbf{n} \times \mathbf{E}|_\Gamma = 0\right\}.$$

In order to derive the weak formulation of (6.2) according to the weighted residual method (see appendix B), the next step is to suppress the residual by enforcing the following equation:

$$\langle \mathbf{F}, \mathscr{L}_E(\lambda)\mathbf{E} - \mathbf{S}\rangle = 0 \qquad \forall \mathbf{F}. \qquad (6.3)$$

The question raised is which functional space the weighting function $\mathbf{F}$ belongs to. The immediate response is to impose $\mathbf{F}$ into the domain of the operator: $\mathbf{F} \in D^D_{\mathscr{L}_E}$. The problem is that it requires the second derivative for the field $\mathbf{F}$, which is too strong and complicated for weight functions which are supposed to be simple and easy to implement in computation. The solution is to 'weaken' the differential condition through the following space:

$$H^D_0(\mathbf{curl}, \Omega) := \{\mathbf{F} : \mathbf{F}, \nabla \times \mathbf{F} \in L^2(\Omega)^3, \mathbf{n} \times \mathbf{F}|_\Gamma = 0\}. \qquad (6.4)$$

Then, (6.3) can be rewritten as:

$$\langle \mathbf{F}, \mathscr{L}_E(\lambda)\mathbf{E} - \mathbf{S} \rangle$$

$$= \int_\Omega \overline{\mathbf{F}} \cdot \left[ \nabla \times \left( \boldsymbol{\mu}^{-1}(\lambda)\nabla \times \mathbf{E} \right) + \lambda^2 \boldsymbol{\varepsilon}(\lambda)\mathbf{E} \right] \, d\Omega - \int_\Omega \overline{\mathbf{F}} \cdot \mathbf{S} \, d\Omega$$

$$= \int_\Omega \left( \nabla \times \overline{\mathbf{F}} \right) \cdot \left( \boldsymbol{\mu}^{-1}(\lambda)\nabla \times \mathbf{E} \right) \, d\Omega + \lambda^2 \int_\Omega \overline{\mathbf{F}} \cdot \left( \boldsymbol{\varepsilon}(\lambda)\mathbf{E} \right) \, d\Omega - \int_\Omega \overline{\mathbf{F}} \cdot \mathbf{S} \, d\Omega$$

$$- \int_\Gamma \left( \mathbf{n} \times \overline{\mathbf{F}} \right) \cdot \left( \boldsymbol{\mu}^{-1}(\lambda)\nabla \times \mathbf{E} \right) \, dS.$$

The second equality is obtained using integration by parts $\int_\Omega \overline{\mathbf{x}} \cdot (\nabla \times \mathbf{y}) \, d\Omega = \int_\Omega (\nabla \times \overline{\mathbf{x}}) \cdot \mathbf{y} \, d\Omega - \int_{\partial\Omega} (\overline{\mathbf{x}} \times \mathbf{y}) \cdot \mathbf{n} \, dS.$

If we set $\mathbf{F} \in H_0^D(\mathbf{curl}, \Omega)$, the boundary term b.t. $= -\int_\Gamma \left( \mathbf{n} \times \overline{\mathbf{F}} \right) \cdot \left( \boldsymbol{\mu}^{-1}(\lambda)\nabla \times \mathbf{E} \right) \, dS$ will vanish and leave the weak formulation:

Find $\mathbf{E} \in H_0^D(\mathbf{curl}, \Omega)$ such that:

$$\int_\Omega \left( \nabla \times \overline{\mathbf{F}} \right) \cdot \left( \boldsymbol{\mu}^{-1}(\lambda)\nabla \times \mathbf{E} \right) \, d\Omega + \lambda^2 \int_\Omega \overline{\mathbf{F}} \cdot \left( \boldsymbol{\varepsilon}(\lambda)\mathbf{E} \right) \, d\Omega - \int_\Omega \overline{\mathbf{F}} \cdot \mathbf{S} \, d\Omega = 0$$

$$\forall \mathbf{F} \in H_0^D(\mathbf{curl}, \Omega) \quad (6.5)$$

We emphasize that the previous equation (6.5) is only the weak version of (6.1). In particular, there is nothing guarantee that the solution $\mathbf{E} \in H_0^D(\mathbf{curl}, \Omega)$ of (6.5) satisfies the differential condition of the operator $\mathscr{L}_E(\lambda)$: $\mathbf{E} \in D_{\mathscr{L}_E}^D$.[1]

It is worth pointing out that since the value of $\boldsymbol{\varepsilon}(\lambda)$ and $\boldsymbol{\mu}(\lambda)$ is not strictly positive: both $\boldsymbol{\varepsilon}(\lambda)$ and $\boldsymbol{\mu}(\lambda)$ are complex function with respect to $\lambda$. Thus, it may be practically impossible to prove the Hermitian property as well as the coercivity of the sesqui-linear form $a(\cdot, \cdot)$:

$$a(\mathbf{F}, \mathbf{E}) = \int_\Omega \left( \nabla \times \overline{\mathbf{F}} \right) \cdot \left( \boldsymbol{\mu}^{-1}(\lambda)\nabla \times \mathbf{E} \right) + \lambda^2 \overline{\mathbf{F}} \cdot \left( \boldsymbol{\varepsilon}(\lambda)\mathbf{E} \right) \, d\Omega, \qquad (6.6)$$

at everywhere in the domain of interest $\Omega$. It will raise the new questions about the well-posedness of the FEA schemes since the Lax-Milgram theorem is no longer valid.[2]

It is also important to note that building the weak formulation is just an intermediate step of Finite Element Modeling. The last step is to choose the finite element subspace $V_h \subset H_0^D(\mathbf{curl}, \Omega)$ in which our solution $\mathbf{E} \in V_h$ is approximated by the expansion:

$$\tilde{\mathbf{E}} = \sum_{i=1}^m c_i \phi_i,$$

where $\text{span}\{\phi_i\} = (V_h)$.[3]

Thus, studying the weak equation (6.5) implies building and solving the matrix system

$$\tilde{\mathscr{L}}_E \cdot c = \ell.$$

1: This indirectly disqualifies the capability to effectively consider each Cartesian component of the vector field as a scalar function using classical elements where the degrees of freedom are nodal values. The further information will be provided in Section 6.3 about the discrete finite element space for vector fields.

2: We will discuss further this ill-posedness property in Chapter 8.

3: The nature of the shape functions $\phi_i$ will be discussed more in Section 6.3.

The matrix $\tilde{\mathscr{L}}_E$ is given by:

$$\tilde{\mathscr{L}}_E = \begin{pmatrix} a(\phi_1, \phi_1) & a(\phi_1, \phi_2) & \dots & a(\phi_1, \phi_m) \\ a(\phi_2, \phi_1) & a(\phi_2, \phi_2) & \dots & a(\phi_2, \phi_m) \\ \vdots & \vdots & \ddots & \vdots \\ a(\phi_m, \phi_1) & a(\phi_m, \phi_2) & \dots & a(\phi_m, \phi_m) \end{pmatrix}, \qquad (6.7)$$

with the sesqui-linear form $a(\cdot, \cdot)$ given by (6.6); and according to (B.4):

$$c = \begin{pmatrix} c_1 \\ c_2 \\ \vdots \\ c_m \end{pmatrix} \qquad \ell = \begin{pmatrix} l(\phi_1) \\ l(\phi_2) \\ \vdots \\ l(\phi_m) \end{pmatrix}.$$

As a result, the infinite-dimensional problem (6.1) is approximated by a matrix system with the finite-dimensional operator (6.7). The spectral properties of the original infinite-dimensional operator $\mathscr{L}_E(\lambda)$ (6.2) in the previous chapter would be transferred to its finite-dimensional counterpart $\tilde{\mathscr{L}}_E(\lambda)$. In the most of the numerical applications, it is implicitly understood that the infinite-dimensional operator $\mathscr{L}_E(\lambda)$ can be discretized and written in terms of a finite-dimensional matrix whose range is given by the degree of freedom $m = \dim(V_h)$: The infinite sum in Lemma 5.1.1, in turn, can be converted into a finite sum according to the degree of freedom $m$.

## Magnetic vector problem

All the previous arguments in the last subsection for electric fields can be extended for the magnetic fields $\mathbf{H}$. Consider the magnetic wave equation:

$$\nabla \times \left( \boldsymbol{\varepsilon}^{-1}(\lambda) \nabla \times \mathbf{H} \right) + \lambda^2 \boldsymbol{\mu}(\lambda) \mathbf{H} = \mathbf{S}_H := \nabla \times \left( \boldsymbol{\varepsilon}^{-1}(\lambda) \mathbf{J} \right), \qquad (6.8)$$

using the following operator:

$$\mathscr{L}_H(\lambda) = \nabla \times \left( \boldsymbol{\varepsilon}^{-1}(\lambda) \nabla \times \cdot \right) + \lambda^2 \boldsymbol{\mu}(\lambda).$$

However, caution must be exercised in order for (6.8) to provide the equivalent solution to (6.1) when converting from magnetic to electric fields. In particular, if the homogeneous Dirichlet boundary condition for electric fields is imposed on the boundary $\Gamma = \partial\Omega$, the homogeneous Neumann boundary condition must be enforced on $\Gamma$ in the case of magnetic fields. As a result, the domain of $\mathscr{L}_H(\lambda)$ is given by:

$$D^N_{\mathscr{L}_H} \equiv \left\{ \mathbf{H} : \mathbf{H}, \nabla \times (\boldsymbol{\varepsilon}^{-1} \nabla \times \mathbf{H}) \in L^2(\Omega)^3, \mathbf{n} \times (\boldsymbol{\mu}^{-1} \nabla \times \mathbf{H})|_\Gamma = 0 \right\}.$$

Then, the weak formulation for the wave equation is obtained as follows:

Find $\mathbf{H} \in H_0^N(\mathbf{curl}, \Omega)$ such that:

$$\int_\Omega \left(\nabla \times \overline{\mathbf{F}}\right) \cdot \left(\boldsymbol{\varepsilon}^{-1}(\lambda)\nabla \times \mathbf{H}\right) \, d\Omega + \lambda^2 \int_\Omega \overline{\mathbf{F}} \cdot \left(\boldsymbol{\mu}(\lambda)\mathbf{H}\right) d\Omega$$

$$- \int_\Omega \overline{\mathbf{F}} \cdot \mathbf{S}_H \, d\Omega = 0 \qquad \forall \mathbf{F} \in H_0^N(\mathbf{curl}, \Omega) \quad (6.9)$$

where

$$H_0^N(\mathbf{curl}, \Omega) := \left\{ \mathbf{F} : \mathbf{F}, \nabla \times \mathbf{F} \in L^2(\Omega)^3, \mathbf{n} \times (\boldsymbol{\mu}^{-1}\nabla \times \mathbf{F})|_\Gamma = 0 \right\}.$$

## Scalar problems

In practice, it is not strictly compulsory to handle the wave equation (6.1) in the vector form. In the case where the material is invariant along one axis, it is possible to simplify the wave equation into two cases:

▶ s-polarization: The electric field is along the axis of invariance.
▶ p-polarization: The magnetic field is along the axis of invariance, which is orthogonal to the electric field.

Let's consider the p-polarization as an example. In such case, the electromagnetic can be decomposed into two separate terms: $\mathbf{H} = H\mathbf{z}$ and $\mathbf{E} = E_x\mathbf{x} + E_y\mathbf{y}$ (see Figure 6.1). As a result, the 3-D problem can be considered as a 2-D equation where the perpendicular out-of-plane magnetic field is given by a scalar quantity:

$$\mathscr{L}_h(\lambda)H := -\nabla \cdot \left(\varepsilon^{-1}\nabla H\right) + \lambda^2(\mu H) = S, \qquad (6.10)$$

where the scalar quantities $\varepsilon^{-1}$ and $\mu$ are given by:

$$\varepsilon^{-1} = \frac{\boldsymbol{\varepsilon}^\mathsf{T}}{\det(\boldsymbol{\varepsilon})} \quad \text{and} \quad \mu = \boldsymbol{\mu}_{zz},$$

Next, we define the Sobolev space for the scalar magnetic field:

$$H_0^N(\mathrm{grad}, \Omega) := \{F : F, \nabla F \in L^2(\Omega), n \cdot (\nabla F)|_\Gamma = 0\}$$

According to (B.3), the weak formulation of (6.10) is deduced as follows:

$$\langle F, \mathscr{L}_h(\lambda)H - S \rangle = \int_\Omega \overline{F} \left[-\nabla \cdot \left(\varepsilon^{-1}(\lambda)\nabla H\right) + \lambda^2\mu(\lambda)H\right] \, d\Omega - \int_\Omega \overline{F}S \, d\Omega$$

$$= \int_\Omega \varepsilon^{-1}(\lambda) \left(\nabla \overline{F}\right) \cdot (\nabla H) \, d\Omega + \lambda^2 \int_\Omega \mu(\lambda)\overline{F}H \, d\Omega$$

$$- \int_\Omega \overline{F}S \, d\Omega - \int_\Gamma \varepsilon^{-1}(\lambda)\overline{F} \left(\mathbf{n} \cdot \nabla H\right) \, dS$$

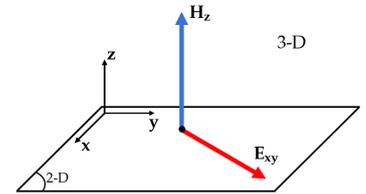

**Figure 6.1:** The magnetic and electric fields in the p-polarized case.

The equality is obtained using $\nabla \cdot (ab) = a(\nabla \cdot b) + b \cdot (\nabla a)$.

Since $H \in H_0^N(\mathrm{grad}, \Omega)$, the boundary term

$$\mathrm{b.t.} = -\int_\Gamma \varepsilon^{-1}(\lambda)\overline{F} \left(\mathbf{n} \cdot \nabla H\right) \, dS$$

will vanish. What remains is the weak formulation:

Find $H \in H_0^N(\text{grad}, \Omega)$ such that:

$$\int_\Omega \varepsilon^{-1}(\lambda) \left(\nabla \overline{F}\right) \cdot (\nabla H) \, d\Omega + \lambda^2 \int_\Omega \mu(\lambda) \overline{F} H \, d\Omega - \int_\Omega \overline{F} S \, d\Omega = 0$$

$$\forall F \in H_0^N(\text{grad}, \Omega) \quad (6.11)$$

As we will explain in Section 6.3, the implementation of the weak formulation of the vector problems (for example (6.5) and (6.9)) under the enforcement of the Sobolev space ($H_0^D(\text{curl}, \Omega)$ and $H_0^N(\text{curl}, \Omega)$ respectively) requires the *curl-conforming* element, i.e. edge elements. However, in the scalar case (6.11), it can be shown that the nodal elements are sufficient to approximate the solution of (6.10) in the finite dimensional subspace $H_0^N(\text{grad}, \Omega)$.

## The finite element solver GetDP

After analytically formulating the discrete problem for our electromagnetic problem, the next issue is how to put Finite Element Analysis (FEA) into practice. There are a number of available software products based on the FEA, such as ANSYS, COMSOL, and MultiPhysics have traditionally been used to model electromagnetic problems. Unfortunately, the main disadvantages of these software systems are the closed source code and high costs. Since the source code is closed, we face the inability to change the configuration of the software for complicated tasks, which may cloud our numerical studies and verification. This leads us to open-source software products with the capability of changing the mathematical models of the investigated processes and a wider choice of algorithms and methods for solving the given problem. A good example of such open-source software is GetDP (`http://getdp.info`) with the support of Gmsh (`http://gmsh.info`), a three-dimensional finite element mesh generator [47, 49] .

An important advantage of GetDP is the visual closeness between the mathematical expression of the electromagnetic problem, i.e. the weak formulation, and the input data defining discrete problems which are written by the user in an ASCII text file. For example, the weak formulation of the vector electric fields (6.5) can be translated into GetDP in the form of an ASCII text as follows:

```
1  Equation {
2      Galerkin { [Inv[mu[]]         *Dof{dE}, {dE}]; In Omega; Jacobian JVol
         ; Integration Int_1;  }
3      Galerkin { [lambda^2*epsilon[]*Dof{E} , {E} ]; In Omega; Jacobian JVol
         ; Integration Int_1;  }
4      Galerkin { [-source[]                 , {E} ]; In Omega; Jacobian JVol
         ; Integration Int_1;  }
5  }
```

**Listing 6.1:** Syntax for the formulation of the electric scattering electric problem.

Besides, there are 2 main features dictating why GetDP is chosen to implement FEA in this thesis:

▶ The versatility of functional space: We can define the basis functions and conditions with their supporting elements simply by listing in the input text file. This allows us to easily construct the discrete finite element space.

▶ The support of open source toolkits: The software can be configured to work with scientific toolkits such as PETSc and SLEPc. For example, the software library in SLEPc provides a variety of solvers for non-linear eigenvalue problem, which plays an essential role in our study of the non-linear electromagnetic operators.

This chapter can be considered as a short guidebook on how to leverage GetDP and related calculation packages on the applications of modal expansion in electromagnetics. Throughout this chapter, we will learn how to construct an electromagnetic problem in GetDP as well as look over several eigensolvers provided in the SLEPc toolkit.

In practice, the discrete problem, which is defined in .pro input files, contains two main input data:

▶ Those including specific data of the given problem, such as geometry, physical characteristics and boundary conditions (i.e., the `Group`, `Function` and `Constraint` objects).

▶ Those defining a resolution method, such as unknowns, equations and related objects (i.e., the `Jacobian`, `Integration`, `FunctionSpace`, `Formulation`, `Resolution` and `PostProcessing` objects).[4]

We will focus on the construction of the `FunctionSpace`, `Formulation` and `Resolution` stages as the main structures to build the discrete problem.

4: For example, the discrete weak formulation (as shown in Listing 6.1) is defined in the `Formulation`.

## 6.2 The eigenvalue problems

The aim of this section is to derive the weak formulation of non-linear eigenvalue problems for our electromagnetic operators and briefly discuss the software library for the solution of these eigenvalue problems.

Firstly, it is worth noting that given the time-harmonic property, all the time dependence of the electromagnetic operators is embedded into the value $\lambda = i\omega$. And since the operators do not involve time derivatives, our numerical system would not explicitly depend on time. In terms of Finite Element, it implies all the degrees of freedom are constants with respect to time and the finite problems will only contain algebraic equations.

Thus, it is a straightforward process to solve the eigenvalue problem for the given operators, which involves looking for the non-zero solutions, i.e. eigenvectors (eigenfields), for appropriate frequencies, i.e eigenvalues, without any source. In the Finite Element Analysis, it means solving the eigenvalue problem of the matrix system $\tilde{\mathscr{L}}(\lambda)$ that is equivalent to the infinite-dimensional wave operators $\mathscr{L}(\lambda)$.

As discussed in Chapter 4 and Chapter 8, the eigenvalue problem of the matrix $\tilde{\mathscr{L}}$ is non-linear with the presence of the frequency-dependent dispersive materials. In particular, the frequency-dispersive property of the permittivity can be generally characterized by an arbitrary rational function:

$$\varepsilon(\lambda) = \frac{\mathscr{N}(\lambda)}{\mathscr{D}(\lambda)},$$

where $\mathcal{N}(\lambda)$ and $\mathcal{D}(\lambda)$ are set to be a polynomial operator degree $N_N$ and $N_D$ respectively.

$$\mathcal{N}(\lambda) = \sum_{i=1}^{N_N} n_i(\lambda)^i \quad \text{and} \quad \mathcal{D}(\lambda) = \sum_{i=1}^{N_D} d_i(\lambda)^i.$$

It is worth reminding that all the coefficients $n_i$ and $d_i$ are real due to the causality property. [5]

### non-linear eigenvalue problems for vector electric fields

Given the electric wave operator according to (6.2)

$$\mathscr{L}_E(\lambda) = \nabla \times \left( \mu^{-1}(\lambda) \nabla \times \cdot \right) + \lambda^2 \varepsilon(\lambda)$$

with the homogeneous Dirichlet boundary condition $\mathbf{n} \times \mathbf{E}|_{\partial\Omega} = 0$, the weak form of the eigenvalue problem is defined as follows:

> Find $(\lambda_n, \mathbf{E}_n) \in \mathbb{C} \times H_0^D(\mathbf{curl}, \Omega)$ such that:
>
> $$\int_\Omega \left( \nabla \times \overline{\mathbf{F}} \right) \cdot \left( \mu^{-1}(\lambda_n) \nabla \times \mathbf{E}_n \right) \, d\Omega + \lambda_n^2 \int_\Omega \overline{\mathbf{F}} \cdot (\varepsilon(\lambda_n) \mathbf{E}_n) \, d\Omega = 0$$
> $$\forall \mathbf{F} \in H_0^D(\mathbf{curl}, \Omega). \quad (6.12)$$

We notice that the previous eigenvalue problem is not linear since both $\varepsilon(\lambda_n)$ and $\mu(\lambda_n)$ depend on the eigenvalue $\lambda_n = i\omega_n$. For the sake of simplicity, in all the numerical examples in this thesis, let's assume that all the materials are isotropic and the permeability is independent of the frequency $\omega$ as well as eigenvalue $\lambda = i\omega$.

Next, we assume that our domain of interest $\Omega$ is filled with air $\varepsilon_a$ and contains $k$ number of subdomain $\Omega_j$ whose permittivity is given by $\varepsilon_j(\lambda) = \frac{\mathcal{N}_j(\lambda)}{\mathcal{D}_j(\lambda)}$ with $j \in \{1, 2, \ldots, k\}$ (see Figure 6.2). Then, (6.12) leads to a rational eigenvalue problem:

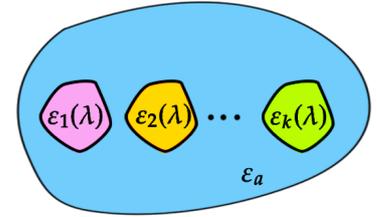

**Figure 6.2:** A closed structure $\Omega$ containing $k$ subdomain $\Omega_j$ with permittivity $\varepsilon_j$

> **Rational eigenvalue problem for vector electric fields**
>
> Find $(\lambda_n, \mathbf{E}_n) \in \mathbb{C} \times H_0^D(\mathbf{curl}, \Omega)$ such that: $\forall \mathbf{F} \in H_0^D(\mathbf{curl}, \Omega)$
>
> $$\int_\Omega \left( \nabla \times \overline{\mathbf{F}} \right) \cdot \left( \mu^{-1} \nabla \times \mathbf{E}_n \right) \, d\Omega + \lambda_n^2 \int_{\Omega_a} \overline{\mathbf{F}} \cdot (\varepsilon_a \mathbf{E}_n) \, d\Omega$$
> $$+ \lambda_n^2 \sum_{j=1}^{k} \left( \frac{\mathcal{N}_j(\lambda_n)}{\mathcal{D}_j(\lambda_n)} \int_{\Omega_j} \overline{\mathbf{F}} \cdot (\mathbf{E}_n) \, d\Omega \right) = 0 \quad (6.13)$$

where $\Omega_a = \Omega \backslash \left( \cup_{j=1}^{k} \Omega_j \right)$.

On the other hand, by multiplication by $\mathcal{D}_j(\lambda_n)$, it is possible to rewrite (6.13) into the form of a polynomial eigenvalue problem:

**Polynomial eigenvalue problem for vector electric fields**

Find $(\lambda_n, \mathbf{E}_n) \in \mathbb{C} \times H_0^D(\mathbf{curl}, \Omega)$ such that: $\forall \mathbf{F} \in H_0^D(\mathbf{curl}, \Omega)$

$$\left(\prod_{j=1}^{k} \mathscr{D}_j(\lambda_n)\right) \int_{\Omega} \left(\nabla \times \overline{\mathbf{F}}\right) \cdot \left(\boldsymbol{\mu}^{-1}\nabla \times \mathbf{E}_n\right) \, d\Omega$$

$$+ \lambda_n^2 \left(\prod_{j=1}^{k} \mathscr{D}_j(\lambda_n)\right) \int_{\Omega_a} \overline{\mathbf{F}} \cdot (\boldsymbol{\varepsilon}_a \mathbf{E}_n) \, d\Omega$$

$$+ \lambda_n^2 \sum_{j=1}^{k} \left(\mathscr{N}_j(\lambda_n) \left(\prod_{\substack{h=1 \\ h \neq j}}^{k} \mathscr{D}_h(\lambda_n)\right) \int_{\Omega_j} \overline{\mathbf{F}} \cdot (\mathbf{E}_n) \, d\Omega\right) = 0, \quad (6.14)$$

if $\mathrm{root}(\mathscr{D}_j(\lambda)) \neq \mathrm{root}(\mathscr{D}_h(\lambda))$ for $\forall j, h \in \{1, 2, \ldots, k\}$. In general, the degree of the polynomial eigenvalue problem is $2 + \sum_{j=1}^{k} \deg(\mathscr{D}_j)$.[6]

We have to pay extra attention to the permittivity Drude model where there exists a pole at zero. As a result, the order of our eigenvalue problem will be reduced accordingly. For example, if our numerical model only contains a single Drude material surrounded by air, the final degree of the polynomial is 3.

## Non-linear eigenvalue problem for scalar magnetic fields

Similar to the scattering problem with the source, in many cases, it is more practical to consider the scalar wave equation, which is much simpler to solve compared to vector equations. It can occur when the material is invariant along one axis, for instance $z$, on which one part of the electromagnetic is oriented.

In this subsection, we will illustrate the non-linear eigenvalue problem for the magnetic field in the p-polarization for the operator $\mathscr{L}_h(\lambda)$ (6.10):

$$\mathscr{L}_h(\lambda) := -\nabla \cdot (\varepsilon^{-1}(\lambda)\nabla) + \lambda^2 \mu(\lambda),$$

under the homogeneous Neumann boundary condition $\mathbf{n} \cdot (\nabla H)|_\Gamma = 0$.

Then, the weak form of the eigvenvalue problem for $\mathscr{L}_h(\lambda)$ is given by:

Find $(\lambda_n, H_n) \in \mathbb{C} \times H_0^N(\mathrm{grad}, \Omega)$ such that:

$$\int_{\Omega} \varepsilon^{-1}(\lambda_n) \left(\nabla\overline{F}\right) \cdot (\nabla H_n) \, d\Omega + \lambda_n^2 \int_{\Omega} \mu(\lambda_n)\overline{F}H_n \, d\Omega 0$$

$$\forall F \in H_0^N(\mathrm{grad}, \Omega). \quad (6.15)$$

Similar to the previous subsection, let's assume that all the materials are isotropic and the permeability is independent the eigenvalue $\lambda = i\omega$. Given the domain of interest $\Omega$ includes $k$ number of subdomain $\Omega_j$ whose permittiviy is given by $\varepsilon_j(\lambda)$ with $j \in \{1, 2, \ldots, k\}$, the weak formula (6.15) can be expressed as follows:

**Rational eigenvalue problem for scalar magnetic fields**

Find $(\lambda_n, H_n) \in \mathbb{C} \times H_0^N(\text{grad}, \Omega)$ such that: $\forall F \in H_0^N(\text{grad}, \Omega)$

$$\int_{\Omega_a} \varepsilon_a^{-1}(\lambda_n) \left(\nabla \overline{F}\right) \cdot (\nabla H_n) \ d\Omega + \sum_{j=1}^{k} \left(\frac{\mathscr{D}_j(\lambda_n)}{\mathscr{N}_j(\lambda_n)} \int_{\Omega_a} \left(\nabla \overline{F}\right) \cdot (\nabla H_n) \ d\Omega\right)$$

$$+ \lambda_n^2 \int_{\Omega} \mu(\lambda_n) \overline{F} H_n \ d\Omega = 0. \quad (6.16)$$

Multiplying by $\mathscr{N}_j(\lambda_n)$, (6.16) can be expressed under the form of a polynomial eigenvalue problem:

**Polynomial eigenvalue problem for vector electric fields**

Find $(\lambda_n, H_n) \in \mathbb{C} \times H_0^N(\text{grad}, \Omega)$ such that: $\forall F \in H_0^N(\text{grad}, \Omega)$

$$\left(\prod_{j=1}^{k} \mathscr{N}_j(\lambda_n)\right) \int_{\Omega_a} \varepsilon_a^{-1}(\lambda_n) \left(\nabla \overline{F}\right) \cdot (\nabla H_n) \ d\Omega$$

$$+ \sum_{j=1}^{k} \left(\mathscr{D}_j(\lambda_n) \left(\prod_{\substack{h=1 \\ h \neq j}}^{k} \mathscr{N}_h(\lambda_n)\right) \int_{\Omega_a} \left(\nabla \overline{F}\right) \cdot (\nabla H_n) \ d\Omega\right)$$

$$+ \lambda_n^2 \left(\prod_{j=1}^{k} \mathscr{N}_j(\lambda_n)\right) \int_{\Omega} \mu(\lambda_n) \overline{F} H_n \ d\Omega = 0, \quad (6.17)$$

for $\text{root}(\mathscr{D}_j(\lambda)) \neq \text{root}(\mathscr{D}_h(\lambda))$ for $\forall j, h \in \{1, 2, \ldots, k\}$. In general, the degree of the polynomial eigenvalue problem is computed by $2 + \sum_{j=1}^{k} \deg(\mathscr{N}_j)$.

## SLEPc: An eigensolver library of GetDP

In the previous section, we have already mentioned GetDP as an intuitive and robust tool to solve the electromagnetic scattering problem. Hereafter, we will explore its ability to solve the eigenvalue problem through the software library SLEPc (https://slepc.upv.es) [44] . In fact, recent progress in sparse matrix eigensolvers allows us to tackle the discrete eigenvalue problem very efficiently. Depending on the given eigenvalue problem, GetDP can call linear, quadratic, general polynomial, or rational compatible eigenvalue solvers of SLEPc. Since one of the challenges of the research of modal expansion is the ability to solve the non-linear eigenvalue problem, the variety and power of eigensolvers in SLEPc library are essential to study the spectral properties of the given non-linear electromagnetic operators. For more information about the application of GetDP with the library SLEPc in solving non-linear eigenvalue problems in photonics, we refer the reader to [50] .

In this thesis, we will focus on two main classes of non-linear eigensolvers in SLEPc:

▶ The PEP class for Polynomial Eigenvalue Problems: Which is intended for addressing discrete polynomial eigenvalue problems of arbitrary degree $N$: $\mathscr{P}_L^N(\lambda_n)\mathbf{v}_n = 0$

▶ The NEP class for non-linear Eigenvalue Problems: Which covers the general case where the eigenvalue problem is non-linear with respect to the eigenvalue, including the case of rational eigenvalue problems.

Pay attention that the operator $\mathscr{P}_L^N(\lambda_n)$ is finite-dimensional.

In each class of non-linear eigensolvers, i.e. PEP and NEP, there are several methods and algorithms to tackle different kinds of eigenvalue problems, which are free to choose by the user. But overall, most of these solvers are based on the idea of linearization. This implies to convert the non-linear eigenvalue problem to the linear one which can be solved with more traditional linear eigensolvers. [7]

7: In the case of polynomial eigenvalue problem, this process of linearization is similar to what we did in Chapter 4. For example, the $d$-order polynomial eigenvalue problem of dimension $n$ can be linearized and rewritten as a linear eigenproblem problem of dimension $d \times n$.

**NEP class**

Similar to the scattering problem, deducing the input data of the discrete problem in GetDP from the weak formulation of the eigenvalue problem is a straightforward process. As an example, let's begin with the NEP class covering the rational eigenvalue problem, which is directly relevant for our electromagnetic operator involving the permittivity given as a rational function of the eigenvalue. For example, let's consider the weak equation (6.12):

$$\int_{\Omega} \left(\nabla \times \overline{\mathbf{F}}\right) \cdot \left(\boldsymbol{\mu}^{-1}(\lambda_n)\nabla \times \mathbf{E}_n\right) d\Omega + \lambda_n^2 \int_{\Omega} \overline{\mathbf{F}} \cdot \left(\boldsymbol{\varepsilon}(\lambda_n)\mathbf{E}_n\right) d\Omega = 0$$
$$\forall \mathbf{F} \in H_0^D(\mathbf{curl}, \Omega), \quad (6.18)$$

in a closed domain $\Omega$ bounded by the homogeneous Dirichlet boundary condition $\mathbf{n} \times \mathbf{E}|_{\partial\Omega} = 0$. Next, we assume that the geometry of $\Omega$ only have one subdomain $\Omega_1$ whose permittivity is given by the Drude model:

$$\varepsilon_d(\omega) = \varepsilon_\infty - \frac{\omega_p^2}{\omega^2 + i\omega\gamma} = \frac{-\varepsilon_\infty(i\omega)^2 + \varepsilon_\infty\gamma(i\omega) - \omega_p^2}{-(i\omega)^2 + \gamma(i\omega)}. \quad (6.19)$$

Then, we have a rational weak eigenvalue problem

$$\int_{\Omega} \left(\nabla \times \overline{\mathbf{F}}\right) \cdot \left(\boldsymbol{\mu}^{-1}(\lambda_n)\nabla \times \mathbf{E}_n\right) d\Omega + \lambda_n^2 \int_{\Omega_0} \overline{\mathbf{F}} \cdot (\boldsymbol{\varepsilon}_0\mathbf{E}_n) d\Omega = 0$$
$$+ \frac{-\varepsilon_\infty\lambda_n^3 + \varepsilon_\infty\gamma\lambda_n^2 - \omega_p^2\lambda_n}{-\lambda_n + \gamma} \int_{\Omega_1} \overline{\mathbf{F}} \cdot (\mathbf{E}_n) d\Omega$$
$$\forall \mathbf{F} \in H_0^D(\mathbf{curl}, \Omega), \quad (6.20)$$

which can be expressed in the `Formulation` object of GetDP via the following text:

```
Equation {
  Galerkin { Eig[ Inv[mur[]]* Dof{d u}, {d u}]; Rational 1; In Omega  ;
    Jacobian JVol; Integration Int_1;}
  Galerkin { Eig[ epsilon[] * Dof{u}  , {u}  ]; Rational 2; In Omega_0;
    Jacobian JVol; Integration Int_1;}
  Galerkin { Eig[ epsilon[] * Dof{u}  , {u}  ]; Rational 3; In Omega_1;
    Jacobian JVol; Integration Int_1;}
```

**Listing 6.2:** Syntax for the formulation of the rational eigenvalue problem for electric fields.

```
5    }
6  }
```

where the `Rational` function are described as a factor of each `Galerkin` term, which will be specified later in the `Resolution` step.

```
1  Resolution {
2    { Name Spectral;
3      System{{ Name M; NameOfFormulation Rational; Type ComplexValue; }}
4      Operation{
5        GenerateSeparate[M];
6        EigenSolve[M,neig,target_real,target_imag,EigFilter[],
7          { {1}, {1,0,0} , {-epsilon_oo,gamma*epsilon_oo,-omega_p^2,0} } ,
8          { {1}, {1}     , {-1,gamma}                                } ];
9        SaveSolutions[M1];
10      }
11    }
12 }
```

It is worth noticing that the position of each numerator (resp. denominator) in the list of numerators (resp. denominators) corresponds to the tag following the `Rational` keyword. A polynomial numerator (resp. denominator), is represented by a list of floats by decreasing power of $\lambda = i\omega$. For example, the term `Rational 2` in Listing 6.2 is corresponding to the rational term $\frac{\{1,0,0\}}{\{1\}}$ in Listing 6.3, which represents the rational function $\frac{\lambda^2}{1}$.

It is also important to understand that in all GetDP eigenvalue solvers the eigenvalue has been chosen to be $i\omega$, which is derived from the Fourier transform with an $\exp(i\omega t)$ time dependence: $\frac{\partial}{\partial t} \to i\omega$. Although this is consistent with our choice of eigenvalue $\lambda = i\omega$, we have to keep in mind that the time dependence throughout this thesis is $\exp(-i\omega t)$, hence $\frac{\partial}{\partial t} \to -i\omega$.

We also want to point out that many eigensolvers in SLEPc provide the tool to solve the 'left' eigensolutions of discrete operators. Unfortunately, these options, which take away our ability to configure metrics, doesn't show its usefulness in our numerical examples. Henceforth, the 'left' eigensolutions will be computed by simply solving the 'right' eigenvalue problem of the adjoint operators.

**PEP class**

Thanks to the variety of eigensolver classes from SLEPc, it is up the user to call the eigensolver that fits the given problem the best. Hereafter, as a demonstration, we will illustrate how to solve the non-linear eigenvalue problem using the PEP class.

Given the same problem as the previous example (6.18) in a closed domain $\Omega$ containing the subdomain $\Omega_{in}$ whose permitivity is given by the Drude model (6.19), we can deduce the following polynomial

eigenvalue problem:

$$(-\lambda_n + \gamma) \int_{\Omega} \left( \nabla \times \overline{\mathbf{F}} \right) \cdot \left( \boldsymbol{\mu}^{-1}(\lambda_n) \nabla \times \mathbf{E}_n \right) \, d\Omega$$

$$+ (-\varepsilon_\infty \lambda_n^3 + \varepsilon_\infty \gamma \lambda_n^2 - \omega_p^2 \lambda_n) \int_{\Omega_1} \overline{\mathbf{F}} \cdot (\boldsymbol{\varepsilon} \mathbf{E}_n) \, d\Omega$$

$$+ (-\lambda_n^3 + \gamma \lambda_n^2) \int_{\Omega_0} \overline{\mathbf{F}} \cdot (\boldsymbol{\varepsilon} \mathbf{E}_n) \, d\Omega = 0 \qquad \forall \mathbf{F} \in H_0^D(\mathbf{curl}, \Omega).$$

The previous weak polynomial eigenvalue problem can be translated into language of GetDP via the following syntax:

```
Equation {
  Galerkin {Eig[-Inv[mu]]                 * Dof{d u}, {d u}]; Order 1;
    In Omega; Jacobian JVol; Integration Int_1;}
  Galerkin {  [ Inv[mu]]*          gamma* Dof{d u}, {d u}];        ;
    In Omega; Jacobian JVol; Integration Int_1;}
  Galerkin {Eig[-epsilon[]*epsilon_oo      * Dof{u}  , {u}  ]; Order 3;
    In Omega_1; Jacobian JVol; Integration Int_1;}
  Galerkin {Eig[-epsilon[]*epsilon_oo*gamma* Dof{u}  , {u}  ]; Order 2;
    In Omega_1; Jacobian JVol; Integration Int_1;}
  Galerkin {Eig[-epsilon[]*omega_p^2       * Dof{u}  , {u}  ]; Order 1;
    In Omega_1; Jacobian JVol; Integration Int_1;}
  Galerkin {Eig[-epsilon[]                 * Dof{u}  , { u} ]; Order 3;
    In Omega_0; Jacobian JVol; Integration Int_1;}
  Galerkin {Eig[ epsilon[]        * gamma* Dof{u}  , { u} ]; Order 2;
    In Omega_0; Jacobian JVol; Integration Int_1;}
}
```

**Listing 6.4**: Syntax for the formulation of the polynomial eigenvalue problem for electric fields.

Since the previous syntax in the `Formulation` object of GetDP already contained all the important information from the weak polynomial eigenvalue problem, we don't have to specify any additional syntax in the `Resolution` part.

```
Resolution {
  { Name Spectral;
    System{{ Name M; NameOfFormulation Polynomial; Type ComplexValue; }}
    Operation{
      GenerateSeparate[M];
      EigenSolve[M,neig,target_real,target_imag,EigFilter[]];
      SaveSolutions[M1];
    }
  }
}
```

**Listing 6.5**: Syntax for the resolution of the polynomial eigenvalue problem for electric fields.

The general SLEPc options for solving of non-linear problems are preset in the source code of GetDP (see `Kernel/EigenSolve_SLEPC.cpp`). Additional or alternative SLEPc options can be passed as command line argument when calling GetDP. There are particularly relevant options that can be passed to SLEPc:

▶ Target: SLEPc eigensolvers will return nev eigenvalues closest to a given target value. The nev parameter can be specified by the user (1 by default), as well as the target value, that represents a point in the complex plane around which the eigenvalues of interest are located. The values can be provided via with the command line arguments `-pep_nev` (or `-nep_nev`), and `-pep_target` (or `-nep_target`).

▶ Regions: The eigenvalues are returned sorted according to their distance to the target. However, only eigenvalues lying inside the region of interest are returned (in other words, eigenvalues outside the region of interest are discarded). The region of interest (which is a rectangle by default) can be specified by the user via the command line argument `-rg_interval_endpoints` (or any other options related to region specification, see SLEPc documentation [51] for details)

## 6.3 Curl-conforming element - Edge Element

After constructing the weak formulation, the next important step is to specify the finite element subspace, the functional space for the shape functions. For scalar quantities, this is a straightforward task: The shape function is the function that interpolates the solution between the discrete values obtained at the mesh nodes, which is usually chosen as low-order polynomials. The study of nodal elements has been well documented and we refer the reader to [52] for more details.

However, for the vector problems, identifying the finite element space $V_h \subset H_0(\text{curl}, \Omega)$ is a complicated task and must be treated with care. In particular, the wrong choice of finite element space $V_h$ could lead to the spectral pollution caused by spurious modes, which has been reported in [53] . Let's consider the eigenvalue problem of the vector electric field as an example.

The strong form of the eigenvalue problem of the operator $\mathscr{L}_E(\lambda)$ is given as follows: Find $(\lambda_n, \mathbf{E}_n)$ such that

$$\mathscr{L}_E(\lambda_n)\mathbf{E}_n = \nabla \times \left( \mu^{-1}(\lambda_n)\nabla \times \mathbf{E}_n \right) + \lambda_n^2 \varepsilon(\lambda_n)\mathbf{E}_n = 0, \qquad (6.21)$$

with the boundary condition $\mathbf{n} \times \mathbf{E}|_\Gamma = 0$.

It is worth pointing out that the previous problem (6.21) implies the weak equation (6.12) but not the other way around. In detail, the equation implicitly contains the divergence condition $\nabla \cdot (\varepsilon \mathbf{E}) = 0$ while the weak (6.12) form does not. Indeed, by taking divergence of (6.21), we obtain:

$$\nabla \cdot \left[ \nabla \times \left( \mu^{-1}(\lambda_n)\nabla \times \mathbf{E}_n \right) \right] + \lambda_n^2 \nabla \cdot (\varepsilon(\lambda_n)\mathbf{E}_n) = 0$$
$$\Rightarrow \nabla \cdot (\varepsilon(\lambda_n)\mathbf{E}_n) = 0. \qquad (6.22)$$

We know $\nabla \cdot \left[ \nabla \times \left( \mu^{-1}(\lambda_n)\nabla \times \mathbf{E}_n \right) \right] = 0$ since $\nabla \cdot (\nabla \times \mathbf{u}) = 0$ (The divergence of the curl is zero).

The spectral solutions of (6.21) automatically satisfy the divergence condition, which makes these solution naturally physical. Unfortunately, this is not the case for the eigensolutions of (6.12), which only satisfy $\mathbf{E}_n \in H_0(\text{curl}, \Omega)$. In other words, since the term $\left( \mu^{-1}(\lambda_n)\nabla \times \mathbf{E}_n \right)$ is not necessarily continuous, the argument $\nabla \cdot \left[ \nabla \times \left( \mu^{-1}(\lambda_n)\nabla \times \mathbf{E}_n \right) \right] = 0$ in (6.22) no longer holds for the weak eigensolutions of (6.12).

The lack of enforcement of the divergence condition is the reason for spurious solutions while solving the weak eigenvalue problem. One simple solution is to enforce the divergence condition through the finite element space $V_h \subset \{H_0(\text{curl}, \Omega) \cap H(\text{div}^0, \Omega)\}$, where $\mathbf{F} \in H(\text{div}^0, \Omega)\}$ satisfies:

$$\int_\Omega \mathbf{F} \cdot (\nabla q)\, d\Omega = 0 \qquad \forall q \in H_0^1(\Omega).$$

This requires to define a whole family of new basis functions constructed by Nédélec [54, 55] for the finite-dimensional subspace of $H(\mathbf{curl}, \Omega)$, which not only satisfy the divergence condition but also possess tangential components continuous across the element boundary. The details of the general way of finding Nédélec's basis functions in affine coordinates of any order and in any dimension can be found in [56] . In the following subsection, we will illustrate the example of the Nédélec element of the first order in 3-D space (The following construction is adapted from [52] ). The finite element properties of 2-D space are similar.

### First order vector tetrahedral elements

Herein, we consider the first-order Nédélec's Finite Element formulation for a single tetrahedral element. Given a tetrahedron with the barycenter coordinates $(r_1, r_2, r_3, r_4)$ (as illustrated in Figure 6.3), let's consider the vector basis functions given by:

$$\mathbf{w}_{ij} = r_i \nabla r_j - r_j \nabla r_i, \tag{6.23}$$

for $i, j \in \{1, 2, 3, 4\}$ and $i < j$.

It is easy to check that $\mathbf{w}_{ij}$ has :

▶ Zero divergence:

$$\nabla \cdot \mathbf{w}_{ij} = \nabla \cdot (r_i \nabla r_j) - \nabla \cdot (r_j \nabla r_i) = 0.$$

▶ Nonzero curl:

$$\nabla \times \mathbf{w}_{ij} = \nabla \times (r_i \nabla r_j) - \nabla \times (r_j \nabla r_i) = 2(\nabla r_i \times \nabla r_j).$$

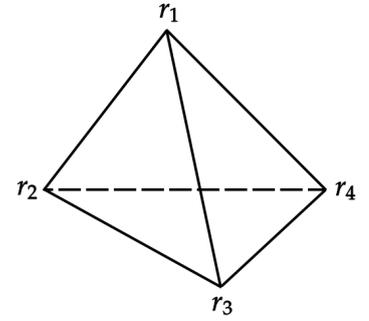

**Figure 6.3:** Tetrahedron with the barycenter coordinates $(r_1, r_2, r_3, r_4)$.

More importantly, it is possible to show that the basis functions maintain tangential continuity across element boundaries. Indeed, given $\mathbf{e}_{ij}$ is the vector pointing from node $i$ to node $j$, the tangential component of the basis function $\mathbf{w}_{ij}$ is given by:

$$\mathbf{e}_{ij} \cdot \mathbf{w}_{ij} = \mathbf{e}_{ij} \cdot (r_i \nabla r_j - r_j \nabla r_i) = (r_i + r_j)/l_{ij} = 1/l_{ij},$$

The second equality is obtained using $\mathbf{e}_{ij} \cdot \nabla r_i = -1/l_{ij}$ and $\mathbf{e}_{ij} \cdot \nabla r_j = 1/l_{ij}$.

where $l_{ij}$ denotes the length of the edge connecting node $i$ and $j$. From here, it is easy to see that $\mathbf{w}_{ij}$ has a constant tangential component along the edge $(r_j, r_j)$ but vanish along the other edges. From here, it is reasonable to redefine the basis function $\mathbf{w}_{ij}$ into $\mathbf{N}_{ij}$:

$$\mathbf{N}_{ij} = \mathbf{w}_{ij} l_{ij},$$

such that the new basis function $\mathbf{N}_{ij}$ is dimensionless.

As the result, the vector field in the given tetrahedral element $(r_1, r_2, r_3, r_4)$ can be approximated by:

$$\mathbf{E}^e = \sum_{\substack{i,j=1 \\ i \neq j}}^{4} \mathbf{N}_{ij} E_{ij}, \tag{6.24}$$

where $E_{ij}$ can be understood as the tangential field along the edge $(r_i, r_j)$ and $\mathbf{E}^e$ represents the approximate field inside the element. By (6.24), we can see there are 6 degrees of freedom that are the line integrals along the 6 edges of the tetrahedron. This is the reason why the first-order Nédélec's element is usually called the edge element and associated with the *curl-conforming* basis functions.

The final approximation of the vector field in the computational domain $\Omega$ is given by:

$$\mathbf{E} = \sum_e \mathbf{E}^e = \sum_e \sum_{\substack{i,j=1 \\ i \neq j}}^{4} \mathbf{N}_{ij} E_{ij}$$

where $\sum_e$ takes into account all the element $e$ in the domain $\Omega$.

It is worth pointing out that the first-order Nédélec elements are exactly the same as the Whitney 1-form element [57] . Similar to the Nédélec family, the elements in the space $W_h^1$ of the Whitney family is chosen such that their normal component is discontinuous while the tangential component is continuous. In particular, it is possible to show that the vectors (6.23) also span the space $W_h^1$ of Whitney element (see [58] for more details).

### Second order vector tetrahedral elements

It is important to emphasize that although the Nédélec space is often referred to as the edge elements, their degrees of freedom are also associated with the faces and volume of the element. Let's consider the second-order Nédélec's basis functions as an illustrative example: There are 12 basis functions associated with the 6 edges of the tetrahedron given by:

$$\mathbf{w}_{ij} = r_i r_j \nabla r_i - r_i^2 \nabla r_j$$
$$\mathbf{w}_{ij} = r_i^2 \nabla r_j - r_i r_j \nabla r_j,$$

for $i, j \in \{1, 2, 3, 4\}$ and $i < j$.

As the same time, there are 8 basis functions associated with the 4 faces of the tetrahedron:

$$\mathbf{w}_{ijk} = r_i r_j \nabla r_k - r_k r_k \nabla r_i$$
$$\mathbf{w}_{ijk} = r_j r_k \nabla r_i - r_k r_i \nabla r_j,$$

for $i, j, k \in \{1, 2, 3, 4\}$ and $i < j < k$.

All 20 degrees of freedom of the second-order Nédélec space are depicted in Figure 6.4. It is clear that the second-order Nédélec requires the contribution of the basis functions not only along the edges but also on the faces of the element. The basis functions associated with the interior of the element will become necessary when the order of Nédélec element reaches 3 [56] .

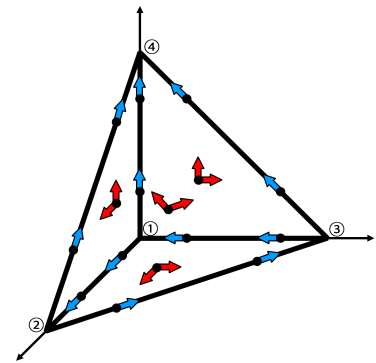

**Figure 6.4:** Degrees of freedom of the second-order Nédélec space for the reference element: Blue arrows refer to the shape functions along the edges while the red arrows indicate the shape functions across the faces of the element.

It is worth reminding that the Nédélec family is not the only way to deduce the higher order 'curl-conforming' finite element. For the developments of other higher order finite elements (for example for Whitney element), we refer the reader to [59, 60] . For example, the expression for the basis vectors of the second-order 1-form Whitney element are given by:

$$\mathbf{w}_{ij} = (8r_i^2 - 4r_i)\nabla r_j + (-8r_i r_j + 2r_j)\nabla r_i$$
$$\mathbf{w}_{ijk} = 16r_i r_j \nabla r_k - 8r_j r_k \nabla r_i - 8r_k r_j \nabla r_i,$$

where $\mathbf{w}_{ij}$ and $\mathbf{w}_{ijk}$ denote the edge and face basis functions respectively. The degrees of freedom of the second-order Whitney 1-form element are illustrated in Figure 6.5.

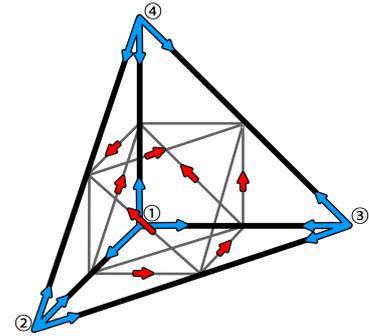

**Figure 6.5:** Degrees of freedom of the second-order Whitney 1-form element.

## Shape function in GetDP

As mentioned at the beginning of this chapter, one of the advantages of GetDp is the flexibility in specifying the finite element subspace, i.e the functional space for the shape functions.

For example, considering the homogeneous Dirichlet boundary condition, the second-order scalar finite element space, which usually refers to the nodal elements, can be set by calling the following syntax:

```
FunctionSpace {
  { Name Egrad; Type Form0;
    BasisFunction {
      { Name sn;  NameOfCoef un;  Function BF_Node    ; Support Region[
  Omega]; Entity NodesOf[Omega]; }
      { Name sn2; NameOfCoef un2; Function BF_Node_2E; Support Region[
  Omega]; Entity EdgesOf[Omega]; }
    }
    Constraint {
      { NameOfCoef un;  EntityType NodesOf ; NameOfConstraint Dirichlet; }
      { NameOfCoef un2; EntityType EdgesOf ; NameOfConstraint Dirichlet; }
    }
  }
}
```

**Listing 6.6:** Syntax for the discrete functional space for second order scalar elements

In the case where only the first-order scalar finite element space is sufficient, we can simply drop the `BasisFunction sn2` in the previous syntax.

Next, we will reveal the syntax for 'Curl-conforming' element in GetDP. It is important to keep in mind that the 'Curl-conforming' element used in GetDP is set in the space $W_h^1$ of Whitney element. From there, defining the discrete function space in GetDp is also very simple:

```
FunctionSpace {
  { Name Ecurl; Type Form1;
    BasisFunction {
      { Name sn;  NameOfCoef un ; Function BF_Edge       ; Support Region[
  Omega]; Entity EdgesOf[All]; }
      { Name sn2; NameOfCoef un2; Function BF_Edge_2E  ; Support Region[
  Omega]; Entity EdgesOf[All]; }
    }
    Constraint {
      { NameOfCoef un;  EntityType EdgesOf  ; NameOfConstraint Dirichlet; }
      { NameOfCoef un2; EntityType EdgesOf  ; NameOfConstraint Dirichlet; }
    }
```

**Listing 6.7:** Syntax for the discrete functional space for second order 'Curl-conforming elements

```
11 |     }
12 | }
```

## 6.4  Numerical examples

In this chapter, we already looked over various numerical techniques that are essential to the numerical simulation of electromagnetic problems: from the theory of finite element analysis to the numerical eigensolver in GetDP. The aim of this section is to set an example where we put all these finite element techniques into practice. In particular, we will demonstrate the correctness and effectiveness of the DQNM expansions for electromagnetic fields in a very simple 3-D bounded model.

Let's consider the geometry of a 3-D object shaped like a sphere $\Omega_1$ inside a perfectly conducting box $\Omega$ (see Figure 6.6) filled with air. The sphere has center O, with the radius $r = 0.2\,\mu m$. The edge of the box has length $d = 6\,\mu m$.

The vacuum electric permittivity and vacuum magnetic permeability are given by $\varepsilon_0 = 8.854{\times}10^{-12} Fm^{-1}$ and $\mu_0 = 1.256{\times}10^{-6}\,NA^{-2}$ respectively. The relative permittivity of air (in the domain $\Omega_0$) is $\varepsilon_a = 1$ while the sphere is made of silver whose relative permittivity is given by the Drude model:

$$\varepsilon_1(\omega) = \varepsilon_\infty - \frac{\omega_p^2}{\omega^2 + i\omega\gamma} \tag{6.25}$$

where $\varepsilon_\infty = 3.36174$, $\omega_p = 1.3388 \times 10^{16} \text{rad.s}^{-1}$, and $\gamma = 7.07592 \times 10^{13} \text{rad.s}^{-1}$.

Our main research subject is the scattered vector electric field **E**, solution of the following direct scattering problem:

$$\mathscr{L}_E(\lambda)\mathbf{E} = \nabla \times \left(\boldsymbol{\mu}^{-1}(\lambda)\nabla \times \mathbf{E}\right) + \lambda^2 \boldsymbol{\varepsilon}(\lambda)\mathbf{E} = \mathbf{S} \tag{6.26}$$

The source is given by a simple oriented Dirac delta function $\mathbf{S} = \delta(x_S, y_S, z_S)\mathbf{z}$ at the coordinates $S(0.24, 0.24, 0.24)\,(\mu m)$ shown in Figure 6.6. Since the boundary of the box is assumed to be perfectly conducting, the electric field **E** must be constrained by the homogeneous Dirichlet boundary condition $\mathbf{n} \times \mathbf{E}|_{\partial\Omega} = 0$.

The whole structure will be discretized using the open-source 3D finite element mesh generator: Gmsh (`https://gmsh.info/`). The mesh of the structure is shown in Figure 6.7. The maximum element size is set to be $0.05\,\mu m$. In this example, we will use the second-order Whitney 1-form finite elements.

Our objectives include:

1. Numerically solving the direct scattering problem (6.26) for a specific real frequency. [8] This implies solving the weak formulation (6.5) in GetDP through the syntax Listing 6.1.

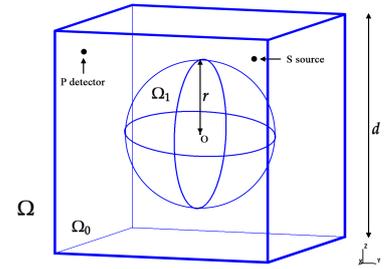

**Figure 6.6:** The geometry of a 3-D square box containing a sphere scatterer $\Omega_1$

[8]: Keep in mind that $\lambda = i\omega$. Thus the value of $\lambda$ in (6.26) is purely imaginary.

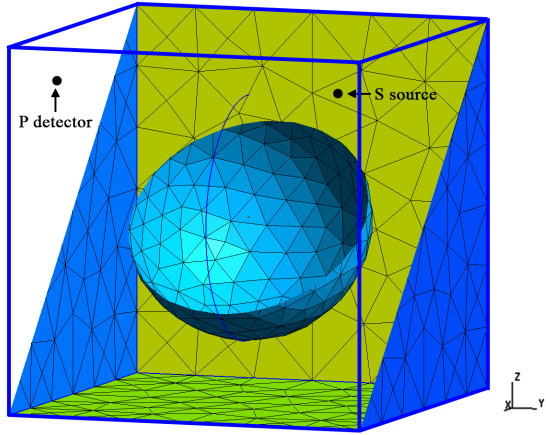

2. Solving the 'left' and 'right' non-linear eigenvalue problem of the operator

$$\langle \mathbf{E}_{ln} | \mathscr{L}_E(\lambda_n) = 0 \qquad \mathscr{L}_E(\lambda_n) | \mathbf{E}_{rn} \rangle = 0$$

using the weak equation (6.20) via the syntax Listing 6.2. [9] Then, we apply the DQNM expansion formula to reconstruct the field $\mathbf{E}$ at the same given value of $\lambda$ as the direct problem. The expansion formula (which is similar to (5.25)) is derived from Lemma 5.1.1:

$$\mathbf{E} = \sum_n \frac{f_p(\lambda_n)}{f_p(\lambda)} \frac{1}{\lambda - \lambda_n} \frac{\int_\Omega \overline{\mathbf{E}_{ln}} \cdot \mathbf{S}}{\langle \mathbf{E}_{ln}, \dot{\mathscr{L}}_E(\lambda_n) \mathbf{E}_{rn} \rangle} \mathbf{E}_{rn}, \qquad (6.27)$$

with

$$\langle \mathbf{E}_{ln}, \dot{\mathscr{L}}_E(\lambda_n) \mathbf{E}_{rn} \rangle =$$

$$\int_\Omega \left[ \overline{\mathbf{E}_{ln}} \cdot \left( \nabla \times ((\mu^{-1}(\lambda_n))' \nabla \times \mathbf{E}_{rn}) \right) + \overline{\mathbf{E}_{ln}} \cdot \left( (\lambda_n^2 \varepsilon(\lambda))' \mathbf{E}_{rn} \right) \right] \, d\Omega.$$

We set three different functions for $f_p(\lambda)$: $f_p(\lambda) = 1$, $f_p(\lambda) = \lambda$, and $f_p(\lambda) = \lambda - \lambda_0$. The value $\lambda_0 = i\omega_0$ is selected such that $\omega_0 = 33$ ($\times 10^{14}$rad.$s^{-1}$), which does not overlap with the zone of influence of other eigenfrequency $\omega_n$.

Next, we notice that our media are reciprocal and the structure is bounded by the homogeneous Dirichlet boundary condition. By Lemma 3.4.1, the 'left' eigenvectors can be computed by simply taking the complex conjugate of 'right' eigenvectors: $\overline{\mathbf{E}_{ln}} = \mathbf{E}_{rn}$.

3. Comparing the results of $\mathbf{E}$ obtained from objective 1 and 2 in a wide range of frequency $\lambda$. Comparative results will be shown through the following quantities:

   ▶ The mean value of the modulus of the electric field inside the object $\Omega_1$: $\int_{\Omega_1} |\mathbf{E}| \, \Omega$.

   ▶ The magnitude of the real part of the field $\mathbf{E}$ at the detector point $(x_P, y_P, z_P)$ with coordinates given by $P(0.24, -0.24, 0.24)$ (see Figure 6.6): $|\Re\{\mathbf{E}(x_P, y_P, z_P)\}|$.

It is worth pointing out that although the non-linear electric wave operator $\mathscr{L}_E(\lambda)$ depends on the variable $\lambda = i\omega$ as a consequence of the causal properties, all the numerical results will be expressed in terms of the

9: In this numerical example, we choose to solve the non-linear eigenvalue problem using rational solver in the NEP class of SLEPc.

frequency $\omega$ as a physical quantity for intuitive reasons.

Figure 6.8 depicts the electric field **E**, solution of the direct problem (6.26) for the fixed frequency $\omega = 35$ ($\times 10^{14}$rad.$s^{-1}$). The strange behavior of the field around the source point is the consequence of the Dirac delta function and doesn't affect the field map on the whole computational domain $\Omega$.

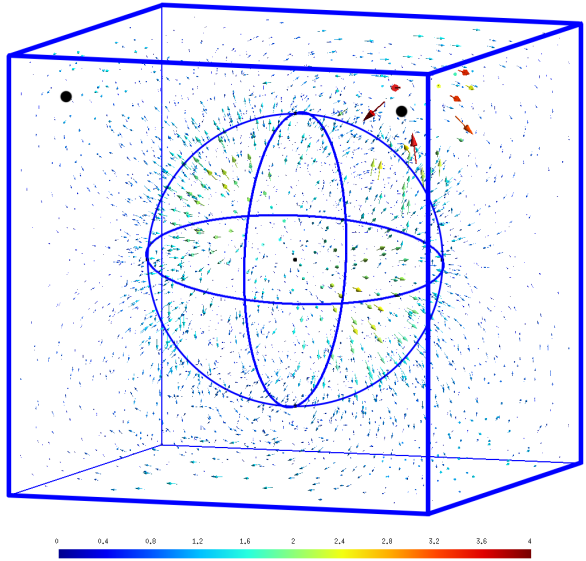

The eigenfrequencies $\omega_n$ (or $\lambda_n$ to be precise), solutions of the eigenvalue problem of the operator $\mathscr{L}_E(\lambda)$, are illustrated in the top panel of Figure 6.9. We notice that a position (for example mode 1, 2, or 3) on the complex plane corresponds to three nearly-equal numerical eigenfrequencies. This implies that the eigenvalues are analytically degenerate: For each individual eigenvalue, there are up to 3 corresponding eigenvectors. This is a consequence of the symmetry of the geometry: The three eigenfields of the same eigenfrequency are actually similar but oriented differently in space. Fortunately, we do not have to pay too much attention to this property in numerical computations: The 3 eigenfields will get 3 nearly-equal but different eigenvalues due to errors in calculations. From the top panel in Figure 6.9, it is easily seen that the locations of these trio eigenvalues are almost identical at low frequency and will begin to separate as the frequency increases.

The middle and bottom panels in Figure 6.9 present the electric field **E** inside and outside the object $\Omega_1$ respectively, through 2 quantities: the norm of the electric field $\int_{\Omega_1} |\mathbf{E}| \, \Omega$ and the magnitude of the real part of the field **E** calculated at the point $(x_P, y_P, z_P)$ $|\Re\{\mathbf{E}(x_P, y_P, z_P)\}|$ in a wide range of frequency $\omega$. We are interested in 4 different values: The green dots refer to the solution obtained by directly solving the scattering problem (6.26), while results reconstructed by the quasinormal modal expansion formula (6.27) with different functions of $f_P(\lambda)$: $f_P(\lambda) = 1$, $f_P(\lambda) = \lambda$, and $f_P(\lambda) = \lambda - \lambda_0$ are represented by blue solid lines, dotted red lines, and purple dashed-dotted lines respectively. In both panels, we notice that the reconstructed fields with $f_P(\lambda) = 1$ (blue solid lines) and $f_P(\lambda) = \lambda$ (dotted red lines) show an incredible agreement with the direct solution (green dots),[10] which numerically confirms our theory on modal expansion for non-linear electromagnetic operators.

10: Note that the data are plotted on a logarithmic scale.

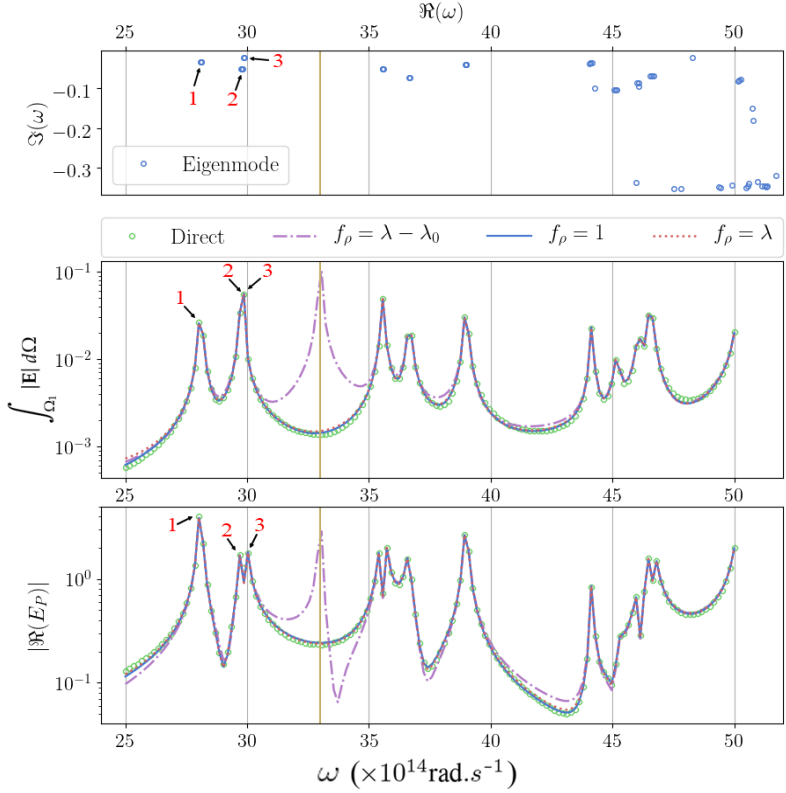

It is also worth pointing out the correspondence between the peaks-/valleys of the norm of the field and the position of the eigenfrequencies in the complex plane. For example, the first three peaks in the middle and bottom panels are clearly dictated by the modes at three locations highlighted by the red numbers in the complex plane. This is the reason for the belief that the optical response of the structure is the synthesis of the excitation of each physical-resonance state in the system. [11]

Moreover, let's pay attention to the case where $f_\rho(\lambda) = \lambda - \lambda_0$ where $\lambda_0 = i\omega_0$ (purple dashed-dotted lines). The frequency $\omega_0 = 33$ ($\times 10^{14} \text{rad.s}^{-1}$) is highlighted by the yellow vertical line. It is clear to see discrepancies at the frequency $\omega_0$ in the middle and bottom panels of Figure 6.9, which confirms our prediction in section Section 5.5 about the singularity at the roots of $f_\rho(\lambda)$. As a result, it is recommended to choose the function $f_\rho(\lambda)$ such that its root is far away from our domain of interest.

Before ending this chapter, it is worth stressing again on the non-uniqueness of the quasinormal modal expansion. In particular, formula (6.27) is not the only modal expansion family for the solution $\mathbf{E}$ of the scattering problem (6.26). Indeed, the modal expansion formula (6.27) is simply derived from Lemma 5.1.1 for the given operator $\mathscr{L}_E(\lambda) = \nabla \times (\boldsymbol{\mu}^{-1}(\lambda)\nabla\times) + \lambda^2 \boldsymbol{\varepsilon}(\lambda)$. This implies for a different choice of operator, we will have a different family of DQNM expansion formulas. This raises the question of what happens if we derive expansion from a polynomial operator. For example, let's rewrite the direct problem (6.26)

11: As we clarified in the previous theoretical chapter, this assumption must be treated with caution.

for the sphere in Figure 6.6 as follows:

$$\mathcal{L}_{PE}(\lambda)\mathbf{E} = (-\lambda + \gamma) \int_{\Omega} \nabla \times \left( \boldsymbol{\mu}^{-1}(\lambda) \nabla \times \mathbf{E} \right) \, d\Omega + (-\lambda^3 + \gamma \lambda^2) \int_{\Omega_0} (\boldsymbol{\varepsilon}_0 \mathbf{E}) \, d\Omega$$

$$+ (-\varepsilon_\infty \lambda^3 + \varepsilon_\infty \gamma \lambda^2 - \omega_p^2 \lambda) \int_{\Omega_1} \mathbf{E} \, d\Omega$$

$$= (-\lambda + \gamma)\mathbf{S} \tag{6.28}$$

It is easy to check that $\mathcal{L}_{PE}$ is a polynomial operator and (6.28) has the same solution as (6.26). Our next step is to deduce the expansion formula for the polynomial operator $\mathcal{L}_{PE}(\lambda)$. By Lemma 4.5.2, we can obtain the quasinormal modal expansion for the solution $\mathbf{E}$ as follows:

$$\mathbf{E} = \sum_n \frac{f_\rho(\lambda_n)}{f_\rho(\lambda)} \frac{1}{\lambda - \lambda_n} \frac{\langle \mathbf{E}_{ln}, (-\lambda + \gamma)\mathbf{S} \rangle}{\langle \mathbf{E}_{ln}, \dot{\mathcal{L}}_{PE}(\lambda_n)\mathbf{E}_{rn} \rangle} \mathbf{E}_{rn} \tag{6.29}$$

with $\dot{\mathcal{L}}_{PE}(\lambda_n)$ is the derivative of the operator $\mathcal{L}_{PE}(\lambda)$ at $\lambda_n$.

It is easily seen that the highest order of $\lambda$ in the operator $\mathcal{L}_{PE}(\lambda)$ is 3. Thus, the function $f_\rho(\lambda)$ can be chosen as an arbitrary polynomial with the degree up to 2. In this example, we will choose the following functions of $f_\rho(\lambda)$: $f_\rho(\lambda) = 1$, $f_\rho(\lambda) = \lambda$, and $f_\rho(\lambda) = \lambda^2$. All numerical results of the quasinormal modal expansion (6.29) for the operator $\mathcal{L}_{PE}(\lambda)$ are depicted in the middle and bottom panels in Figure 6.10:[12]

12: It is important to distinguish Figure 6.10 from Figure 6.9.

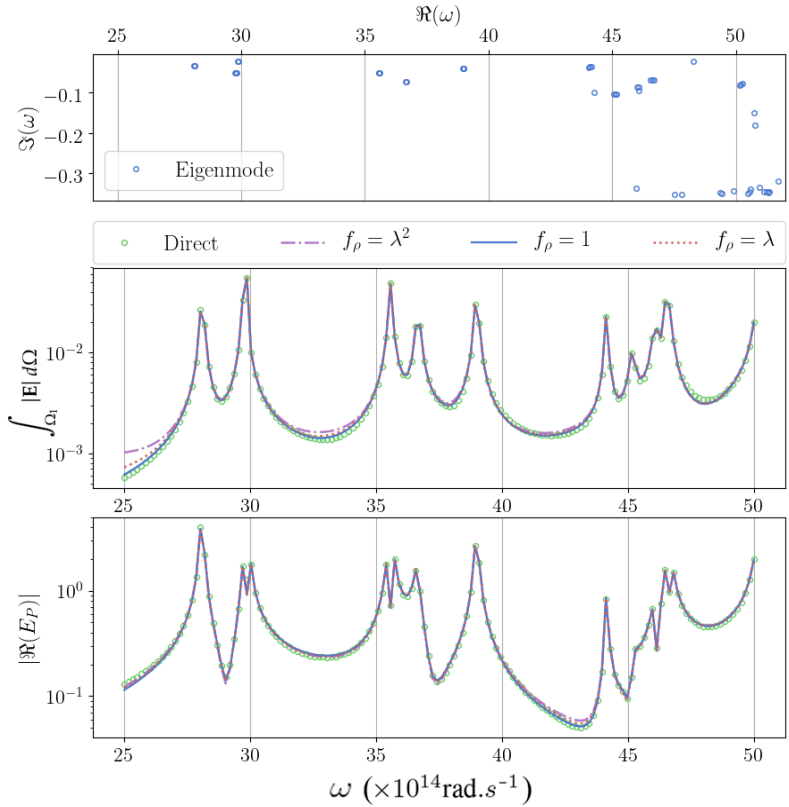

**Figure 6.10:** Modal expansion for the polynomial operator $\mathcal{L}_{PE}(\lambda)$ with 3 different functions $f_\rho(\lambda)$: $f_\rho(\lambda) = 1$ (solid blue lines), $f_\rho(\lambda) = \lambda$ (dotted red lines), and $f_\rho(\lambda) = \lambda^2$ (purple dashed-dotted lines)
(TOP) Spectrum of complex eigenfrequencies.
(MIDDLE) The mean value of the modulus of the electric field inside the object $\Omega_1$: $\int_{\Omega_1} |\mathbf{E}| \, d\Omega$.
(BOTTOM) The magnitude of the real part of the field $\mathbf{E}$ at the detector point $(x_P, y_P, z_P)$: $|\Re\{\mathbf{E}(x_P, y_P, z_P)\}|$.

Unsurprisingly, we again witness a beautiful agreement between the direct solution (green dots) with the three results obtained from the modal expansion (6.29), i.e blue, red, purple lines. It emphasizes the fact that the quasinormal modal expansion formula in electromagnetics is not unique and heavily depends on the way of constructing the electromagnetic operator.

# Perfectly Matched Layer (PML) for open structures

# 7

In the previous chapter, we have provided a simple modeling example of the application of QNM expansion using Finite Element. In particular, we study the 3-D bounded structures given by a closed box containing a metallic sphere. In terms of mathematics, it is equivalent to study the Maxwell operator constrained by the following boundary conditions:

▶ homogeneous Dirichlet boundary condition:

$$D_L^D \equiv \left\{ \mathbf{u} : \mathbf{u}, \nabla \times (\mu^{-1} \nabla \times \mathbf{u}) \in L^2(\Omega)^3, \mathbf{n} \times \mathbf{u}|_\Gamma = 0 \right\},$$

▶ homogeneous Neumannn boundary condition:

$$D_L^N \equiv \left\{ \mathbf{u} : \mathbf{u}, \nabla \times (\mu^{-1} \nabla \times \mathbf{u}) \in L^2(\Omega)^3, \mathbf{n} \times \left( \mu^{-1} \nabla \times \mathbf{u} \right)|_\Gamma = 0 \right\}.$$

In such bounded structures, the energy dissipation, which is indicated by the imaginary part of the frequencies, only results from the energy absorption of the materials. For the electromagnetic problems in open regions, the previous argument is no longer true since the energy freely radiates (leaks) into the surrounding medium. This modifies the characteristics of our resonant modes (QNMs) [61] where the energy dissipation issues from not only the absorption of material but also the loss to the surrounding environment. For easier understanding, we will call them 'leaky modes', the term is commonly used in the research of waveguides [62] . In addition, there also exists a new kind of mode, which resonant mainly in the surrounding medium, called radiation modes. As mentioned in chapter 3, these radiation modes constitute the continuous spectrum of our Maxwell operators.

The numerical modeling of open structures encounters two difficulties: the computational domain is unbounded and the amplitude of leaky modes exponentially increases. In order to overcome these difficulties, the classical Finite Element method must be associated with other techniques to truncate the computational region:

One of the most commonly used grid truncation techniques are so-called Absorbing Boundary Conditions (ABC) B. Engquist and A. Majda in 1977 [63] . The idea is to replace the Dirichlet or Neumann boundary conditions, which are responsible for the reflection of an outgoing wave on the artificial boundary of a computational domain, by the designed absorbing conditions, which can yield a small reflection coefficient on the truncation boundary.

The main drawbacks of the ABC are that it can degrade the sparsity of the system matrix and only works at a single frequency. As a result, we need new techniques that can be applied to a wide frequency range. In 1994, Bérénger suggested an elegant technique, which consists of analytically extending real coordinates of physical equations into the complex ones, called the Perfectly Matched Layer (PML) method [64] . The main principle of the PML technique is to border the computational domain by a dissipative zone which can damp the incoming waves [65] . In particular, the PML is designed such that:

▶ There is no reflection at the interface between the PML and non-PML regions for all frequencies and all angles of incidence and polarization.

▶ The PML domains, made of lossy materials, strongly absorb outgoing waves from the interior of a computational region without reflecting them back into the interior. In other words, in the PML, the propagating (oscillating) waves are replaced by exponentially decaying waves.

In this chapter, we will focus our study on the technique of Perfectly Matched Layer to truncate the open computational domain. The ultimate goal is to get a better understanding of the influence of PML parameters upon the eigenmodes as well as the numerical results of the QNM expansion. The construction in this chapter is inspired by [52] .

## 7.1 The Cartesian PML techniques

In this section, we will introduce the basic concepts of PML in the Cartesian coordinate system. Given an unbounded region $\Omega$, we assume that outside the region of the main resonators, the media are homogeneous and isotropic when $r \to \infty$. Thus, the permittivity and permeability at infinity can be expressed as scalar quantities. The basic idea consists of closing the resonant section by replacing the unbounded homogeneous region with a PML of finite thickness, which can attenuate waves in the surrounding medium. Figure 7.1 illustrates the application of the Cartesian PML in a 2-D open domain. Since the computational domain is truncated by the PML, the problem becomes closed and can be implemented via Finite Element Analysis as in the previous chapter.

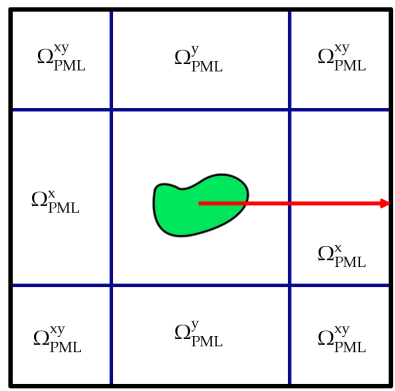

**Figure 7.1:** An open 2-D geometric structure truncated by the Cartesian PML

### Complex stretch of coordinates

In order to illustrate the effect of PML, let's consider a one-dimensional situation along the $x$-axis. In particular, the wave travels along the $x$-axis can be expressed as an exponential function $\exp(ik_x x)$ where $k_x$ is the wavevector along the $x$-axis . Figure 7.2 illustrate how the wave is transmitted from the region of resonators $|x| < d_x$, where the media can be inhomogeneous, anisotropic and dispersive, to the homogeneous surrounding medium $d_x < |x|$.[1] .

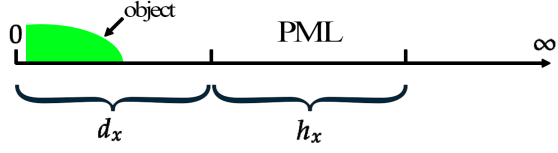

**Figure 7.2:** A 2-D geometric structure truncated by the Cartesian PML

Instead of the real value $x$, we can choose a particular path $\tilde{x}(x)$ in the complex plane parametrized by a real variable x such that $\exp(ik_x\tilde{x}(x))$ decays exponentially as $x \to \infty$. For example, we can define:

$$\tilde{x}(x) = \int_0^x s_x(x')dx',$$

where $s_x$ is the complex function with respect to the real variable $x'$ satisfying:

► $s_x(x) = 1$ for $|x| < d_x$.
► $\mathfrak{I}(s_x) > 0$ for $d_x < |x|$.

Then, it is easily seen that for any function $\tilde{f}(\tilde{x})$, we have the change of variable $\tilde{x} \to x$ as follows:

$$\frac{\partial \tilde{f}}{\partial \tilde{x}} = \frac{1}{s_x}\frac{\partial f}{\partial x} \quad \text{and} \quad d\tilde{x} = s_x dx.$$

In order words, the $x$-axis is stretched by a factor $s_x$. For this particular choice of function $s_x$, we will have $\tilde{x} = x$ for $|x| < d_x$, which implies that the electromagnetic fields remain the same in the region of resonators. It also can be prove that if $\mathfrak{I}(s_x) > 0$ for $d_x < |x|$, the fields in the surrounding medium will converge exponentially towards zero when $x \to \infty$. Thus, it is possible to truncate this region by a layer with the value $h_x$, which is called the thickness of PML along the $x$-direction.[2]

The same considerations apply in the directions of $y$ and $z$. From here, we can define the Cartesian PML as the method consisting in extending the original equations to complex coordinates $\tilde{x}$, $\tilde{y}$ and $\tilde{z}$ [66] . In particular, we have:

$$\frac{\partial}{\partial \tilde{x}} = \frac{1}{s_x}\frac{\partial}{\partial x} \qquad \frac{\partial}{\partial \tilde{y}} = \frac{1}{s_y}\frac{\partial}{\partial y} \qquad \frac{\partial}{\partial \tilde{z}} = \frac{1}{s_z}\frac{\partial}{\partial z},$$

for

$$\tilde{x}(x) = \int_0^x s_x(x')dx'$$
$$\tilde{y}(y) = \int_0^x s_y(y')dy'$$
$$\tilde{z}(z) = \int_0^x s_z(z')dz'.$$

Our classical Maxwell equations in the new complex coordinates are

given by

$$\tilde{\nabla} \times \mathbf{H}_s = -\lambda \boldsymbol{\varepsilon} \mathbf{E}_s$$
$$\tilde{\nabla} \times \mathbf{E}_s = \lambda \boldsymbol{\mu} \mathbf{H}_s, \tag{7.1}$$

with the help of the modified operator $\tilde{\nabla}$:

$$\tilde{\nabla} = \tilde{x} \frac{\partial}{\partial \tilde{x}} + \tilde{y} \frac{\partial}{\partial \tilde{y}} + \tilde{z} \frac{\partial}{\partial \tilde{z}}. \tag{7.2}$$

It is easily seen that the solutions of (7.1), i.e. $\begin{pmatrix} \mathbf{E}_s \\ \mathbf{H}_s \end{pmatrix}$ are identical to $\begin{pmatrix} \mathbf{E} \\ \mathbf{H} \end{pmatrix}$, solution of the original Maxwell equation in the region around the resonators ($|x| < d_x$). The only difference is that along the $x$-axis the fields $\begin{pmatrix} \mathbf{E}_s \\ \mathbf{H}_s \end{pmatrix}$ attenuate and converge to zero in the PML ($d_x < |x| < d_x + h_x$), while the physical counter part $\begin{pmatrix} \mathbf{E} \\ \mathbf{H} \end{pmatrix}$ do not. The similar considerations apply for the $y$ and $z$ directions.

## The absorption property of PML

The next question arises is how to choose the complex function $s_x$, $s_y$, and $s_z$ to maximize the attenuation of waves with minimal reflections in the region of PML. In order to answer the problem, let's reconsider the effect of PML along the $x$-axis as shown in Figure 7.2. For the wave $\exp(ik_x \tilde{x})$, it can be shown that the total attenuation across a layer of finite thickness $h_x$ is proportional to the exponential function

$$\exp(-\Im(k \gamma_x)) = \exp(-|k_x| |\gamma_x| \sin(\arg(k_x) + \arg(\gamma_x))),$$

where $\gamma_x$ is given by:

$$\gamma_x = \int_{d_x}^{d_x + h_x} s_x(x') dx'.$$

It is worth reminding in the open structure, the field of leaky modes grow exponentially in space at infinity. This implies $\Im(k_x) < 0$, which is equivalent to $\arg(k_x) < 0$. The intuitive response is to choose the function $s_x$ to compromise this 'space explosion' effect. Indeed, it is easily seen that for the choice $\arg(\gamma_x) > -\arg(k_x)$, the leaky modes can be completely attenuated by the PML. In other words, increasing $\arg(\gamma_x)$ will enlarge the region of the complex plane where leaky modes can be computed. It is important to see that the absorption property of PML increases with the increment of the value $|\gamma_x|$. In particular, if we choose $\Re(\gamma_x) > h_x$, the PML can also enhance the natural decay of the modes.

At first glance, it is reasonable to choose $s_x$ such that the value $|\gamma_x|$ is as high as possible. Unfortunately, it leads to the very fast attenuation of the electromagnetic fields, which may not be fully captured by the FE discretization [67] .

A simple version of $s_x$ is constant functions, such as $s_x = 1 + i$. In scattering problems, since the media can be dispersive, it is wiser to set the absorption function $s_x$ frequency dependent, for instance, $s_x = 1 + i\sigma_x(x)/\omega$ where $\sigma_x(x)$ is a continuous function, parabolic inside the PML region [67] [68] . Obviously, the frequency-dependent absorption function $s_x$ would be a better choice to attenuate the fields in the PML domain. Unfortunately, since the effects of dispersive PML can obscure spectral characteristics of our study of modal expansion, we prioritize the use of fixed absorption function $s_x$ in this thesis. [3]

3: The impact of the dispersive PML on the modal expansion is out of scope of this thesis and needs a in-depth separate study in the future.

### Interpretation of complex stretch as absorbing materials

Although the introduction of the complex stretch is necessary to attenuate the electromagnetic wave in the region of PML, the complex coordinate is definitely not an easily receptive concept. Fortunately, the operating principle of PML can be easily explained through the concept of absorbing materials. In particular, the PML can be considered to be made of absorbing material in the real space, which can completely absorb any electromagnetic wave coming from the internal region [69] . This interpretation can be shown via the change of coordinate from the complex space to the real space.

We have the Jacobian associated to these changes of coordinates are given by

$$\mathbf{J} = \mathbf{diag}(\frac{\partial \tilde{x}}{\partial x}, \frac{\partial \tilde{y}}{\partial y}, \frac{\partial \tilde{z}}{\partial z}) = \mathbf{diag}(s_x, s_y, s_z).$$

Then, the Maxwell equations in the complex coordinates (7.1) can be proved to be equivalent to the following equations:

$$\nabla \times \mathbf{H}_s = -\lambda \boldsymbol{\varepsilon}_s \mathbf{E}_s$$
$$\nabla \times \mathbf{E}_s = \lambda \boldsymbol{\mu}_s \mathbf{H}_s, \tag{7.3}$$

where the new tensors $\boldsymbol{\varepsilon}_s$ and $\boldsymbol{\mu}_s$ are defined as follows:

$$\boldsymbol{\delta}_s := \mathbf{J}^{-1} \boldsymbol{\delta} \mathbf{J}^{-\top} \det(\mathbf{J}) \quad \text{for} \quad \boldsymbol{\delta} = \{\boldsymbol{\varepsilon}, \boldsymbol{\mu}\}. \tag{7.4}$$

It is worth noting that the modified tensors of permittivity and permeability $\boldsymbol{\varepsilon}_s$ and $\boldsymbol{\mu}_s$ remain unchanged in the region around the resonators. On the other hand, in the PMLs, the media are no longer homogeneous but replaced by absorbing materials $\boldsymbol{\varepsilon}_s$ and $\boldsymbol{\mu}_s$.

In the case where the permittivity $\boldsymbol{\varepsilon}$ and permeability $\boldsymbol{\mu}$ can be expressed in terms of diagonal tensors

$$\boldsymbol{\varepsilon} = \begin{pmatrix} \varepsilon_{xx} & 0 & 0 \\ 0 & \varepsilon_{yy} & 0 \\ 0 & 0 & \varepsilon_{zz} \end{pmatrix} \quad \text{and} \quad \boldsymbol{\mu} = \begin{pmatrix} \mu_{xx} & 0 & 0 \\ 0 & \mu_{yy} & 0 \\ 0 & 0 & \mu_{zz} \end{pmatrix},$$

the new tensors for the permittivity and permeability can be written as follows:

$$\boldsymbol{\delta}_s = \boldsymbol{\delta} \boldsymbol{\Lambda} \quad \text{for} \quad \boldsymbol{\delta} = \{\boldsymbol{\varepsilon}, \boldsymbol{\mu}\}$$

with

$$\mathbf{\Lambda} = \mathbf{diag}(\frac{s_y s_z}{s_x}, \frac{s_z s_x}{s_y}, \frac{s_x s_y}{s_z})$$

Henceforth, the implementation of the Cartesian PML technique is as simple as replacing the media in the appropriate media given by (7.4). For example, the parameters for the Cartesian PML in 2-D in Figure 7.2 are given as follows:

▶ $s_x \in \mathbb{C}^+$ and $s_y = s_z = 1$ in the region $\Omega_{\text{PML}}^x$.
▶ $s_y \in \mathbb{C}^+$ and $s_x = s_z = 1$ in the region $\Omega_{\text{PML}}^y$.
▶ $s_x = s_y \in \mathbb{C}^+$ and $s_z = 1$ in the region $\Omega_{\text{PML}}^{xy}$.

## 7.2 The radial PML techniques

The PML technique is not only limited by the Cartesian coordinate. In fact, it is straightforward to develop the PML in the cylindrical or spherical system [66] . In particular, instead of stretching the 3 coordinates $x$, $y$, and $z$ in the Cartesian system $(x, y, z)$, the radial coordinate $\rho$ in the cylindrical system $(\rho, \phi, z)$ can be changed to:

$$\tilde{\rho} = \int_0^\rho s_\rho(\rho')d\rho',$$

and similarly, the radial coordinate $r$ in the spherical system $(r, \theta, \phi)$ can be modified as:

$$\tilde{r} = \int_0^r s_r(r')dr',$$

to attenuate the wave traveling along the radial direction. The other coordinates, i.e. $\phi$ and $z$ in the cylindrical system $(\rho, \phi, z)$; or $\theta$ and $\phi$ in the spherical system $(r, \theta, \phi)$, usually remain untouched by the complex stretch.

By carrying out the same procedure as in the Cartesian system, we see that the electromagnetic wave is attenuated when entering the region of PML in the cylindrical/spherical system. This results from the absorption properties of the surrounding materials: $\delta_s = \delta \mathbf{\Lambda}$ for $\delta = \{\varepsilon, \mu\}$ where $\mathbf{\Lambda}$ is given by

$$\mathbf{\Lambda} = \mathbf{diag}\left(\left(\frac{\tilde{\rho}}{\rho}\right)\left(\frac{s_z}{s_\rho}\right), \left(\frac{\rho}{\tilde{\rho}}\right)(s_z s_\rho), \left(\frac{\tilde{\rho}}{\rho}\right)\left(\frac{s_\rho}{s_z}\right)\right),$$

for the cylindrical PML; and

$$\mathbf{\Lambda} = \mathbf{diag}\left(\left(\frac{\tilde{\rho}}{\rho}\right)^2\left(\frac{1}{s_r}\right), s_r, s_r\right),$$

for the spherical PML.

[66]: Chew et al. (1997), 'Complex coordinate system as a generalized absorbing boundary condition'

## 7.3 Quasinormal modal expansion for a sphere in the open space

As a proof of concept, in this section, let's apply the technique of PML in the DQNM expansion for electromagnetic fields in a simple 3-D open system.

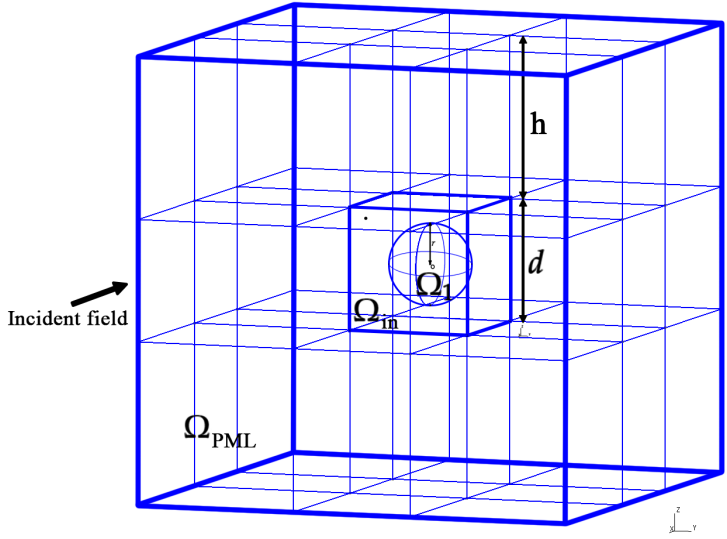

We consider the geometry of a 3-D object shaped like a sphere $\Omega_1$, which has center O, with the radius $r$, in an open space $\Omega$ (see Figure 7.3). In order to cope with the infinity of space, we will close the domain of interest by replacing the unbounded homogeneous region $\Omega$ with Perfectly Matched Layers (PMLs) of finite thickness $h$, which can attenuate waves in the surrounding medium. The distance from the spherical center to the PML layer is denoted by $d/2$. Figure 7.3 demonstrates the application of the Cartesian PML on our 3-D computational domain. As an extension of the PML application of the 2-D geometric structure (depicted in Figure 7.1), the Cartesian PML $\Omega_{PML}$ in 3-D contains 26 subregions. In each region, the coordinate is complex-stretched to the direction along which the field must decay:

- ▶ $s_x = s \in \mathbb{C}^+$ and $s_y = s_z = 1$ in the 2 subregion $\Omega_{PML}^x$.
- ▶ $s_y = s \in \mathbb{C}^+$ and $s_x = s_z = 1$ in the 2 subregion $\Omega_{PML}^y$.
- ▶ $s_z = s \in \mathbb{C}^+$ and $s_x = s_y = 1$ in the 2 subregion $\Omega_{PML}^z$.
- ▶ $s_x = s_y = s \in \mathbb{C}^+$ and $s_z = 1$ in the 4 subregion $\Omega_{PML}^{xy}$.
- ▶ $s_x = s_z = s \in \mathbb{C}^+$ and $s_y = 1$ in the 4 subregion $\Omega_{PML}^{xz}$.
- ▶ $s_y = s_z = s \in \mathbb{C}^+$ and $s_x = 1$ in the 4 subregion $\Omega_{PML}^{yz}$.
- ▶ $s_y = s_y = s_z = s \in \mathbb{C}^+$ in the 8 subregion $\Omega_{PML}^{xyz}$.

In this example, the geometrical parameters are chosen as follows: $r = 0.1\,\mu m$, $d = 0.3\,\mu m$, $h = 0.4\,\mu m$. The complex stretching function is given by $s = 1 + 0.2i$.

The relative permittivity of air is fixed to be constant $\varepsilon_a = 1$ while the relative permittivity of the sphere is given by the Drude model:

$$\varepsilon_1(\omega) = \varepsilon_\infty - \frac{\omega_p^2}{\omega^2 + i\omega\gamma},$$

where $\varepsilon_\infty = 6$, $\omega_p = 2.286 \times 10^{15}\,\mathrm{rad.s^{-1}}$, and $\gamma = 1.332 \times 10^{15}\,\mathrm{rad.s^{-1}}$ [41].

The whole structure will be discretized using Gmsh (`https://gmsh.info/`) with the maximum element size is set to be $0.015\,\mu m$ inside the sphere, which will increase to $0.1\,\mu m$ at the outer boundary of PMLs. The second-order Whitney 1-form finite elements are used to discretize the problem. We will impose the Dirichlet boundary condition on the outer boundary of PMLs. The mesh of the structure is shown in Figure 7.4.

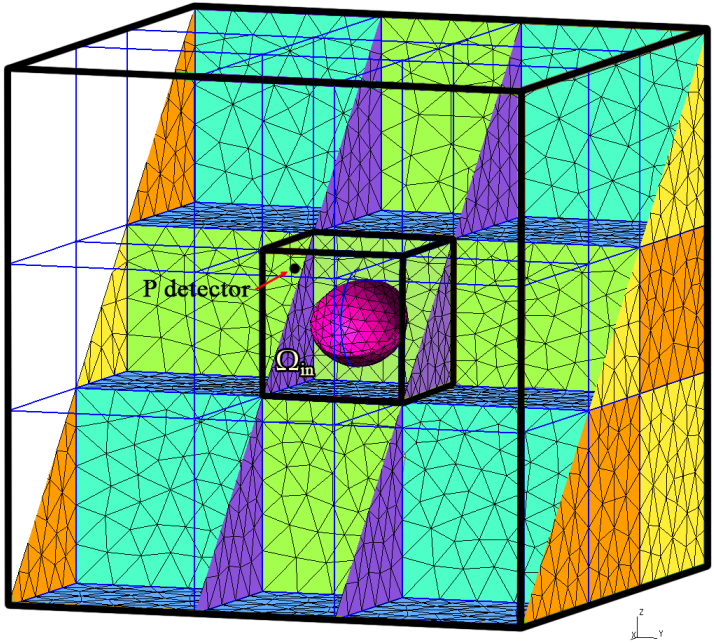

**Figure 7.4:** The 3-D mesh (displayed via the 2-D surface) of the sphere in an open domain.

In this numerical example, the Dirac delta source is replaced: The geometrical structure will be illuminated by an incident plane wave oriented in $x$-direction and polarized along $z$-direction.[4]

$$\mathbf{E}_0 = \exp(ikx)\mathbf{e}_z,$$

where $k$ stands for the $x$-component of wavevector in air $k = \omega\sqrt{\varepsilon}$.

Then, the scattered vector electric field $\mathbf{E}$ is the solution of the following direct radiation problem:

$$\mathcal{L}_{\mu,\varepsilon}(\lambda)\mathbf{E}_s = \nabla \times \left(\mu^{-1}(\lambda)\nabla \times \mathbf{E}_s\right) + \lambda^2 \varepsilon(\lambda)\mathbf{E}_s = \mathbf{S}_0, \qquad (7.5)$$

where the source is given by

$$\begin{aligned}
\mathbf{S}_0 &= \mathcal{L}_{\mu_0-\mu,\varepsilon_0-\varepsilon}(\lambda)\mathbf{E}_0 \\
&= \nabla \times \left((\mu_0^{-1} - \mu^{-1}(\lambda))\nabla \times \mathbf{E}_0\right) + \lambda^2(\varepsilon_0 - \varepsilon(\lambda))\mathbf{E}_0.
\end{aligned}$$

4: The direction of the incident wave is indicated by the arrow in Figure 7.3.

The tensors $\boldsymbol{\mu}_0$ and $\boldsymbol{\varepsilon}_0$ denote the permeability and permittivity of the background:

$$\boldsymbol{\mu}_0 = \mu_0 \mathcal{I} \quad \text{and} \quad \boldsymbol{\varepsilon}_0 = \varepsilon_0 \varepsilon_a \mathcal{I}$$

where $\varepsilon_0$ and $\mu_0$ are the vacuum electric permittivity and vacuum magnetic permeability; while $\mathcal{I}$ stands for the identity tensor.

Similar to the previous example, our objectives include:

1. Numerically solving the direct scattering problem (7.5) for a wide range of incident frequency $\lambda$.

2. Solving the eigenvalue problem [5]

$$\mathcal{L}_{\boldsymbol{\mu},\boldsymbol{\varepsilon}}(\lambda_n)\mathbf{E}_{rn} = 0.$$

Then, applying the DQNM expansion formula to reconstruct the field $\mathbf{E}_s$ at the same given value of $\lambda$ as the direct problem. The expansion formula is derived from Lemma 5.1.1:

$$\mathbf{E}_s = \sum_n \frac{f_p(\lambda_n)}{f_p(\lambda)} \frac{1}{\lambda - \lambda_n} \frac{\int_\Omega \mathbf{E}_{rn} \cdot \mathbf{S}_0}{\langle \overline{\mathbf{E}_{rn}}, \dot{\mathcal{L}}_E(\lambda_n)\mathbf{E}_{rn} \rangle} \mathbf{E}_{rn}, \qquad (7.6)$$

with

$$\langle \overline{\mathbf{E}_{rn}}, \dot{\mathcal{L}}_E(\lambda_n)\mathbf{E}_{rn} \rangle =$$
$$\int_\Omega \left[ \mathbf{E}_{rn} \cdot \left( \nabla \times ((\boldsymbol{\mu}^{-1}(\lambda_n))' \nabla \times \mathbf{E}_{rn}) \right) + \mathbf{E}_{rn} \cdot \left( (\lambda_n^2 \boldsymbol{\varepsilon}(\lambda))' \mathbf{E}_{rn} \right) \right] \, d\Omega.$$

3. Comparing the total electric field $\mathbf{E} = \mathbf{E}_s + \mathbf{E}_0$ obtained from objective 1 and 2. Comparative results will be shown through the following quantities:

   ▶ The mean value of the electric field in the non-PML region $\Omega_{\text{in}}$ (which also contains the sphere domain $\Omega_1$): $\int_{\Omega_{\text{in}}} |\mathbf{E}| \, \Omega$.
   ▶ The magnitude of the real part of the field $\mathbf{E}$ at the detector point $(x_P, y_P, z_P)$ with coordinates given by $P(0.12, -0.12, 0.12)$ (see Figure 7.4): $|\Re\{\mathbf{E}(x_P, y_P, z_P)\}|$.

The top panel of Figure 7.5 illustrates the eigenfrequencies $\omega_n$ (or $\lambda_n$ to be precise), solutions of the eigenvalue problem of the operator $\mathcal{L}_{\boldsymbol{\mu},\boldsymbol{\varepsilon}}(\lambda)$ in the complex plane. It is worth reminding that each blue dot in the complex plane indeed corresponds to three nearly-equal numerical eigenfrequencies. This is the consequence of the symmetry of the geometry: For each eigenfrequency, there are three eigen-fields that are actually similar but oriented differently. In the case of a closed structure (see Figure 6.9), the position of these trio eigenvalues are almost indistinguishable. On the other hand, in this example of an open resonator, the location of each frequency in the trio become more recognizable (The trio eigenvalues begin to separate from each other). For example, the three modes 1, 2, and 3 in the top panel of Figure 7.5 are supposed to share the same analytical eigenfrequency. Their eigen-fields (depicted in Figure 7.6) are indeed the same but have different orientations in space.

We recognize there are two kinds of eigenmodes: The first kind has their eigen-fields distributed around the sphere (for example, see modes 1, 2, and 3 in Figure 7.6) and seems to influence the 'physical' property of

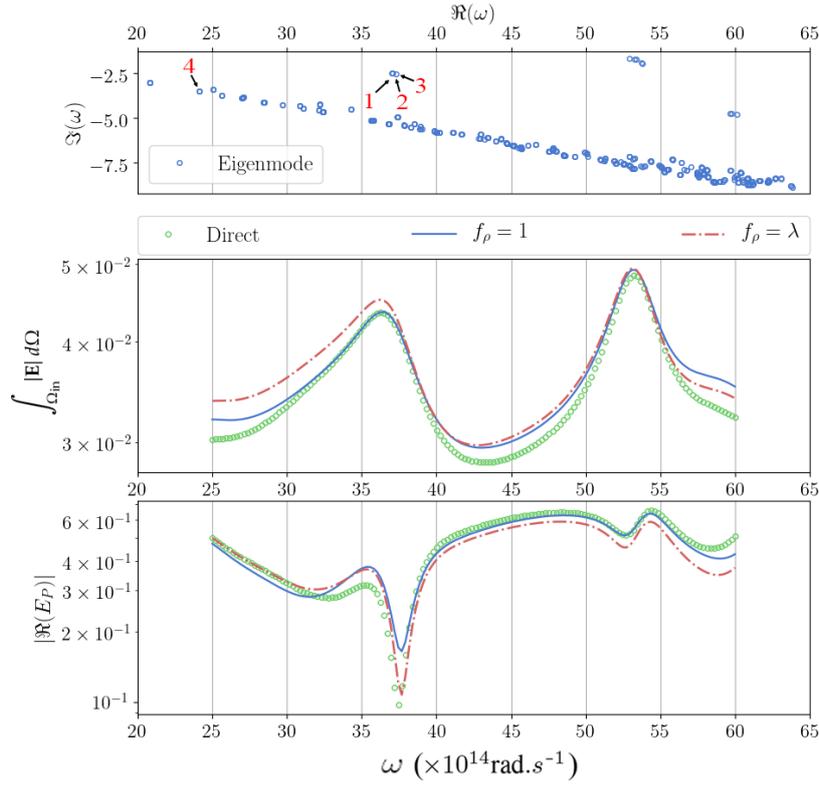

**Figure 7.5:** Modal expansion for a sphere in an open space.
(TOP) Spectrum of complex eigenfrequencies.
(MIDDLE) The mean value of the electric field outside of the PMLs: $\int_{\Omega_{in}} |\mathbf{E}| \, \Omega$.
(BOTTOM) The magnitude of the real part of the field $\mathbf{E}$ at the detector point $(x_P, y_P, z_P)$: $|\Re\{\mathbf{E}(x_P, y_P, z_P)\}|$.
The yellow vertical lines in the middle and bottom panels indicate the value $\Re(\omega_0)$.

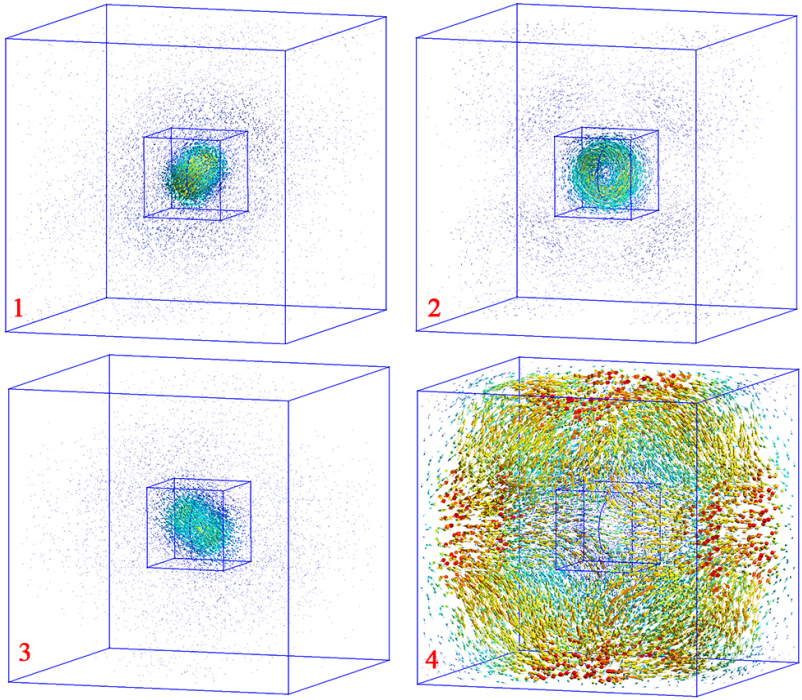

**Figure 7.6:** The 3-D field of 4 eigenmodes from the top panel of Figure 7.5. The blue color of the field maps indicates the minimum value while the red is the maximum.

the system. These eigenmodes can be considered as 'resonant modes' of the structure. On the other hand, the second kind possesses certain characteristics related to the PML:

▶ Their eigen-fields (for example see modes 4 in Figure 7.6) are scattered throughout the domain of PMLs.

▶ Their position evolves according to the change of the PML parame-ters: The thickness $h$ of PML and the complex stretch parameter $s$.

▶ Their locations are neatly aligned around a fixed line.

For now, without going deeper into details, let's call the second kind of eigenmodes by the name 'PML' modes.

The middle and bottom panels in Figure 7.5 present the total electric field $\mathbf{E}$ through 2 quantities: the norm of the electric field $\int_{\Omega_{in}} |\mathbf{E}| \, \Omega$ and the magnitude of the real part of the field $\mathbf{E}$ calculated at the point $(x_P, y_P, z_P)$, namely $|\mathfrak{R}\{\mathbf{E}(x_P, y_P, z_P)\}|$. We notice that, compared to the closed resonator, the discrepancies between the green dots, which refer to the solution obtained by directly solving the scattering problem (7.5), and blue solid (resp. dotted red) lines, which are reconstructed by the quasinormal modal expansion formula (7.6) with $f_p(\lambda) = 1$ (resp. $f_p(\lambda) = \lambda$), are quite significant. These errors are of no surprise considering the presence of 3-D Cartesian PMLs to simulate open space. [6] Eventually, the DQNM expansion still succeeds in a relatively accurate reconstruction of the optical properties of the structure.

6: Keep in mind that the data are plotted on a logarithmic scale.

Finally, it is worth noting that among all the eigenmodes, 'resonant' modes have a direct impact on the optical characteristics of the given structure. For example, it is easily seen that the modes 1, 2 and 3 are directly responsible for the peak and valley of the middle and bottom panels in Figure 7.5, respectively, around the frequency $\omega = 36.5 \times 10^{14} \, \text{rad.s}^{-1}$. On the other hand, the contribution of each 'PML' mode is insignificant. However, numerical computations point out that the joint contribution of 'PML' modes in the DQNM expansion is too big to be negligible: We have to take into account both the 'resonant' and 'PML' eigenmodes in the DQNM expansion.

## 7.4 Quasinormal modal expansion in the diffraction grating

In the previous section, we have pointed out the importance of PML modes in the DQNM expansion. However, the nature and characteristics of there PML modes still remain to be elucidated. In order to simplify the problem and to speed up the numerical research of PML, it is convenient to reduce the dimension from 3-D to 2-D model.

In addition, in the previous two numerical examples, we have not really used the concept of 'left' eigenvectors. Therefore, it is of interest to demonstrate the irreplaceable importance of 'left' eigenvectors in the DQNM expansion in practical applications. Thus, in this section, we will study a 2-D structure of the diffraction grating problem.

### The 2-D model

We provide a simple grating structure, made of a periodic slit array etched in a silver membrane. This diffraction grating structure can be modeled by a 2-D geometry on the O**xy** plane filled with dispersive Drude-like scatterers as shown in Figure 7.7. For the sake of simplification, only $z$-anisotropic materials are considered; no mixing longitudinal and transverse. The studied structure is invariant along the $z$-axis and contains an infinitely $d$-periodic chain of scatterers ($\varepsilon^d$) along the $x$-axis, whose relative permittivity is given by the Drude model similar to (6.25):

$$\varepsilon_1(\omega) = \varepsilon_\infty - \frac{\omega_p^2}{\omega^2 + i\omega\gamma}$$

where $\varepsilon_\infty = 3.36174$, $\omega_p = 1.3388 \times 10^{16} \text{rad.s}^{-1}$, and $\gamma = 7.07592 \times 10^{13} \text{rad.s}^{-1}$, and surrounded by air $\varepsilon^+ = \varepsilon^- = 1$.

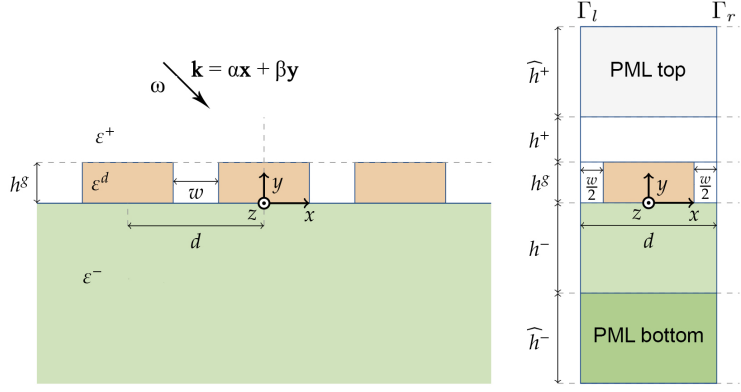

We will illustrate numerical results in the case of the diffraction grating with the geometry proposed by [70][71] where $h^g = 0.13\mu m$, $d = 0.4825\mu m$, $w = 0.135\mu m$, $h^+ = h^- = 2\mu m$, and $\widehat{h^+} = \widehat{h^-} = 25\mu m$. The parameters of the structure are chosen in such a way that the problem is well-posed in our domain of computation, (i.e. our eigenfrequencies are computed far away from the complex-frequency region where the Drude model has a purely real negative value [72]).[7] Thus we can perform our modal expansion without concerning the issue of corner modes.

7: We will study further this properties in Chapter 8

The difficulties of simulation lie not only in the dispersive material but also in the periodic and open computational domain. In order to handle the infinite $d$-periodicity along $x$-axis, a computational cell (the right-hand panel of Figure 7.7) is introduced. Let us denote by $\Gamma_l$, $\Gamma_r$ the two parallel boundaries of such cell orthogonal to the direction of periodicity $x$ and separated by $d$. For more details about the modeling of the 2-D diffraction grating, we refer the reader to appendix C.

In this paper, we will pay attention to the case where the magnetic field **H** is linearly polarized along the $z$-axis (i.e. p-polarization case). The whole structure is illuminated by an incident plane wave $\mathbf{H}_0 = H_0\mathbf{e}_z = \exp(i\mathbf{k} \cdot \mathbf{r})\mathbf{e}_z$, with wave vector **k** defined by the angle $\theta$ with respect to O**y** axis: $\mathbf{k} = \alpha\mathbf{x} - \beta\mathbf{y} = k(\sin\theta\mathbf{x} - \cos\theta\mathbf{y})$.[8]

8: see the Bloch-Floquet quasi-periodicity conditions in Section 3.4 and Section C.1 for more details.

To solve the diffraction problem of the given structure is equivalent to find the solution of the equation:

$$\mathscr{L}_{\varepsilon,\mu}(\lambda)H = -\nabla \cdot (\varepsilon^{-1}\nabla H) + \lambda^2\mu H = 0. \tag{7.7}$$

with the incident field $H_0$, and the diffracted field $H^d := H - H_0$ must satisfy an outgoing wave condition (OWC).

The field $H_1$ is defined as a solution of a simple reference diopter problem $\mathscr{L}_{\varepsilon_r,\mu_r}(\omega)H_1 = 0$ such that $H_1^d = H_1 - H_0$ satisfies an OWC. $\varepsilon_r$, and $\mu_r$ are chosen such that $H_1$ is considered to be known in a closed form (i.e. the single interface problem). Then, the problem (7.7) can be rewritten through an equivalent radiation problem:

$$\mathscr{L}_{\varepsilon,\mu}(\lambda)H^d = S_r, \qquad (7.8)$$

with $H^d = H - H_1 = H^d - H_1^d$. The source $S_r$ is supposed to be known $S_r = \mathscr{T}_{\varepsilon_r - \varepsilon, \mu_r - \mu}(\omega)H_1$. According to Bloch theorem, we have $H^d(x + d) = H^d(x)\exp(i\alpha d)$.

We apply PML by complex stretching the coordinates in the PML top and bottom domains as the extension part of superstratum and substratum [9] to the direction along which the field must decay ($y$-direction). In particular, $s_y(y) = \sigma\exp(i\phi)$ is chosen as a complex valued function of $y$ that is taken as a constant function (with $\sigma = 1$ and $0 < \phi < \pi/2$). By replacing $\varepsilon$ and $\mu$ in (7.8) the PML domain by new tensors $\delta_s = \text{diag}(s_y\delta_{xx}, s_y^{-1}\delta_{yy}, s_y\delta_{zz})$ for $\delta \in \{\varepsilon, \mu\}$ we will get a new field $H_{\text{PML}}$ which is identical to $H^d$, the solution of (7.8) outside the PML domain.

## Numerical results for the case $\alpha = 0$

### Eigen-solutions

By solving the eigenvalue problem at $\alpha = 0$ with different values of $\phi$, we obtain a map of eigenfrequencies in the complex plane as depicted in Figure 7.8.

It is easy to see that the original theoretical continuous spectrum, which is supposed to be located on $\mathbb{R}^+$ axis, is rotated through different angles. The precise value of these angles are $-\arg(h^+ + \widehat{h^+}\exp(i\phi)) \approx -\phi$ (For detailed instructions of calculation of the angle of rotation, we refer the reader to [73] ). This results in a large number of so-called discretized Bérenger 'PML' modes [64] whose position numerically changes based on PML parameters (i.e. the thickness $\widehat{h^+}$, $\widehat{h^-}$ and angle of rotation $\phi$). All the PML parameters ($\phi$, $\widehat{h^+}$ and $\widehat{h^-}$) must be chosen such that the absorption of the PML:

$$|\gamma_y| = \left| \max\left( \int_{h^+}^{\widehat{h^+}} s_y(y')dy', \int_{h^-}^{\widehat{h^-}} s_y(y')dy' \right) \right| \qquad (7.9)$$

is large enough to guarantee the complete absorption property of the PMLs, but at the same time, still able to be captured by the FE discretization.

Empirically, we can draw the following conclusion:

▶ The smaller the angle $\phi$, the denser the density of PML modes in the complex plane.
▶ The thinner the layer of PML $\widehat{h^+}$ and $\widehat{h^-}$, the denser the density of PML modes in the complex plane.

It should be noticed that the neatly aligned points corresponding to the continuous spectrum (i.e. PML modes) are clearly becoming numerically unstable further on the curve as explained by the pseudo-spectrum theory of L. Trefethen [74] . The field-map of these "PML" modes is concentrated in the PML region and far away from the scatterers (as can be seen from the field map of mode 3 in Figure 7.8). On the other hand, the field map of 'natural' resonant modes (i.e. mode 1 and mode 2) is distributed around the scatterers in the center of the structure. The position of these modes remains untouched when we change the angle $\phi$ of PML (see red sidebar Figure 7.8b).

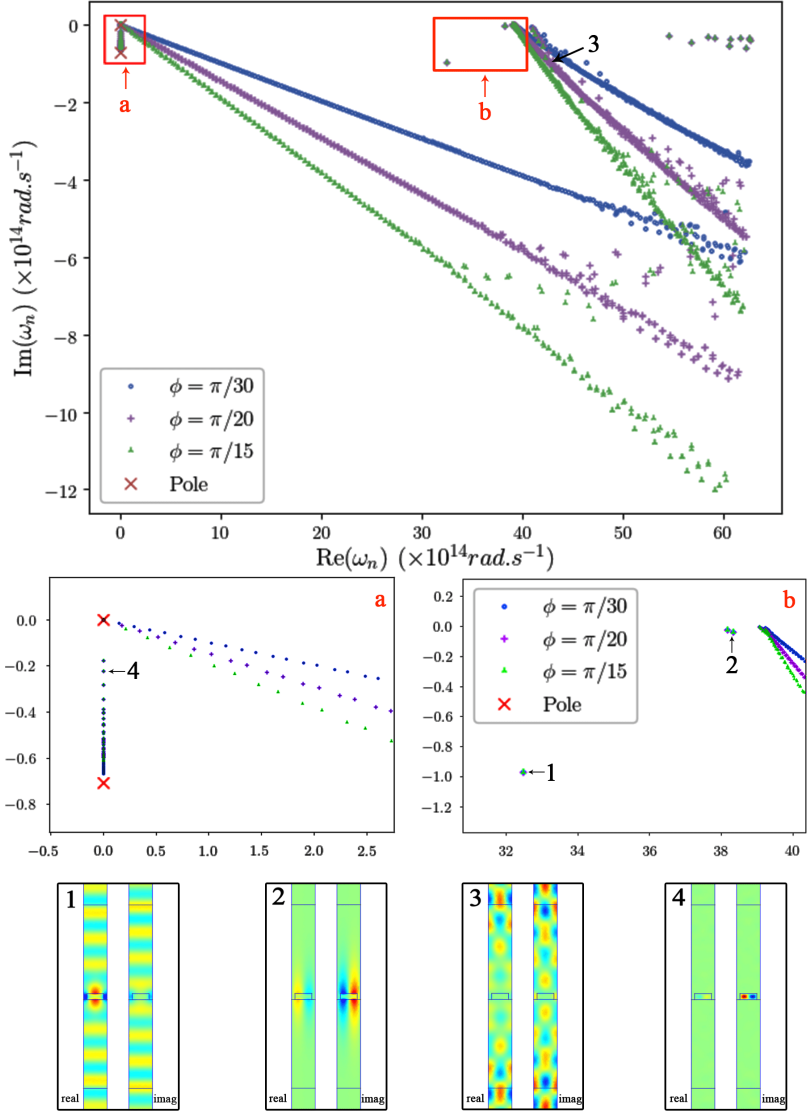

**Figure 7.8:**
(TOP)The spectrum of complex eigenfrequencies (computed at $\alpha = 0$) in the complex plan.
(MIDDLE) The two below sidebar a are b are the zoomed images at the areas indicated by the red boxes.
(BOTTOM) The Four eigen-fields are also depicted at the bottom (the blue color of the field maps indicates the minimum value and the red is the maximum).

Finally, it is worth pointing out that there are 2 accumulation points corresponding to 2 poles (i.e. $\omega = 0$ and $\omega = -i\gamma$) of the Drude model (see red sidebar Figure 7.8a), where the value of permittivity goes to infinity. The modes around these poles possess purely imaginary frequencies and concentrate inside the scatterers (for instance mode 4 in Figure 7.8). Numerical experiments show that these modes can overwhelm other DQNMs in the modal expansion, thus will be eliminated from the reconstruction of scattered fields in the present case.[10]

10: The issue of accumulation points around the complex pole of permittivity model will be addressed in Chapter 8.

**Quasinormal Modal expansion**

The next step is to compare the radiation solution $H^d$ computed by solving directly the problem (7.8) and the one reconstructed using the QNM expansion. By Lemma 5.1.1, the QNM expansion formula for the rational operator $\mathscr{L}_{\varepsilon,\mu}(\lambda)$ (7.7) is expressed as follows:

$$H^d = \sum_n \frac{f_\rho(\lambda_n)}{f_\rho(\lambda)} \frac{1}{(\lambda - \lambda_n)} \frac{\langle H_{ln}, S_r \rangle}{\langle H_{ln}, \dot{\mathscr{L}}_{\varepsilon,\mu}(\lambda_n) H_{rn} \rangle} |H_{rn}\rangle, \qquad (7.10)$$

where the inner product $\langle H_{ln}, \dot{\mathscr{L}}_{\varepsilon,\mu}(\lambda_n) H_{rn} \rangle$ is given by

$$\langle H_{ln}, \dot{\mathscr{L}}_{\varepsilon,\mu}(\lambda_n) H_{rn} \rangle = \int_\Omega \overline{H_{ln}} \left( -\nabla \cdot ((\varepsilon^{-1})' \nabla H_{rn}) + (\lambda^2 \mu(\lambda))' H_{rn} \right) \, d\Omega$$

$(\lambda_n, \langle \mathbf{H}_{ln}|, |\mathbf{H}_{rn}\rangle)$ are the eigen-triplets of the operator $\mathscr{L}_{\varepsilon,\mu}(\lambda)$:

$$\langle H_{ln} | \mathscr{L}_{\varepsilon,\mu}(\lambda_n) = 0 \qquad \mathscr{L}_{\varepsilon,\mu}(\lambda_n) | H_{rn} \rangle = 0.$$

In this numerical example, we want to focus on stuyding the impact of PML on the QNM expansion. Thus, the function $f_\rho(\lambda)$ will be chosen to be constant: $f_\rho(\lambda) = 1$.[11]

It is worth reminding that for $\alpha = 0$, the 'left' eigenvector is the complex conjugate of the 'right' counterpart: $H_{ln} = \overline{H_{rn}}$.[12] Thus, we only have to solve the eigenvalue problem once.

The diffraction efficiency is computed from the scattered fields reconstructed using (7.10) for all the incident frequencies in the domain of interest (red lines in Figure 7.9.1) and is compared with the exact direct data (blue lines). It is easy to see that the diffraction properties of the given structure are fully captured by our modal expansion technique with acceptable discrepancies at high frequencies. It is possible to reduce the error by lowering the value of $\phi$ (see Figure 7.9.2). It is also worth pointing out that all the computations at different $\phi$ are performed with the same thickness of PML $(\widehat{h^+}, \widehat{h^-})$ for the sake of comparison. In practice, it is recommended to increase the PML thickness while lowering the value of $\phi$ in order to maintain the value $|\gamma_y|$. This, in turn, requires increasing the number of modes in the modal expansion to cover the same range of frequencies, since the distribution of PML modes in the complex plane is denser.

The biggest difficulty in using the PML in the simulation of DQNM expansion comes from the non-dispersive-ness of the absorption value $|\gamma_y|$.[13] In particular, different frequencies require different absorption levels of PML. Since our main goal is to reconstruct the diffraction efficiency in the wide range of frequency, several issues occur:

▶ At low frequencies, i.e. high wavelengths, the absorption level of PML must be set to be high in order to completely absorb all the incoming waves from inside. This requires a high thickness of PML.

▶ At high frequencies, i.e. low wavelengths, the FE mesh size has to be chosen to be small enough in order to capture all the high-frequency oscillation of the electromagnetic field.

11: The numerical results of the QNM expansion for other functions $f_\rho(\lambda)$ are similar.

12: see Section 3.4 for more details

13: the complex stretching function $s_y$ to be more precise

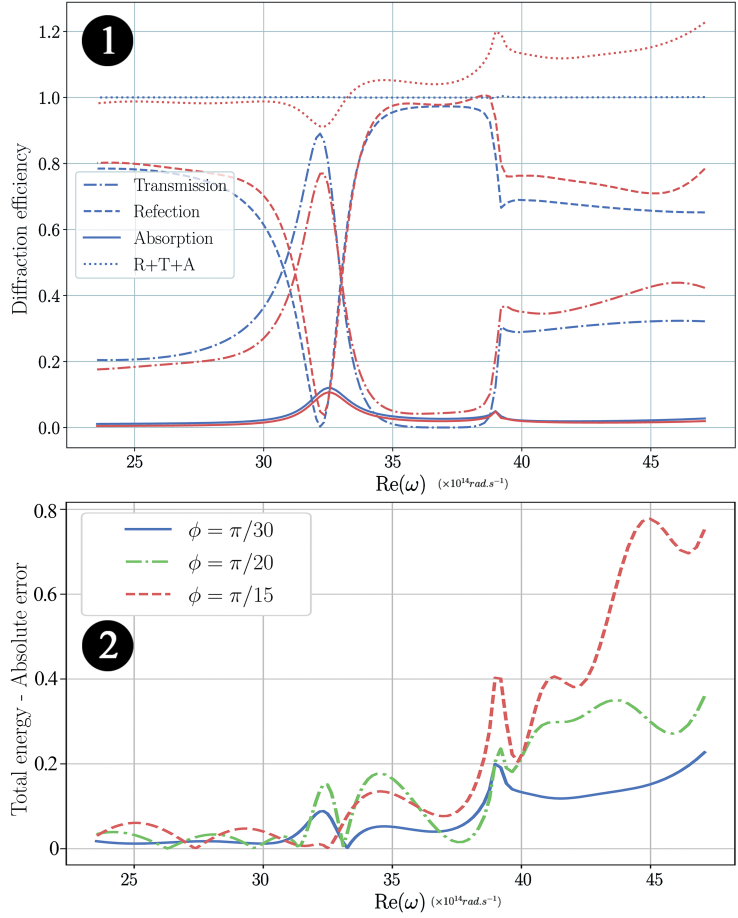

**Figure 7.9:** 1/ Diffraction efficiency computed at $\alpha = 0$: The blue lines refer to the direct computation, the red lines are reconstructed by the DQNM expansion from 700 modes with $\phi = \pi/30$.
2/ Absolute errors on total energy ($R + T + A$ in the upper subfigure) rebuilt from 700 modes for different values $\phi$.

▶ Increasing the value $\phi$ could enhance the instability of the PML modes according to the pseudo-spectrum theory of L. Trefethen. This could highly reduce the numerical accuracy of QNM expansion as we pointed out in Figure 7.9.2.

As a result, we face the problem of high computational cost for simulating a large computational domain with a refined FE mesh.

## Numerical results for the case $\alpha \neq 0$

Before ending this chapter, it is of interest to study the diffraction grating problem at $\alpha \neq 0$, one of the few electromagnetic problems which require solving the 'left' eigenvalue problem explicitly. The parameters of the geometric structure are given as follows: $\alpha = \pi/(2d)$, $\phi = \pi/20$, $h^+ = h^- = 1\mu m$, $\widehat{h^+} = \widehat{h^-} = 11\mu m$, which is slightly different from the previous example.

Figure 7.10.1 demonstrates the effectiveness of QNM expansion in calculating the diffraction efficiency.

It is easy to notice that the diffraction efficiency is dramatically affected by the 'resonant' mode (which is highlighted in the top panel of Figure 7.10). Thus, it is natural to wonder how many eigenmodes are required to reconstruct the diffraction efficiency around this 'resonant' mode. For example, the absorption efficiency, which is rebuilt using a different number of modes, is shown in Figure 7.10.2. It is interesting to witness that only one 'resonant' mode can reproduce the absorption characteristic of diffraction grating really well around the resonant mode $\omega \simeq 27.7$ $\times 10^{14}$ rad.$s^{-1}$. The accuracy of the reconstruction of the field map at such 'natural resonant mode' is also depicted in Figure 7.10.3. This is indeed a wonderful example to point out the usefulness of DQNM in understanding the intrinsic spectral properties of electromagnetic structures. With a small number of modes, we can recreate some important optical response of certain systems with acceptable errors.

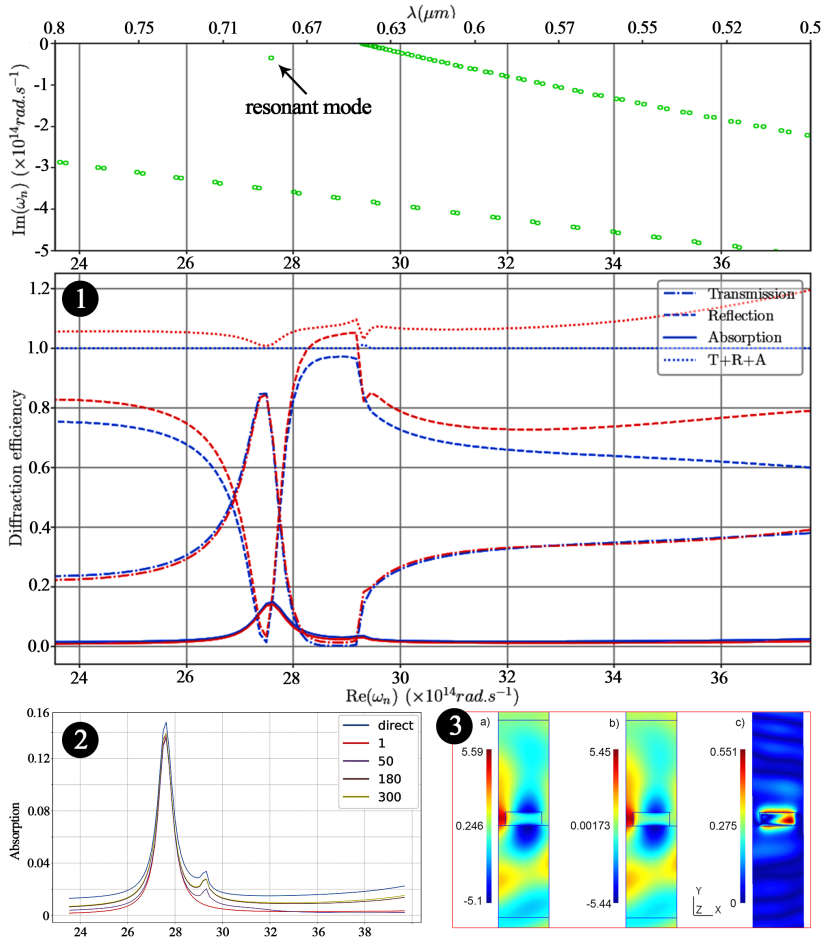

**Figure 7.10:** 1/ Diffraction efficiency computed at $\alpha = \pi/(2d)$, the upper figure presents several positions of these eigenmodes in the vicinity the 'resonant' DQNM mode in the complex plane.
2/ Absorption efficiency at $\alpha = \pi/(2d)$ reconstructed by the QNM expansion from different numbers of modes (from 1 to 300).
3/ The real part of the field at the 'resonant' frequency with $\alpha = \pi/(2d)$ calculated from the direct problem (a), rebuilt by the QNM expansion from 300 modes (b), and absolute error (c).

# Rational operators - Limitations of modal expansion  8

Before beginning the last chapter of this thesis, it is worth spending some time to review what we have achieved thus far.

The second part of the thesis explains the process deriving the Dispersive Quasinormal Modal expansion (DQNM expansion) in electromagnetics: We start by re-studying the spectral properties of linear operators, in particular, the Fourier expansion of self-adjoint linear operators. From that foundation, we have succeeded in extending the modal expansion to the case of non-self-adjoint operators with the help of the 'left 'eigenvectors. Next, we explore spectral characteristics of non-linear operators and formalize the modal expansion equations for 'rational 'operators. Since Maxwell's equations can be re-written in terms of 'rational ', the electromagnetic fields can be decomposed (expanded) as the DQNM expansion into a vector space formed by the resonant states (QNMs). Afterward, in the third part of the thesis, the correctness of the expansion has been numerically verified in both bounded and unbounded (open) structures. In particular, the non-uniqueness of DQNM expansions has been demonstrated: We have shown that there exists not just one, but a continuous family of DQNM expansion formulas. Then, in the previous chapter, we introduced the concept of PMLs to truncate the infinite computational domain for unbounded structures and studied the impact of PML on the modal expansion.

Until now, it seems arbitrary to construct the expansion based on 'rational' operators. Thus, the reader may wonder why we don't develop spectral theorems based on other non-linear operators. In fact, rational electromagnetic operators are derived from the 'rational' functions for the permittivity and permeability. In other words, querying the nature of 'rational' operators is equivalent to question the essence of the constitutive relations between the displacement field $\mathbf{D}$ and electric field $\mathbf{E}$, (or between the magnetic field $\mathbf{H}$ and $\mathbf{B}$). The situation boils down to the question of which function best represents the physical properties of permittivity and permeability. This question becomes more essential when the numerical modeling requires to include the actual measurement data of permittivity and permeability.

As a result, in this chapter, we will study how to extract a 'rational' mathematical model of permittivity from experimental data. This opens up opportunities applying the DQNM not only to hypothetical pre-modeled materials but also to realistic 'physical' materials. Through several examples, we also reveal the limitations of the modal expansion based on rational operators.

## 8.1 Extracting an accurate model of permittivity from experimental data

In this section, for the sake of simplicity, we only discuss the problem of extracting the formula of the permittivity from experimental data. The same consideration and procedure can be applied for the permeability. The construction of this section follows [43] .

In practice, the experimental values of complex permittivity and permeability are often given as tabulated data [75]. When solving the direct scattering problem in the frequency domain, the process of using these tabulated data is quite straightforward: For the given frequency, we can extract the complex value of the permittivity or permeability (or up to a simple interpolation) from the tabulate. Unfortunately, the research of spectral properties of dispersive structures requires an analytical expression of relative permittivity and permeability, which has to be extracted from the data. Indeed, this problem is well known in electromagnetics and several closed forms have already been proposed: The Drude (or Drude-Lorentz) model, the Debye model, the critical points model [76] ... The problem of these models is that they all try to pose a hypothetical mathematical model to permittivity. Thus, they are material-dependent and not flexible enough for a wide range of media.

The purpose of this section is to deduce a general representation of the frequency-dependence of the permittivities of media in terms of rational functions. In order to do this, we have to enforce the causality principle via the constitutive relation between the electric field $\mathbf{E}$ and polarization vector $\mathbf{P}_e$.

### The constitutive relation

For simplicity, in this section, we will assume materials are isotropic. Therefore, the second-order tensor of permittivity $\varepsilon$ can be expressed in terms of a scalar $\varepsilon$. In order to derive the model of the relative permittivity in the frequency domain, we have to come back to the constitutive relations (1.1) in the time domain:

$$\mathbf{H} = \mu_0^{-1}\mathbf{B} - \mathbf{M} \quad \text{and} \quad \mathbf{D} = \varepsilon_0 \mathbf{E} + \mathbf{P}_e,$$

where $\mathbf{M}$ and $\mathbf{P}_e$ are the magnetization and polarization vectors respectively.

We remind that due to time-varying media, the electric polarization of the materials, characterized by vector $\mathbf{P}$, can be expressed by the following equation:[1]

$$\mathbf{P}_e(\mathbf{r}, t) = \frac{\varepsilon_0}{2\pi} \int_{-\infty}^{+\infty} \chi_e(t - \tau)\mathbf{E}(\mathbf{r}, \tau) \, d\tau. \tag{8.1}$$

Similarly, due to hysteresis, the magnetic magnetization is formalized as follows:

$$\mathbf{M}(\mathbf{r}, t) = \frac{1}{2\pi\mu_0} \int_{-\infty}^{+\infty} \chi_m(t - \tau)\mathbf{B}(\mathbf{r}, \tau) \, d\tau. \tag{8.2}$$

1: Note that the factor before the integral is chosen to be consistent with our Fourier Transform convention (1.3).

The equation (8.1) is interpreted as follows: The polarization at the instant $t$ depends on all the past history $\tau$ of the electric field, modulated by the function $\chi_e(t)$ which is called the electric susceptibility of the material. Thus according to the causality, $\chi_e(t)$ has a positive support: $\chi_e(t) = 0$ for $t < 0$. The same argument applies to the magnetization. However, in this thesis, we will assume $\chi_m(t) = 0$ for simplicity.

From here, finding the permittivity model from experimental data is equivalent to formulating the electric susceptibility.

## An accurate model of the electric susceptibility from experimental data

The constitutive relation (8.1) implies that the electric susceptibility $\chi_e(t)$ can be considered as the Green function of a differential equation connecting the electric field and polarization vector, which is indeed the only requirement of the function $\chi_e(t)$. Thus, in order to keep $\chi_e(t)$ as general as possible, we start with the following constitutive relation:

$$\sum_{i=0}^{N_p} a_i \frac{\partial^i \mathbf{P}_e(t)}{\partial t^i} = \varepsilon_0 \sum_{j=0}^{N_e} b_j \frac{\partial^k \mathbf{E}(t)}{\partial t^k}. \tag{8.3}$$

Since the previous equation refers to realistic problems with physical quantities, all the coefficients $a_j$, $b_i$ must be real. Then, the frequency-dependent electric susceptibility $\chi_e(\omega)$ is given by carrying out the Fourier transform (1.3):[2]

$$\left(\sum_{i=0}^{N_p} a_i(-i\omega)^i\right) \mathbf{P}(\omega) = \varepsilon_0 \left(\sum_{j=0}^{N_e} b_j(-i\omega)^j\right) \mathbf{E}(\omega).$$

Thus, the electric susceptibility $\chi_e(\omega)$ reads:

$$\chi_e(\omega) = \frac{\sum_{j=0}^{N_e} b_j(-i\omega)^j}{\sum_{i=0}^{N_p} a_i(-i\omega)^i}.$$

It is easily seen that the susceptibility is represented by a rational function. Adopting the notation $\lambda = i\omega$ from Chapter 4 and Chapter 5, the electric susceptibility can be viewed as a rational function of the variable $\lambda$:

$$\chi_e(\lambda) = \frac{\mathcal{N}^{N_e}(\lambda)}{\mathcal{D}^{N_p}(\lambda)} = \frac{\sum_{j=0}^{N_e} c_j \lambda^j}{\sum_{i=0}^{N_p} d_i \lambda^i}, \tag{8.4}$$

where the denominator $\mathcal{D}^{N_p}(\lambda)$ (resp. numerator $\mathcal{N}^{N_e}(\lambda)$) is described as a polynomial degree $N_p$ (resp. $N_e$) of the variable $\lambda$. All the coefficients $c_j$ and $d_i$ of real value. As a result, the permittivity

$$\varepsilon(\lambda) = \varepsilon_0(1 + \chi_e(\lambda)),$$

must also be a rational function with respect to $\lambda$. [3]

With a closed form of the susceptibility, our next step is try to fit the model (8.4) with actual experimental data. In practice, the data [75] are

2: It is worth pointing out all the causality properties are reserved after the Fourier transform.

3: Keep in mind that with the assumption of isotropic media, the permittivity, in this case, is written in forms of a scalar. In the general case, the permittivity must be seen as a second-order tensor whose entries are rational functions of the variable $\lambda$.

often given by a set of corresponding points $(\omega^{\text{data}}, \chi^{\text{data}})$ in a tabulate.[4] These data can be re-organized in terms of vectors as follows:

$$\underline{\lambda} = \lambda_m = (\lambda_1^{\text{data}}, \lambda_2^{\text{data}}, \dots, \lambda_M^{\text{data}})^\mathsf{T}$$
$$\underline{\chi} = \chi_m^{\text{data}} = (\chi_1^{\text{data}}, \chi_2^{\text{data}}, \dots, \chi_M^{\text{data}})^\mathsf{T} \tag{8.5}$$

where $\lambda^{\text{data}} = i\omega^{\text{data}}$. The range $M$ of these vectors is only limited by the granularity of experimental data.

In order to extract a rational function from the data, we firstly rewrite the susceptibility by fixing $d_0 = 1$ as follows:

$$\chi_e(\lambda) = \frac{\sum_{j=0}^{N_e} c_j \lambda^j}{1 + \sum_{i=1}^{N_p} d_i \lambda^i}. \tag{8.6}$$

Next, we suppose the data in (8.5) satisfies the the rational form (8.6):

$$\chi_m^{\text{data}} = \frac{\sum_{j=0}^{N_e} c_j \lambda_m^j}{1 + \sum_{i=1}^{N_p} d_i \lambda_m^i}. \tag{8.7}$$

The previous equation can be rewritten as follows:

$$\chi_m^{\text{data}} = \sum_{n=0}^{N_e+N_p} r_n \xi_{mn},$$

where

$$r_n = \begin{cases} c_n & \text{if} \quad n = 0, \dots, N_e \\ d_{n-N_e} & \text{if} \quad n = N_e + 1, \dots, N_e + N_p \end{cases}$$

and

$$\xi_{mn} = \begin{cases} \lambda_m^n & \text{if} \quad n = 0, \dots, N_e \\ -\chi_m^{\text{data}} \lambda_m^{n-N_e} & \text{if} \quad n = N_e + 1, \dots, N_e + N_p \end{cases}.$$

Then, (8.7) can be expressed in the matrix form:

$$\underline{\chi} = \xi \underline{r}, \tag{8.8}$$

where $\xi$ is the $M \times (N_e + N_p + 1)$ matrix with entries $\xi_{mn}$ and $\underline{r}$ is the vector with entries $r_n$.

It is clear that the system (8.8) is over-determined. Therefore, the value of $\underline{r}$ can be solved by minimizing the following squared Euclidean 2-norm (see [77] for more details of the least-squares method):

Find the vector $\underline{R}$ such that:

$$\|\underline{\chi} - \xi\underline{R}\|_2^2 = \min\|\underline{\chi} - \xi\underline{r}.\|_2^2$$

It is worth reminding that the value $r_n$ must be real. However, in numerical computations, it is more convenient to set $r_n$ to be complex in order to relax our numerical scheme for the complex-valued susceptibility. This process, nevertheless, does not affect the causality assumption because the magnitude of the imaginary part of $r_n$ will be found out to be much smaller than the magnitude of the real part. Thus the imaginary part of these numbers will simply be omitted in the final results.

**Complex-valued electric poles**

Once the values $c_j$ and $d_i$ of (8.4) are extracted from the experimental data, it is straightforward to express the susceptibility as a rational function of the variable $\lambda$. Unfortunately, (8.4) does not provide much insight into the material properties. Thus, we want to look for the electric poles $\omega_i^\varepsilon$ of the susceptibility by finding the root of $\sum_{i=0}^{N_p} d_i \lambda^i$. As a result, the susceptibility will be re-written in forms of a multi-pole model:

$$\chi_e = \sum_{i=1}^{N_p} \frac{A_i}{\omega - \omega_i^\varepsilon} + g(\omega), \tag{8.9}$$

where $g$ is a holomorphic function representing a non-resonant term of $\chi_e$. By assuming that $g$ is negligible, the amplitude coefficients $A_i$ are computed by using another least-squares procedure. The electric poles $\omega_i^\varepsilon$ are written in terms of frequencies for intuitive reasons.

By the Titchmarsh theorem [78, 79] , the imaginary part of $\omega_i^\varepsilon$ must be non-negative to ensure causality. At the same time, the susceptibility is also required to be Hermitian symmetric. Therefore, it is more convenient to pair each 'physical' electric poles with its corresponding symmetric $-\overline{\omega_i^\varepsilon}$. Thus, the susceptibility reads:

$$\chi_e = \sum_{i=1}^{N_p} \left( \frac{A_i}{\omega - \omega_i^\varepsilon} - \frac{\overline{A_i}}{\omega + \overline{\omega_i^\varepsilon}} \right), \tag{8.10}$$

The previous equation indeed guarantees the Hermitian symmetry of the electric susceptibility.

It is easy to see that by adding more poles in (8.10), which is equivalent to increasing the order of the polynomial $\mathfrak{D}^{N_p}$, we can draw a better approximation of the electric susceptibility of realistic materials from measurement data. For more details about the procedure of hunting complex poles for the permittivity model from experimental data, we refer the reader to [43] .

As an example of this technique, we will try to extract the rational model for the permittivity of silicon from experimental data provided in [80]. The final result are depicted in the following tabular up to $N_p = 4$:[5]

5: The frequency unit is $\times 10^{14} (\text{rad.s}^{-1})$

**Table 8.1:** Electric poles of the rational model for the permittivity of silicon

| $i$ | 1 | 2 | 3 | 4 |
|---|---|---|---|---|
| $A_i(\mathbf{r})$ | $-165.959 - 20.199i$ | $-113.424 + 89.872i$ | $-41.362 + 41.091i$ | $-34.218 - 47.163i$ |
| $\omega_i^\varepsilon$ | $64.605 - 4.127i$ | $72.079 - 14.16i$ | $51.186 - 2.109i$ | $59.553 - 4.219i$ |

Figure 8.1 shows the real and imaginary part of the permittivity of silicon calculated based on (8.10) using the electric poles given by Table 8.1. According to Figure 8.1, it is clear that the more number of poles, the more our model of permittivity resembles realistic data.

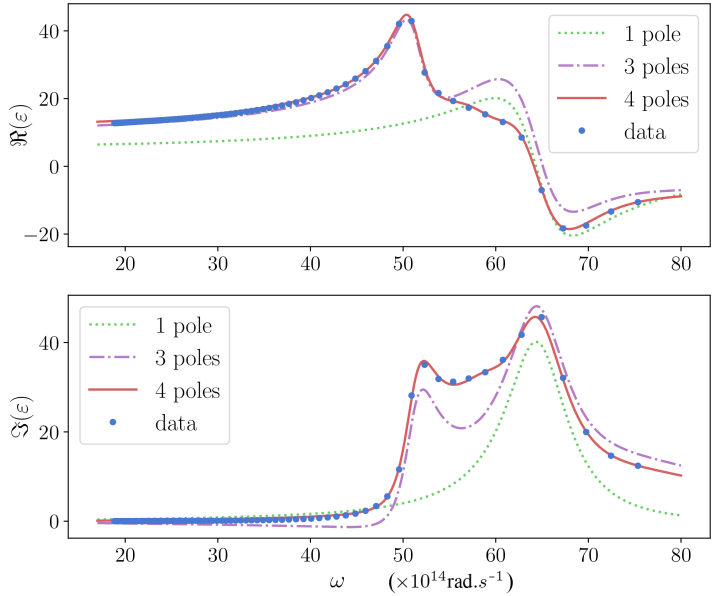

## Relation between $N_e$ and $N_p$

Before ending this section, it is worth spending a little time to reveal the analytical relation between $N_e$ and $N_p$. Based on numerical computations, a good choice is to set $N_e = N_p$, which is indeed the numerical approach we follow in this chapter. However, we may wonder if there is any physical explanation for this approach. The answer to this equation may be found in the nature of the electric polarization.

In practice, for very high frequencies, the permittivity of materials is shown to converge to a constant value, called 'high-frequency permittivity limit'. This phenomenon is mentioned and explained in several models of permittivity such as the Drude-Lorentz model or the Debye model [81] :

▶ In the Drude-Lorentz model, the polarization is associated with bounded electrons or lattice interacting with the electromagnetic field, which generally oscillate around their equilibrium position. Under the influence of the electric field, these electrons oscillate with damping frequencies, which is responsible for the electric susceptibility $\chi_e$. At very high frequencies, the susceptibility $\chi_e$ decreases asymptotically as $\chi_e \propto (1/\omega^2)$. This comes from the fact that the electrons can not react fast enough to the incident electric field. In the case of the Drude model, a metal becomes transparent since there is no more absorption.

▶ In the Debye model, the polarization is determined by permanent electric dipoles. Such permanent dipoles already exist in the materials but are randomly oriented. Only when an electromagnetic field is applied, these dipoles will be realigned towards the direction of the electromagnetic field. Unlike the Lorentz model, in which springs will drag a polarized oscillator back to its equilibrium position, the dipole orientation is randomized by thermal motion. Regardless, at high frequencies, the molecules can not follow the fast oscillation of the electromagnetic fields and the susceptibility $\chi_e$ decreases asymptotically as $\chi_e \propto (1/\omega)$.

Through the two previous models of susceptibility and the existence of the 'high-frequency permittivity limit', it is reasonable to conclude that $N_e \leq N_p$. Since the permittivity is given by $\varepsilon(\lambda) = \varepsilon_0(1 + \chi_e(\lambda))$, it is safe to assume Proposition 5.2.1:

> If the frequency-dependence of permittivity is expressed as a rational function, the order of polynomial of the numerator and denominator of such rational function are equal.

## 8.2 Complex-values poles of rational eigenvalue problem

It is clear that the dispersion of permittivity can be obtained through an interpolation method [43] that is very accurate on a large range of frequencies and thrifty with the number of poles. The obtained rational functions are naturally causal (following Kramers-Kronig relations) and provide a natural analytic continuation of permittivities in the complex plane [79] . The same argument can be applied to the permeability. Thus, the constructed electromagnetic operators $\mathscr{L}(\lambda)$ must also be rational.

It is worth pointing out that the domain of $\mathscr{L}(\lambda)$ must exclude all the values $\lambda$ which make the denominator zero. If we ignore the effect of magnetic magnetization, this implies that the operator $\mathscr{L}(\lambda)$ is not well-defined at the electric poles of the permittivity. It raises questions about the influence of those electric poles on the spectral properties of the 'rational' operator $\mathscr{L}(\lambda)$. To answer this question, let's set up some simple models made of materials whose permittivity is given by a multi-pole rational function (8.10).

### Silicon scatterer in a closed structure

We will illustrate numerical results in the geometry of an object shaped like an ellipse $\Omega_1$ inside a perfectly conducting vacuum square $\Omega_0$ (see Figure 8.2). The parameters of the structure are chosen in such a way that the material and geometric resonances highly interact with each other. In particular, we want to exhibit the problem where the geometric resonances are in the vicinity of the electric poles of the permittivity.

The elliptic scatterer is made of silicon whose dispersive relative permittivity $\varepsilon_{Si}$ is given by the multi-pole rational model (8.10) and Table 8.1. At the same time, the relative permittivity of vacuum is set to be constant $\varepsilon_{vac} = 1$. Then, the whole structure will be illuminated by the Dirac delta source $\mathbf{S} = \delta(\mathbf{r}_S)$ whose coordinates is given by $\mathbf{r}_S = (-2.4, 0.8) \, (\times 10^{-1} \mu m)$, see Figure 8.2. The maximum element size is set to be $0.03 \, \mu m$, in comparison to the smallest wavelength in vacuum $0.2355 \, \mu m$ (equivalent to the highest frequency of the spectrum).

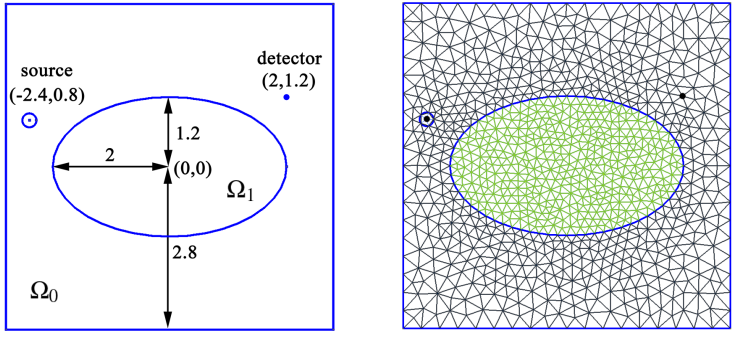

## DQNM expansion

In this example, we reconsider the scattered vector electric field $\mathbf{E}$, the solution of the following direct scattering problem:

$$\mathscr{L}_{\boldsymbol{\mu},\boldsymbol{\varepsilon}}(\lambda)\mathbf{E} = \nabla \times \left(\boldsymbol{\mu}^{-1}(\lambda)\nabla \times \mathbf{E}\right) + \lambda^2 \boldsymbol{\varepsilon}(\lambda)\mathbf{E} = \mathbf{S}. \quad (8.11)$$

Since the entries of the tensor $\boldsymbol{\varepsilon}(\lambda)$ are rational functions of the variable $\lambda$, while $\boldsymbol{\varepsilon}$ is assumed to be constant, $\mathscr{L}_{\boldsymbol{\mu},\boldsymbol{\varepsilon}}(\lambda)$ can be seen as a rational operator:

$$\mathscr{L}_{\boldsymbol{\mu},\boldsymbol{\varepsilon}}(\lambda) = \frac{\mathscr{N}_L^{N_N}(\lambda)}{\mathscr{D}^{N_D}(\lambda)}. \quad (8.12)$$

where the numerator $\mathscr{N}_L^{N_N}(\lambda)$ is set to be a polynomial operator of degree $N_N$, while the denominator $\mathscr{D}^{N_D}(\lambda)$ is described as a polynomial of degree $N_D$.

Similar to the previous chapters, with the help of the eigentriplets $(\lambda_n, \langle \mathbf{E}_{ln}|, |\mathbf{E}_{rn}\rangle)$ satisfying:

$$\langle \mathbf{E}_{rn}|\mathscr{L}_{\boldsymbol{\mu},\boldsymbol{\varepsilon}}(\lambda_n) = 0 \qquad \text{and} \qquad \mathscr{L}_{\boldsymbol{\mu},\boldsymbol{\varepsilon}}(\lambda_n)|\mathbf{E}_{rn}\rangle = 0,$$

the DQNM expansion of the solution $\mathbf{E}$ appears to be:

$$\mathbf{E} = \sum_n \frac{f_\rho(\lambda_n)}{f_\rho(\lambda)} \frac{1}{\lambda - \lambda_n} \frac{\langle \mathbf{E}_{ln}, \mathbf{S}\rangle}{\langle \mathbf{E}_{ln}, \mathscr{L}_{\boldsymbol{\mu},\boldsymbol{\varepsilon}}(\lambda_n)\mathbf{E}_{rn}\rangle} \mathbf{E}_{rn}, \quad (8.13)$$

where the inner product in the denominator can be computed explicitly as follows:

$$\langle \mathbf{E}_{ln}, \mathscr{L}_{\boldsymbol{\mu},\boldsymbol{\varepsilon}}(\lambda_n)\mathbf{E}_{rn}\rangle =$$
$$\int_\Omega \left[ \overline{\mathbf{E}_{ln}} \cdot \left(\nabla \times ((\boldsymbol{\mu}^{-1}(\lambda_n))'\nabla \times \mathbf{E}_{rn})\right) + \overline{\mathbf{E}_{ln}} \cdot \left((\lambda_n^2 \boldsymbol{\varepsilon}(\lambda_n))'\mathbf{E}_{rn}\right) \right] d\Omega.$$

The next important step is to identify the value of degree $\rho$ in (8.13) by finding out the relation between $N_N$ and $N_D$. In the previous section, we have already explained that the frequency-dependence of permittivity can be efficiently represented by a rational function, whose numerator and denominator are polynomials of the same degree (see Proposition 5.2.1). The same argument is applied to the permeability. Then, by (5.4), the values of $N_N$ and $N_D$ in (8.11) are given by:

$$N_N - N_D = 2. \tag{8.14}$$

It is worth pointing out that the previous conclusion holds for all dispersive, anisotropic and even non-reciprocal materials.

As a result, $f_p(\lambda)$ in (8.13) can be any arbitrary polynomial up to degree 1. This means $f_p(\lambda)$ would take the form $f_p(\lambda) = \alpha + \lambda\beta$ with $\forall \alpha, \beta \in \mathbb{C}$ and $|\alpha| + |\beta| \neq 0$. As a consequence, there exists a continuous family of expansion formulas for the electric field $\mathbf{E}$ of the operator $\mathscr{L}_{\boldsymbol{\mu},\boldsymbol{\varepsilon}}(\lambda)$ in (8.11).

## Accumulation points at complex electric poles

Now, we are ready to apply the DQNM expansion to solve the scattering problem (8.11) (TE polarization) in the structure given by Figure 8.2. It is also worth reminding that the 3-D electrodynamic eigenvalue computations require genuine edge elements to avoid spurious modes but, in our 2-D case, we use longitudinal fields $E_z\mathbf{e}_z$ or $H_z\mathbf{e}_z$ and the associated edge elements reduce to the Lagrange basis elements, here second-order, for the (scalar) component $E_z$ or $H_z$. For the sake of clarification, let's begin with a 1-pole model of permittivity, i.e. $N_p = 1$.

### 1-pole model of permittivity

The complex eigenfrequencies are shown in the bottom-left panel of Figure 8.3. It is easily seen that the function of 1-pole permittivity (in the top-left panel) changes dramatically at the frequency of the electric pole $\omega_1^e$ (red cross in the bottom-left panel). We emphasize the existence of an accumulation point in the vicinity of the electric pole where $\varepsilon_{Si} \to \infty$. The modes around the pole concentrate inside the scatterer and have spatial frequency tending to infinity (for instance mode 2 in Figure 8.3), which distinguishes them from conventional modes whose eigenfield is located around the scatterer (see mode 1 and mode 3 in Figure 8.3).

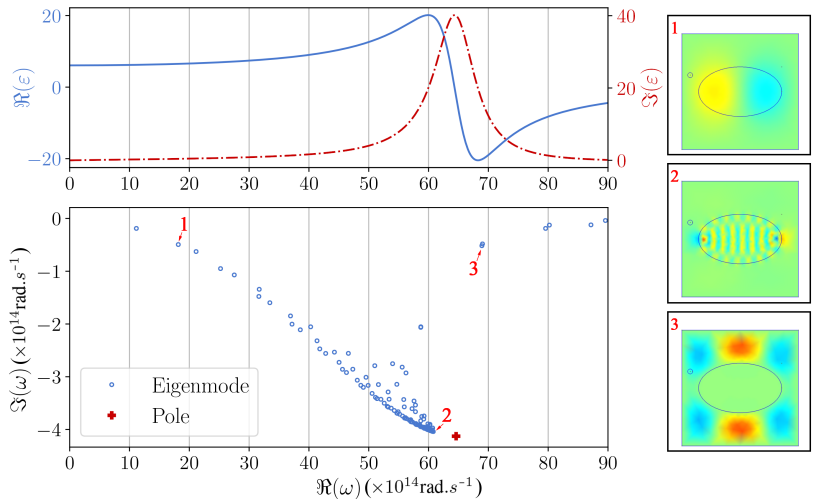

**Figure 8.3:**
(TOP LEFT) The analytical real (blue line) and imaginary part (red line) of the permittivity of silicon computed using 1 electric pole.
(BOTTOM LEFT) Spectrum of complex eigenfrequencies (left bottom) corresponding to the 1-pole relative permittivity (in the left top panel).
(RIGHT) Three eigenfields (real part) are depicted (the blue color of the field maps indicates the minimum value and the red is the maximum).

In fact, we have already witnessed the existence of the accumulation points around poles of the Drude model in Figure 8.3 of Section 7.4. In that case, it is fortunate that the modes around these poles possess purely imaginary frequencies. Thus, these accumulation points do not affect the 'physical' properties of the diffraction grating and are simply eliminated from the reconstruction of scattered fields. On the other hand, in the current example, the accumulation point is in the middle of our domain of interest. Therefore, it is hard to precisely distinguish the eigenmodes accumulating around the electric pole from the 'physical' conventional modes (for example mode 1). As a result, we have to take into account all the above eigenmodes in the DQNM expansion. We have to keep in mind that, in the case of closed structures, the spectrum of eigenfrequencies is indeed symmetric through the imaginary axis. Since the contribution of the modes on the left half of the complex plane is numerically insignificant,[6] we do not include them in the computation.

[6]: The factor $\frac{1}{\lambda - \lambda_n}$ of these modes is relatively small compared to their counterparts on the right half of the complex plane

The numerical comparison between the direct computation of electric field **E** and the reconstruction based on the DQNM expansion (8.13) where $f_p(\lambda) = 1$, and $f_p(\lambda) = \lambda$ is displayed in Figure 8.4. The results are represented in three figures corresponding to three quantities: the norm of the electric field in the domain $\Omega_1$: $\int_{\Omega_1} |\mathbf{E}| d\Omega$ (displayed in a logarithmic scale at the top panel of Figure 8.4); the real and imaginary part of the electric field $\mathbf{E}_1$ calculated at the detector point in Figure 8.2 (represented by the middle and bottom panel of Figure 8.4). The DQNM expansion in the cases where $f_p(\lambda) = 1$ (blue lines) and $f_p(\lambda) = \lambda$ (red lines) demonstrates the excellent agreement with green dots obtained by directly solving the scattering problem except in the vicinity of $\Re(\omega_1^{\varepsilon})$.

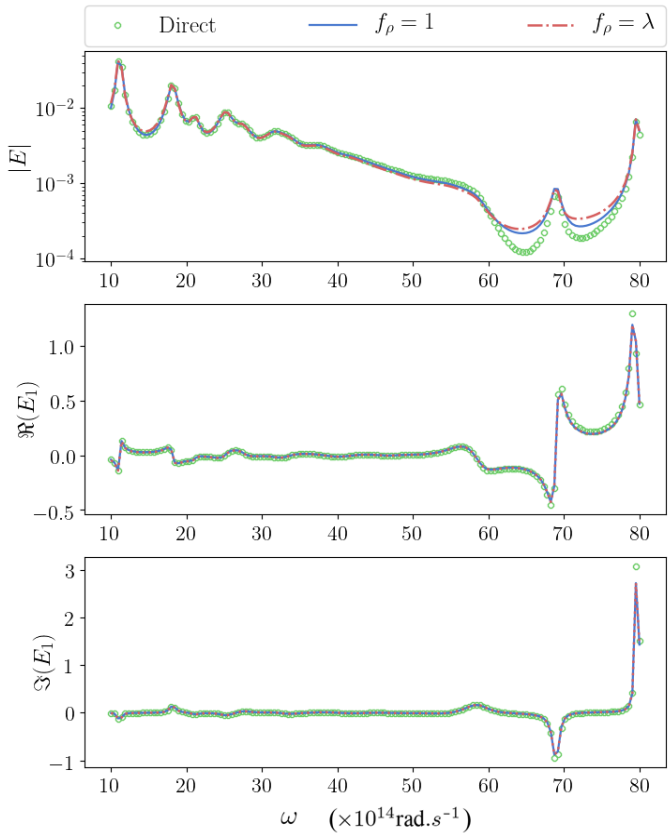

**Figure 8.4:** Scattered field **E** obtained by different expansion formulas (8.13) or by solving a direct problem classically (green dots) corresponding to the 1-pole permittivity:
(TOP) Integral over $\Omega_1$ of the norm of the electric field $\int_{\Omega_1} |\mathbf{E}| d\Omega$.
(MIDDLE and BOTTOM) The real and imaginary part of the electric field calculated at the detector point.

In addition, the norm of the field $\int_{\Omega_1} |\mathbf{E}| d\Omega$ reconstructed using the DQNM expansion formula with $f_\rho(\lambda) = \lambda - \lambda_0$ and $f_\rho(\lambda) = \lambda^2$ are shown in the top panel of Figure 8.5. The value $\lambda_0 = i\omega_0$ is selected such that $\omega_0 = 31.628 - 1.478i$ ($\times 10^{14}$rad.$s^{-1}$), which is almost coincide with an eigenfrequency. With $f_\rho(\lambda) = \lambda - \lambda_0$ (purple lines), we notice discrepancies around the frequency $\mathfrak{R}(\omega_0)$ (highlighted by the yellow vertical line), which remind us that the singularity at the roots of $f_\rho(\lambda)$ must be treated with caution[7] . As a result, it is recommended to choose $f_\rho(\lambda) = \alpha + \lambda\beta$ such that the value $-\alpha/\beta$ is far away from our domain of interest. Unsurprisingly, when the degree of the polynomial $f_\rho(\lambda)$ is higher than 1, i.e. $f_\rho(\lambda) = \lambda^2$ (orange line), we see the less accurate in the numerical results of the DQNM expansion since it is not an appropriate formula.

7: see Section 5.5 for more details

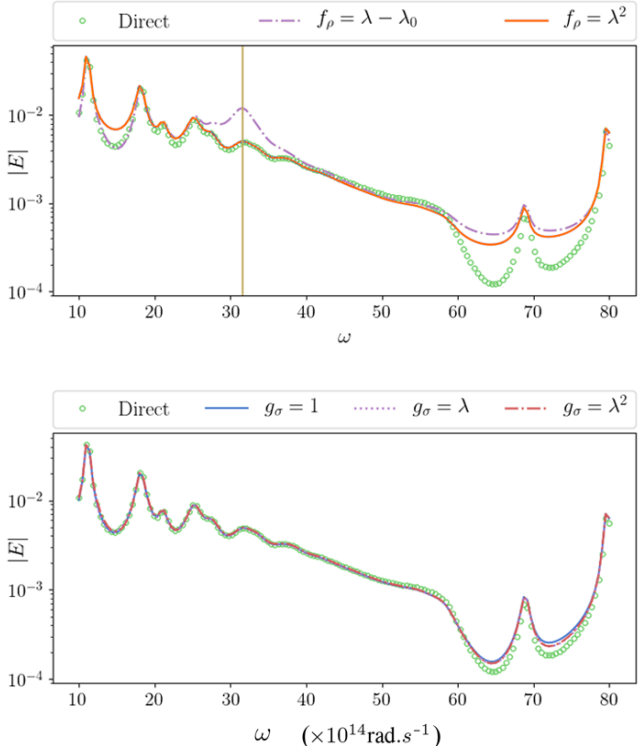

**Figure 8.5:** Integral over $\Omega_1$ of the norm of the electric field $\int_{\Omega_1} |\mathbf{E}| d\Omega$ for:
(TOP) Scattered field $\mathbf{E}$ reconstructed by (8.13) with $f_\rho(\lambda) = \lambda - \lambda_0$ and $f_\rho(\lambda) = \lambda^2$. The orange vertical line indicates the value $\mathfrak{R}(\omega_0)$.
(BOTTOM) Scattered field $\mathbf{E}$ rebuilt by (8.16) for $g_\sigma(\lambda) = 1$, $g_\sigma(\lambda) = \lambda$ and $g_\sigma(\lambda) = \lambda^2$.

Finally, we emphasize again on the non-uniqueness of the quasinormal modal expansion. In particular, formula (8.13) is not the only modal expansion family for the solution $\mathbf{E}$ of the scattering problem (8.11). Indeed, since $\mathcal{L}_{\mu,\varepsilon}(\lambda)$ is a rational operator (8.12), $\mathbf{E}$ is also the solution of the following equation:

$$\mathcal{N}_L^{N_N}(\lambda)\mathbf{E} = \mathscr{D}^{N_D}(\lambda)\mathbf{S}. \tag{8.15}$$

By Lemma 4.5.2, we can obtain another family of DQNM expansion for the solution $\mathbf{E}$ as follows:

$$\mathbf{E} = \sum_n \frac{g_\sigma(\lambda_n)}{g_\sigma(\lambda)} \frac{1}{\lambda - \lambda_n} \frac{\langle \mathbf{E}_{ln}, \mathscr{D}^{N_D}(\lambda)\mathbf{S} \rangle}{\langle \mathbf{E}_{ln}, \dot{\mathcal{N}}_L^{N_N}(\lambda_n)\mathbf{E}_{rn} \rangle} \mathbf{E}_{rn}, \tag{8.16}$$

with $\dot{\mathcal{N}}_L^{N_N}(\lambda_n)$ is the derivative of the operator $\mathcal{N}_L^{N_N}(\lambda)$ at $\lambda_n$.

We notice that the degree of the polynomial $g_\sigma(\lambda)$ can be set to be larger than 1 (as demonstrated by the bottom panel of Figure 8.5) and only be limited by the degree $N_N$ of the polynomial operator $\mathcal{N}_L^{N_N}(\lambda)$. This implies that the more poles the permittivity model has (the larger the value $N_p$ is), the higher the degree of the polynomial $g_\sigma(\lambda)$ can be.

## Multi-pole model of permittivity

From the previous subsection, it is clear that electric poles of the permittivity may cause discrepancies around the accumulation point in the DQNM expansion. The question arises of what happens to the DQNM expansion if we try to increase the number of poles in the permittivity model (8.10).

The spectrum of complex eigenfrequencies for the 4-pole permittivity model $N_p = 4$ is shown in the bottom-left panel of Figure 8.6. For the sake of comparison, the function of the 4-pole permittivity is plotted in the top-left panel of Figure 8.6.

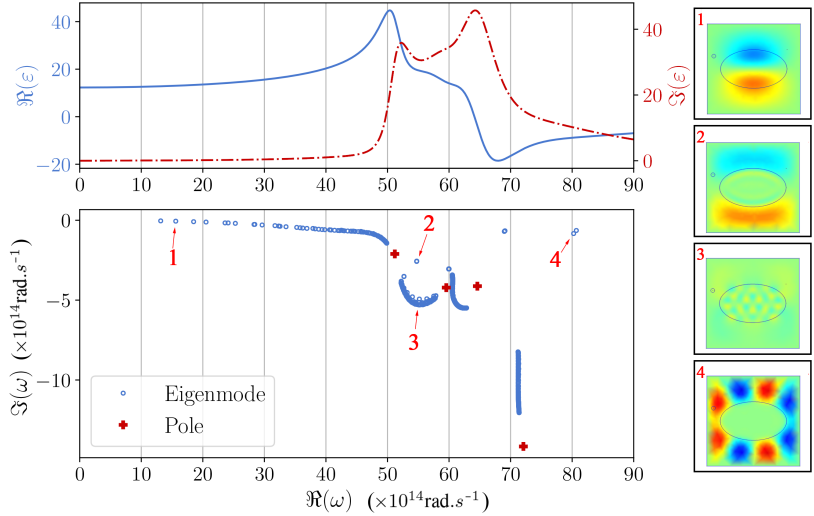

**Figure 8.6:**
(TOP LEFT) The analytical real (blue line) and imaginary part (red line) of the permittivity of silicon computed using 4 electric poles.
(BOTTOM LEFT) Spectrum of complex eigenfrequencies (left bottom) corresponding to the 4-pole relative permittivity (in the left top panel).
(RIGHT) Eigenfields (real part) are depicted (the blue color of the field maps indicates the minimum value and the red is the maximum).

In Figure 8.6, we especially focus on the 4 electric poles (red crosses), which divide the complex plane into several sub-regions. In those subregions, we can distinguish two types of eigenfrequencies:

▶ Some are distributed separately (for example mode 1, 2, and 4 illustrated on the right side of Figure 8.6) which are responsible for the physical properties of the system.
▶ Some gather into separate clusters (for example mode 3 on the right side of Figure 8.6) whose eigenfields oscillate with very high spatial frequencies. These modes result from the accumulation points around the poles of the permittivity.

Numerical computations show that the second kind of eigenmodes, which we will call 'cluster' modes, affects the performance of the DQNM expansion around the corresponding frequencies (see Figure 8.7).

We again compare electric field **E** between the direct computation and the reconstruction based on the DQNM expansion (8.13) where $f_p(\lambda) = 1$, and $f_\rho(\lambda) = \lambda$ through three quantities: the norm of electric field in the domain $\Omega_1$: $\int_{\Omega_1} |\mathbf{E}| d\Omega$ (displayed in a logarithmic scale at the top panel of Figure 8.7); the real and imaginary part of the electric field $\mathbf{E}_1$ calculated at the detector point in Figure 8.7 (represented by the middle and bottom panel of Figure 8.7). Indeed, the DQNM expansion shows good agreement with respect to the direct data (green dots) except around the frequencies of the 'cluster' modes and the electric poles of the permittivity. It is also worth noting that when $f_\rho(\lambda) = \lambda - \lambda_0$, the expansion (8.13) endures discrepancies around $\mathfrak{R}(\omega_0)$ where $\omega_0 = 29.633 - 0.295i$ ($\times 10^{14}$rad.$s^{-1}$).

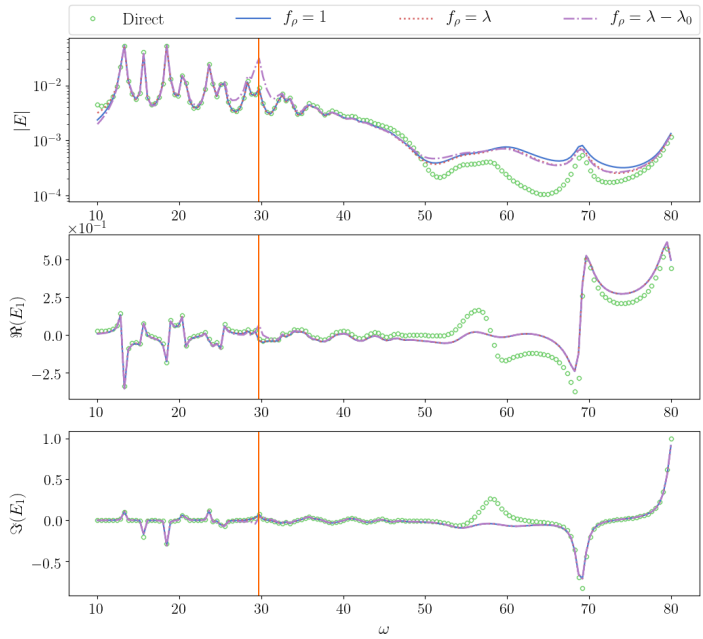

**Figure 8.7:** Scattered field $\mathbf{E}$ obtained by expansion for different functions of $f_\rho$ (blue, red and purple curves) or by solving a direct problem classically (green dots) corresponding to the 4-pole permittivity:
(TOP) Integral over $\Omega_1$ of the norm of the electric field $\int_{\Omega_1} |\mathbf{E}| d\Omega$.
(MIDDLE and BOTTOM) The real and imaginary part of the electric field calculated at the detector point. The orange vertical line indicates the value $\mathfrak{R}(\omega_0)$.

The last two examples indicate the principal difficulty of DQNM expansion based on rational operators. In particular, the electromagnetic rational operators are not well-defined around the electric poles of the permittivity. These electric poles, in turn, fragment the spectrum of eigenfrequencies in the complex plane into many sub-regions, thus, limit the effective frequency range of our DQNM expansion. For example, in the previous example of the 4-pole permittivity, Figure 8.7 shows that the DQNM expansion can no longer reproduce exactly the electric field in the frequency range from 50 to 75 ($\times 10^{14}$ rad.$s^{-1}$). In short, we face a trade-off situation:

► By raising the number of poles, the permittivity model fits better with realistic materials.
► Increasing the number of electric poles results in increasing ambiguous domains (ill-defined domains) of the rational operators. Although these ambiguous domains do not affect the numerical solution of the direct scattering problem, they cause the accumulation points in the spectrum of complex resonances in the complex plane. These accumulation points around the electric poles, if not handled with care, can cause uncontrolled errors in the DQNM expansion.

In this section, we have mentioned the problem of the electric poles of the permittivity. The spectrum of eigenfrequencies of the electromagnetic operators is fragmented and spoiled by accumulation points around these electric poles. It is important to keep in mind that these accumulation points are not related to any realistic 'physical' characteristics of the electromagnetic system: They are just numerical consequences of our choice of the mathematical model. In particular, the electric poles result from the use of rational functions to represent the electric susceptibility of the material.

In fact, the electric poles of the permittivity are not the only value at which the rational operator is not defined. In the next section, we will study another source of accumulation points: The regions where the finite element analysis of the electromagnetic scattering problem is no longer well-posed.

## 8.3 Sign-changing coefficient in finite element analysis

In order to understand the problem of the ill-posedness of the finite element analysis of electromagnetic systems, let's consider the magnetic scattering problem (TM polarization) in the same geometric structure of Figure 8.2:

$$\mathcal{L}_{\varepsilon,\mu}(\lambda)\mathbf{H} = \nabla \times \left(\varepsilon^{-1}(\lambda)\nabla \times \mathbf{H}\right) + \lambda^2 \mu(\lambda)\mathbf{H} = \mathbf{S}. \tag{8.17}$$

with the homogeneous Dirichlet boundary condition $\mathbf{H} \in H_0^D(\mathbf{curl}, \Omega)$ where

$$H_0^D(\mathbf{curl}, \Omega) := \{\mathbf{F} : \mathbf{F}, \nabla \times \mathbf{F} \in L^2(\Omega)^3, \mathbf{n} \times \mathbf{F}|_\Gamma = 0\}.$$

From Section 6.1 and Appendix B, the weak formulation of (8.17) reads:

$$a(\mathbf{F}, \mathbf{H}) = \langle \mathbf{F}, \mathbf{S} \rangle \qquad \forall \mathbf{F} \in H_0^D(\mathbf{curl}, \Omega) \tag{8.18}$$

where the sesqui-linear form $a(\cdot, \cdot)$ is given by

$$a(\mathbf{F}, \mathbf{H}) = \int_\Omega \left(\nabla \times \overline{\mathbf{F}}\right) \cdot \left(\varepsilon^{-1}(\lambda)\nabla \times \mathbf{H}\right) + \lambda^2 \overline{\mathbf{F}} \cdot \left(\mu(\lambda)\mathbf{H}\right) d\Omega. \tag{8.19}$$

It is worth noting that materials do not always have positive permittivity. In particular, a negative material can be metal (where the permittivity is given by the Drude model as numerical examples in Chapter 6 and Chapter 7) or dielectric at high frequencies (for example silicon in Figure 8.1) or metamaterial [82] . Since the permittivity of the scatterer can be negative, there will be the sign shift of $\varepsilon$ across the interface $\Sigma$, dividing the scatterer $\Omega_1$ and the surrounding background $\Omega_0$. Thus, the coercivity of the sesqui-linear form $a(\cdot, \cdot)$ is not guaranteed. As a result, Theorem B.2.1 can no longer be applied to establish the well-posedness of (8.18).

## Surface plasmon resonance

The difficult related to sign-changing coefficients has been well known in both direct problems and spectral problems [83, 84] . The solution is an abstract mathematical approach called T-coercivity. In particular, under some assumptions, the problem (8.18) can be shown to be well-posed as long as the contrast, i.e. the ratio of the values of $\varepsilon$ across the interface, lies outside some interval $I_c$, called critical interval.

Let's denote $\kappa_\varepsilon$ the ratio of the values of the permittivity across the interface $\Sigma$ between the scatterer $\Omega_1$ (which is made of silicon in this example) and the vacuum background:

$$\kappa_\varepsilon = \frac{\varepsilon_{\mathsf{Si}}}{\varepsilon_0}.$$

According to [84], since $\Sigma$ is smooth, $I_c = \{-1\}$. From a physical point of view, the critical point $I_c = \{-1\}$ associated with special resonant state so-called surface plasmon. Surface plasmons have many applications in practice, including guiding or confining light in nano-photonic devices [85, 86] .

In the direct problem, with a real and fixed frequency, the problem of well-posedness is hidden by the fact that most of the physical problems are dissipative (i.e. the real-part changing coefficient has a non-vanishing imaginary part). Thus, the contrast $\kappa_\varepsilon$ can not be exactly equal to $-1$. However, in spectral problems with complex frequencies, the complex plane of frequencies can contain regions where the sign-changing coefficient is purely real. This is indeed the case of our example of the magnetic fields in TM polarization (8.17).

Figure 8.8 illustrates the spectrum of eigenfrequencies of magnetic fields **H** for the 4-pole permittivity. Besides the 2 types of modes we mentioned in the previous section, there exists the third kind of eigenmodes which accumulates around the plasmon branch where $\varepsilon_{\mathsf{Si}}(\omega_2) = -1$ (green crosses in Figure 8.8). These modes are indeed plasmonic resonances that distribute on the interface $\Sigma$ between $\Omega_0$ and $\Omega_1$ with the spatial frequencies tending to infinity (for example mode 1, 3, and 4 in Figure 8.8). They must be distinguished from the 'cluster' modes caused by electric poles of the permittivity (mode 2 in Figure 8.8).

It is worth pointing out that the locations of these plasmonic resonances do not conspicuously converge (see inset C of Figure 8.8 for example) to the analytical position $\omega_2$ where $\varepsilon_{\mathsf{Si}}(\omega_2) = -1$. This can be explained by the ill-posedness of the finite element scheme around the critical point $I_c = \{-1\}$. This problem can be fixed by imposing the T-coercivity on the sesquilinear form $a(\cdot, \cdot)$. In the T-coercivity framework, it is possible to prove that the discretized problem of (8.18) is well-posed and its solution converges to the solution of the continuous problem. In detail, the technique requires to build a structured symmetric mesh around the interface $\Sigma$ to stabilize the numerical discretization of the plasmonic accumulation points [50] . Although the construction of symmetric mesh in the case of polygonal sign-changing interfaces has been well documented in [83], the symmetry requirements with respect to a curved boundary for this example remain to be elucidated.

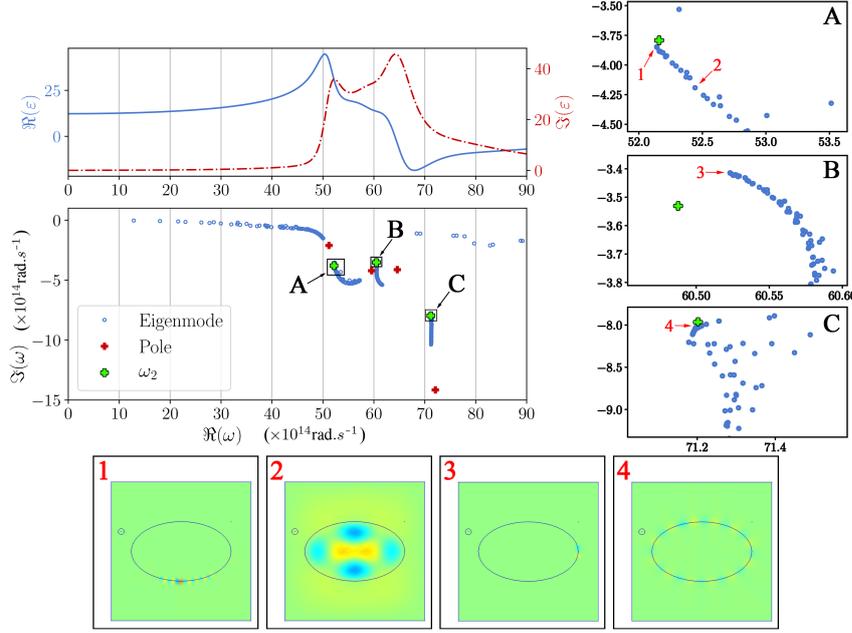

**Figure 8.8:**
(TOP LEFT) The analytical real (blue line) and imaginary part (red line) of the permittivity of silicon computed using 4 electric poles.
(MIDDLE LEFT) Spectrum of complex eigenfrequencies of magnetic field **H** corresponding to the 4-pole permittivity. The green crosses refer to the location of plasmons, solutions of $\varepsilon_{Si}(\omega_2) = -1$ (There should be 4 green crosses but the forth one is out of our domain of interest).
(A, B and C) Enlarged images of the spectrum around $\omega_2$.
(BOTTOM) Four eigenfields (real part): The blue color of the field maps indicates the minimum value and the red is the maximum.

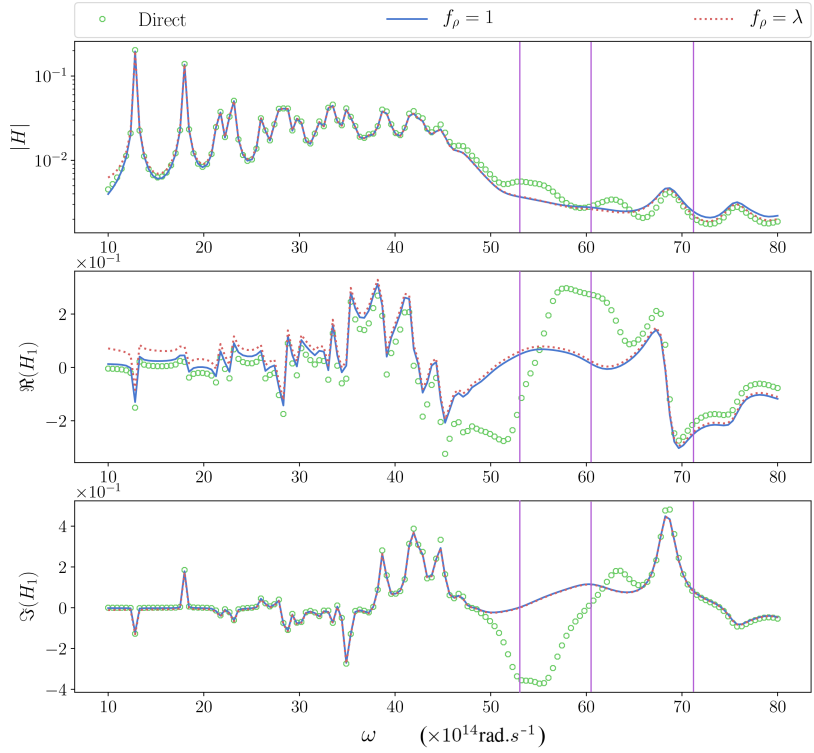

**Figure 8.9:** Scattered field **H** obtained by solving a direct problem classically (green dots) or rebuilt by the DQNM expansion formulas (8.20) for different functions $f_\rho(\lambda)$. for the 4-pole permittivity. The purple vertical lines indicate the positions of $\Re(\omega_2)$
(TOP) Integral over $\Omega_1$ of the norm of the magnetic field $\int_{\Omega_1} |\mathbf{H}| d\Omega$.
(MIDDLE and BOTTOM) The real and imaginary part of the magnetic field calculated at the detector point.

Since the problem of structured symmetric mesh for arbitrary interfaces is out of the scope of this thesis, our numerical example for magnetic fields is carried out with the same mesh in Figure 8.2. Figure 8.9 shows the comparison of the magnetic field between the direct computation and the reconstruction based on the DQNM expansion:

$$\mathbf{H} = \sum_n \frac{f_p(\lambda_n)}{f_p(\lambda)} \frac{1}{\lambda - \lambda_n} \frac{\langle \mathbf{H}_{ln}, \mathbf{S} \rangle}{\langle \mathbf{H}_{ln}, \dot{\mathcal{L}}_{\varepsilon,\mu}(\lambda_n)\mathbf{H}_{rn} \rangle} \mathbf{H}_{rn}, \quad (8.20)$$

where the inner product in the denominator can be computed explicitly as follows:

$$\langle \mathbf{H}_{ln}, \dot{\mathcal{L}}_{\varepsilon,\mu}(\lambda_n)\mathbf{H}_{rn} \rangle$$
$$= \int_\Omega \left[ \overline{\mathbf{H}_{ln}} \cdot \left( \nabla \times ((\varepsilon^{-1}(\lambda_n))' \nabla \times \mathbf{H}_{rn}) \right) + \overline{\mathbf{H}_{ln}} \cdot \left( (\lambda_n^2 \mu(\lambda_n))' \mathbf{H}_{rn} \right) \right] d\Omega.$$

Unsurprisingly, the plasmonic accumulation points exacerbate the error of the DQNM expansion (8.20) in the frequency range from 50 to 75 ($\times 10^{14}$ rad.s$^{-1}$) (see Figure 8.6). Since both electric poles and the critical points $\omega_2$, i.e. surface plasmon frequencies, are located in the vicinity of each other, it is hard to tell which is the cause of discrepancy of the DQNM expansion at certain frequencies. In fact, these errors can come from the accumulation points associated with both the electric poles and surface plasmon resonance.

## 8.4 Rational eigenvalue problem in open structure

In the previous computations, we have illustrated the expansion formalisms for multi-pole rational operators in bounded structures. Thus, for the last numerical example of this thesis, it is of interest to apply the DQNM expansion of multi-pole rational operators in open structures. We remind that in the open electromagnetic system, there exist the leaky resonant modes, which grow exponentially in space at infinity. A solution is to use the Perfectly Matched Layers (PML) to truncate and damp the electromagnetic fields in free space. As a result, in this example, we have to take into account the effects of both the PML modes and the electric poles of the permittivity.

In order to impose the out-going wave condition for the electromagnetic field, we will cover the closed structure in Figure 8.2 by the PML layer $\Omega_{\mathrm{PML}}$. The final geometry of the structure is described in Figure 8.10.

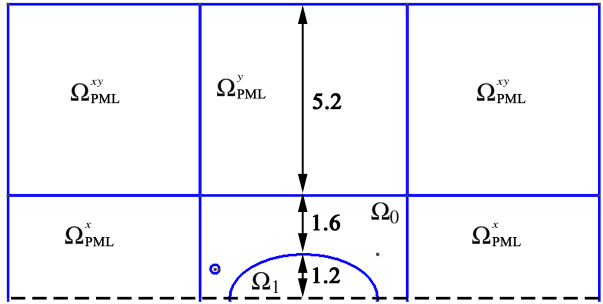

**Figure 8.10:** The upper half of the geometry for the unbounded structure.

We again apply the DQNM expansion to the electric field, the solution of the scattering problem (8.11) (TE polarization) in the open structure

Figure 8.10. The Dirac delta source $\mathbf{S} = \delta(\mathbf{r}_S)$, whose coordinates is given by $\mathbf{r}_S = (-2.4, 0.8)$ $(\times 10^{-1} \mu m)$, remains unchanged.

In order to replicate the open space, we follow Chapter 7 in replacing the initial material properties $\boldsymbol{\varepsilon}$ and $\boldsymbol{\mu}$ in the PML domain $\Omega_{\text{PML}}$ (vacuum in this case) by equivalent medium $\boldsymbol{\varepsilon}^s$ and $\boldsymbol{\mu}^s$ given by the following rule:

$$\boldsymbol{\delta}_s := \mathbf{J}_s^{-1} \boldsymbol{\delta} \mathbf{J}_s^{-\top} \det(\mathbf{J}_s) \quad \text{for} \quad \boldsymbol{\delta} = \{\boldsymbol{\varepsilon}, \boldsymbol{\mu}\},$$

where $\mathbf{J}_s$ is the stretched Jacobian matrix such that:

$$\mathbf{J}_s = \begin{cases} \text{diag}(s_x, 1, 1) & \text{in} \quad \Omega_{\text{PML}}^x \\ \text{diag}(1, s_y, 1) & \text{in} \quad \Omega_{\text{PML}}^y \\ \text{diag}(s_x, s_y, 1) & \text{in} \quad \Omega_{\text{PML}}^{xy} \end{cases},$$

where $s_x = s_y = s = \sigma \exp(i\phi)$ with $\sigma = 1$ and $\phi = \pi/10$. The location of the regions $\Omega_{\text{PML}}^x$, $\Omega_{\text{PML}}^y$, and $\Omega_{\text{PML}}^{xy}$ is shown as in Figure 8.10.

After solving the eigenvalue problem, eigenfrequencies are shown in Figure 8.11. It is easy to see that the theoretical continuous spectrum, which is supposed to be located on $\mathbb{R}^+$, is rotated of an angle $\theta = -\arg(2.8 + 5.2 \exp(i\phi)) \approx -0.20456$ rad (see [73] for detailed instructions of calculation of the angle $\theta$). This results in a large number of discretized Bérenger 'PML' modes [64, 87], whose eigenfield is concentrated in the PML region $\Omega_{\text{PML}}$ and far away from the scatterers (as can be seen from the field map of mode 1 in Figure 8.11). It is easily seen that, at low frequencies, these PML modes are arranged neatly along a straight line. However, when the frequency increases, the location of PML modes are clearly becoming numerically unstable further on the line, which has been predicted by the pseudo-spectrum theory of L. Trefethen [74].

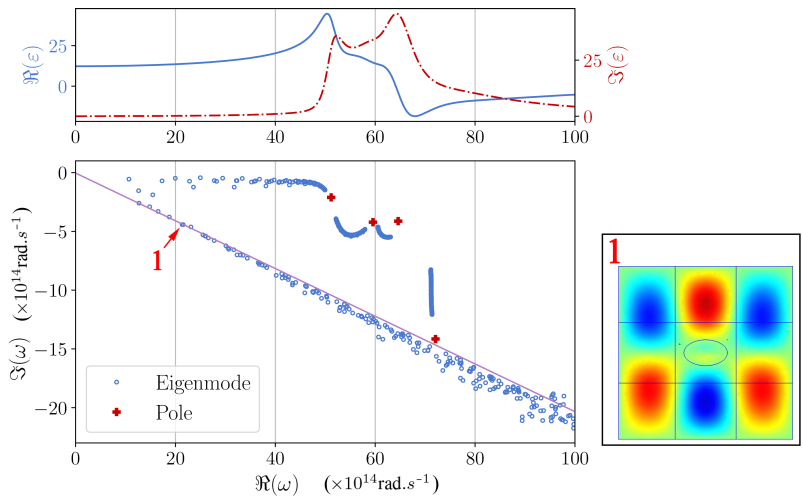

**Figure 8.11:** (BOTTOM LEFT) Spectrum of complex eigenfrequencies of electric field in the unbounded structure.
(RIGHT) The eigenfield of PML mode .
The purple line illustrates the slope $\theta = -0.20456$.

Outside the straight line formed by PML modes, we also notice the existence of the electric poles of the permittivity, which break the complex plane into several sections. Similar to previous examples, around the electric poles, there are eigenfrequencies, whose eigenfields oscillate with very high spatial frequencies, accumulates into clusters. It is worth mentioning that the imaginary part of the complex stretching function $s$ must be chosen to be large enough to reveal all the electric poles as well as the eigenmodes caused by these electric poles.

Finally, the numerical results of the DQNM expansion of electric field **E** are shown in Figure 8.12. We can see that the optical properties of the given open structure are fully captured by our DQNM expansion technique at low frequencies. When the frequency is larger, the discrepancies of the norm of electric fields inside the scatterer become noticeable since there is almost no field in the domain $\Omega_1$.[8] Numerical experiences show that the instability of PML modes (at high frequencies) may add noise to the DQNM expansion. In addition, we notice that the expansion with $f_\rho = \lambda$ provides a better approximation of the field inside the scatterer comparing to the case $f_\rho = 1$.

8: Pay attention that the figure is drawn in logarithmic scale and the values are really low

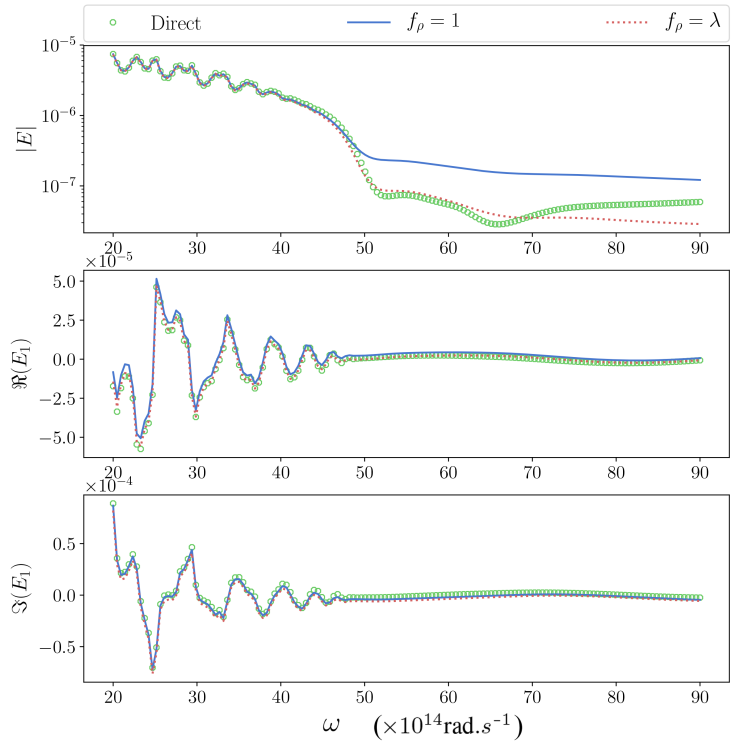

**Figure 8.12:** Electric field **E** obtained by solving a direct problem classically (green dots) or rebuilt by the DQNM expansion formulas (8.13) for 4-pole permittivity.
(TOP) Integral over $\Omega_1$ of the norm of the magnetic field $\int_{\Omega_1} |\mathbf{E}| d\Omega$.
(MIDDLE and BOTTOM) The real and imaginary part of the eletric field calculated at the detector point.

# General conclusions | 9

In this thesis, we have provided several spectral theories and numerical techniques regarding the Quasinormal Modal expansion in electromagnetics. Important points to be mentioned are:

▶ In chapter 1, we introduce the definition of Quasinormal Mode in the electromagnetic system. It was intended to motivate the study of DQNM expansion as a technique to connect the concept of optical resonance to the physical behaviors of the optical structure. In addition, it poses many important questions about the spectral properties of optical resonance such as the normalization of eigensolutions, the over-completeness and non-uniqueness of the modal expansion... which will be answered in the following chapters.

▶ In chapter 2, we studied the Fourier expansion for self-adjoint and linear operators of electromagnetic fields. It is set as a foundation on which other complicated spectral theories of non-linear operators are built. Most importantly, the normalization of electromagnetic fields based on energy is solely explained as the mathematical consequence of the construction of self-adjoint operators. At the end of this chapter, some signs of the non-uniqueness of the modal expansion are also pointed out.

▶ In chapter 3, we developed the modal expansion for linear operators. The aim of this chapter is to bring the non-self-adjointness into the previous Fourier expansion. This is done by imposing different boundary conditions from Dirichlet boundary condition to outgoing wave condition and Bloch-Floquet quasi-periodicity conditions... From there, we show the necessity of the 'left' eigenvectors in the expansion formulas. Finally, the non-uniqueness of both the normalization and the modal expansion are also mentioned.

▶ In chapter 4, we introduce the concept of polynomial operators. Through the process of linearization, i.e rewriting polynomial operators in terms of a system of linear problems, the QNM expansion for polynomial operators is formulated. We proved that formula of the QNM expansion is not unique. This non-uniqueness of the QNM expansion can be explained by the fact that the set of eigenvectors is not linear independent.

▶ Chapter 5 was devoted to derive the QNM expansion formalism Lemma 5.1.1 for rational operators based on the result obtained from the previous chapters. Since the electromagnetic problem in dispersive media can be represented by rational operators, it is straightforward to decompose the electromagnetic fields onto the resonant-state basis by using Lemma 5.1.1. Accordingly, the formulation of the Dispersive Quasinormal Modal expansion (DQNM expansion) for electromagnetic fields is not unique. One of these formulas is proved to be the general expansion of other well-known results in the literature.

▶ Chapter 6 is intended to be a brief guidebook for numerical modeling in electromagnetics. After a short reminder of Finite Element

Analysis in electromagnetics, we introduce the finite element solver GetDP and its eigensolver library SLEPc. We also list some basic syntax as well as examples, which can help readers get acquainted with this software and are able to reproduce the numerical results in this thesis. In order to put all these finite element tools into practice, an example of the DQNM expansion in a 3-D closed structure is given. The results of this example numerically confirm the existence of a continuous family of DQNM expansion formulas.

▶ In chapter 7, we studied the Perfectly Matched Layer as a tool to impose the outgoing wave condition in numerical computations. The technique involves truncating the computational domain by a dissipative zone which can damp the incoming waves. The goal of this chapter is to get a better understanding of the influence of PML parameters upon the eigen-solutions of electromagnetic operators. In order to do that, a 3-D model of a sphere and a 2-D structure of diffraction grating are given. Moreover, the example of the diffraction grating also demonstrates the importance of the 'left' eigenvectors in the DQNM expansion.

▶ Chapter 8 indicates the limitations of the DQNM expansion for rational operators when considering 'actual' materials.

In the first section of the chapter, we work on the extraction of an accurate model of permittivity from experimental data: The permittivity in our calculations is no longer limited to simple models (for example the Drude-Lorentz model) but is drawn directly from actual measurement data. The result is multi-pole rational functions of permittivity which leads to the construction of rational operators.

The next section provides two numerical examples where the permittivity is given by 1-pole and multi-pole rational functions. We discover that these electric poles result in accumulation points in the complex plane of the spectrum. These accumulation points, in turn, cause errors in calculations of the DQNM expansion.

The third section of the chapter mentions the surface plasmon resonance for magnetic fields. This happens when the value of the permittivity changes sign when crossing the interface between 2 media. The plasmonic resonances also cause another kind of accumulation points, which reduces the accuracy of DQNM expansion. Finally, we provide an example of the DQNM expansion for multipole rational operators in an open structure.

# A

# Linear Operator

The construction in the this appendix follows [42] .

## A.1 Basis of Hilbert space

This section serves as a review of basic spectral theory.

First, we define a three-dimensional space $\Omega \subset \mathbf{R}^3$. A Lebesgue space $\mathbf{L}^2(\Omega)^3$ is the set of vector fields $\mathbf{a}(\mathbf{r}) = (a_x(\mathbf{r}), a_y(\mathbf{r}), a_z(\mathbf{r}))$ such that: [1]

$$\int_\Omega |\mathbf{a}(\mathbf{r})|^2 d\Omega < \infty.$$

1: In terms of physics, this implies that the energy associated with the field in $\Omega$ is finite.

Given $\mathbf{x}, \mathbf{y} \in \mathbf{L}^2(\Omega)^3$, the inner product $\langle \mathbf{x}, \mathbf{y} \rangle$ is defined as follows:

$$\langle \mathbf{x}, \mathbf{y} \rangle = \int_\Omega \overline{\mathbf{x}(\mathbf{r})} \cdot \mathbf{y}(\mathbf{r}) d\Omega.$$

We call the inner product space $\langle \mathbf{x}, \mathbf{y} \rangle$ in $\mathbf{L}^2(\Omega)^3$ complete in the norm $\|\mathbf{x}\| = \langle \mathbf{x}, \mathbf{x} \rangle^{1/2}$ a Hilbert space $H$.

**Definition A.1.1** *An orthonormal set of elements $\{v_n\}$ is said to be a basis for a Hilbert space $H$ if each element $u \in H$ can be written in a unique way as*

$$u = \sum_{n=1}^{\infty} \langle v_n, u \rangle v_n.$$

In electromagnetics, Hilbert spaces typically arise as infinite-dimensional function spaces. That is the reason why the sum runs from one to infinity. Thus, this must be understood as a Hilbert basis (or countable basis) with infinite sums.

## A.2 Self-adjoint operator

**Definition A.2.1** *A mapping $A$ from $X$ to $Y$, denoted as $A : X \rightarrow Y$, is called a linear operator (linear mapping, linear transformation) if for all $x$ and $y$ in the domain of $A$ (defined below) and $\alpha, \beta \in \mathbf{C}$,*

$$A(\alpha x + \beta y) = \alpha A x + \beta A y,$$

*where $Ax, Ay \in Y$.*

**Definition A.2.2** *The domain of a linear operator $A : X \rightarrow Y$, denoted as $D_A$, is simply the set of elements for which the mapping $A$ is defined.*

*Example:* The differential operator $A : L^2(\Omega)^3 \rightarrow L^2(\Omega)^3$ defined by $(A\mathbf{x})(\mathbf{r}) = \nabla \times \nabla \times \mathbf{x}(\mathbf{r})$ cannot have its domain all of $L^2(\Omega)^3$ since many functions in this space are not differentiable, or even continuous. Instead, we have to specify the domain $D_A$: $D_A = \{\mathbf{x} : \mathbf{x}(\mathbf{r}), \nabla \times \nabla \times \mathbf{x}(\mathbf{r}) \in L^2(\Omega)^3, \mathbf{n} \times \mathbf{x}(\mathbf{r})|_{\partial\Omega} = 0\}$, where the last term refers to the boundary conditions.

**Definition A.2.3** *A linear operator $A : H \rightarrow H$ is bounded on its domain if, for all $\mathbf{x} \in H$, there exist., a number $k > 0$ such that $\|A\mathbf{x}\|_H \leq k\|\mathbf{x}\|_H$.*

It is important to point out that differential operators are unbounded (not bounded) based on the definitions above. Hence, all electromagnetic operators in the thesis have to be treated carefully as unbounded ones.

**Definition A.2.4** *Given an operator $A : H \rightarrow H$ with the domain $D_A$. The domain $D_{A^\dagger}$ of the adjoint operator $A^\dagger$ is defined such that:*

$$D_{A^\dagger} \equiv \{\mathbf{y} \in H : \exists A^\dagger \text{ such that } \langle \mathbf{y}, A\mathbf{x} \rangle = \langle A^\dagger \mathbf{y}, \mathbf{x} \rangle, \forall \mathbf{x} \in D_A\}.$$

We will assume that all the operators in this thesis possess adjoints. It is clear that $D_{A^\dagger}$ is not unique and can be chosen depending on $D_A$, which, in turn, is related to our choice of boundary conditions.

*Example:* Let's consider operator $A : L^2(\Omega)^3 \rightarrow L^2(\Omega)^3$:

$$(A\mathbf{x})(\mathbf{r}) \equiv \nabla \times \nabla \times \mathbf{x}(\mathbf{r})$$
$$D_A \equiv \{\mathbf{x} : \mathbf{x}(\mathbf{r}), \nabla \times \nabla \times \mathbf{x}(\mathbf{r}) \in L^2(\Omega)^3, \mathbf{n} \times \mathbf{x}(\mathbf{r})|_\Gamma = 0\}.$$

The adjoint operator can be found using integration by parts:

$$\langle \mathbf{y}, A\mathbf{x} \rangle = \int_\Omega \overline{\mathbf{y}} \cdot (\nabla \times \nabla \times \mathbf{x}) \, d\Omega$$
$$= \int_\Omega (\nabla \times \nabla \times \overline{\mathbf{y}}) \cdot \mathbf{x} \, d\Omega + \int_\Gamma ((\nabla \times \overline{\mathbf{y}}) \cdot (\mathbf{n} \times \mathbf{x}) - (n \times \overline{\mathbf{y}}) \cdot (\nabla \times \mathbf{x})) \, dS$$
$$= \langle A^\dagger \mathbf{y}, \mathbf{x} \rangle$$

where $\Gamma$ is the boundary of $\Omega$.

The operator $A^\dagger$ is called the formal adjoint of $A$. The boundary term (i.e. the integrated term) is known as the conjunct (see [42] for more details). The domain of the adjoint operator is determined by requiring that the conjunct vanish so that we have $\langle \mathbf{y}, A\mathbf{x} \rangle = \langle A^\dagger \mathbf{y}, \mathbf{x} \rangle$.

We proceed to see that:

$$(A^\dagger \mathbf{y})(\mathbf{r}) = \nabla \times \nabla \times \mathbf{y}(\mathbf{r}) = (A\mathbf{y})(\mathbf{r})$$
$$D_{A^\dagger} = \{\mathbf{y} : \mathbf{y}(\mathbf{r}), \nabla \times \nabla \times \mathbf{y}(\mathbf{r}) \in L^2(\Omega)^3, \mathbf{n} \times \mathbf{y}(\mathbf{r})|_\Gamma = 0\} = D_A.$$

**Definition A.2.5** *A operator $A$ is called self-adjoint if $D_A = D_{A^\dagger}$ and $A^\dagger \mathbf{x} = A\mathbf{x}$ for all $\mathbf{x} \in D_A = D_{A^\dagger}$.*

It is vital to keep in mind that the self-adjointness of an operator (especially unbounded operator) not only requires $A^\dagger \mathbf{x} = A\mathbf{x}$ but also the domains $D_A = D_{A^\dagger}$. Thus, the self-adjointness of operator $A$ with respect to the domain $D_A$ may no longer hold if we change the boundary conditions, which in turn, modifies $D_A$ and $D_{A^\dagger}$, even though the equality $A^\dagger \mathbf{x} = A\mathbf{x}$ remains unchanged.

***Example:*** It is easily seen that the operator $A : L^2(\Omega)^3 \to L^2(\Omega)^3$:

$$(A\mathbf{x})(\mathbf{r}) \equiv \nabla \times \nabla \times \mathbf{x}(\mathbf{r})$$

$$D_A \equiv \{\mathbf{x} : \mathbf{x}(\mathbf{r}), \nabla \times \nabla \times \mathbf{x}(\mathbf{r}) \in L^2(\Omega)^3, \mathbf{n} \times \mathbf{x}(\mathbf{r})|_\Gamma = 0\},$$

is self-adjoint, since $D_A = D_{A^\dagger}$ and $A^\dagger \mathbf{x} = A\mathbf{x}$.

However, if we change the domain from the Dirichlet boundary condition to the outgoing radiation condition, the operator is no longer self-adjoint. Let's consider the operator $B : L^2(\Omega)^3 \to L^2(\Omega)^3$:

$$(B\mathbf{x})(\mathbf{r}) \equiv \nabla \times \nabla \times \mathbf{x}(\mathbf{r})$$

$$D_B \equiv \{\mathbf{x} : \mathbf{x}(\mathbf{r}), \nabla \times \nabla \times \mathbf{x}(\mathbf{r}) \in L^2(\Omega)^3, \lim_{r \to \infty} r\,(\nabla \times \mathbf{E} + ik\hat{\mathbf{r}} \times \mathbf{E}) = 0\},$$

$B$ is not a self-adjoint operator. [2]

2: See Section 3.3 for more details.

The condition for a linear operator on a Hilbert space to be self-adjoint is sometimes too strong. There is indeed a weaker property call formally self-adjoint (i.e. symmetric).

> **Definition A.2.6** *An operator $A$ is called symmetric if $D_A = D_{A^\dagger}$ and $A^\dagger x = Ax$ for all $x \in D_A \subseteq D_{A^\dagger}$.*

We emphasize that the symmetric property of operator is different from our common understanding of symmetric matrix. In this case, the operator's symmetry is just the "weaker" self-adjointness which allows the domain of the operator $A^\dagger$ contains the domain of the operator $A$.

## A.3 Linear eigenvalue problem

The spectral problem consists of examining solutions of the eigenvalue problems. In this thesis, a generalized linear eigenvalue problem is defined as follows:

> **Definition A.3.1** (Generalized linear eigenvalue problem) *Given two operators $A, B : H \to H$, with the domain $D_L$; the vector $\boldsymbol{v}_n \in D_L$ such that $\boldsymbol{v}_n \neq 0$ and are called an eigenvectors with a corresponding eigenvalues $\lambda_n \in \mathbb{C}$.*
>
> $$L(\lambda_n)\boldsymbol{v}_n = (A + \lambda_n B)\boldsymbol{v}_n = 0. \tag{A.1}$$

From (A.1), it is useful to examine properties of the resolvent operator

$$R_\lambda(A, B) \equiv (A + \lambda B)^{-1}. \tag{A.2}$$

It is clear that if $R_\lambda(A, B)$ exists for a particular $\tilde{\lambda}$, then that $\tilde{\lambda}$ cannot be an eigenvalue since (A.1) would only have trivial solutions. Therefore, spectral properties of the operator $L(\lambda)$ can be determined by considering properties of $R_\lambda(A, B)$. In particular, the total spectrum can be decomposed as follows:

**Point spectrum** $\sigma_p(L)$ : Values of $\lambda$ for which $R_\lambda(A, B)$ does not exist. This consists of all the discrete eigenvalues of $L(\lambda)$. That is why in many literature, eigenvalues of $L(\lambda)$ can be considered to be poles of the resolvent operator $R_\lambda(A, B)$.

**Continuous spectrum** $\sigma_c(L)$ : Values of $\lambda$ for which $R_\lambda(A, B)$ exists and is unbounded, meanwhile make the range of $L(\lambda)$ is dense in $H$.

**Residual spectrum** $\sigma_r(L)$ : Values of $\lambda$ for which $R_\lambda(A, B)$ exists and the closure of the range of $L(\lambda)$ is a proper subset of $H$.

**Resolvent set** $\rho(L)$ : Values of $\lambda$ for which $R_\lambda(A, B)$ exists and bounded, while the range of $L(\lambda)$ is dense in $H$. This is not part of the spectrum of $L(\lambda)$.

Unlike finite-dimensional spaces where the spectrum are associated with point spectrum, continuous and residual spectrum only arise in the infinite-dimensional case. Fortunately, the residual spectrum does not usually occur in electromagnetic applications. Therefore, our remaining work is to verify the existence of $R_\lambda(A, B)$ to distinguish between point and continuous spectrum.

**Theorem A.3.1** *Let $A, B : H \to H$ be self-adjoint (or symmetric) linear operators such that $\langle v_n, Av_n \rangle \neq 0$ or $\langle v_n, Bv_n \rangle \neq 0$. Then eigenvalues $\lambda_n$ corresponding to $L(\lambda_n)v_n = (A + \lambda_n B)v_n = 0$ are real-valued.*

**Proof:** Firstly, we assume that $\langle \mathbf{v}_n, B\mathbf{v}_n \rangle \neq 0$. From $(A + \lambda B)\mathbf{v}_n = 0$, it is clear that $\langle \mathbf{v}_n, A\mathbf{v}_n \rangle + \langle \mathbf{v}_n, \lambda B\mathbf{v}_n \rangle = \langle A\mathbf{v}_n, \mathbf{v}_n \rangle + \langle \lambda B\mathbf{v}_n, \mathbf{v}_n \rangle = 0$. Given both $A$ and $B$ symmetric, we have $\langle \mathbf{v}_n, \lambda_n B\mathbf{v}_n \rangle = \langle \lambda_n B\mathbf{v}_n, \mathbf{v}_n \rangle$, which implies $(\lambda_n - \overline{\lambda_n})\langle \mathbf{v}_n, B\mathbf{v}_n \rangle = 0$. This concludes that if $\langle \mathbf{v}_n, B\mathbf{v}_n \rangle \neq 0$, then $\lambda_n = \overline{\lambda_n}$, which proves $\lambda_n$ is real. The proof for $\langle \mathbf{v}_n, A\mathbf{v}_n \rangle \neq 0$ is similar. Given $(A + \lambda_n B)\mathbf{v}_n = 0$ and the condition $\lambda_n \neq 0$, we have $\langle \mathbf{v}_n, (1/\lambda_n)A\mathbf{v}_n \rangle + \langle \mathbf{v}_n, \mathbf{v}_n \rangle = \langle (1/\lambda_n)A\mathbf{v}_n, \mathbf{v}_n \rangle + \langle B\mathbf{v}_n, \mathbf{v}_n \rangle = 0$. Since $A$ and $B$ symmetric, we have $\langle \mathbf{v}_n, (1/\lambda_n)A\mathbf{v} \rangle = \langle (1/\lambda_n)A\mathbf{v}_n, \mathbf{v}_n \rangle$, which implies $(1/\lambda_n - 1/\overline{\lambda_n})\langle \mathbf{v}_n, A\mathbf{v}_n \rangle = 0$. This concludes that if $\langle \mathbf{v}_n, A\mathbf{v}_n \rangle \neq 0$, then $1/\lambda_n = 1/\overline{\lambda_n}$, which proves $\lambda_n$ is real.

**Theorem A.3.2** *Let $A, B : H \to H$ be self-adjoint (or symmetric) operators. Then eigenvectors corresponding to $(A + \lambda_n B)v_n = 0$ satisfy the orthogonality relationships.*

$$(\lambda_m - \overline{\lambda_n})\langle v_n, Bv_m \rangle = 0$$

**Proof:** Given $A$ self-adjoint (or symmetric), we have $\langle \mathbf{v}_n, A\mathbf{v}_m \rangle = \langle A\mathbf{v}_n, \mathbf{v}_m \rangle$. Then, it is easily seen that $\langle \mathbf{v}_n, \lambda_m B\mathbf{v}_m \rangle = \langle \lambda_n B\mathbf{v}_n, \mathbf{v}_m \rangle$. Since B is self-adjoint (or symmetric), we obtain $(\lambda_m - \overline{\lambda_n})$ $\langle \mathbf{v}_n, B\mathbf{v}_m \rangle = 0$. If $A$ or $B$ is definite (i.e. $\langle \mathbf{v}, A\mathbf{v} \rangle \neq 0$ or $\langle \mathbf{v}, B\mathbf{v} \rangle \neq 0$), we get $\lambda_m, \lambda_n \in \mathbf{R}$ and $(\lambda_m - \lambda_n)\langle \mathbf{v}_n, B\mathbf{v}_m \rangle = 0$.

**Remark A.3.1** The self-adjointness of the operators $A, B : H \to H$ and $\lambda \in \mathbf{R}$ do not imply that the operator $L(\lambda) = A + \lambda B$ is also self-adjoint.

**Theorem A.3.3** (Spectral theorem for compact self-adjoint operators)
*Let $L : H \rightarrow H$ be a compact, self-adjoint linear operator acting on an infinite-dimensional Hilbert space $H$. Then there exists an orthonormal basis for $H$ of eigenvectors $\boldsymbol{v}_n$ with corresponding eigenvalues $\lambda_n$. For every $\boldsymbol{u} \in H$, we have*

$$\boldsymbol{u} = \sum_n \langle \boldsymbol{v}_n, \boldsymbol{u} \rangle v_n$$

It is clear that our electromagnetic operator, i.e. differential operators, are indeed unbounded, hence not compact. The previous theorem no longer holds for unbounded operators with boundary conditions.

However, there is one way to bypass the requirement of compactness by considering the inverse operator.

**Theorem A.3.4** *Let $A : H \rightarrow H$ be an invertible linear operator with eigenvalues $\lambda$ and corresponding eigenvectors $v$ such that $Av = \lambda v$. Then, $A^{-1} : H \rightarrow H$ has eigenvalues $1/\lambda$ and corresponding eigenvectors $v$.*

The proof is straightforward. Given $A\mathbf{v} = \lambda\mathbf{v}$, we can deduce $A^{-1}A\mathbf{v} = \lambda A^{-1}\mathbf{v}$, which leads to $A^{-1}\mathbf{v} = (1/\lambda)\mathbf{v}$.

Given operators $A, B : H \rightarrow H$ self-adjoint and invertible, then the eigen-problem $L(\lambda)\mathbf{v} = (A + \lambda B)\mathbf{v} = 0$ shares the same eigenvalues $\lambda$ and eigenfunctions $\mathbf{v}$ with $C\mathbf{v} = \lambda\mathbf{v}$ where $C = (-B)^{-1}A$. It is possible to prove that the operator $C$ has a compact self-adjoint inverse operator $C^{-1}$ on $H$, with an associated orthonormal eigenbasis by Theorem A.3.4 (In most of the cases of differential operators, the inverse operator would be the integral bounded by boundary conditions, hence be bounded and compact). Then by Theorem A.3.3, the eigenfunctions of the inverse operator $C^{-1}$ are eigenfunctions of $L(\lambda)$, and therefore $L(\lambda)$ possesses an eigenbasis in H.

**Proposition A.3.5** *The eigenfunctions of a self-adjoint boundary value problem $L(\lambda_n)\boldsymbol{v}_n = (A + \lambda_n B)\boldsymbol{v}_n = 0$ on a Hilbert space $H$ form an orthonormal basis.*

This is a very important observation, which allows us to expand any $\mathbf{u} \in H$ as a linear combination of eigenfunctions $\mathbf{v}$ even without the compactness of the operator $L(\lambda)$. We can write

$$\mathbf{u} = \sum_n \langle \mathbf{v}_n, \mathbf{u} \rangle \mathbf{v}_n \qquad (A.3)$$

# B

# Finite Element Analysis (FEA)

## B.1 Introduction of Finite Element Analysis

The technique of Finite Element Analysis involves finding the approximation of continuous unknown solutions of boundary-value problems. It consists of splitting the numerical solution into pieces, each defined over a small subset (elements) of the computational domain. On each of these elements, the approximate solution is assumed to have a particularly simple form, usually a linear combination of a finite number of predefined functions. By selecting the 'right' linear combination, i.e. choosing the finite element space, it is possible to prove that the numerical approximation converges to the analytic solution if the size of the subdomain (element) is refined to be small enough.

The main idea of Finite Element Analysis can be encapsulated in three steps:

- ▶ Weak formulation: Convert the given continuous problem into a weak form.
- ▶ Discretization: Divide the computational continuous domain into several sub-domains, i.e. finite elements and define the finite-dimensional subspace on which the solution function can be decomposed as a linear combination.
- ▶ Linearization: Convert the weak formulation into a linear system, commonly by using the Galerkin method. The linear system (expressed in terms of matrices) can then be solved numerically which leads to a global approximate solution of the problem.

We will give a brief exposition of the basic concepts of FEA as follows.

## B.2 The Lax-Milgram theorem

Without loss of generality, we will apply the Finite Element process to solve the following differential problem with appropriate boundary conditions:

$$Au = f. \tag{B.1}$$

for the operator $A : H \rightarrow H$, which is assumed to be complex-valued unless otherwise specified. [1]

1: It is worth reminding that $H$ is a infinite-dimensional space.

The sesqui-linear form $a(\cdot, \cdot) : H \times H \rightarrow \mathbb{C}$ is then given by:

$$a(w, v) := \langle w, Av \rangle.$$

It is immediate that if $A$ is self-adjoint, $\langle w, Av \rangle = \langle Aw, v \rangle$, the sesqui-linear form $a(\cdot, \cdot)$ is Hermitian:

$$a(w, v) = \overline{a(v, w)}.$$

Next, we will state an important theorem in FEA.

---

**Theorem B.2.1** (The Lax-Milgram theorem) *Given the sesqui-linear form $a(\cdot, \cdot) : H \times H \to \mathbb{C}$ and the functional $l(\cdot) : H \to \mathbb{C}$*

$$l(v) = \langle v, f \rangle$$

*Then, if there exists $c_1, c_2, c_3 > 0$ such that*

- ▶ $|a(w, v)| \leq c_1 \|w\| \|v\| \quad \forall u, v \in H$ *(continuity of a)*
- ▶ $\Re(a(v, v)) \geq c_2 \|v\|^2 \quad \forall v \in H$ *(coercivity of a)*
- ▶ $|l(v)| \leq c_3 \|v\|^2 \quad \forall v \in H$ *(continuity of l)*

*then there exists a unique element $u \in H$ such that*

$$a(v, u) = l(v) \quad \forall v \in H$$

---

If $a$ is Hermitian, $u$ is the unique solution of the minimization problem:

$$J(u) = \min_{v \in H} J(v),$$

where the functional $J$ is defined by:

$$J(v) = \frac{1}{2} a(v, v) - l(v).$$

## B.3 The Galerkin method

Our goal is to deduce the weak formulation of (B.1) using the sesqui-linear form $a(\cdot, \cdot)$. It is worth point out that since the operator $A$ is not necessarily self-adjoint, the mathematical deduction based on the Rayleigh-Ritz method is invalid here. In fact, the weak formulation is obtained through the annulation of the weighted residuals where the shape and weight functions are in the same space: The Galerkin method. The procedure is as follows:

Given the domain $\Omega$, we choose the finite dimensional space $V_h \subset H_0^1(\Omega)$ with the dimension $m = dim(V_h)$ and the basis $\mathrm{span}\{\phi_i\} = (V_h)$. Then a function $\tilde{u} \in V_h$ can be approximated by the expansion:

$$\tilde{u} = \sum_{i=1}^{m} c_i \phi_i, \tag{B.2}$$

where $c_i$ are constant coefficient to be determined.

The idea is to find $c_i$ such that $\tilde{u}$ an approximate solution to (B.1). Since $\tilde{u}$ is just an approximation, it is easily seen that the following residual is nonzero:

$$R = A\tilde{u} - f \neq 0.$$

The best solution we can end up is the one that reduces the residual to the least value at all points of $\Omega$. The next condition is the spirit of the weighted residual method:

$$\langle w_i, R \rangle = \langle w_i, A\tilde{u} - f \rangle = 0 \quad \forall w_i \in V_h \tag{B.3}$$

where $w_i$ refer to chosen weighting functions. The previous equation (B.3) is also called the weak formulation of (B.1).

In the Galerkin method, the weighting functions $w_i$ are selected to be the same as those used for expansion of the approximation solution, i.e $\phi_i$: $w_i = \phi_i$. As a result, (B.3) yields:

$$\langle \phi_i, A\tilde{u} - f \rangle = 0,$$

which can be written:

$$a(\phi_j, \sum_{i=1}^{m} c_i \phi_i) = l(\phi_j),$$

using $l(\phi_j) = \langle \phi_i, f \rangle$ and $a(\phi_j, \sum_{i=1}^{m} c_i \phi_i) = \langle \phi_i, A \sum_{i=1}^{m} c_i \phi_i \rangle$

Since $a$ has a sesqui-linear form, it is possible to take the sum out of the integral in eq as follows:

$$\sum_{i=1}^{m} a(\phi_j, \phi_i) c_i = l(\phi_j)$$

The previous equation is nothing more than a set of linear equations which can be expressed in terms of matrix form.

$$\begin{pmatrix} a(\phi_1, \phi_1) & a(\phi_1, \phi_2) & \dots & a(\phi_1, \phi_m) \\ a(\phi_2, \phi_1) & a(\phi_2, \phi_2) & \dots & a(\phi_m, \phi_2) \\ \vdots & \vdots & \ddots & \vdots \\ a(\phi_m, \phi_1) & a(\phi_m, \phi_2) & \dots & a(\phi_m, \phi_m) \end{pmatrix} \begin{pmatrix} c_1 \\ c_2 \\ \vdots \\ c_m \end{pmatrix} = \begin{pmatrix} l(\phi_1) \\ l(\phi_2) \\ \vdots \\ l(\phi_m) \end{pmatrix} \tag{B.4}$$

By solving the previous linear system to obtain $c_i$, we can reconstruct the numerical solution $\tilde{u}$, the approximation of the continuous solution $u$ of (B.1).

Up until now, we claim very little restriction on $V_h$ or the basis $\{\phi_i\}$ used. However, care should be taken since the numerical solution heavily depends on the construction of $V_h$. The wrong choice of $\{\phi_i\}$ could lead to a non-physical solution.

For example, the domain $\Omega$ can be divided into a set of triangles which pairwise have in common either a full edge (see Figure B.1) or a node or nothing in the two-dimensional case and into a set of tetrahedra which pairwise have in common either a full triangular facet (see Figure B.2), or a full edge or a node or nothing in the three-dimensional case. The triangles in the two-dimensional case and tetrahedra in the three-dimensional case are called the elements, the keystone in FEA.

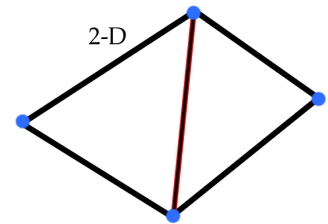

**Figure B.1:** Two 2-D triangles have in common a full edge.

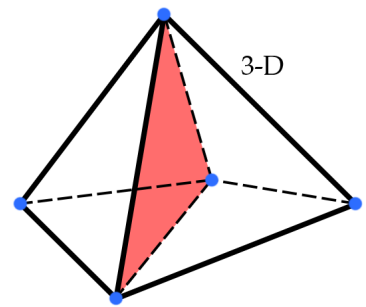

**Figure B.2:** Two 3-D tetrahedra have in common a full facet.

## B.4  Error estimation

The final issue arises is how good is the numerical solution $\tilde{u}$ (B.2) in approximating the actual function $u$. Thus, it is important to discuss the convergence property of the Galerkin method.

As most other numerical methods, the Galerkin method only finds an approximate solution to our problem. Given $u$ as the exact solution to the continuous problem (B.1) and $\tilde{u}$ the solution of (B.3), our goal is to bound on the error $\|u - \tilde{u}\|$ by a certain norm $\|\cdot\|$. The process can be done by Céa's lemma:

**Lemma B.4.1** (Céa's lemma)  *Given the sesqui-linear form $a(\cdot, \cdot)$ satisfies the conditions of Theorem B.2.1 with $H^m$-norm, assume $u \in H^m(\Omega)$ and $\tilde{u} \in V_h \subset H^m(\Omega)$ are the solutions of (B.1) and (B.3) respectively; then there exists a constant $c(\Omega)$ such that*

$$\|u - \tilde{u}\|_{H^m(\Omega)} \leq c(\Omega) \inf_{v \in V_h} \|u - v\|_{H^m(\Omega)}. \qquad (B.5)$$

It is worth noting that the constant $c(\Omega)$ is independent of $h$ but only depending on $\Omega$. In fact, the quantity $\inf_{v \in V_h} \|u - v\|_{H^m(\Omega)}$ is known as the best approximation error since it is the error of the best possible approximation for all $v \in V_h$.

Remark: Using Céa's lemma implies that we assumed $a$ is coercive which is not true for many cases where the permittivity is negative. There are generalizations to Céa's lemma where we replace the assumption of coercivity with the one that $a$ satisfies a discrete inf-sup condition. One of consequence of these generalizations leads to the non-conforming methods of FEA which is outside the scope of this thesis

# C

# The diffraction grating problem

In optics, a diffraction grating is an optical component with a periodic structure that splits and diffracts light into several waves traveling in different directions. The directions of these diffracted waves depend on the geometric structure of the grating and the frequency of incident waves that the grating acts as a dispersive element. Thanks to its dispersion relation, the diffraction grating has a wide range of applications in optics [88, 89] .

In this thesis, we will choose to study a metallic grating structure because of its ability to support a wide range of resonant modes and provide a good platform to study the modal expansion. The study of direct scattering problem consists of simulating diffracted waves in given grating structures under the illumination of specific incident waves. The aim of this chapter is to introduce the numerical modeling of the diffraction grating and describe important physical quantities related to the grating system (see [90] for more details).

## C.1 Direct grating problem

We remind that our optical problem is time harmonic with the frequency $\omega$: The electric and magnetic fields can be represented by complex vector $\mathbf{E}$ and $\mathbf{H}$ with Fourier transform is set as:

$$\hat{f}(\omega) = \int f(t) \exp(-i\omega t) dt.$$

Next, we define the wave vector $k = \omega.$

For the diffraction grating problem, $z$-anisotropic materials will be used with relative permittivity $\boldsymbol{\varepsilon}$ and relative permeability $\boldsymbol{\mu}$ as follows:

$$\boldsymbol{\varepsilon} = \begin{pmatrix} \varepsilon_{xx} & \bar{\varepsilon}_a & 0 \\ \varepsilon_a & \varepsilon_{yy} & 0 \\ 0 & 0 & \varepsilon_{zz} \end{pmatrix} \quad \text{and} \quad \boldsymbol{\mu} = \begin{pmatrix} \mu_{xx} & \bar{\mu}_a & 0 \\ \mu_a & \mu_{yy} & 0 \\ 0 & 0 & \mu_{zz} \end{pmatrix}.$$

where $\varepsilon_{xx}, \varepsilon_a, \dots \mu_{zz}$ are complex functions of variables x, y as well as frequency $\omega$ and $\bar{\varepsilon}_a$ represents the conjugate complex of $\varepsilon_a$.

### Geometric structure

Hereby, the geometric structure of the grating problem is presented in Figure C.1. For simplicity, we will restrict our attention to the numerical

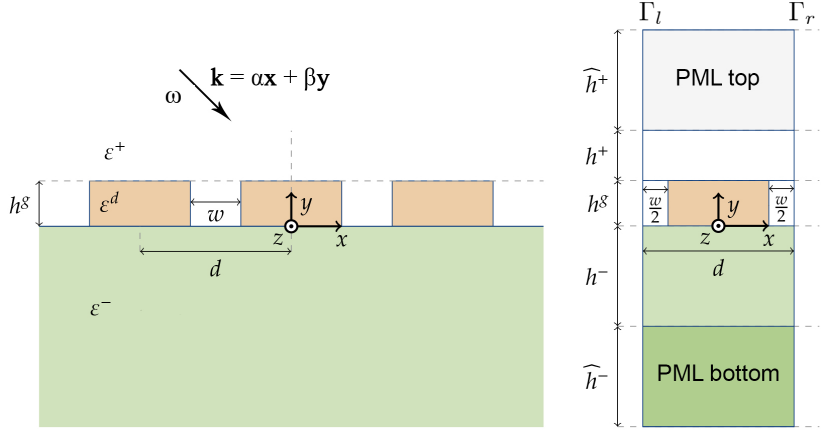

modeling of 2-D structures, consisting of 5 regions:

▶ The groove region ($0 < y \le h^g$) characterized by the 2 tensors $\boldsymbol{\varepsilon}^g(x, y, \omega)$ and $\boldsymbol{\mu}^g(x, y, \omega)$. In the groove region, there are scatterers with the domain $\Omega_1$ as shown in the picture

▶ The superstratum ($h^g < y \le h^g + h^+$) which is homogeneous, isotropic and lossless characterized by fixed permittivity $\varepsilon^+$ and permeability $\mu^+$. Let's denotes $k^+ = k_0 \sqrt{\varepsilon^+ \mu^+}$.

▶ The substratum ($-h^- < y \le 0$) which is also homogeneous, isotropic and lossless characterized by fixed permittivity $\varepsilon^-$ and permeability $\mu^-$. Let's denotes $k^- = k_0 \sqrt{\varepsilon^- \mu^-}$.

▶ The top PML domain ($h^g + h^+ < y$) is computational domain which is used later when discretizing the problem. Here, the permittivity, and permeability will be modified later in section C.5.

▶ The bottom PML domain ($y \le -h^-$).

It is easy to see that the grating geometric structure is periodic along x-axis with the period $d$.

## Equation of the diffraction grating problem

The geometric structure is illuminated by an incident plane wave from the superstratum domain. It is common to divided diffraction problems into two polarization cases based on the geometrical orientation of the oscillations of incident electromagnetic fields as follows:

$$\mathbf{E}_e^0 = A^0 \exp(i\mathbf{k}^+ \cdot \mathbf{r})\mathbf{z} \quad \text{for TE (p-polarization) case} \tag{C.1a}$$

$$\mathbf{H}_p^0 = A^0 \exp(i\mathbf{k}^+ \cdot \mathbf{r})\mathbf{z} \quad \text{for TH (s-polarization) case} \tag{C.1b}$$

with the wave vector $\mathbf{k}^+$ defined as:

$$\mathbf{k}^+ = a\mathbf{x} - b\mathbf{y} = k^+(\sin(\theta)\mathbf{x} - \cos(\theta)\mathbf{y}), \tag{C.2}$$

where $\theta$ is the incidence angle to the $y$-axis

For TE case (resp. TM case), the magnetic (resp. electric) field derived from $\mathbf{E}_e^0$ (resp. $\mathbf{H}_p^0$) is denoted by $\mathbf{H}_e^0$ (resp. $\mathbf{H}_p^0$) to form the incident electromagnetic field ($\mathbf{E}_e^0, \mathbf{H}_e^0$) (resp. ($\mathbf{E}_p^0, \mathbf{H}_p^0$)).

In order to deduce the equation of diffraction problem, we recall Maxwell's equations:

$$\nabla \times \mathbf{E} = i\omega \boldsymbol{\mu} \mathbf{H}$$

$$\nabla \times \mathbf{H} = -i\omega \boldsymbol{\varepsilon} \mathbf{E},$$

where the unique solution $(\mathbf{E}, \mathbf{H})$ must satisfy the Outgoing Waves Condition (OWC).

From these Maxwell's equations, we define the wave operator $\mathcal{L}_{\xi,\chi}$ such that:

$$\mathcal{L}_{\xi,\chi}\mathbf{u} = \nabla \times (\xi^{-1}\nabla \times \mathbf{u}) + \lambda^2 \chi \mathbf{u} = 0, \qquad (C.3)$$

with $\lambda = i\omega$. The tensors $\xi, \chi$ stand for the permeability $\mu$ (resp. the permittivity $\varepsilon$) and the permittivity $\varepsilon$ (resp. the permeability $\mu$) in the TE case (resp. TM case). $u$ is the unknown solution for $\mathbf{E}$ field (resp. $\mathbf{H}$ field ) in the TE case (resp. TM case). In fact, the value of tensors $\xi, \chi$ will characterized by the geometric structures of the grating.

Similar to Section 6.1, we can simplify the vector equation (C.3) by the following scalar quantity:

$$\mathcal{L}_{\xi,\chi}u = -\nabla \cdot (\xi^{-1}\nabla u) + \lambda^2 \chi u = 0. \qquad (C.4)$$

The corresponding scalar values $\xi, \chi$ in the previous equation are re-defined as:

$$\xi^{-1} = \frac{\widetilde{\xi}^{\intercal}}{\det(\widetilde{\xi})}, \quad \chi = \chi_{zz}, \qquad (C.5)$$

where $\widetilde{\xi} = \begin{pmatrix} \xi_{zz} & \bar{\xi}_a \\ \xi_a & \xi_{yy} \end{pmatrix}$ of the tensor $\xi$ and $\chi_{zz}$ is the zz-component of the tensor $\chi$.

In this chapter, (C.4) will be set as the master equation for diffraction grating problem.

## C.2  The equivalent radiation problem

Before moving forward, let's contemplate some problems when simulating the diffraction grating:

▶ In this example, the plane wave is set as the incident field (C.1). The problem is that in order to generate the plane wave, the source must be infinitely far. This is quite different from the case of Dirac delta source, where we can model the source inside our domain of computation.

▶ The structure is open: The domain of the superstratum and substratum is unbounded, in which the fields must satisfy the Outgoing Waves Condition OWC. On the other hand, the grating is both periodic and infinite.

As a result, the structure of the diffraction problem is not well suited to the Finite Element Method (FEM) in a finite domain. In order to cope with these difficulties, an equivalent radiation problem should be introduced.

The idea is to split the diffraction problem into two parts: The first part consists of the classical calculation of the total field solution of a simple reference problem. The second one is to look for a radiated field with sources confined within the diffractive scatterers and deduced from the previous reference problem. From this viewpoint, the later radiated field can be interpreted as an exact perturbation of the total field.

Since in the equivalent radiation problem, the sources are inside the scatterers, it is a straightforward process to use PMLs to truncate the unbounded domain at a finite distance. Then, we can simplify the whole grating structure by a single-period cell which is bounded by the quasi-periodicity conditions.

The radiation problem is constructed with the decomposition of the solution $u$ into 2 functions $u = u_1 + u_2^d$, whose meaning will be explained later. First, we define $u_1$ as the solution of "reference" diffraction problem

$$\mathscr{L}_{\xi_r, \chi_r}(u_1) = 0,$$

where $\mathscr{L}_{\xi_r, \chi_r}$ chosen to be simple such that $u_1$ is considered to be known in the closed form.

The next step is to take the action of operator $\mathscr{L}_{\xi, \chi}$ onto function $u_2^d$ to obtain:

$$\mathscr{L}_{\xi, \chi}(u_2^d) = \mathscr{L}_{\xi, \chi}(u - u_1) = -\mathscr{L}_{\xi, \chi}(u_1). \tag{C.6}$$

If the RHS of (C.6) $-\mathscr{L}_{\xi, \chi}(u_1)$ is considered as a known source, $u_2^d$ will be the solution of radiation problem of the following source:

$$S = -\mathscr{L}_{\xi, \chi}(u_1) = \mathscr{L}_{\xi_r, \chi_r}(u_1) - \mathscr{L}_{\xi, \chi}(u_1) = \mathscr{L}_{\xi_r - \xi, \chi_r - \chi}(u_1). \tag{C.7}$$

Equation (C.6) is indeed the radiation problem transformed from the direct diffraction problem (C.4).

## Reference diffraction problem

In order to understand the source $S = -\mathscr{L}_{\xi, \chi}(u_1)$, it is necessary to define precisely our "reference" diffraction problem. Empirically, it is advantageous to set the permittivity and permeability in the groove region to be equal value of the permittivity and permeability in the superstratum:

$$\xi_r = \begin{cases} \xi^+ & \text{for} \quad 0 < y \\ \xi^- & \text{for} \quad y \le 0 \end{cases} \qquad \chi_r = \begin{cases} \chi^+ & \text{for} \quad 0 < y \\ \chi^- & \text{for} \quad y \le 0 \end{cases}.$$

We will notice that the permittivity and permeability of the reference problem are equal to the ones of the direct problem everywhere except inside the scatterers ($\xi = \xi_r$, and $\chi = \chi_r$ in $\Omega \backslash \Omega_1$). Therefore, the source $S = \mathscr{L}_{\xi_r - \xi, \chi_r - \chi}(u_1)$ is equal to zero outside the scatterers. This has a great advantage considering that the source doesn't extend to the boundary but is localized only within the region of scatterers $\Omega_1$.

The 'reference' diffraction problem now becomes a simple refection problem where we can decompose again the "reference" solution $u_1$ into $u_1 = u_0 + u_1^d$ where $u_1^d$ satisfies an OWC. In this case, the incident field: $u_0$ is defined as follows:

$$u_0 = \begin{cases} E_0^+ & \text{for TE case and} \quad h^g < y \leq h^g + h^+ \\ H_0^+ & \text{for TM case and} \quad h^g < y \leq h^g + h^+ \end{cases}.$$

From (C.7), we can decompose the source $S$ into 2 parts $S = S_1 + S_2$, where

$$S_1 = \mathcal{L}_{\xi_r - \xi, \chi_r - \chi}(u_0)$$
$$= -i\nabla \cdot \left[ (\xi_r - \xi) \mathbf{k}^+ \exp(i\mathbf{k}^+ \cdot \mathbf{r}) \right] + \lambda^2 (\chi_r - \chi) \exp(i\mathbf{k}^+ \cdot \mathbf{r}).$$

$$S_2 = \mathcal{L}_{\xi_r - \xi, \chi_r - \chi}(u_1^d)$$
$$= \rho \left\{ -i\nabla \cdot \left[ (\xi_r - \xi) \mathbf{k}^- \exp(i\mathbf{k}^- \cdot \mathbf{r}) \right] + \lambda^2 (\chi_r - \chi) \exp(i\mathbf{k}^- \cdot \mathbf{r}) \right\}.$$

with $\mathbf{k}^+, \mathbf{k}^-$ are wave vectors from the incident and reflected field (for the "reference" problem) in the superstratum respectively; and $\rho$ is the complex reflection coefficient.

$$\rho = \frac{\rho^+ - \rho^-}{\rho^+ + \rho^-} \quad \rho^\pm = \begin{cases} \beta^\pm & : \text{TE case} \\ \frac{\beta^\pm}{\varepsilon^\pm} & : \text{TM case} \end{cases}.$$

## C.3 Diffraction efficiency

There is no better way to understand diffractive optical phenomena, especially the diffraction grating problem than to study the power throughput. Roughly speaking, we want to measure how much optical power is diffracted into a designated direction compared to the power incident. Thus it is important to deduce formulas for diffraction efficiency. The construction is adapted from [91] . From (C.1), we can re-write the scalar value of incident wave as follows:

$$u_0(x, y) = A^0 \exp(iax - iby).$$

For geometric structure with the period $d$, the previous equation implies:

$$u_0(x + d, y) = \exp(iad)u_0(x, y).$$

It is reasonable to expect that the diffraction solution $u^d := u - u_0$ of (C.4) will share the same property:

$$u^d(x + d, y) = \exp(iad)u^d(x, y).$$

Without loss of generality, we can define a new intermediate function $v(x, y)$ such that:

$$v(x, y) := \frac{u^d(x, y)}{\exp(iad)}. \tag{C.8}$$

The new function $v(x, y)$ can be shown to be periodic in x:

$$v(x + d, y) = \frac{u^d(x + d, y)}{\exp(iad)\exp(iax)} = \frac{u^d(x, y)}{\exp(iad)} = v(x, y).$$

Thus, it is easy to check that the function $v(x, y)$ can be expressed by Fourier series:

$$v(x, y) = \frac{u^d(x, y)}{\exp(iad)} = \sum_n u_n^d(y)\exp(i\frac{2\pi}{d}nx).$$

By combining the last equation with (C.8), we get:

$$u^d(x, y) = \sum_{n \in \mathbb{Z}} u_n^d(y)\exp(ia_nx) \quad \text{with} \quad a_n = a + \frac{2\pi}{d}n, \qquad \text{(C.9)}$$

where

$$u_n^d(y) = \frac{1}{d}\int_{-d/2}^{d/2} u^d(x, y)\exp(-ia_nx)\, dx.$$

(C.9) is known as the Rayleigh expansion for $u^d(x, y)$.

Finally, by substituting (C.9) into (C.4) gives us the Rayleigh coefficients:

$$u_n^d(y) = \begin{cases} s_n\exp(-ib_n^+y) + r_n\exp(ib_n^+y) & \text{for} \quad h^g < y \\ \\ u_n\exp(-ib_n^-y) + t_n\exp(ib_n^-y) & \text{for} \quad y \leq 0 \end{cases} \quad \text{with} \quad b_n^{\pm^2} = k^{\pm^2} - a_n^2$$

It suffices to reveal 4 coefficients $s_n, r_n, u_n, t_n$ in the previous equation. By recalling the OWC, it is straightforward to impose $s_n = u_n = 0$. The value of $r_n$ and $u_n$ can be deduced from the following equations:

$$\begin{cases} r_n = \frac{1}{d}\int_{-d/2}^{d/2} u^d(x, y_0)\exp(-ia_nx - ib_n^+y_0)\, dx & \text{for} \quad h^g < y \\ \\ t_n = \frac{1}{d}\int_{-d/2}^{d/2} u^d(x, y_0)\exp(-ia_nx + ib_n^+y_0)\, dx & \text{for} \quad y \leq 0 \end{cases}$$

at a fixed $y_0$.

## Reflection and transmission efficiency

The reflected and transmitted diffraction efficiency of propagative orders $T_n$ and $R_n$ are deduced from the Rayleigh coefficients as follows:

$$\begin{cases} R_n = r_n\overline{r_n}\frac{b_n^+}{b^+} & \text{for} \quad h^g < y \\ \\ T_n = t_n\overline{t_n}\frac{b_n^-}{b^+}\frac{\gamma^+}{\gamma^-} & \text{for} \quad y < 0 \end{cases} \quad \text{where} \quad \gamma^{\pm} = \begin{cases} \mu^{\pm} & \text{for TE case} \\ \\ \varepsilon^{\pm} & \text{for TM case} \end{cases}$$

### Absorption efficiency

In the harmonic frequency domain, we recall the Poynting vector:

$$\mathbf{P} = \frac{1}{2}(\mathbf{E} \times \mathbf{H}).$$

For the grating geometric structure, in one d-periodic cell, the power of the incident wave traveling though the surface of the superstratum $\Sigma := \{y = h^g + h^+\}$ can be expressed as:

$$Q_0 = \Re\left(\int_\Sigma \mathbf{P} \cdot \mathbf{n}\, dl\right) = \frac{1}{2}|A^0|^2 d\eta \cos\theta \quad \text{with} \quad \eta = \begin{cases} \sqrt{\dfrac{\varepsilon_0 \varepsilon}{\mu_0 \mu}} & \text{for TE case} \\[3mm] \sqrt{\dfrac{\mu_0 \mu}{\varepsilon_0 \varepsilon}} & \text{for TM case} \end{cases}$$

with $n$ is the normal vector of $\Sigma$.

Next, we can calculate the dissipated power in the lost domain $\Omega_p$ by the following equation:

$$Q_p = \Re\left(\int_{\Omega_p} \nabla \cdot \mathbf{P}\, dS\right) = \frac{1}{2}\omega\varepsilon_0 \int_{\Omega_p} \Im(\varepsilon)|\mathbf{E}|^2 dS.$$

The absorption efficiency is expressed as the ratio of the dissipated power and the total power of the incident wave.

$$A = \frac{Q_p}{Q_0}.$$

### Conservation of energy

In the studying of diffraction gratings, the diffraction efficiency actually provides us a powerful tool to verify the consistency of our numerical models. In particular, according to the conservation law of energy, we obtain the following equations for the energy En:

$$\text{En} := \sum_{n \in \mathbb{Z}} R_n + \sum_{n \in \mathbb{Z}} T_n + A = 1$$

It is necessary to check if the numerical value of the quantity En is approximately equal to 1; in order to confirm that the conservation of total energy conserves in our geometric model.

## C.4  Weak formulation

After understanding all the physical properties of the grating structure, our next step is to construct the weak formulation of the diffraction grating problem. Firstly, the surface of 1 cell $\Omega$ in Figure C.1 is divided into 4 parts $\partial\Omega = \Gamma_u + \Gamma_d + \Gamma_l + \Gamma_r$, where $\Gamma_l$ and $\Gamma_r$ refer to the lines parallel to the y-axis delimiting a cell of the grating from its left and right neighbors respectively; $\Gamma_u$ and $\Gamma_d$ denote the lines parallel to the x-axis bounding the cell from above and below.

Next, we define a Sobolev space $H_{d,0}(\Omega, \nabla)$ such that:

$$H_{d,0}(\text{grad}, \Omega) = \left\{ u \in L^2(\Omega)^2 : \nabla u \in L^2(\Omega)^2, u|_{\Gamma_r} = e^{iad} u|_{\Gamma_l}, u|_{\Gamma_u} = u|_{\Gamma_d} = 0 \right\}$$

Then the radiation problem $\mathscr{L}_{\xi,\chi}(u_2^d) = S$ with $S := \mathscr{L}_{\xi_r - \xi, \chi_r - \chi}(u_1)$ can be expressed in the form of the following weak formulation:

Find $u \in H_{d,0}(\text{grad}, \Omega)$ such that

$$\int_\Omega \xi^{-1}(\nabla \overline{u'}) \cdot (\nabla u) + \lambda^2 \chi \overline{u'} u \, d\Omega - \int_\Omega u' S \, d\Omega$$

$$- \int_{\partial \Omega} \overline{u'}(\xi^{-1}\nabla u) \cdot \mathbf{n} \, dl = 0 \quad \forall u' \in H_{d,0}(\text{grad}, \Omega) \quad \text{(C.10)}$$

We can see that if the boundary term b.t. $= \int_{\partial \Omega} \overline{u'}(\xi^{-1}\nabla u) \cdot \mathbf{n} \, dl$ vanishes, solution $u$ of (C.10) will satisfy the radiation problem $\mathscr{L}_{\xi,\chi}(u_2^d) = S$. In particular, the boundary term can be set to be zero by the following boundary conditions

- ▶ On the boundaries $\Gamma_u$, $\Gamma_d$, we impose the Dirichlet conditions by choosing $u'|_{\Gamma_u, \Gamma_d} = 0$ or the homogeneous Neumann condition $(\xi^{-1}\nabla u) \cdot \mathbf{n} = 0$.
- ▶ On the boundaries $\Gamma_l$, $\Gamma_r$, by applying the Bloch-Floquet quasi-periodicity conditions,[1] , we have :     

$$\int_{\Gamma_l \cup \Gamma_r} \overline{u'}(\xi \nabla u) \cdot \mathbf{n} dl = \int_{\Gamma_l \cup \Gamma_r} \overline{u'}_\sharp e^{-iax}(\xi^{-1}\nabla u_\sharp e^{iax}) \cdot \mathbf{n} dl$$

$$= \int_{\Gamma_l \cup \Gamma_r} \overline{u'}_\sharp \left( \xi^{-1}(\nabla u_\sharp + iau_\sharp) \right) \cdot \mathbf{n} dl = 0$$

where $u(x, y) = u_\sharp(x, y) \exp(iax)$.

Because the term $\overline{u'}_\sharp \left( \xi(\nabla u_\sharp + iau_\sharp) \right) \cdot \mathbf{n}$ is periodic along x-axis, the the normal $\mathbf{n}$ has opposite direction on $\Gamma_l$ and $\Gamma_r$, so the contributions of these two boundaries have the same absolute value with opposite signs. As a result, the contribution of 2 boundaries $\Gamma_l$, $\Gamma_r$ in the boundary term is zero.

## C.5 Perfectly Matched Layer (PML)

The final step is to tackle the unbounded characteristic of the diffraction grating. Indeed, the propagating modes outside the main structure (in our case the electromagnetic fields in superstratum and substratum) have non-decreasing behavior, which need to be handled by Perfectly Matched Layer (PML). In this 2D diffraction grating, we cover the top and bottom by 2 layers of PML. The PML method will transform the initial operator $\mathscr{L}_{\xi,\chi}$ into new operator on a mapped space; whose eigen-modes are exponentially damped inside the PMLs.

Then, the complex stretching the coordinate is applied in the PML top and bottom domain (as the extension part of superstratum and substratum) to the direction along which the field must decay (y-direction in our case). In particular, we introduce new tensors $\varepsilon_s$ and $\mu_s$:

$$\delta_s := \mathbf{J}_s^{-1} \delta \mathbf{J}_s^{-T} \text{def}(\mathbf{J}_s)| \quad \text{for} \quad \delta = \{\varepsilon, \mu\}$$

where $J_s$ is stretched Jacobian matrix. Since the stretch should be applied along $y$-direction, we have:

$$J_s = \text{diag}(1, s_y(y), 1)$$

with $s_y(y)$ is a complex valued function of y.[2]

By replacing $\varepsilon$ and $\mu$ in the PML domain by new tensors $\varepsilon_s$ and $\mu_s$

$$\delta_s = \begin{pmatrix} s_y \delta_{xx} & \bar{\delta}_d & 0 \\ \delta_d & s_y^{-1} \delta_{yy} & 0 \\ 0 & 0 & s_y \delta zz \end{pmatrix}$$

we will get a new electromagnetic field $(\mathbf{E}_s, \mathbf{H}_s)$ which is identical to $(\mathbf{E}, \mathbf{H})$ solution of equation outside the PML domain.

2:  As discussed in Section 7.1, $s_y(y) \in \mathbb{C}^+$.